\documentclass[fleqn,10pt]{wlscirep}
\usepackage[utf8]{inputenc}
\usepackage[T1]{fontenc}

\usepackage{amsmath}
\usepackage{fixmath}
\usepackage{subcaption}
\definecolor{myc1}{HTML}{003049}
\definecolor{myc2}{HTML}{d62828}
\definecolor{myc3}{HTML}{f77f00}
\definecolor{myc4}{HTML}{6ca13b}

\definecolor{Q2}{HTML}{1f77b4}
\definecolor{Q1}{HTML}{ff7f0e}
\definecolor{Q3}{HTML}{2ca02c}

\definecolor{Q2}{HTML}{6ca13b}
\definecolor{Q1}{HTML}{6ca13b}
\definecolor{Q3}{HTML}{6ca13b}

\usepackage{tikz}
\usepackage{tikz-3dplot}
\usetikzlibrary{decorations.markings,arrows.meta,calc,shapes,intersections,backgrounds,patterns.meta,bending,angles,decorations.pathreplacing,decorations.pathmorphing}
\usepackage{xifthen}
\usepackage[normalem]{ulem}
\usepackage{pdflscape}

\usepackage{pgfplots}
\usepgfplotslibrary{fillbetween}
\pgfplotsset{compat=1.12} 

\usepackage{pifont}
\newcommand{\cmark}{\ding{51}}
\newcommand{\xmark}{\ding{55}}

\newcommand{\alvaro}[1]{{#1}}

\let\tw\textwidth

\def\beq{\begin{equation}}
\def\eeq{\end{equation}}

\def\beq{\begin{equation}}
\def\eeq{\end{equation}}

\def\dis{\varepsilon}

\definecolor{olivegreen}{rgb}{0,0.6,0}

\def\drawline#1#2{\raise 2.5pt\vbox{\hrule width #1pt height #2pt}}

\def\trian{\raise 1.25pt\hbox{$\scriptstyle\triangle$}\nobreak}

\def\dtrian{\raise 1.25pt\hbox%
{$\scriptscriptstyle\bigtriangledown$}\nobreak}

\def\squar{\raise 1.25pt\hbox{$\scriptstyle\Box$}\nobreak}

\def\diamon{\raise 1.25pt\hbox{$\scriptstyle\diamond$}\nobreak}

\def\beq{\begin{equation}}
\def\eeq{\end{equation}}

\definecolor{C0}{HTML}{1F77B4} 
\definecolor{C1}{HTML}{FF7F0E}
\definecolor{C2}{HTML}{2CA02C}
\definecolor{C3}{HTML}{D62728}
\definecolor{C4}{HTML}{9467BD}
\definecolor{C5}{HTML}{8C564B}
\definecolor{C6}{HTML}{E377C2}
\definecolor{C7}{HTML}{7F7F7F}
\definecolor{C8}{HTML}{BCBD22}
\definecolor{C9}{HTML}{17BECF}

\let\bs\mathbold
\let\provc\providecommand
\newcommand{\mybq}{\bs{q}}        
\newcommand{\bq}{\bs{q}}        
\newcommand{\bQ}{\bs{Q}}        
\newcommand{\bi}{\bs{i}}        
\newcommand{\Is}{\tilde{\imath}}        

\newcommand{\myindexvar}[3]{%
  \ifthenelse{\isempty{#2}\and\isempty{#3}}%
  {#1}{#1_{#2}^{#3}}} 

\provc{\sq}[2]{\myindexvar{q}{#1}{#2}}       
\provc{\bsq}[2]{\myindexvar{\mybq}{#1}{#2}}  

\newcommand{\mybS}{\bs{S}}        
\newcommand{\mybA}{\bs{A}}        
\newcommand{\mybth}{\bs{\theta}}  
\newcommand{\mybJ}{\bs{J}}        

\provc{\bS}[2]{\myindexvar{\mybS}{#1}{#2}}     
\provc{\bA}[2]{\myindexvar{\mybA}{#1}{#2}}     
\provc{\bth}[2]{\myindexvar{\mybth}{#1}{#2}}   

\provc{\bWs}[2]{\myindexvar{\bs{W}}{#1}{#2}}               
\provc{\bWa}[2]{\myindexvar{\bs{V}}{#1}{#2}}   

\provc{\bJ}[2]{\myindexvar{\mybJ}{#1}{#2}}                  
\provc{\bJtilde}[2]{\myindexvar{\widetilde{\mybJ}}{#1}{#2}} 
\provc{\bJhat}[2]{\myindexvar{\widehat{\mybJ}}{#1}{#2}}     

\provc{\J}[2]{\myindexvar{J}{#1}{#2}}                  
\provc{\Jtilde}[2]{\myindexvar{\widetilde{J}}{#1}{#2}} 
\provc{\Jhat}[2]{\myindexvar{\widehat{J}}{#1}{#2}}     

\provc{\mun}{\mu}
\provc{\sgn}{\Xi}
\provc{\bmun}{\bs{\mu}}
\provc{\bsgn}{\bs{\Xi}}
\provc{\muopt}{\mu^*}
\provc{\sgopt}{\Xi^*}
\provc{\bmuopt}{\bs{\mu}^*}
\provc{\bsgopt}{\bs{\Xi}^*}
\provc{\mutar}{\hat{\mu}}
\provc{\sgtar}{\widehat{\Xi}}
\provc{\bmutar}{\hat{\bs{\mu}}}
\provc{\bsgtar}{\widehat{\bs{\Xi}}}

\provc{\relf}{\alpha}
\provc{\relfmu}{\relf_\mu}
\provc{\relfsg}{\relf_\xi}

\provc{\bths}{\mybth_s} 
\provc{\bthpa}{\mybth_p}
\provc{\bthaa}{\mybth_a}

\provc{\ths}{\theta_s} 
\provc{\thpa}{\theta_p}
\provc{\thaa}{\theta_a}

\provc{\shiftM}{\bs{a}}
\provc{\shiftmi}{a}

\definecolor{myc1}{HTML}{003049}
\definecolor{myc2}{HTML}{d62828}
\definecolor{myc3}{HTML}{f77f00}
\definecolor{myc4}{HTML}{6ca13b}

\title{Decomposing causality into its synergistic, unique, and redundant components}

\author[1,*]{Álvaro Martínez-Sánchez}
\author[1]{Gonzalo Arranz}
\author[1]{Adrián Lozano-Durán}
\affil[1]{Department of Aeronautics and Astronautics, Massachusetts Institute of Technology, Cambridge, MA 02139}

\affil[*]{alvaroms@mit.edu}

\begin{abstract}

Causality lies at the heart of scientific inquiry, serving as the
fundamental basis for understanding interactions among variables in
physical systems. Despite its central role, current methods for causal
inference face significant challenges due to nonlinear dependencies,
stochastic interactions, self-causation, collider effects, and
influences from exogenous factors, among others. While existing
methods can effectively address some of these challenges, no single
approach has successfully integrated all these aspects. Here, we
address these challenges with SURD: Synergistic-Unique-Redundant
Decomposition of causality. SURD quantifies causality as the
increments of redundant, unique, and synergistic information gained
about future events from past observations. The formulation is
non-intrusive and applicable to both computational and experimental
investigations, even when samples are scarce. We benchmark SURD in
scenarios that pose significant challenges for causal inference and
demonstrate that it offers a more reliable quantification of causality
compared to previous methods.

\end{abstract}

\begin{document}

\flushbottom
\maketitle
%
%
\thispagestyle{empty}

\section*{Introduction}

The quest for understanding causality is the cornerstone of scientific
discovery\cite{pearl2000}. It is through the exploration of
cause-and-effect relationships that we are able to understand a given
phenomenon and shape the course of events through deliberate
actions\cite{bunge1979}. This has accelerated the proliferation of
methods for causal inference, as they hold the potential to drive
progress across multiple scientific and engineering domains, such as
climate research\cite{Runge2023}, neuroscience\cite{neuroscience2016},
economics\cite{economic2008}, epidemiology\cite{epidemiology2005},
social sciences\cite{social2010}, and fluid
dynamics\cite{lozano2020,martinez2023}, among others.

A central aspect of causality is the concept of physical
influence\cite{Eichler2013}: manipulation of the cause manifests as
changes in the effects\cite{pearl2000, barndorff2000,
  Spirtes2001, Dawid2002}. For example, prolonged exposure to elevated
air pollution levels has a causal connection to a higher incidence of
chronic respiratory conditions\cite{kampa2008}.  The precise
definition of causality remains elusive, yet it must be distinguished
from the concepts of association and correlation. Association
indicates a statistical relationship between two variables in which
they have a tendency to co-occur more often than would be expected by
random chance. Yet, association does not automatically imply
causation\cite{Altman2015}. Association may arise from shared causes,
statistical coincidences, or the influence of confounding factors. An
example of association can be observed in the increased rates of
chronic respiratory diseases in regions undergoing significant
deforestation. Although it may seem that deforestation directly
contributes to respiratory health issues, this might primarily be due
to the confounding factor of air pollution.  Correlation, on the other
hand, refers to a particular type of association that measures the monotonic
strength and direction of
variables\cite{pearson1895,Spearman1987, Altman2015,
  agresti2007art}. Correlation implies association but not causation;
causation implies association but not correlation\cite{Altman2015}.
Discerning between causality, association, and correlation poses a
significant challenge in the development of methods for causal
discovery. Here, we introduce an approach for causal inference that
facilitates the study of complex systems in a manner that surpasses
simple correlational and associational analyses.

The first factor to consider is the nature of interaction among
variables. Three building blocks serve as the
foundations of causal interactions\cite{pearl2000}: mediator,
confounder, and collider effects. These interactions can intertwine
and manifest concurrently, leading to more complex causal
networks. Therefore, accurately capturing these interactions is key to
faithfully characterizing more general causal patterns. Consider the
three events denoted by $A$, $B$, and $C$:
\begin{itemize}
    \item Mediator variables ($A\rightarrow B \rightarrow C$) emerge
      in the causal chain between variable $A$ to variable $C$, with
      variable $B$ acting as a bridge.  In this scenario, $B$ is often
      viewed as the mechanism or mediator responsible for transmitting
      the influence of $A$ to $C$. Mediator variables help explain the
      underlying mechanisms by which an independent variable
      influences a dependent variable. A simple example is $\uparrow$
      education level $\rightarrow$ $\uparrow$ job skills
      $\rightarrow$ $\uparrow$ income.
    \item Confounding variables ($A \leftarrow B\rightarrow C$) act as
      the common cause for two variables: $B \rightarrow A$ and $B
      \rightarrow C$. They have the potential to create a statistical
      correlation between $A$ and $C$, even if there is no direct
      causal link between them. Consequently, confounding variables
      can obscure or distort the genuine relationship between
      variables. Following the example above, air pollution
      $\rightarrow$ $\uparrow$ deforestation\cite{Horn2018}, and air pollution
      $\rightarrow$ $\uparrow$ respiratory health conditions\cite{duan2020}.
    \item Collider variables ($A \rightarrow B\leftarrow C$)
      represent the effect of multiple factors acting on the same
      variable: $A \rightarrow B$ and $C \rightarrow B$.  This
      scenario is particularly relevant in nonlinear dynamical systems,
      where most variables are affected by multiple causes due to
      their coupling.  A collider exhibits
      \emph{redundant} causes when both $A$ and $C$ contribute to the
      same effect or outcome of $B$, creating overlapping or
      duplicative influences on the outcome.  Consequently, redundant
      causes result in multiple pathways to the same effect. For
      instance, both hard work and high intelligence can independently
      contribute to the good grades of a student. Note that $A$ and
      $C$ may not necessarily be independent. A collider is
      \emph{synergistic} if the combined effect of $A$ and $C$ on $B$
      surpasses their individual effects on $B$ when considered
      separately. For example, two drugs, $A$ and $C$, may be required
      in tandem to effectively treat a condition $B$, when each drug
      alone is ineffective.
\end{itemize}
%

The search for mathematical definitions of causality that accurately
identify mediator, confounder, and collider effects remains an active
area of research\cite{camps2023}. One of the most intuitive
formulations of causality relies on the concept of
interventions~\cite{pearl2000, Eichler2013}. The approach offers a
pathway for evaluating the causal effect that the process $A$ exerts
on another process $B$ by setting $A$ at a modified value $\tilde{A}$
and observing the consequences of the post-intervention in
$B$. Despite its intuitiveness, interventional studies are not without
limitations~\cite{eberhardt2007, Runge2023}. Causality with
interventions is intrusive (i.e., it requires modification of the
system) and costly (the experiments or simulations need to be
repeated). When data are gathered from physical experiments,
establishing causality through interventions may become highly
challenging or impractical (e.g., we cannot use interventions to
assess the causality in the stock market in 2008). Additionally, the
notion of causality with interventions prompts questions about the
type of intervention that should be introduced and whether this
intervention could affect the outcome of the exercise as a consequence
of forcing the system out of its natural state. Interventional studies
can also pose ethical problems in fields such as neuroscience or
climate science~\cite{runge2019pcmci}.  For example, they might
involve manipulating neural functions in living organisms or altering
natural environmental conditions, potentially leading to irreversible
changes or damage.

The alternative approach to interventions involves discovering causal
links through observations. Observational methods are predominantly
data-driven and do not require alterations to the original system.  In
the recent years, the steady advancements in computational power
coupled with the exponential growth of big data have significantly
contributed to the widespread adoption of observational
techniques. One of the pioneering approaches is rooted in the use of
forecasting models. The concept was initially proposed by
Wiener\cite{wiener1956} and later quantified by
Granger\cite{granger1969}. Granger causality (GC) measures the
causality from the process $B$ to $A$ by evaluating how the inclusion
of $B$ in an autoregressive model reduces the forecast error for
$A$. Originally developed for linear bivariate relationships, GC has
since expanded to encompass nonlinear and multivariate
scenarios\cite{geweke1984, barnett2009, barnett2014, Barnett2015},
finding applications in diverse fields ranging from
econometrics\cite{Hiemstra1994, Bell9967, abhyankar1998} to fluid
dynamics\cite{tissot2014}, and biology\cite{ancona2004, bueso2020}.

Model-free approaches for causal discovery have also been proposed to
overcome the limitations of GC. A leading method in this domain is
convergent cross-mapping (CCM)\cite{sugihara2012} and its
variants\cite{pai2014,mccm2015, extendedccm2015,pcm2020,lccm2021},
which apply Takens' embedding theorem\cite{Takens1981} to establish
connections between variables and the attractor of the
system. \alvaro{An alternative approach, known as continuity
  scaling\cite{cont_scaling_2022}, directly assesses causal
  relationships by examining the scaling laws governing the continuity
  of the system.}

Information theory\alvaro{, the science of message
  communication\cite{shannon1948},} has also served as a framework for
model-free causality quantification. \alvaro{The success of
  information theory relies on the notion of information as a
  fundamental property of physical systems, closely tied to the
  restrictions and possibilities of the laws of
  physics\cite{Lozano2022, yuan2024}. The grounds for causality as information
  are rooted in the intimate connection between information and the
  arrow of time. Time-asymmetries present in the system at a
  macroscopic level can be leveraged to measure the causality of
  events using information-theoretic metrics based on the Shannon
  entropy\cite{shannon1948}.} The initial applications of information
theory for causality were formally established through the use of
conditional entropies, employing what is known as directed
information\cite{massey1990, kramer1998}. Among the most recognized
contributions is transfer entropy (TE)\cite{schreiber2000}, which
measures the reduction in entropy about the future state of a variable
by knowing the past states of another. Various improvements have been
proposed to address the inherent limitations of TE. Among them, we can
cite conditional transfer entropy (CTE) \cite{verdes2005, lizier2008,
  lizier2010, cte2016}, which stands as the nonlinear, nonparametric
extension of conditional GC\cite{barnett2009}. Subsequent advancements
of the method include multivariate formulations of
CTE\cite{Lozano2022} \alvaro{and momentary information
  transfer\cite{pompe2011}, which extends TE by examining the transfer
  of information at each time step. Other information-theoretic
  methods, derived from dynamical system theory\cite{liang2006,
    liang2016if, liang2008if,liang2013if}, quantify causality as the
  amount of information that flows from one process to another as
  dictated by the governing equations.}

Another family of methods for causal inference relies on conducting
conditional independence tests. This approach was popularized by the
Peter-Clark algorithm (PC)\cite{spirtes1991}, with subsequent
extensions incorporating tests for momentary conditional independence
(PCMCI)\cite{runge2018cmi, runge2019pcmci}. PCMCI aims to optimally
identify a reduced conditioning set that includes the parents of the
target variable\cite{runge2023comment}. This method has been shown to
be effective in accurately detecting causal relationships while
controlling for false positives\cite{runge2019pcmci}. Recently, new
PCMCI variants have been developed for identifying contemporaneous
links\cite{runge2020pcmci+}, latent
confounders\cite{Gerhardus2020LPCMCI}, and regime-dependent
relationships\cite{saggioro2020RPCMCI}.

The methods for causal inference discussed above have significantly
advanced our understanding of cause-effect interactions in complex
systems. Despite the progress, current approaches face limitations in
the presence of nonlinear dependencies, stochastic interactions (i.e.,
noise), self-causation, mediator, confounder, and collider effects, to
name a few.  Moreover, they are not capable of classifying causal
interactions as redundant, unique, and synergistic, which is crucial
to identify the fundamental relationships within the system. Another
gap in existing methodologies is their inability to quantify causality
that remains unaccounted for due to unobserved variables. To address
these shortcomings, we propose SURD: Synergistic-Unique-Redundant
Decomposition of causality. SURD offers causal quantification in terms
of redundant, unique, and synergistic contributions and provides a
measure of the causality from hidden variables. The approach can be
used to detect causal relationships in systems with multiple
variables, dependencies at different time lags, and instantaneous
links. We demonstrate the performance of SURD across a large
collection of scenarios that have proven challenging for causal
inference and compare the results to previous approaches.

\section*{Results}
\subsection*{Theoretical background} 

Consider the collection of $N$ time-dependent variables given by the
vector $\bQ = [Q_1(t), Q_2(t), \ldots, Q_N(t)]$. For example, $Q_i$
may represent the regional average of climatological variables (e.g.,
temperature, pressure,...) or the evolution of human heart rate. The
components of $\bQ$ are the observables and are treated as random
variables. Our objective is to quantify the causality from the
components of $\bQ$ to the future of the target variable $Q_j$,
denoted by $Q_j^+ = Q_j(t + \Delta T)$, where $\Delta T > 0$ is an
arbitrary time increment. The vector $\bQ$ can include variables at
times less or equal to $t+\Delta T$, which allows us to identify both
lagged and instantaneous dependencies.

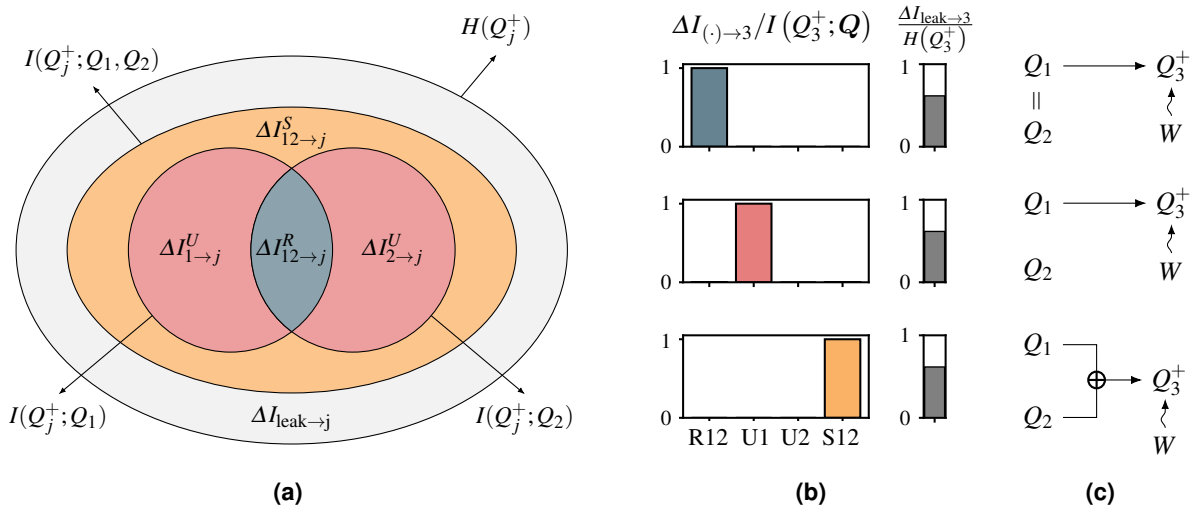
\begin{figure}[t!]
\vspace{1cm}
    \centering
    \subfloat[]{\vspace{0.1in}\begin{tikzpicture}[scale=1.35,thin,>={Latex[length=.1cm]}]
    \colorlet{c2}{myc2}
    \colorlet{c1}{myc3}

    \colorlet{MI}{myc1}
    \pgfmathsetmacro{\dis}{.6}
    \begin{small}

        \draw[->] (-\dis,0) ++ (0:1)  --+ (50:2.5)  node[anchor=south,pos=1] {$H(Q^+_j)$};
        
        \draw[fill=black!5] (0,0) ellipse (2.7 and 1.9);
        
        \draw[->] (-\dis,0) ++ (130:1)  --+ (130:1.1)  node[anchor=south,pos=1] {$I(Q^+_j;Q_1,Q_2)$};

        \draw[fill=c1!45] (0,0) ellipse (2.2 and 1.4);

        \draw[fill=c2!45] (-\dis,0) circle (1);
        \draw[fill=c2!45] (\dis,0) circle (1);
        
        \begin{scope}
            \clip (-\dis,0) circle (1);
            \draw[fill=MI!45] (\dis,0) circle (1);
        \end{scope}
        \begin{scope}
            \clip (\dis,0) circle (1);
            \draw (-\dis,0) circle (1);
        \end{scope}

        \node[anchor=east] at (-\dis,0) {$\Delta I_{1\to j}^U$};
        \node[anchor=west] at (+\dis,0) {$\Delta I_{2\to j}^U$};
        \node              at (    0,0) {$\Delta I_{12\to j}^R$};
        \node              at ( 0, 1.15) {$\Delta I_{12\to j}^S$};
        \node              at ( 0, -1.65) {$\Delta I_{\rm{leak}\to j}$};


        \draw[->] (+\dis,0) ++ (-40:1)  --+ (-40:1.2)  node[anchor=north,pos=1] {$I(Q^+_j;Q_2)$};
        \draw[->] (-\dis,0) ++ (220:1)  --+ (220:1.2)  node[anchor=north,pos=1] {$I(Q^+_j;Q_1)$};

    \end{small}
\end{tikzpicture}}
    \hspace{0.04\tw}
    \subfloat[]{
    \begin{minipage}{0.25\tw}
        \includegraphics[width=\tw,trim={0 0.855cm 0 0},clip]{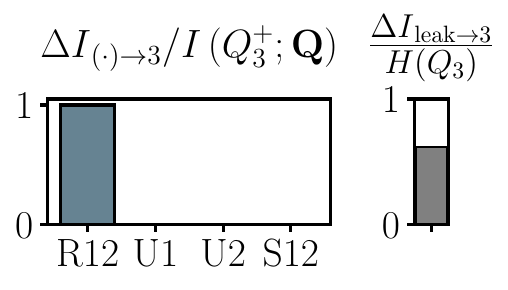}
        \begin{tikzpicture}[overlay, remember picture]
            \node[fill=white] at (3.75,2.1) {$\frac{{\Delta I}_{\mathrm{leak} \rightarrow 3}}{H \left(Q_{3}^+ \right)}$};

            \node[fill=white] at (1.6,2.125) {${{\Delta I}}_{(\cdot) \rightarrow {3}} / I \left(Q_{3}^+ ; \mathrm{\bQ} \right)$};

            {\footnotesize
            \node[fill=white] at (0.19,1.6) {$1$};
            \node[fill=white] at (3.35,1.6) {$1$};
            \node[fill=white] at (0.19,0.55) {$0$};
            \node[fill=white] at (3.35,0.55) {$0$};
            }
            
        \end{tikzpicture}
        \vspace{0.01mm} 
    
        \includegraphics[width=\tw,trim={0 0.855cm 0 1.44cm},clip]{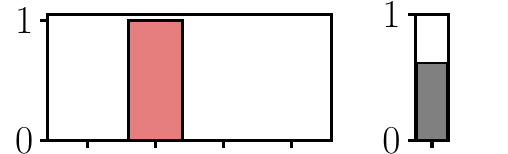}
        \begin{tikzpicture}[overlay, remember picture]
            {\footnotesize
            \node[fill=white] at (0.19,1.6) {$1$};
            \node[fill=white] at (3.35,1.6) {$1$};
            \node[fill=white] at (0.19,0.55) {$0$};
            \node[fill=white] at (3.35,0.55) {$0$};
            }
        \end{tikzpicture}
        \vspace{0.01mm} 
    
        \includegraphics[width=\tw,trim={0 0 0 1.44cm},clip]{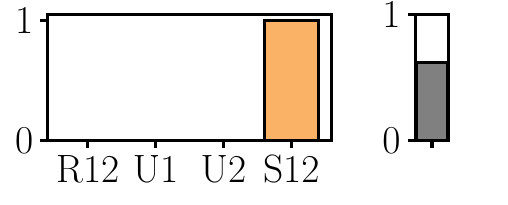}
        \begin{tikzpicture}[overlay, remember picture]
            {\small
            \node[fill=white] at (0.75,0.725) {$\rm{R}12$};
            \node[fill=white] at (1.375,0.725) {$\rm{U}1$};
            \node[fill=white] at (1.9,0.725) {$\rm{U}2$};
            \node[fill=white] at (2.5,0.725) {$\rm{S}12$};
            }
            {\footnotesize
            \node[fill=white] at (0.19,2) {$1$};
            \node[fill=white] at (3.35,2.075) {$1$};
            \node[fill=white] at (0.19,0.975) {$0$};
            \node[fill=white] at (3.35,0.975) {$0$};
            }
        \end{tikzpicture}
        \vspace{-0.375cm} 
    \end{minipage}}
    \hspace{0.01\tw}
    \subfloat[]{\vspace{0.in} \begin{tikzpicture}[
 			bar/.style={thick,black!70,fill=black!5,
            rounded corners=.1mm},
 			barline/.style={thick,black!70,rounded corners=.1mm},
 			myfill/.style={thick,rounded corners=.1mm},
            >={Latex[length=.2cm]},
            scale=0.55
    ]
    


    \def\ths{{1.,1.,1.}}
    \def\lbs{{1,2,12}}


    \pgfmathsetmacro{\ypaneltwo}{+3.3}
        

    \draw[->,>={Latex[length=.15cm]}] (4.6,\ths[1]+\ypaneltwo+1.1) node[anchor=east] {$Q_1$} --++ (2,0) node[anchor=west] {$Q_3^+$};
    
    \node[anchor=south] at (4,\ths[1]+\ypaneltwo-0.2) {\rotatebox{90}{$=$}};
    \node[anchor=south] at (4,\ths[1]+\ypaneltwo-1.) {$Q_2$};

    \draw[decorate,decoration={snake,post length=0pt,amplitude=.15em},->,>={Latex[length=.135cm]}] (7.2,\ths[1]+\ypaneltwo-0.1) node[anchor=north] {$W$} --++ (0,0.7);

    \pgfmathsetmacro{\ypaneltwo}{0}

    \draw[->,>={Latex[length=.15cm]}] (4.6,\ths[1]+\ypaneltwo+1.1) node[anchor=east] {$Q_1$} --++ (2,0) node[anchor=west] {$Q_3^+$};
    \node[anchor=south] at (4,\ths[1]+\ypaneltwo-1) {$Q_2$};

    \draw[decorate,decoration={snake,post length=0pt,amplitude=.15em},->,>={Latex[length=.135cm]}] (7.2,\ths[1]+\ypaneltwo-0.1) node[anchor=north] {$W$} --++ (0,0.7);
    
	
    

    \pgfmathsetmacro{\ypaneltwo}{-3.3}

    \node[anchor=south] (q1) at (4,\ths[1]+\ypaneltwo+0.5) {$Q_1$};
    \node[anchor=south] (q2) at (4,\ths[1]+\ypaneltwo-1.25) {$Q_2$};
    
    \coordinate (circ) at (5.4,\ths[1]+\ypaneltwo+0.15);
    \draw (q1.east) -| (circ) |- (q2) -- (q2.east);
    
    \draw[->,>={Latex[length=.15cm]}] (circ) --+ (1.1,0) node[anchor=west,pos=1] {$Q_3^+$};
            
    \draw[fill=white,anchor=south,thick] (circ) circle (0.2);
    \draw[thick] ($(circ)-(90:.2)$) --++ (90:.4);
    \draw[thick] ($(circ)-(0:.2)$) --++ (0:.4);

    \draw[decorate,decoration={snake,post length=0pt,amplitude=.15em},->,>={Latex[length=.135cm]}] (7.05,\ths[1]+\ypaneltwo-1.) node[anchor=north] {$W$} --++ (0,0.7);
    

\end{tikzpicture}}
    \caption{\textbf{SURD: Synergistic-Unique-Redundant Decomposition
        of causality}. (a) Diagram of the decomposition of causal
      dependencies between a vector of observed variables $\bQ =
      [Q_1,Q_2]$ (past) and a target variable $Q_j^+$ (future) into
      their synergistic (S), unique (U) and redundant (R) components
      (in yellow, red, and blue, respectively) and contributions to
      the total, $I(Q_j^+; Q_1,Q_2)$, and individual, $I(Q_j^+; Q_i)$, 
      mutual information. The causality leak is
      represented in gray. A version of this diagram for three
      variables is shown in the Supplementary Materials.  (b)
      Redundant, unique, and synergistic causalities for the simple
      examples of (c) a duplicated input (top panel), an output equal
      to the first input (middle panel), and an exclusive-OR output
      (bottom panel). The notation used is such that $[R12,U1,U2,S12]
      \equiv [\Delta I_{12\rightarrow 3}^R, \Delta I_{1\rightarrow
          3}^U, \Delta I_{2\rightarrow 3}^U, \Delta I_{12\rightarrow
          3}^S]$. The target variable $Q_3^+$ is affected by external
      stochastic forcing $W$, which is independent of the observed
      variables in $\bQ$. The effect of $W$ is measured by the
      causality leak, represented by the gray bar.}
    \label{fig:method}
\end{figure}

SURD quantifies causality as the increase in information ($\Delta I$)
about $Q_j^+$ obtained from observing individual components or groups
of components from $\bQ$. The information in $Q_j^+$ is measured by
the Shannon entropy\cite{shannon1948}, denoted by $H(Q_j^+)$, which
represents the average number of bits required to unambiguously
determine $Q_j^+$. \alvaro{It is also useful to interpret Shannon
  entropy as a measure of uncertainty. Processes that are highly
  uncertain (high entropy) are also the ones from which we gain the
  most information when their states are determined. Conversely,
  uncertainty is zero when the process is completely deterministic,
  indicating no information is gained when the outcome is revealed.}
Using the principle of forward-in-time propagation of information
(i.e., information only flows toward the future)\cite{Lozano2022},
$H(Q_j^+)$ can be decomposed as the sum of all causal contributions
from the past and present:
\begin{equation}
  \label{eq:conservation_info}
  H(Q_j^+) = \sum_{\mathbold{i} \in \mathcal{C}} \Delta I_{\bi
    \rightarrow j}^R + \sum_{i=1}^N \Delta I_{i\rightarrow j}^U +
  \sum_{\bi\in \mathcal{C}} \Delta I_{\bi \rightarrow j}^S + \Delta
  I_{\text{leak}\rightarrow j},
\end{equation}
where $\Delta I_{\bi \rightarrow j}^R$, $\Delta I_{i\rightarrow j}^U$,
and $\Delta I_{\bi \rightarrow j}^S$ are the redundant, unique, and
synergistic causalities, respectively, from the observed variables to
$Q_j^+$, and $\Delta I_{\text{leak}\rightarrow j}$ is the causality
from unobserved variables, referred to as the causality leak. Unique
causalities are associated with individual components of $\bQ$,
whereas redundant and synergistic causalities arise from groups of
variables from $\bQ$. Consequently, the set $\mathcal{C}$ contains all
combinations involving more than one variable. For instance, for
$N=2$, Equation~(\ref{eq:conservation_info}) reduces to $H(Q_j^+) =
\Delta I_{12\rightarrow j}^R + \Delta I_{1\rightarrow j}^U + \Delta
I_{2\rightarrow j}^U + \Delta I_{12 \rightarrow j}^S + \Delta
I_{\text{leak} \rightarrow j}$.
%
Figure~\ref{fig:method} shows the diagram of the redundant, unique,
and synergistic causalities for $N=2$. The formal definitions of
causality can be found in the Supplementary Materials. Here, we offer
an interpretation of each term:
 \begin{itemize}
\setlength\itemsep{-0.5em}
\item Redundant causality from $\bQ_{\bi}=[Q_{i_1},Q_{i_2},\ldots]$ to
  $Q_j^+$ (denoted by $\Delta I^R_{\bi \rightarrow j}$) is the common
  causality shared among all the components of $\bQ_{\bi}$, where
  $\bQ_{\bi}$ is a subset of $\bQ$. \alvaro{Redundant causality occurs
    when all the variables in $\bQ_{\bi}$ contain the same amount of
    information about $Q_j^+$. Therefore, any component of $\bQ_{\bi}$
    offers identical insight into the outcome of $Q_j^+$.}
 \item Unique causality from $Q_i$ to $Q_j^+$ (denoted by $\Delta
   I^U_{i \rightarrow j}$) is the causality from $Q_i$ that cannot be
   obtained from any other individual variable $Q_k\neq
   Q_i$. \alvaro{This causality occurs when observing $Q_i$ yields
     more information about some outcomes of $Q_j^+$ than observing any
     other isolated variable.}
 \item Synergistic causality from $\bQ_{\bi}=[Q_{i_1},Q_{i_2},\ldots]$
   to $Q^+_j$ (denoted by $\Delta I^S_{\bi \rightarrow j}$) is the
   causality arising from the joint effect of the variables in
   $\bQ_{\bi}$. \alvaro{This causality occurs when more information
     about $Q_j^+$ is gained by observing a collection of variables
     simultaneously than by observing each variable individually.}
 \item Causality leak represents the effect from unobserved variables
   that influence $Q_j^+$ but are not contained in $\bQ$.
   \alvaro{This is the amount of information missing that would be
     required to unambiguously determine the future of $Q_j$ after
     considering all observable variables collectively}.
 \end{itemize}
%
SURD exhibits several key properties that facilitate the precise
identification of interactions by preventing the duplication of
causality. This is illustrated in Figure~\ref{fig:method}. First, the
terms in Equation~(\ref{eq:conservation_info}) are non-negative and
such that the sum of redundant, unique, and synergistic causalities
equals the information shared between $Q_j^+$ and $\mathbf{Q}$,
referred to as the mutual information $I(Q_j^+; \mathbf{Q})$
\cite{shannon1948, kullback1951, Kreer1957}. SURD also satisfies that
the mutual information between individual variables $Q_i$ and $Q_j^+$,
denoted as $I(Q_j^+;Q_i)$, is represented by the sum of unique and
redundant causalities involving $Q_i$. This condition is consistent
with the notion that causality from an individual variable to $Q_j^+$
is composed solely of unique and redundant causalities, while
synergistic causalities emerge from the combined effects of two or
more variables\cite{Robin2017}. The information-theoretic formulation
of SURD is also well-suited for capturing nonlinear dependencies, as
well as deterministic and stochastic interactions, and self-causation.

The forward propagation of information from
Equation~(\ref{eq:conservation_info}) also lays the foundation for
normalizing causality within SURD. Unique, redundant, and synergistic
causalities to $Q_j^+$ are normalized by $I(Q_j^+; \mathbf{Q})$, such
that their sum equals 1. Similarly, the causality leak is normalized
by $H(Q_j^+)$, which bounds its values between 0 (indicating that all
causalities to $Q_j^+$ are accounted for by $\bQ$) and 1 (indicating
that none of the causalities are accounted for by $\bQ$).
Figure~\ref{fig:method} includes the results of SURD for three simple
examples. Each case represents a system characterized exclusively by
redundant, unique, or synergistic causality, respectively. These
examples also allow us to introduce the notation used in the following
figures: $[R12, U1, U2, S12] \equiv [\Delta I_{12\rightarrow 3}^R,
  \Delta I_{1\rightarrow 3}^U, \Delta I_{2\rightarrow 3}^U, \Delta
  I_{12\rightarrow 3}^S]$, where the index of the target variable
$Q_3$ is omitted but it will be unambiguous from the context.

\subsection*{Validation}

\begin{table}[t!]
\centering
\begin{tabular}{|l|c|c|c|c|c|c|c|}
\hline
\textbf{Case}  & \textbf{CGC} & \textbf{CTE} & \textbf{CCM} & \textbf{PCMCI} & \textbf{SURD} \\
\hline
Mediator variable              & \xmark & \cmark & \xmark & \cmark  & \cmark \\
Confounder variable            & \cmark & \cmark & \cmark & \cmark$^\ddag$ &  \cmark \\ 
Synergistic collider variable  &  \xmark &  \cmark$^\dag$ &  \xmark & \cmark$^\dag$  &  \cmark \\
Redundant collider variable    &  \xmark &  \xmark &  \xmark & \xmark  &  \cmark \\
Turbulent energy cascade           & \xmark  & \cmark$^\ddag$ & \xmark & \cmark$^\ddag$  & \cmark \\
\alvaro{Experimental turbulent boundary layer}           & \cmark  & \cmark & \xmark & \xmark  & \cmark \\
Lotka--Volterra prey-predator model\cite{lotka1925,volterra1926} & \cmark & \cmark & \cmark & \xmark  &  \cmark \\
Three-interacting species system\cite{pcm2020} & \xmark & \cmark$^\dag$ & \xmark & \xmark  &  \cmark \\
Moran effect model\cite{moran1953}                 & \cmark & \cmark & \cmark & \cmark  & \cmark \\
One-way coupling nonlinear logistic difference system\cite{sugihara2012} & \xmark & \cmark & \cmark & \xmark &  \cmark \\
Two-way coupling nonlinear logistic difference system\cite{sugihara2012} & \xmark & \cmark  & \cmark & \xmark & \cmark \\
Stochastic system with linear time-lagged dependencies\cite{ding2006} & \cmark & \cmark & \xmark  & \cmark$^\ddag$ & \cmark \\
Stochastic system with non-linear time-lagged dependencies\cite{bueso2020} & \xmark  & \cmark & \xmark & \cmark  & \cmark \\
Synchronization of two variables in logistic maps\cite{may1976}   & \xmark & \xmark & \xmark & \cmark$^{\dag\dag}$ & \cmark \\
Synchronization of three variables in logistic maps\cite{may1976}  & \xmark & \xmark  & \xmark & \xmark & \cmark \\
Uncoupled Rössler--Lorenz system\cite{lorenz1963, rossler1977} & \xmark &  \cmark & \xmark & \cmark &  \cmark \\
One-way coupled Rössler--Lorenz system\cite{lorenz1963, rossler1977} & \xmark & \cmark  & \cmark &  \cmark$^\ddag$  &  \cmark \\
\hline
\end{tabular}
\caption{\textbf{{Summary of the performance of different methods for
      causal inference.}} The markers \cmark\ and \xmark\ denote
  consistent and inconsistent identification of causal links,
  respectively, according to the functional dependency of variables
  within the system.  The methods considered are conditional Granger
  causality (CGC), conditional transfer entropy (CTE), convergent
  cross-mapping (CCM), Peter and Clark momentary conditional
  independence (PCMCI) based on conditional mutual information tests
  with a $k$-nearest neighbor estimator, and
  synergistic-unique-redundant decomposition of causality (SURD). A
  summary of the results for PCMCI with different independence tests
  is provided in the Supplementary Materials. $^\dag$Causalities are
  detected but the method cannot discern whether they are synergistic
  or unique. $^\ddag$The causality detected is consistent with the
  interactions in the system, although the causal strength of the
  links is weak. $^{\dag\dag}$The method cannot detect duplicated
  variables and redundant causalities.}
\label{tab:summary}
\end{table}

We validate SURD across multiple scenarios that pose significant
challenges in causal inference. These include systems with mediator,
confounder and colliders effects, Lotka--Volterra prey-predator
model\cite{lotka1925, volterra1926}, three-interacting species
system\cite{pcm2020}, the Moran effect model\cite{moran1953},
turbulent energy cascade\cite{richardson1922, obukhov1941,
  kolmogorov1941}, \alvaro{experimental data for a turbulent boundary
  layer\cite{Baars2015,Baars2017,marusic2020}}, deterministic and
stochastic systems with time-lagged dependencies proposed by Sugihara
et al.\cite{sugihara2012}, Ding et al.\cite{ding2006}, and Bueso et
al.\cite{bueso2020}, logic gates, synchronization of logistic
maps\cite{may1976}, and the coupled Rössler--Lorenz
system\cite{lorenz1963, rossler1977, Quiroga2000, Krakovska2018}.  A
summary of the results is shown in Table~\ref{tab:summary}, where the
metric for success is based on whether the results are consistent with
the functional dependencies of the system, rather than on the concrete
value of the causal strength provided by each method.

The ability of SURD to identify causal relationships is compared with
other methods for causal inference, which are also included in
Table~\ref{tab:summary}. The approaches considered are conditional
Granger causality (CGC)\cite{geweke1984}, conditional transfer entropy
(CTE)\cite{cte2016}, convergent cross-mapping (CCM)\cite{sugihara2012,
  causalccm}, and Peter and Clark momentary conditional independence
(PCMCI)\cite{runge2019pcmci}.  Each method brings distinct
capabilities to address specific challenges in the field of causal
inference. To facilitate comparison, we have summarized the key
properties associated with each approach in Table
\ref{tab:comparison}. This classification outlines the ability of each
method to handle multivariate relationships, nonlinear dependencies,
stochastic (nondeterministic) processes, contemporaneous links (i.e.,
those occurring at a smaller time-scale than the time resolution of
the data), estimation of the causality leak (i.e., causality from
unobserved variables), time-delayed dependencies, and
self-causation. The table also highlights the ability of SURD to
account for all the scenarios described above.
\begin{table}[t!]
\centering
\begin{tabular}{|l|c|c|c|c|c|c|c|}
\hline
\textbf{Method} & \textbf{Multivariate} & \textbf{Nonlinear} & \textbf{Stochastic} & \textbf{Contemporaneous} & \textbf{Leak} & \textbf{Time-delay} & \textbf{Self-causation} \\
\hline
CGC\cite{geweke1984}     & \cmark & \xmark & \cmark & \xmark & \xmark & \cmark & \cmark \\
CTE\cite{cte2016}     & \cmark & \cmark & \cmark & \xmark & \xmark & \cmark & \cmark \\
CCM\cite{sugihara2012}    & \xmark & \cmark & \xmark$^1$ & \cmark & \xmark & \xmark$^2$ & \xmark \\
PCMCI\cite{runge2019pcmci}  & \cmark & \cmark & \cmark & \xmark$^3$ & \xmark & \cmark & \cmark \\
SURD   & \cmark & \cmark & \cmark & \cmark & \cmark & \cmark & \cmark \\
\hline
\end{tabular}
\caption{\textbf{Method for causal inference}. List of methods for
  causal inference investigated, along with the methodological
  challenges each method is capable of addressing: multivariate
  relationships, nonlinear dependencies, stochastic (nondeterministic)
  processes, contemporaneous links, estimation of the causality leak,
  time-delayed dependencies, and self-causation. $^1$CCM aims to
  reconstruct the attractor manifold associated with two given
  variables, making it potentially effective for stochastic
  systems. However, the presence of increased dynamical noise can
  complicate the reconstruction process. \alvaro{$^2$An extension of
    the CCM method, extended CCM\cite{extendedccm2015}, introduces the
    concept of time-delayed causal interactions.} $^3$A recent variant
  of the PCMCI method, PCMCI+\cite{runge2020pcmci+}, accounts for
  contemporaneous links.}
\label{tab:comparison}
\end{table}

The findings in Table~\ref{tab:summary} show that, although there are
methods capable of effectively tackling certain situations, SURD is
consistently successful across all the cases considered. In
particular, SURD offers a distinct advantage in the presence of
redundant variables and synchronization phenomena.
%
In the following, we focus our discussion on the cases from
Table~\ref{tab:summary} involving fundamental causal interactions:
mediators, confounders, and colliders, as these are key to understand
the success of SURD. We also discuss the application of SURD to unveil
the causality in two turbulent flow scenarios. Readers are
referred to the Supplementary Materials for a comprehensive discussion
of the results presented in Table~\ref{tab:summary}, as well as a
detailed overview of the causal inference methods utilized and their
implementation. Furthermore, the Supplementary Materials also contain a section on the application of SURD to select the most effective input variables for temporal forecasting, as well as a section that discusses the non-separability problem for nonlinear dynamical systems and demonstrates the robustness of SURD under such conditions.

\subsubsection*{Mediator variable} 
\begin{figure}[t!]
    \centering
    \hspace{-0.02\textwidth}
    \begin{minipage}{0.175\textwidth}
    \vspace{-1.7\textwidth}
        $\hspace{0.5cm}{\hbox{


\begin{tikzpicture}[cir/.style={circle,draw=black!70!white,,thick,inner sep=.5em},
    >={Latex[length=.2cm]},scale=.8]
    

    \colorlet{cA}{Q1!30!white}
    \colorlet{cB}{Q2!30!white}
    \colorlet{cC}{Q3!30!white}


    \node [cir,fill=cB] (q1) at (.73,0)   {$Q_1$}; 
    \node [cir,fill=cA] (q2) at (-1,-1) {$Q_2$}; 
    \node [cir,fill=cC] (q3) at (-1,+1) {$Q_3$}; 
    
    \node [anchor=north] at (q2.south) {\scriptsize mediator};

    \draw[decorate,decoration={snake,post length=4pt,amplitude=.2em}, thick,->]
    (-2.3,1) node[anchor=east] {$W_3$} -- (q3);
    \draw[decorate,decoration={snake,post length=4pt,amplitude=.2em}, thick,->]
    (-2.3,-1) node[anchor=east] {$W_2$} -- (q2);
    \draw[decorate,decoration={snake,post length=4pt,amplitude=.2em}, thick,->]
    (.73,-1.3) node[anchor=north] {$W_1$} -- (q1);

    \path[thick,->]
        (q3) edge[bend right] node [left] {} (q2)
        (q2) edge[bend right] node [left] {} (q1);
    \path[thick,<-] (q3) edge[loop above] node {} (q3);
    
\end{tikzpicture}

}}$
        {\footnotesize
          \begin{flalign*}
            Q_1^{n+1} &= \sin \left(Q_2^{n}\right) + 0.001W_1^{n} \\
            Q_2^{n+1} &= \cos \left(Q_3^{n}\right) + 0.01W_2^{n} \\
            Q_3^{n+1} &= 0.5Q_3^{n} + 0.1W_3^{n}
          \end{flalign*}}
    \end{minipage}
    \hspace{0.02\textwidth}
    \begin{minipage}{0.3352\textwidth}
        \vspace{-0.86\textwidth}
        {\includegraphics[width=\textwidth]{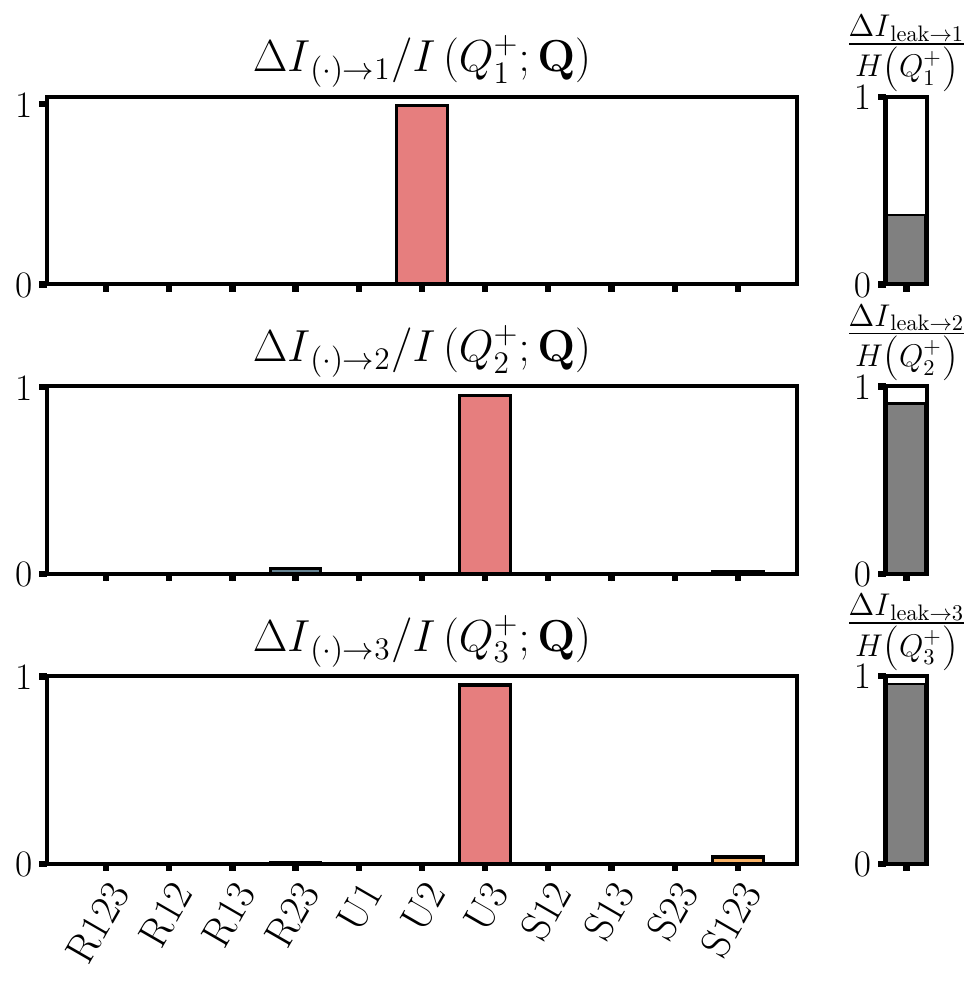}}
    \end{minipage}
    \hspace*{-0.002\textwidth}
    {\includegraphics[width=0.4\textwidth]{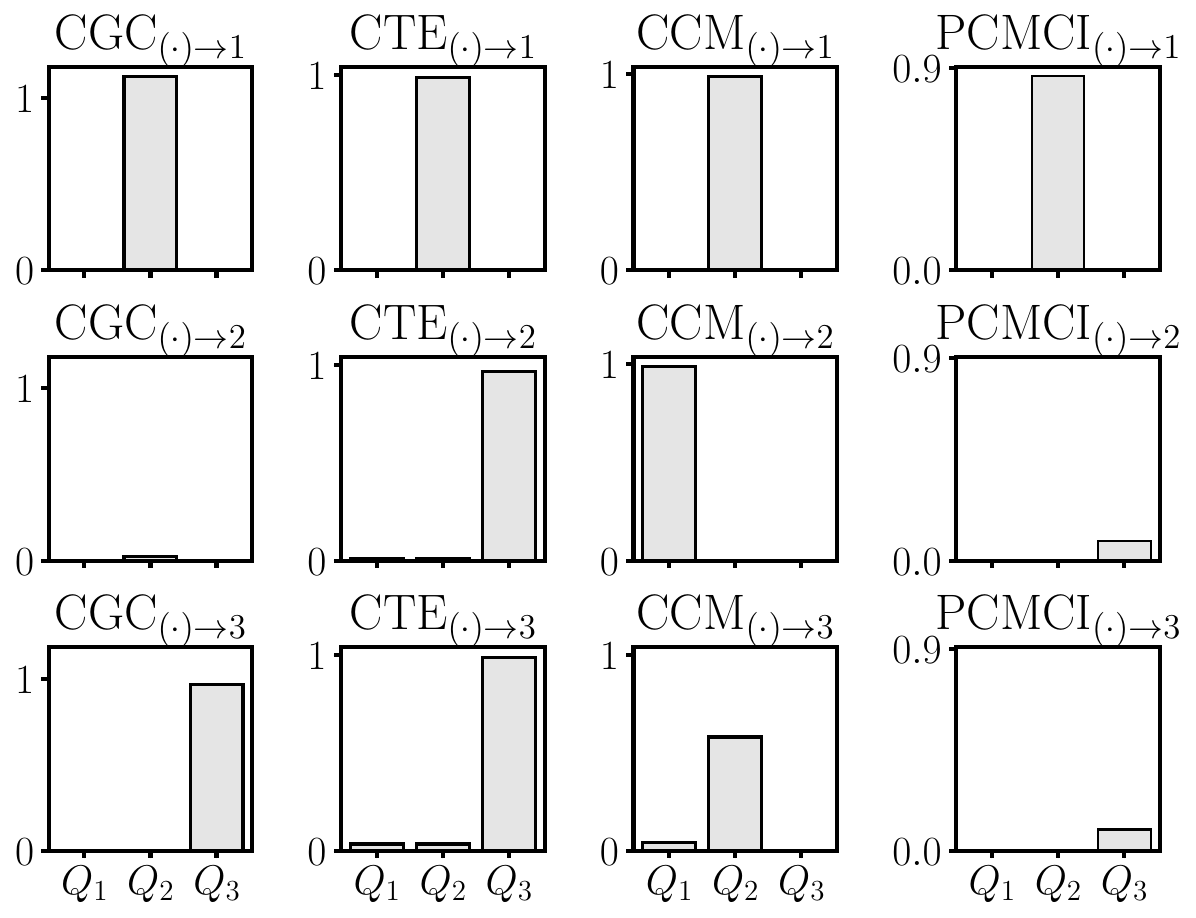}}
    \caption{\textbf{System with mediator variable}. (Left panel)
      Schematic of the functional dependence among variables and
      system equations, where $W_i$ represents unobserved, stochastic
      forcing on the variable $Q_i$. We use the notation
      $Q_i^n=Q_i(t_n)$, where $n$ indicates the time step. (Center and
      right panels) Results from SURD with redundant (R), unique (U)
      and synergistic (S) causalities in blue, red and yellow,
      respectively. The notation employed is such that R123 denotes
      $\Delta I_{123\rightarrow j}^R$ and so on. The gray bar is the
      causality leak. The results from CGC, CTE, PCMCI, and CCM are
      depicted on the right.  In all methods but CCM, the value of the
      bar represent the strength of the causal link.  In CCM, a causal
      link is detected only when the value converges to 1 as the
      length of the time series increases, but not otherwise.  CGC and
      CTE use the same normalization as SURD.  The values for SURD,
      CTE and CCM are upper bounded by 1.  The values for PCMCI
      represent conditional mutual information and are unbounded. The
      equations to quantify causality by each of the methods are in
      the Supplementary Materials.}
    \label{fig:mediator}
\end{figure}
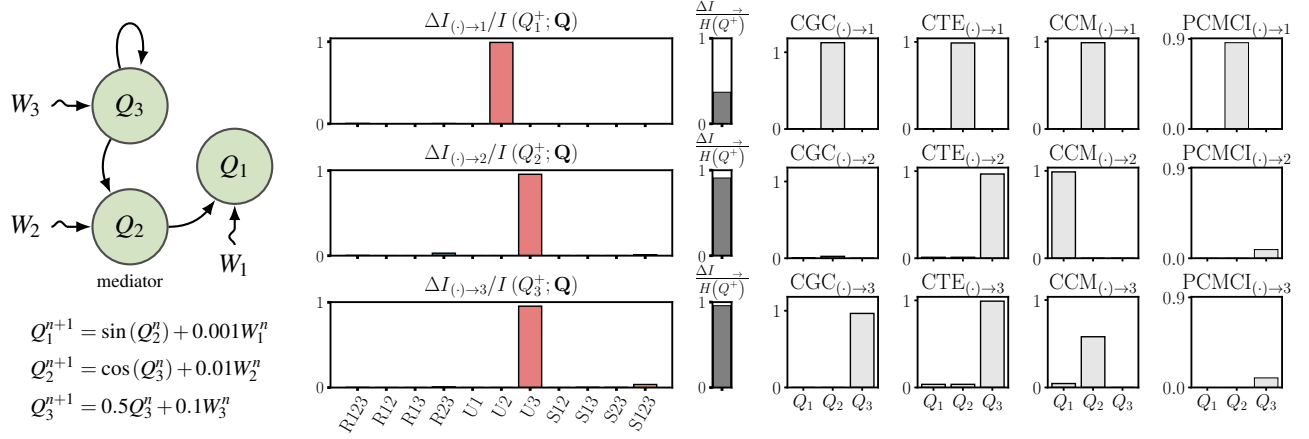

The first case investigated corresponds to the system $Q_3 \rightarrow
Q_2 \rightarrow Q_1$, where $Q_3$ influences $Q_1$ through the
mediator variable $Q_2$. Figure~\ref{fig:mediator} displays a diagram
illustrating the relationships among the variables, along with the
results derived from SURD and other causal analysis methods. Note that
while the exact value of the causal strength may be subject to debate,
the predicted causal links by each method should maintain consistency
with the arrows depicted in the diagram.

The causal contributions detected by SURD are the unique causalities
${\Delta I}_{3\rightarrow 3}^U$, ${\Delta I}_{3\rightarrow 2}^U$, and
${\Delta I}_{2\rightarrow 1}^U$, which are consistent with the
dependency of variables in the system. PCMCI and CTE yield similar
results; however, CGC cannot identify the link $Q_3 \rightarrow
Q_2$. CCM also fails to unambiguously capture the causal links, as the
only causality converging to one as the length of the time series
increases is $Q_1 \rightarrow Q_2$, which is inconsistent with the
equations of the system.

SURD also offers an estimate of the causality leak, which exceeds 95\%
for $Q_2$ and $Q_3$. This is attributed to the influence of the
stochastic forcing terms $W_2$ and $W_3$, respectively, which are
assumed to be unknown. The causality leak for $Q_1$ is the lowest
(below 50\%), as the noise from $W_1$ is the smallest among the three
$W_i$. Note that none of the methods offers any insight into the
causality leak, and this is also the case for all the subsequent
benchmarks.

\subsubsection*{Confounder variable} 

\begin{figure}[t!]
    \centering
    \hspace{-0.5cm}
    \begin{minipage}{0.175\textwidth}
    \vspace{-1.7\textwidth}
        $\hspace{0.55cm}{\hbox{


\begin{tikzpicture}[cir/.style={circle,draw=black!70!white,,thick,inner sep=.5em},
    >={Latex[length=.2cm]},scale=.8]
    

    \colorlet{cA}{Q1!30!white}
    \colorlet{cB}{Q2!30!white}
    \colorlet{cC}{Q3!30!white}

    \node [cir,fill=cB] (q1) at (.73,0) {$Q_1$}; 
    \node [cir,fill=cA] (q2) at (-1,-1) {$Q_2$}; 
    \node [cir,fill=cC] (q3) at (-1,+1) {$Q_3$}; 

    \node [anchor=west] at (q3.north) {\quad \scriptsize confounder};

    \draw[decorate,decoration={snake,post length=4pt,amplitude=.2em}, thick,->]
    (-2.3,1) node[anchor=east] {$W_3$} -- (q3);
    \draw[decorate,decoration={snake,post length=4pt,amplitude=.2em}, thick,->]
    (-2.3,-1) node[anchor=east] {$W_2$} -- (q2);
    \draw[decorate,decoration={snake,post length=4pt,amplitude=.2em}, thick,->]
    (.73,-1.3) node[anchor=north] {$W_1$} -- (q1);

    \path[thick,->]
        (q3) edge[bend left] node [right] {} (q1)
        (q3) edge[bend right] node [left] {} (q2);
    \path[thick,<-] (q1) edge[loop right] node {} (q1);
    \path[thick,<-] (q2) edge[loop below] node {} (q2);
    \path[thick,<-] (q3) edge[loop above] node {} (q3);

\end{tikzpicture}

}}$
        {\footnotesize
          \begin{flalign*}
            Q_1^{n+1} &= \sin\left(Q_1^n + Q_3^n\right) + 0.01W_1^n\\
            Q_2^{n+1} &= \cos\left(Q_2^n - Q_3^n\right) + 0.01W_2^n\\
            Q_3^{n+1} &= 0.5Q_3^n + 0.1W_3^n
          \end{flalign*}}
    \end{minipage}
    \hspace{0.0075\textwidth}
    \begin{minipage}{0.3395\textwidth}
        \vspace{-0.85\textwidth}
        {\includegraphics[width=\textwidth]{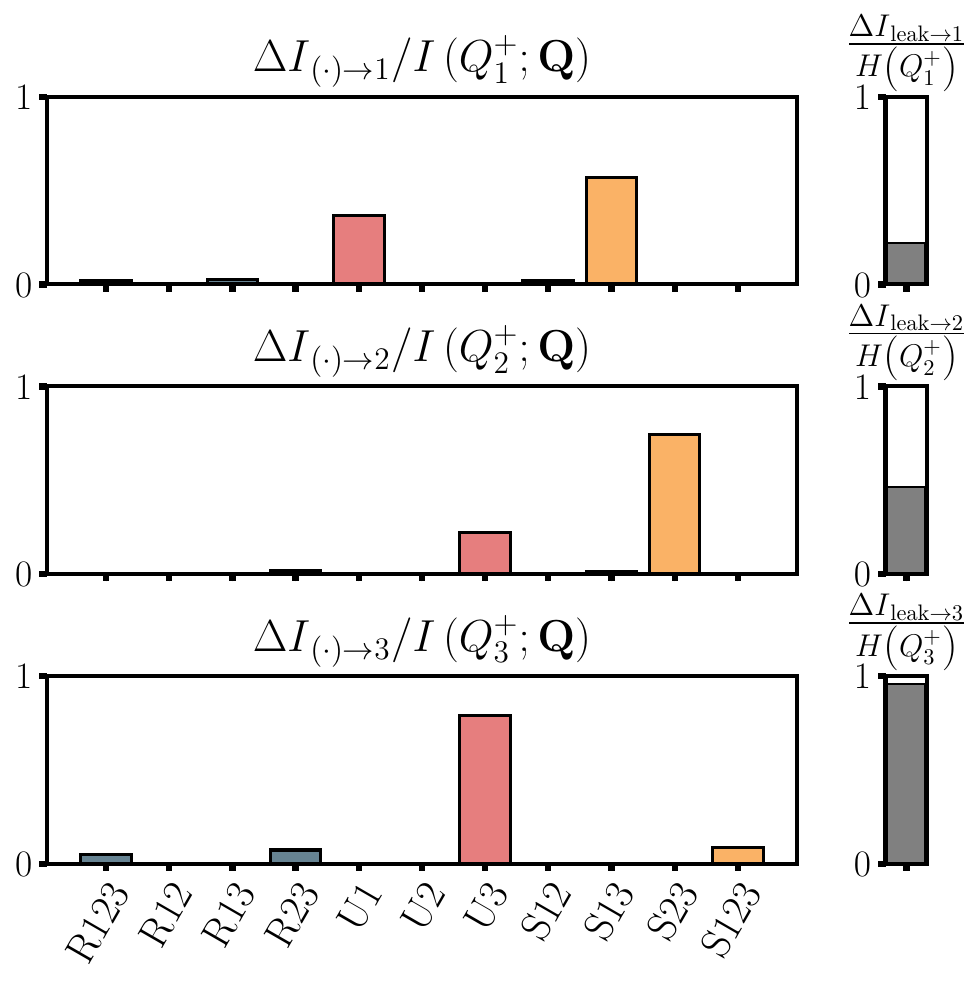}}
    \end{minipage}
    \hspace*{-0.002\textwidth}
    {\includegraphics[width=0.4\textwidth]{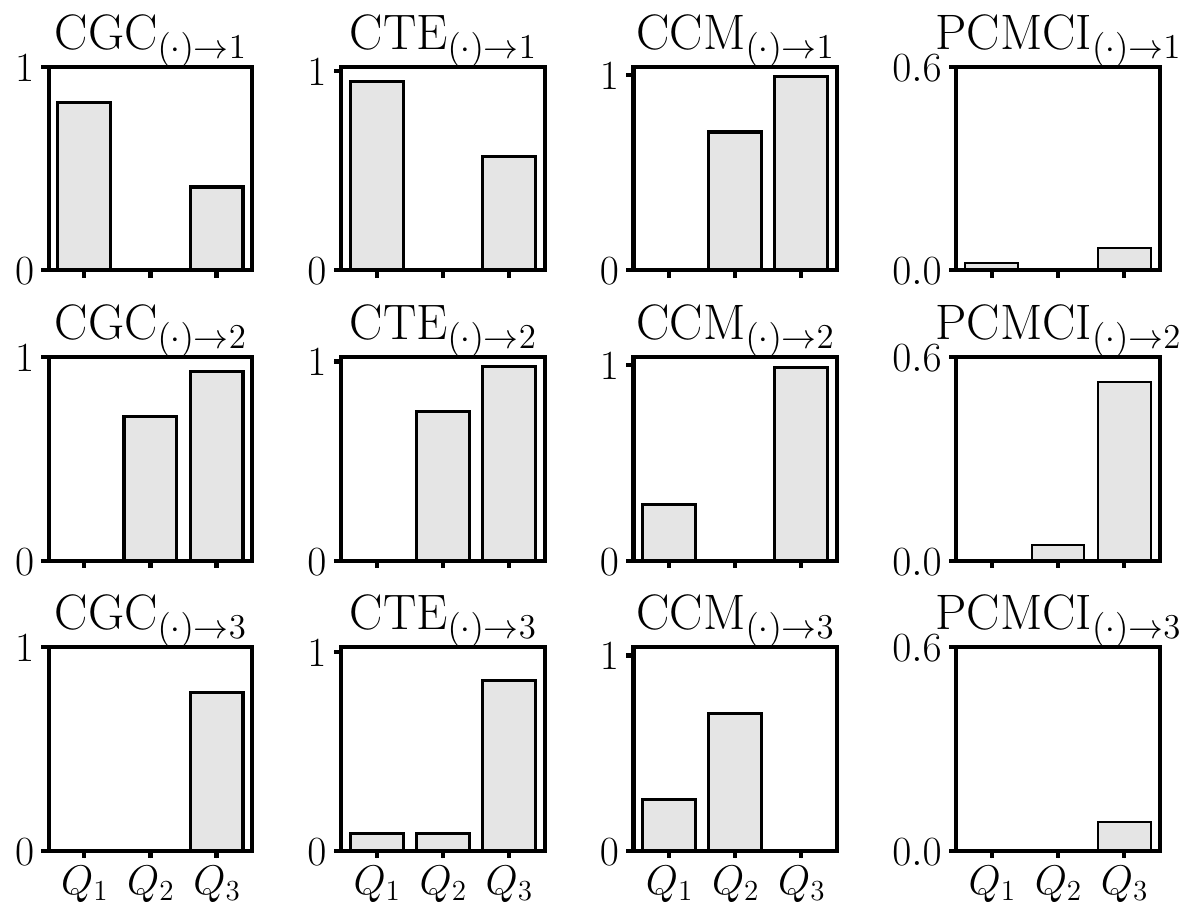}}
    \caption{\textbf{System with confounder variable}. Same as
      Figure~\ref{fig:mediator}.}
    \label{fig:confounder}
\end{figure}
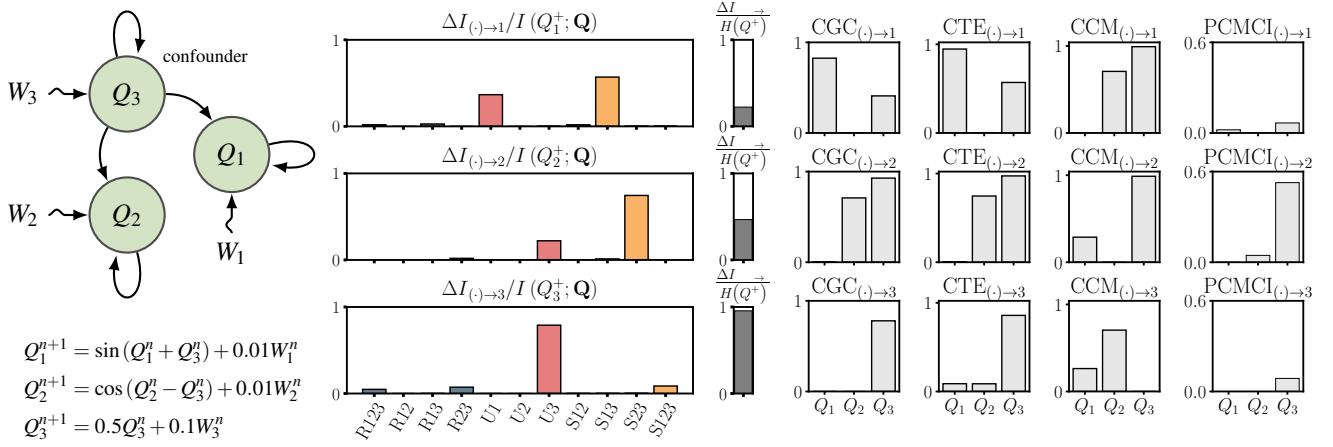

The second case (Figure~\ref{fig:confounder}) corresponds to a system
where $Q_3$ acts as a confounding variable for $Q_1$ and $Q_2$, i.e.,
$Q_1 \leftarrow Q_3 \rightarrow Q_2$. The presence of confounding
effects is captured in SURD by the synergistic causalities $\Delta
I_{13\rightarrow 1}^S$ and $\Delta I_{23\rightarrow 2}^S$, while the
self-induced causality from $Q_3$ is detected by $\Delta
I_{3\rightarrow 3}^U$. Other causalities manifest as $\Delta
I_{1\rightarrow 1}^U$ and $\Delta I_{3\rightarrow 2}^U$, which are
also consistent with the causal structure of the system. Regarding
other methods, all of them correctly identified the confounding
effects, although the link $Q_3 \rightarrow Q_1$ and the self-induced
causalities identified by PCMCI are barely significant.  This
highlights another advantage of SURD: the relative importance of the
causalities is easier to interpret, since the sum of their normalized
values must always equal one. Also note that CCM cannot detect
self-induced causalities by construction.  The largest causality leak
occurs for $Q_3$, which is consistent with the fact that the
(unobserved) stochastic term acting on $Q_3$ is ten times larger than
the (unobserved) stochastic terms acting on $Q_1$ and $Q_2$.

%

\subsubsection*{Collider with synergistic variables} 

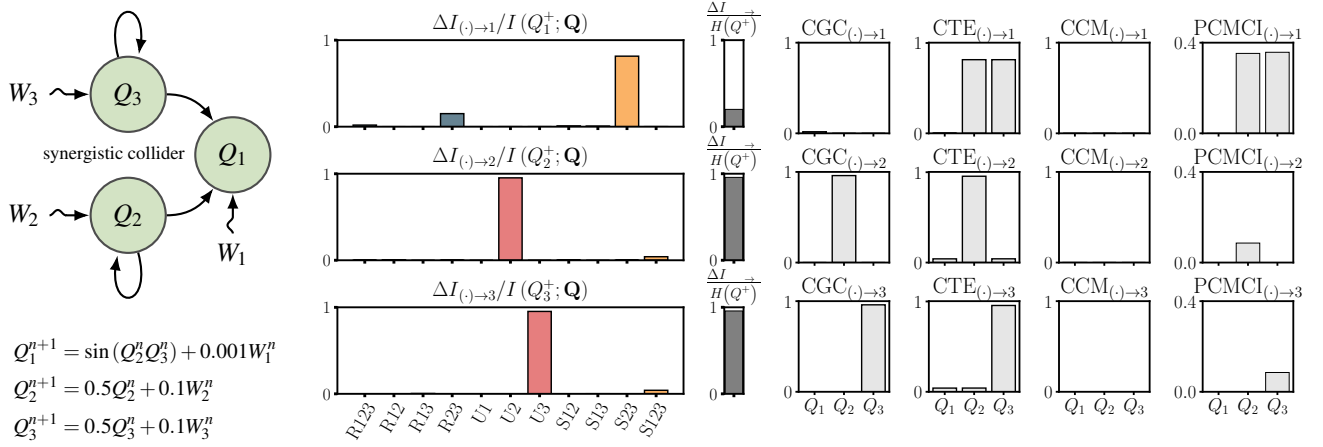
\begin{figure}[t!]
    \centering
    \hspace{-0.5cm}
    \begin{minipage}{0.175\textwidth}
    \vspace{-1.7\textwidth}
        $\hspace{0.7cm}{\hbox{


\begin{tikzpicture}[cir/.style={circle,draw=black!70!white,,thick,inner sep=.5em},
    >={Latex[length=.2cm]},scale=.8]
    

    \colorlet{cA}{Q1!30!white}
    \colorlet{cB}{Q2!30!white}
    \colorlet{cC}{Q3!30!white}

    \node [cir,fill=cB] (q1) at (.73,0)   {$Q_1$}; 
    \node [cir,fill=cA] (q2) at (-1,-1) {$Q_2$}; 
    \node [cir,fill=cC] (q3) at (-1,+1) {$Q_3$}; 
    
    \node [anchor=east] at (q1.west) {\scriptsize synergistic collider}; 

    \draw[decorate,decoration={snake,post length=4pt,amplitude=.2em}, thick,->]
    (-2.3,1) node[anchor=east] {$W_3$} -- (q3);
    \draw[decorate,decoration={snake,post length=4pt,amplitude=.2em}, thick,->]
    (-2.3,-1) node[anchor=east] {$W_2$} -- (q2);
    \draw[decorate,decoration={snake,post length=4pt,amplitude=.2em}, thick,->]
    (.73,-1.3) node[anchor=north] {$W_1$} -- (q1);

    \path[thick,->]
    (q3) edge[bend left] node [left] {} (q1)
    (q2) edge[bend right] node  [left] {} (q1);

    \path[thick,<-] (q2) edge[loop below] node {} (q2);
    \path[thick,<-] (q3) edge[loop above] node {} (q3);

\end{tikzpicture}

}}$
        {\footnotesize
          \begin{flalign*}
            Q_1^{n+1} &= \sin\left(Q_2^nQ_3^n\right) + 0.001W_1^n\\
            Q_2^{n+1} &= 0.5Q_2^n + 0.1W_2^n\\
            Q_3^{n+1} &= 0.5Q_3^n + 0.1W_3^n
          \end{flalign*}}
    \end{minipage}
    \hspace{0.02\textwidth}
    \begin{minipage}{0.3395\textwidth}
        \vspace{-0.85\textwidth}
        {\includegraphics[width=\textwidth]{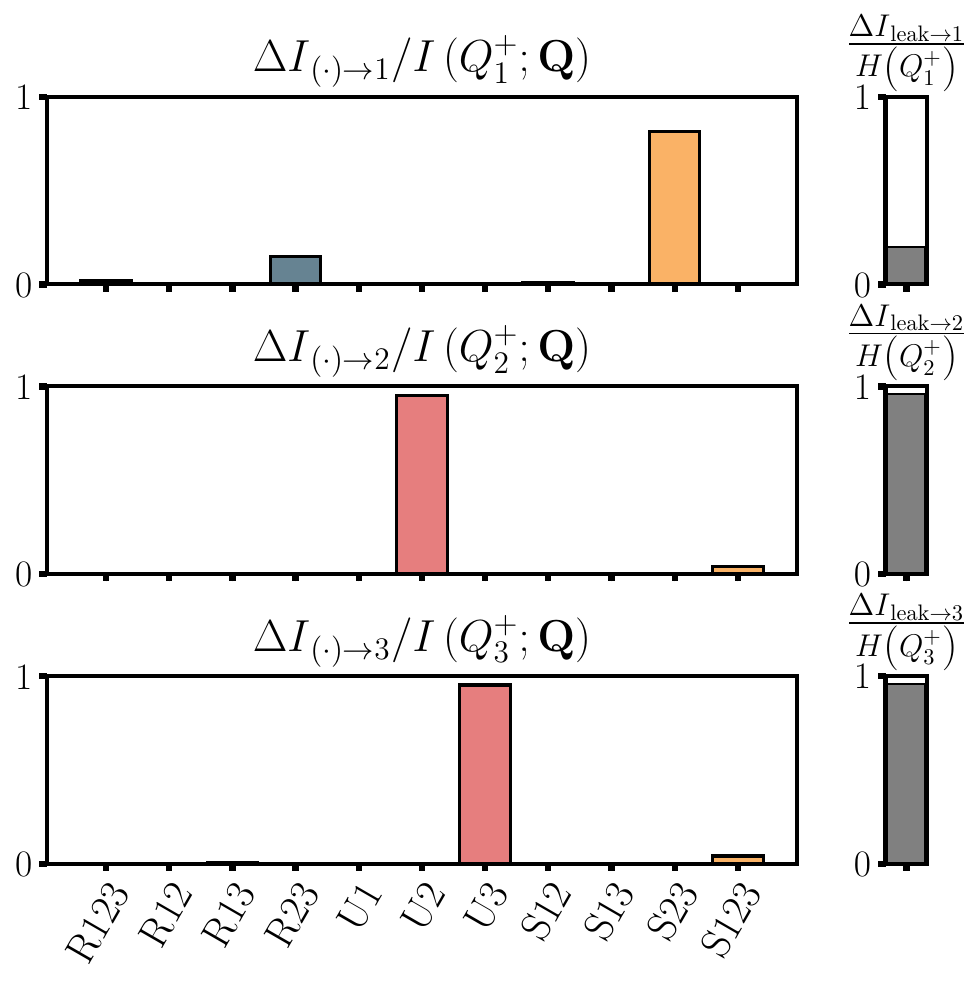}}
    \end{minipage}
    \hspace*{-0.002\textwidth}
    {\includegraphics[width=0.4\textwidth]{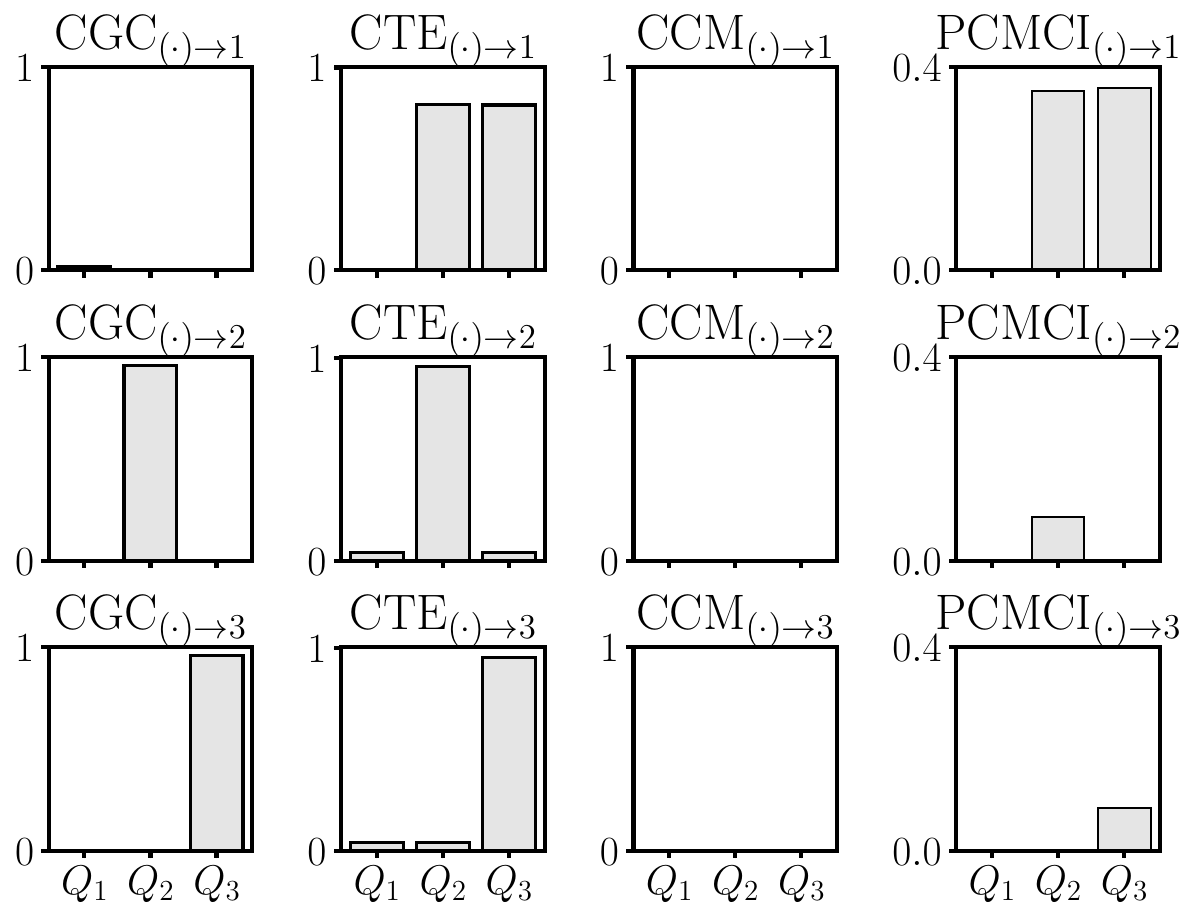}}
    \caption{\textbf{System with synergistic collider variables}.
      Same as Figure~\ref{fig:mediator}.}
    \label{fig:synergistic}
\end{figure}

Next, we consider the system $[Q_2,Q_3] \rightarrow Q_1$, where $Q_2$
and $Q_3$ act together to influence $Q_1$. In reality, $Q_2$ and $Q_3$
behave as a single random variable that drives $Q_1$. The results,
presented in Figure~\ref{fig:synergistic}, demonstrate that SURD is
able to detect the dominant synergistic effect of $Q_2$ and $Q_3$ on
$Q_1$ through $\Delta I_{23\rightarrow 1}^S$ along with the
self-induced causalities $\Delta I_{2\rightarrow 2}^U$ and $\Delta
I_{3\rightarrow 3}^U$. The smallest causality leak is associated with
$Q_1$, as it is affected by the lowest stochastic forcing while being
strongly influenced by the (observed) variables $Q_2$ and $Q_3$.

PCMCI and CTE also identify the self-influence of $Q_2$ and $Q_3$ and
the effect of $Q_2$ and $Q_3$ on $Q_1$. However, neither of the two
methods can label the interaction as synergistic and, hence, cannot
show that both variables are required in combination, rather than
individually, to affect $Q_1$. On the other hand, CGC and CCM are
unsuccessful in identifying the interactions of $Q_2$ and $Q_3$ with $Q_1$.

\subsubsection*{Collider with redundant variables} 

\begin{figure}[t!]
    \centering
    \hspace{-1cm}
    \begin{minipage}{0.175\textwidth}
    \vspace{-1.55\textwidth}
        $\hspace{0.65cm}{\hbox{


\begin{tikzpicture}[cir/.style={circle,draw=black!70!white,,thick,inner sep=.5em},
    >={Latex[length=.2cm]},scale=.8]
    

    \colorlet{cA}{Q1!30!white}
    \colorlet{cB}{Q2!30!white}
    \colorlet{cC}{Q3!30!white}

    \node [cir,fill=cB] (q1) at (1.73,0)   {$Q_1$}; 
    \node [cir,fill=cA] (q2) at (0,-1)     {$Q_2$}; 
    \node [cir,fill=cC] (q3) at (0,+1)     {$Q_3$}; 

    \node[anchor=south] at (q3.north) {\scriptsize redundant collider};
    \draw[decorate,decoration={snake,post length=4pt,amplitude=.2em}, thick,->]
    (-1.3,-1) node[anchor=east] {$W_2$} -- (q2);
    \draw[decorate,decoration={snake,post length=4pt,amplitude=.2em}, thick,->]
    (1.73,-1.3) node[anchor=north] {$W_1$} -- (q1);

    \path[thick,->]
    (q3) edge[bend left] node [left] {} (q1)
    (q2) edge[bend right] node  [left] {} (q1);

    \path[thick,<-] (q2) edge[loop below] node {} (q2);
    \path[thick,<-] (q1) edge[loop right] node {} (q1);
    
    \draw[thick,double,<->,double distance=.1em] (q3) -- (q2) node [pos=.6,anchor=east] {\scriptsize };
\end{tikzpicture}

}}$
        {\footnotesize
          \begin{flalign*}
            Q_1^{n+1} &=  0.3Q_1^n + \sin\left(Q_2^nQ_3^n\right) + 0.001 W_1^n\\
            Q_2^{n+1} &= 0.5 Q_2^n + 0.1W_2^n\\
            Q_3^{n+1} &\equiv Q_2^{n+1}
          \end{flalign*}}
    \end{minipage}
    \begin{minipage}{0.3362\textwidth}
        \vspace{-0.85\textwidth}
        {\includegraphics[width=\textwidth]{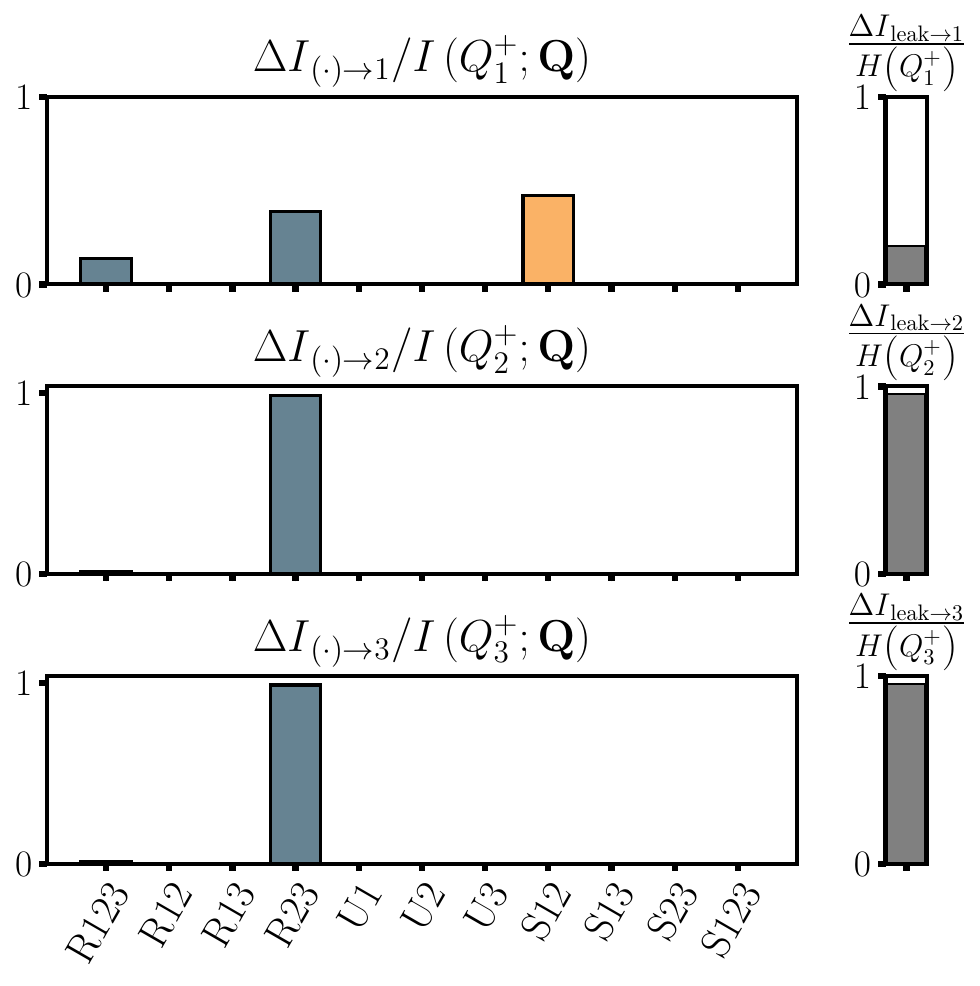}}
    \end{minipage}
    \hspace*{-0.002\textwidth}
    {\includegraphics[width=0.396\textwidth]{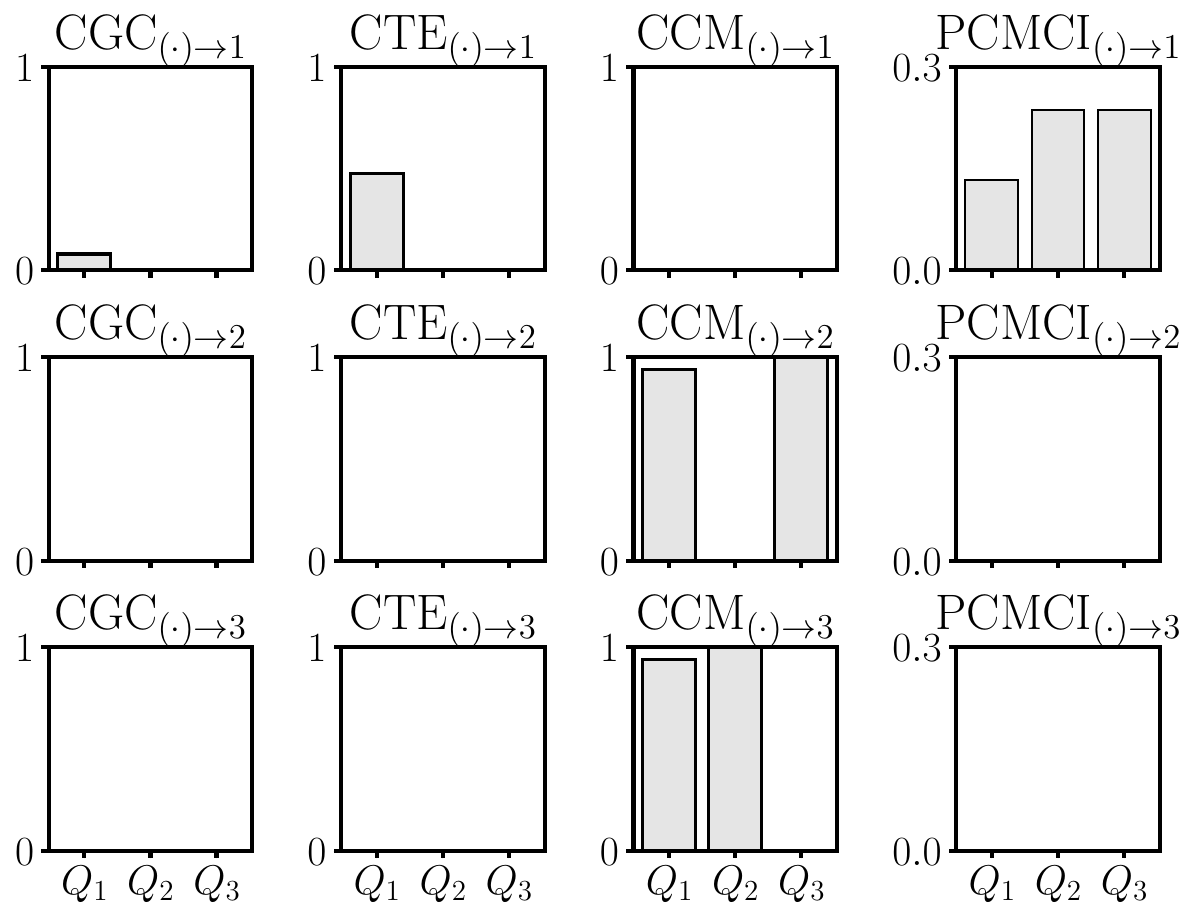}}

    \caption{\textbf{System with redundant collider variables}.  Same
      as Figure~\ref{fig:mediator}. The symbol $\Leftrightarrow$
      indicates that variables $Q_2$ and $Q_3$ are the identical.}
    \label{fig:redundant}
\end{figure}
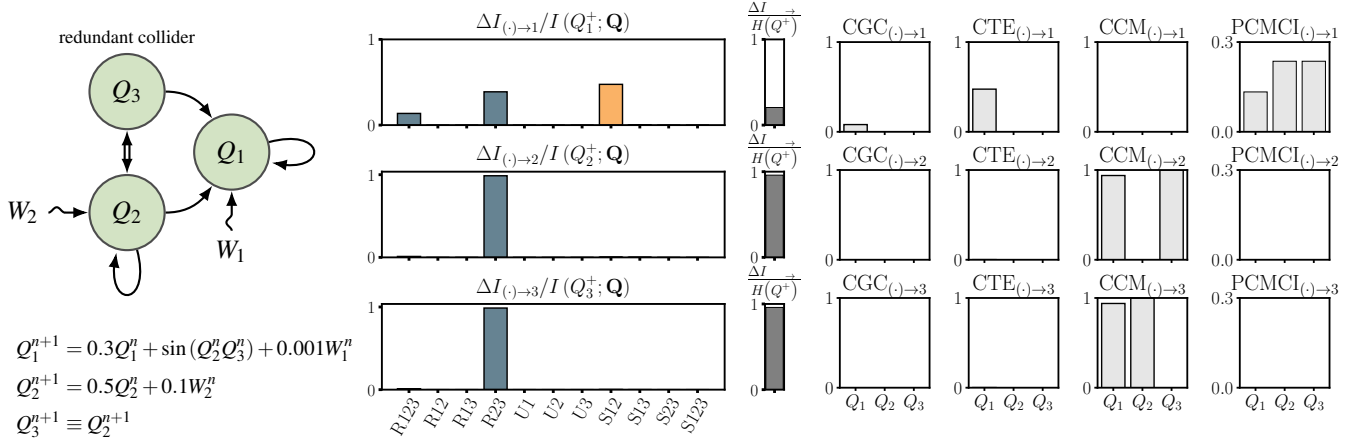

The fourth case under examination explores the fundamental interaction
$Q_2 \equiv Q_3 \rightarrow Q_1$, where $Q_3$ is identical to
$Q_2$. In this scenario, $Q_2$ and $Q_3$ equally influence the future
outcomes of $Q_1$. SURD identifies $\Delta I^R_{23 \rightarrow 2} =
\Delta I^R_{23 \rightarrow 3}$ as the most significant causalities
associated with $Q_2$ and $Q_3$, respectively, as shown in Figure
\ref{fig:redundant}. Moreover, the identical causalities (and
causality leaks) for both $Q_2$ and $Q_3$ suggest that they represent
the same variable. SURD also identifies the influence of $Q_2$ and
$Q_3$ on $Q_1$ mostly via $\Delta I^R_{23 \rightarrow 1}$ and $\Delta
I^S_{12 \rightarrow 1}$. Given that $Q_2$ and $Q_3$ are identical,
SURD assigns a nonzero value only to $\Delta I^S_{12 \rightarrow 1}$,
but not to $\Delta I^S_{13 \rightarrow 1}$, to prevent the duplication
of causality in compliance with Equation (\ref{eq:conservation_info}).

For the rest of the methods, CGC and CTE are unable to identify any
causal links between different variables. CCM identifies a
bidirectional causal connection between $Q_2$ and $Q_3$, but cannot
identify their influence to $Q_1$. Finally, PCMCI exhibits consistent
results for $Q_1$; however, it does not offer a mechanism to identify
the redundancy between $Q_2$ and $Q_3$.

\subsection*{Application to the energy cascade in turbulence}

\begin{figure}[t!]
    \centering
    
    \subfloat[]{
    \begin{minipage}{\tw}
        \centering
        \hspace*{0.03\textwidth}
        \includegraphics[width=0.19\textwidth, trim={3.6cm 0.4cm 4.7cm 0},clip]{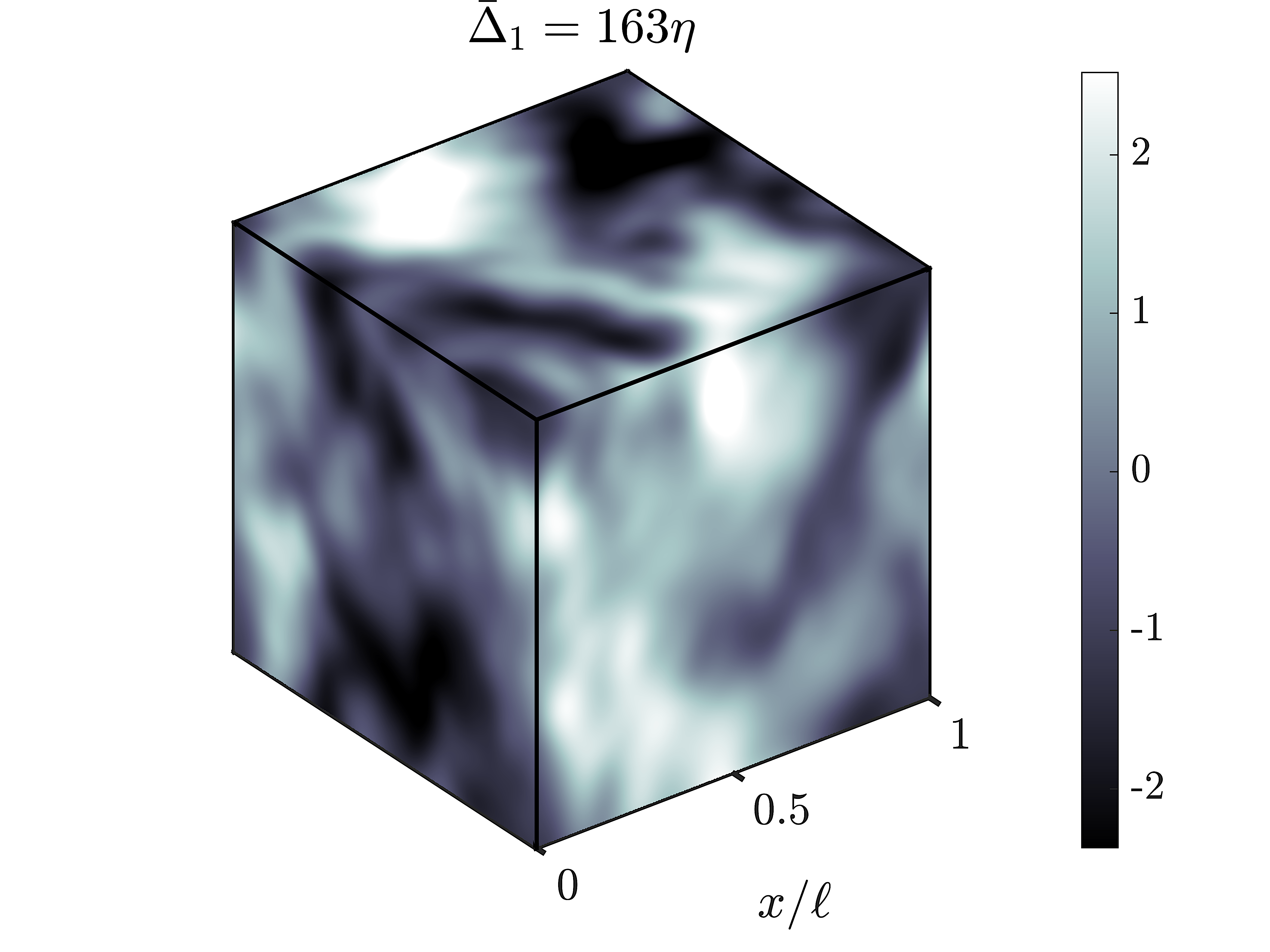}
        \begin{tikzpicture}[overlay, remember picture]
            \node[fill=white] at (-1.8,4.25) {$\bar{\Delta}_1=163\eta$};
            \node[fill=white] at (-0.7,0.005) {$x/\ell$};
            {\footnotesize
            \node[fill=white] at (-0.1,0.8) {$1$};
            \node[fill=white] at (-0.9,0.45) {$0.5$};
            \node[fill=white] at (-1.8,0.17) {$0$};
            }
        \end{tikzpicture}
        \hspace*{0.039\textwidth}
        \includegraphics[width=0.19\textwidth, trim={3.6cm 0.4cm 4.7cm 0},clip]{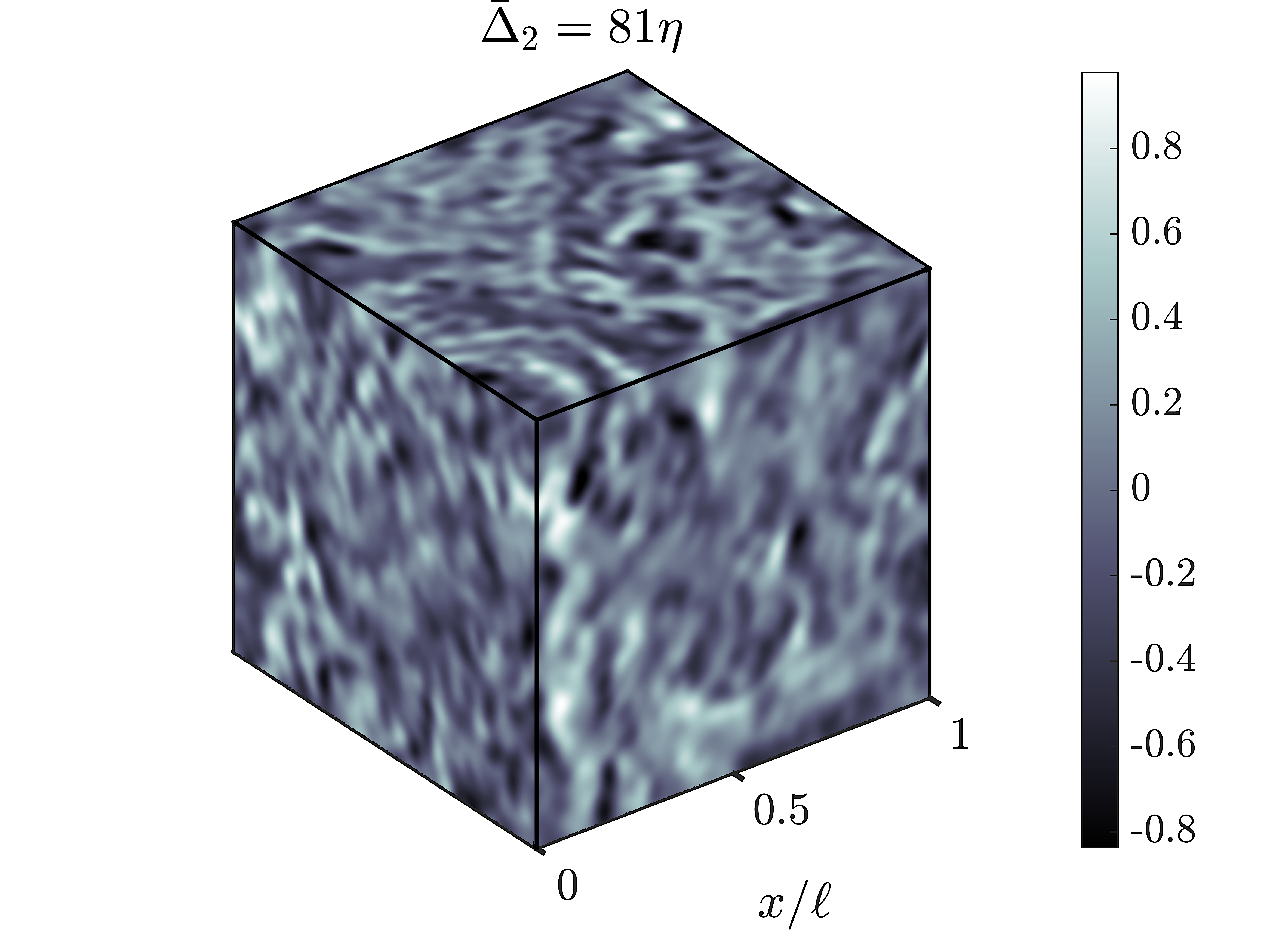}
        \begin{tikzpicture}[overlay, remember picture]
            \node[fill=white] at (-1.8,4.25) {$\bar{\Delta}_1=81\eta$};
            \node[fill=white] at (-0.7,0.005) {$x/\ell$};
            {\footnotesize
            \node[fill=white] at (-0.1,0.8) {$1$};
            \node[fill=white] at (-0.9,0.45) {$0.5$};
            \node[fill=white] at (-1.8,0.17) {$0$};
            }
        \end{tikzpicture}
        \hspace*{0.039\textwidth}
        \includegraphics[width=0.19\textwidth, trim={3.6cm 0.4cm 4.7cm 0},clip]{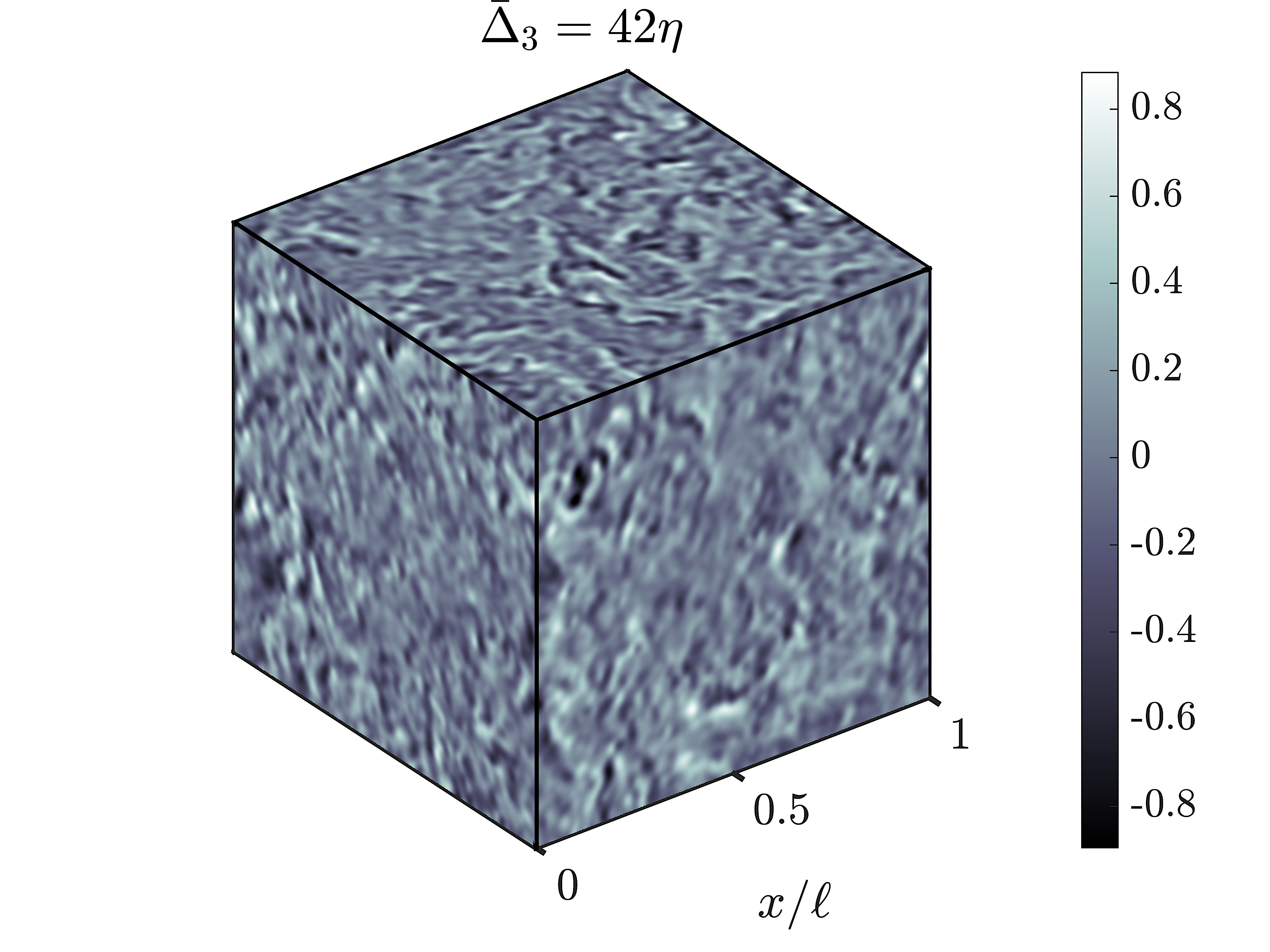}
        \begin{tikzpicture}[overlay, remember picture]
            \node[fill=white] at (-1.8,4.25) {$\bar{\Delta}_1=42\eta$};
            \node[fill=white] at (-0.7,0.005) {$x/\ell$};
            {\footnotesize
            \node[fill=white] at (-0.1,0.8) {$1$};
            \node[fill=white] at (-0.9,0.45) {$0.5$};
            \node[fill=white] at (-1.8,0.17) {$0$};
            }
        \end{tikzpicture}
        \hspace*{0.039\textwidth}
        \includegraphics[width=0.19\textwidth, trim={3.6cm 0.4cm 4.7cm 0},clip]{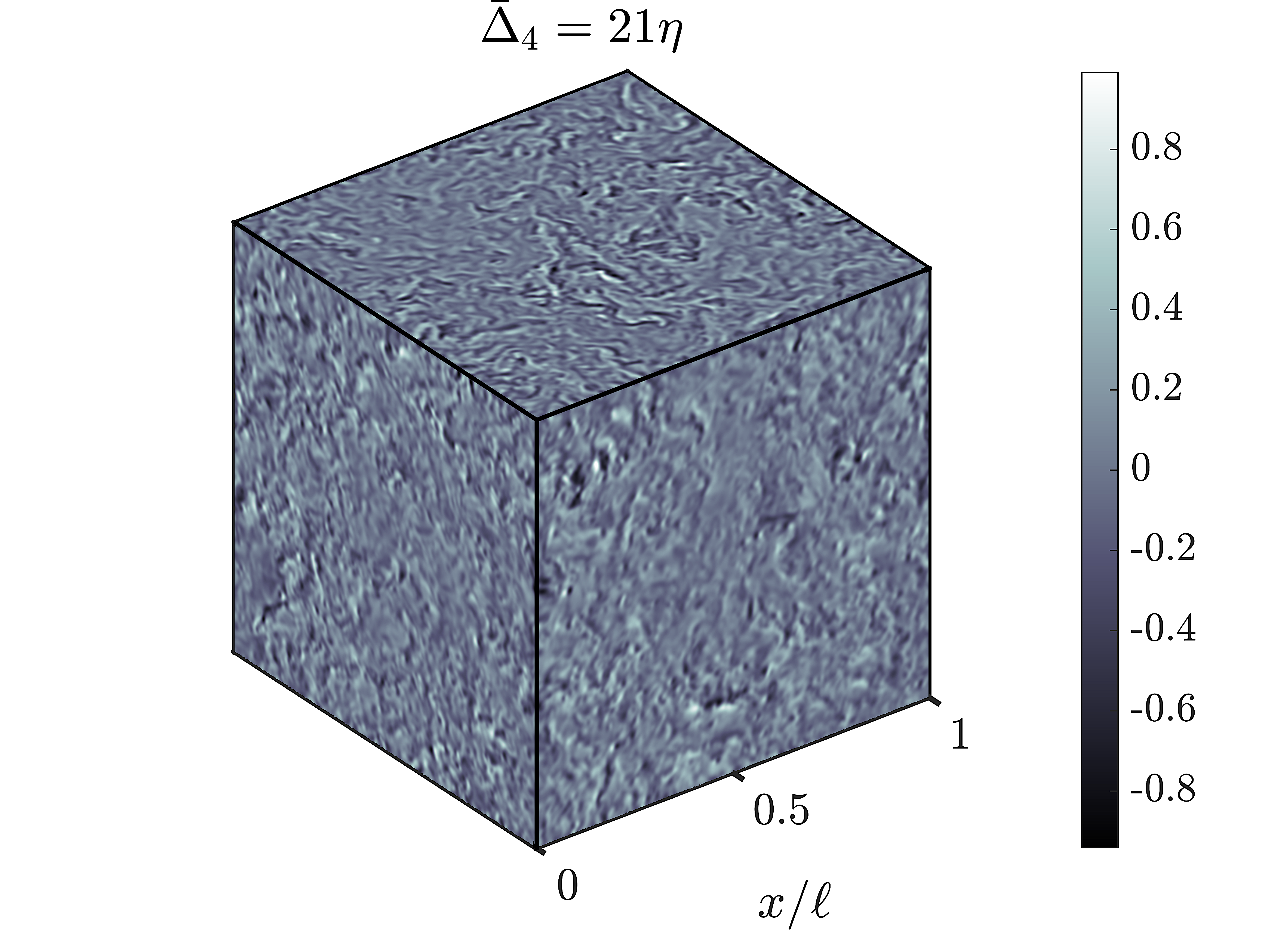}
        \begin{tikzpicture}[overlay, remember picture]
            \node[fill=white] at (-1.8,4.25) {$\bar{\Delta}_1=21\eta$};
            \node[fill=white] at (-0.7,0.005) {$x/\ell$};
            {\footnotesize
            \node[fill=white] at (-0.1,0.8) {$1$};
            \node[fill=white] at (-0.9,0.45) {$0.5$};
            \node[fill=white] at (-1.8,0.17) {$0$};
            }
        \end{tikzpicture}
        \vspace{0.15cm} 
    \end{minipage}
    }

    \vspace*{0.005\textwidth}
    \subfloat[]{
    \includegraphics[width=0.23\textwidth]{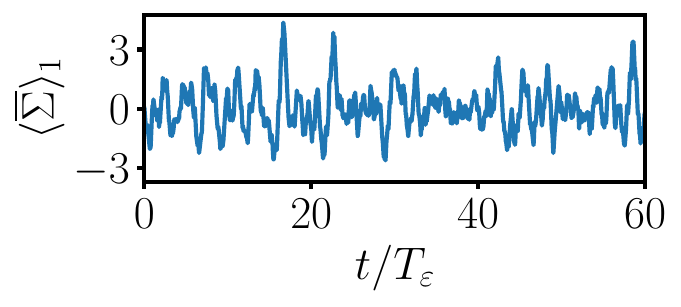}
    \hspace*{0.006\textwidth}
    \includegraphics[width=0.23\textwidth]{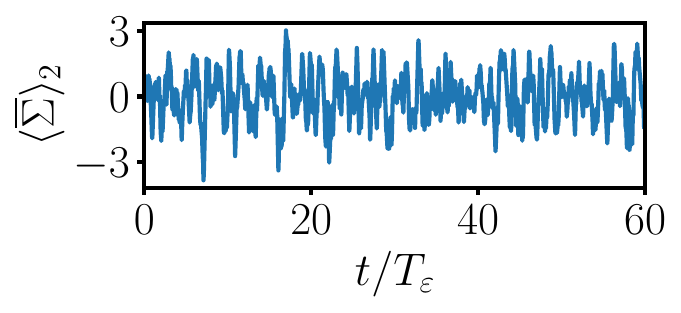}
    \hspace*{0.006\textwidth}
    \includegraphics[width=0.23\textwidth]{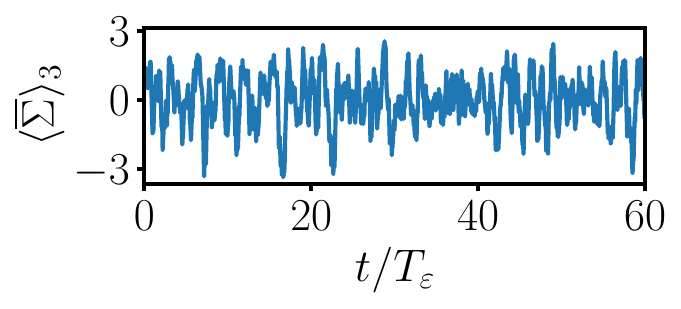}
    \hspace*{0.006\textwidth}
    \includegraphics[width=0.23\textwidth]{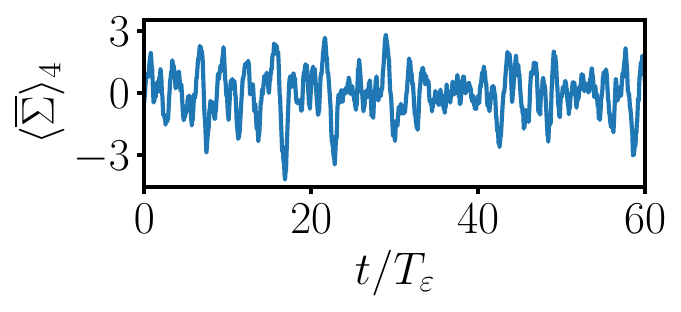}
    }

    \subfloat[]{
    \includegraphics[width=0.368\textwidth]{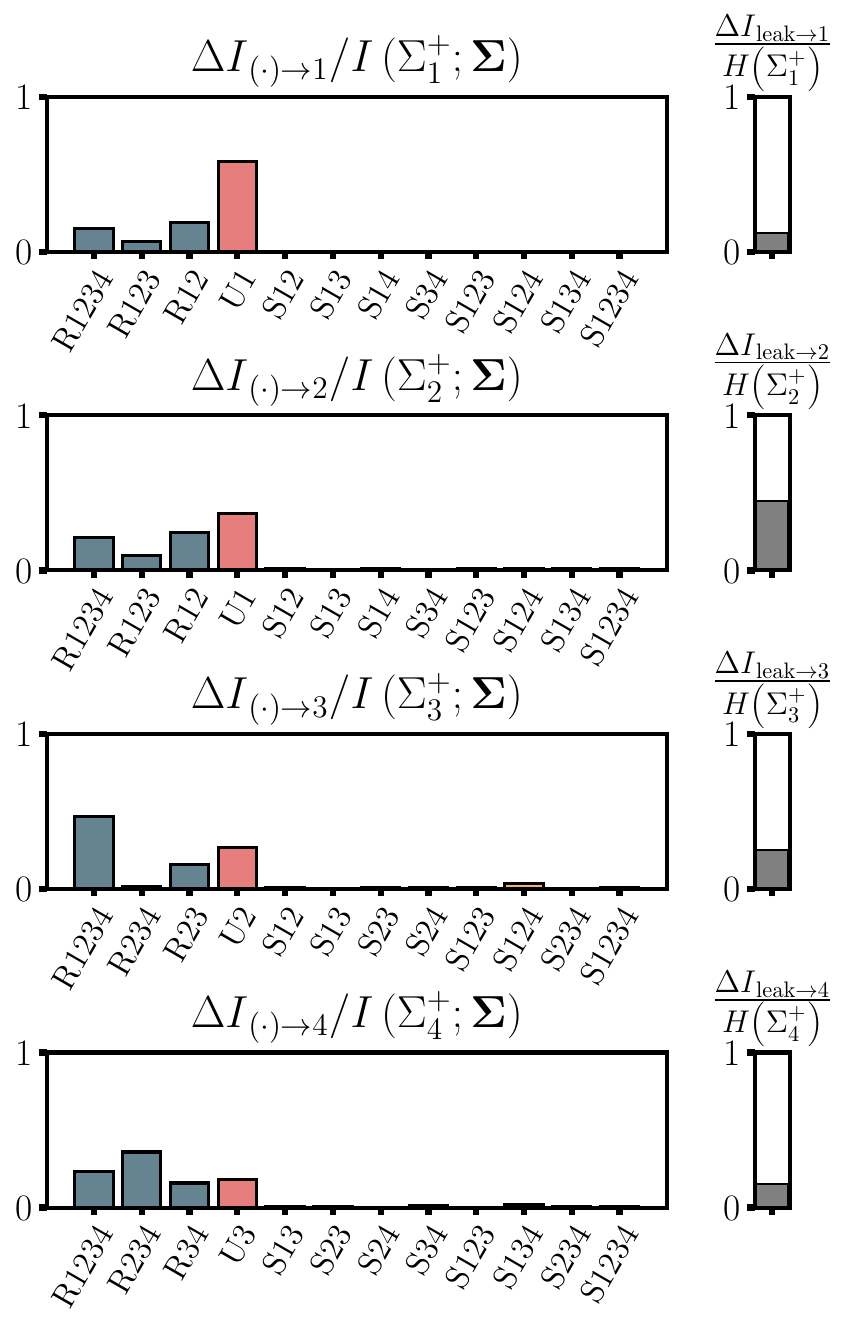}
    \hspace*{-0.01\textwidth}
    \begin{minipage}{0.142\textwidth}
        \vspace{-4.1\textwidth}
        {\includegraphics[width=\textwidth]{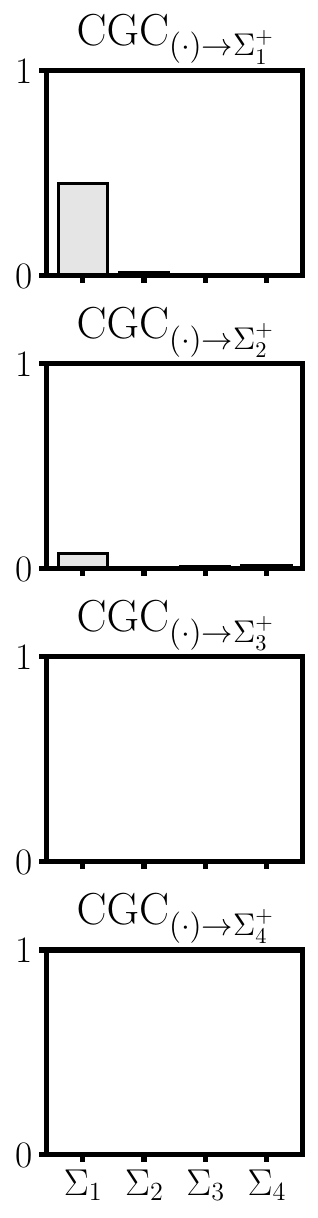}}
    \end{minipage}
    \hfill
    \begin{minipage}{0.142\textwidth}
        \vspace{-4.1\textwidth}
        {\includegraphics[width=\textwidth]{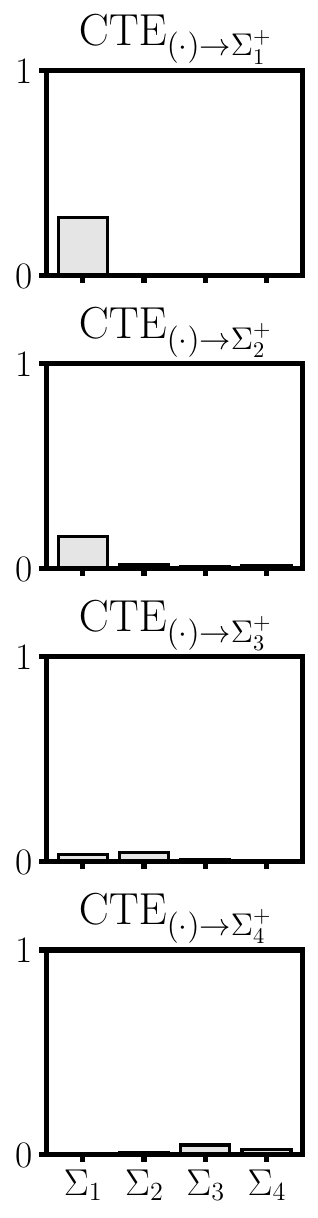}}
    \end{minipage}
    \hfill
    \begin{minipage}{0.142\textwidth}
        \vspace{-4.1\textwidth}
        {\includegraphics[width=\textwidth]{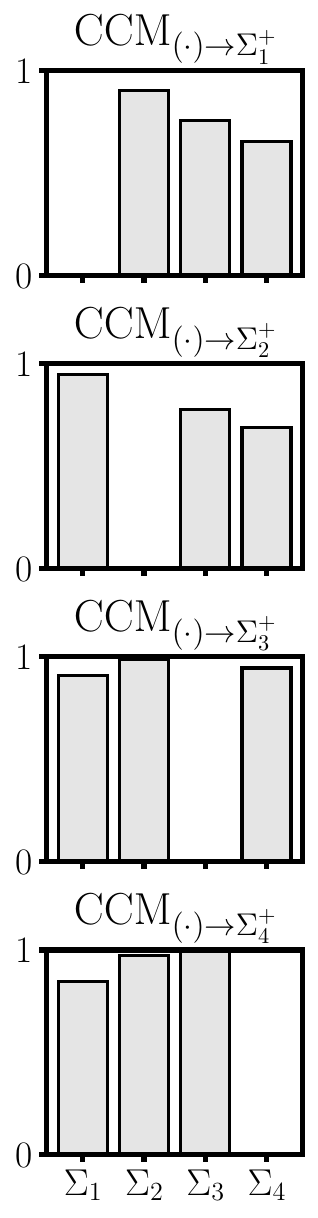}}
    \end{minipage}
    \begin{minipage}{0.154\textwidth}
        \vspace{-3.775\textwidth}
        {\includegraphics[width=\textwidth]{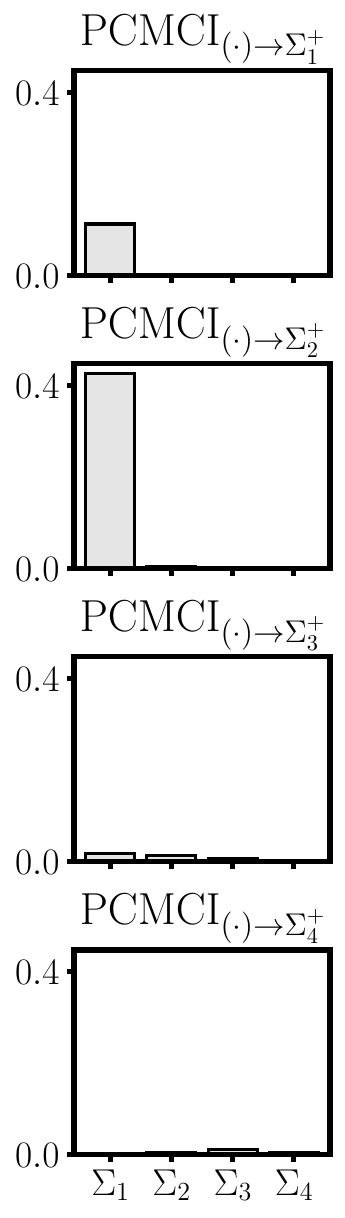}}
    \end{minipage}
    }
    \caption{\textbf{Causality in the turbulent energy cascade}. (a)
      Visualization of the magnitude of the velocity field for four
      filter sizes $\Delta_i$ at the same instant. The parameter
      $\eta$ is the Kolmogorov length-scale and represents the
      smallest scale in the flow.  (b) An extract of the time history
      of $\Sigma_1$, $\Sigma_2$, $\Sigma_3$ and $\Sigma_4$. The
      notation $\langle \overline{{\Sigma}}_i \rangle$ is used to
      denote standardization using the mean and standard deviation of
      the signals. The time is non-dimensionalized by $T_\epsilon$,
      which is the integral time-scale of the flow.  (c) Redundant
      (R), unique (U), and synergistic (S) causal contributions from
      SURD. Only the top $12$ contributions, satisfying the condition
      ${\Delta I}_{(\cdot)\rightarrow j} / I(\Sigma_j^+; \bs{\Sigma})
      \geq 10^{-3}$, are represented, where $\bs{\Sigma}=[\Sigma_1,
        \Sigma_2, \Sigma_3, \Sigma_4]$. The gray bar is the causality
      leak. The results of CGC, CTE, CCM, and PCMCI are depicted on
      the right.}
    \label{fig:cascade}
\end{figure}
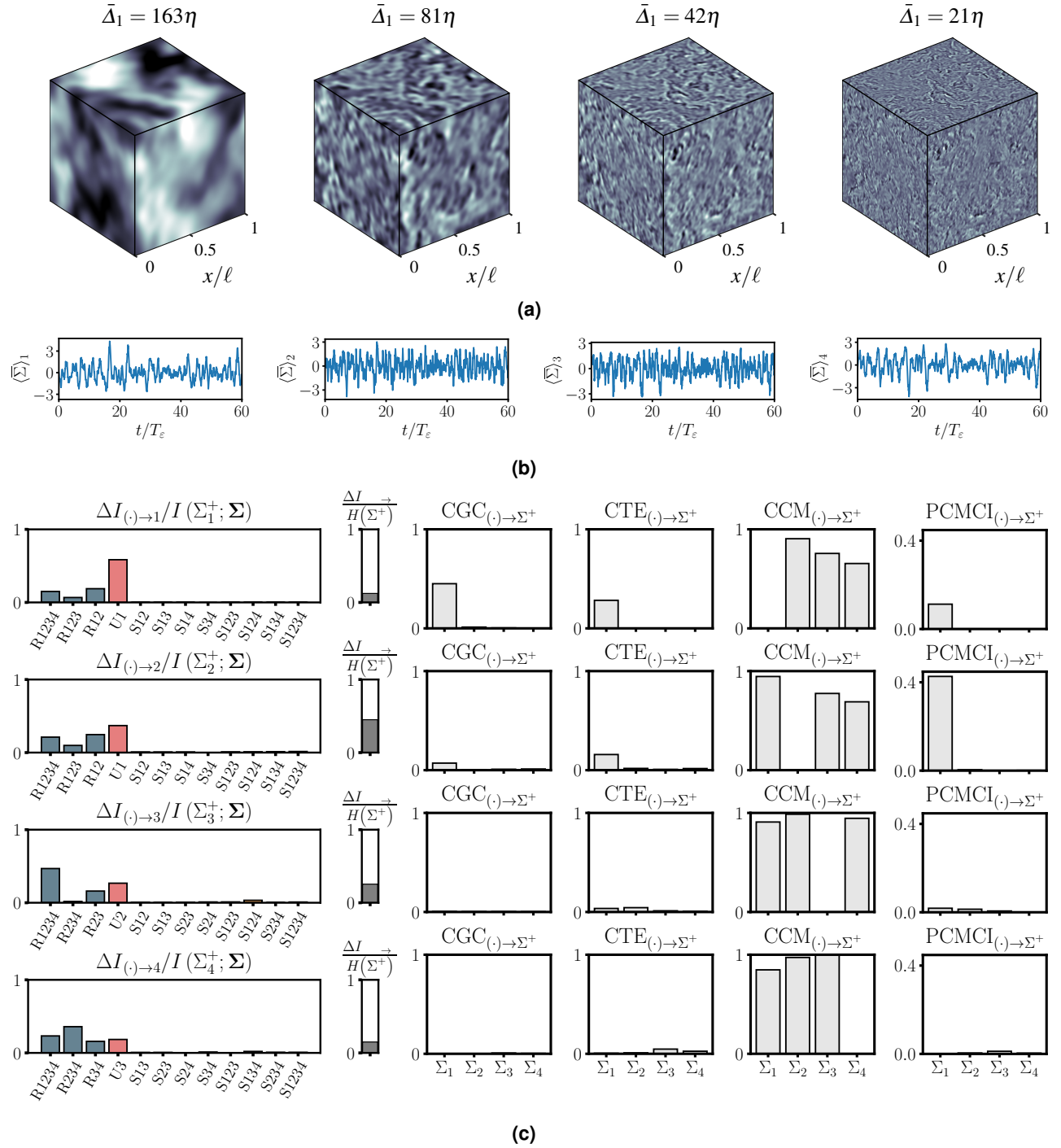

We apply SURD to investigate the causality of the energy cascade in
turbulence, which serves as a primary example of a chaotic,
multi-scale, high-dimensional system.  The energy cascade is the
transfer of kinetic energy from large to small scales in the flow
(forward cascade), or vice versa (backward cascade), and has been the
cornerstone of most theories and models of turbulence since the
1940s\cite{richardson1922, obukhov1941, kolmogorov1941,
  kolmogorov1962, aoyama2005, falkovich2009, cardesa2017}. However,
understanding the dynamics of the energy transfer across scales
remains an outstanding challenge. Given the ubiquity of turbulence, a
deeper understanding of the energy transfer among flow scales could
enable significant progress across various fields, ranging from plasma
physics\cite{Yamada2008}, combustion\cite{veynante2002},
climate\cite{bodenschatz2015}, and astrophysics\cite{young2017} to
engineering applications in aero/hydrodynamics\cite{sirovich1997,
  hof2010, marusic2010, kuhnen2018}. Despite the progress made in
recent decades, the causal interactions of energy among scales in
turbulent flows have received less attention. Here, we investigate the
redundant, unique, and synergistic causalities involved in the energy
transfer.

We use data from a high-fidelity simulation of isotropic turbulence in
a triply periodic domain\cite{cardesa2017}. This case is the testbed
used by the community to understand the fundamental physical processes
of the energy cascade. The total number of degrees of freedom of the
system is of the order of $10^9$. The kinetic energy transfer was
obtained by filtering the velocity field at four different length
scales (denoted by $\Delta_i$, for $i=1,2,3,4$) and calculating the
energy flux across those scales. Causality is computed among the time
signals of the volume-averaged energy transfer, denoted by $\Sigma_i$
for $i=1,2,3,4$, where the index signifies energy transfer at scale
$\Delta_i$. The top panels of Figure \ref{fig:cascade} depicts a
visualization of the filtered velocity from the largest to the
smallest flow scale, together with the time evolution of the energy
transfer signals at that scale. The causal relationships among energy
transfers identified by SURD are shown in the left panels of
Figure~\ref{fig:cascade}. The dominant contributions come from
redundant and unique causalities, whereas synergistic causalities play
a minor role. The unique causalities (depicted in red) vividly capture
the forward energy cascade of causality from large to smaller scales,
inferred from the non-zero terms $\Delta I^U_{1\rightarrow 2}$,
$\Delta I^U_{2\rightarrow 3}$, and $\Delta I^U_{3\rightarrow
  4}$. Curiously, no unique causality is observed from smaller to
larger scales, and any causality from the backward cascade arises
solely through redundant relationships. In the context of SURD, this
implies that no new information is conveyed from the smaller scales to
the larger scales, which is consistent with recent views of the
backward energy cascade in the literature\cite{Vela2021, Vela2022}.
\alvaro{From the modeling perspective, this justifies the success of
  subgrid-scale modeling in large-eddy simulation, as the information
  contained in the smaller scales is redundant and does not constitute
  a key ingredient in solving the closure model problem. The results
  obtained from SURD also provide support for classic hypotheses about
  the energy cascade from a new causal-effect perspective. Among them,
  we can cite Taylor's dissipation surrogate
  assumption~\cite{taylor1935} and the dissipation
  anomaly~\cite{frisch1995}. The former posits that the dissipation
  rate can be determined by large-scale dynamics, even if dissipation
  is formally a small-scale feature of the flow. SURD clearly supports
  this assumption due to the lack of unique and synergistic causality
  from small to large scales. The results from SURD are also
  consistent with the dissipation anomaly (i.e., the constant rate of
  energy dissipation despite decreasing viscosity), which is enabled
  by the forward directionality of the energy cascade process.}

SURD also provides the causality leak ($\Delta I_{\mathrm{leak}
  \rightarrow j}$) that measures the amount of causality unaccounted
for due to unobserved variables. The largest causality leak occurs for
$\Sigma_2$, where approximately 47\% of the causality is carried by
variables not included within $[\Sigma_1, \Sigma_2, \Sigma_3,
  \Sigma_4]$. This implies that there are other factors affecting
$\Sigma_2$ that have not been accounted for and that explain the
remaining 53\% of the causality of the variable. Conversely, the
energy transfer at the largest scale $\Sigma_1$ bears the smallest
leak of 14\%, which is due to the high value of the unique causality
$\Delta I^U_{1 \rightarrow 1}$. The latter implies that the future of
the largest scales is mostly determined by its own past.

Finally, the results from SURD are compared with the other
methods. CGC and CCM do not support the hypothesis of a forward energy
cascade, which disagrees with the consensus within the fluid dynamics
community~\cite{zhou1993a, aoyama2005, eyink2005, mininni2006,
  aluie2009, domaradzki2009, cardesa2017}. \alvaro{The formulation of
  CCM used in this study adheres to the original work by Sugihara et
  al.\cite{sugihara2012} However, more recent iterations of CCM, such
  as Extended CCM \cite{extendedccm2015}, which explicitly account for
  time delays, have demonstrated efficacy in accurately detecting
  causality in systems with strongly synchronized variables. Hence,
  these and other improved versions of CCM might be more suitable for
  analyzing the turbulent energy cascade, where smaller scales are
  enslaved to the larger ones.} CTE and PCMCI are consistent with the
forward propagation of energy, but the strength of the causal links
detected is extremely weak.  Beyond the failure of some methods to
support the forward energy cascade hypothesis, different approaches
also yield conflicting outcomes regarding the path followed by the
energy across scales and the significance of the backward energy
cascade.  Additionally, none of the other previous methods offer
quantification of missing causality due to unobserved variables, in
contrast to the causality leak provided by SURD.
%

\section*{{Application to experimental data from a turbulent boundary layer}}

\begin{figure}[t!]
    \centering
    \begin{minipage}{\tw}
    \centering
    \subfloat[]{
        \hspace{0.01\textwidth}
        \includegraphics[width=0.85\textwidth]{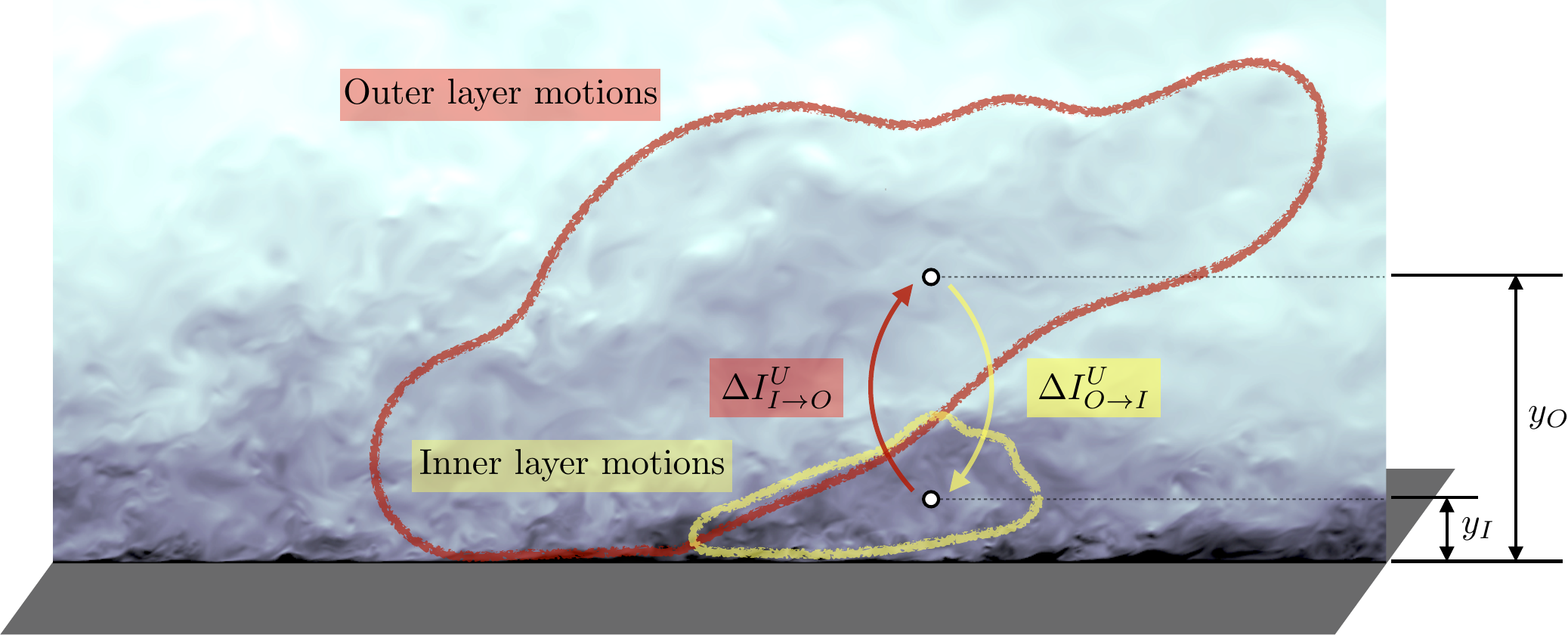}
    }
    \end{minipage}
    \begin{minipage}{\tw}
    \subfloat[]{
    \centering
    \begin{minipage}{0.28\tw}
    \vspace{-4.325cm}
        \includegraphics[width=\textwidth]{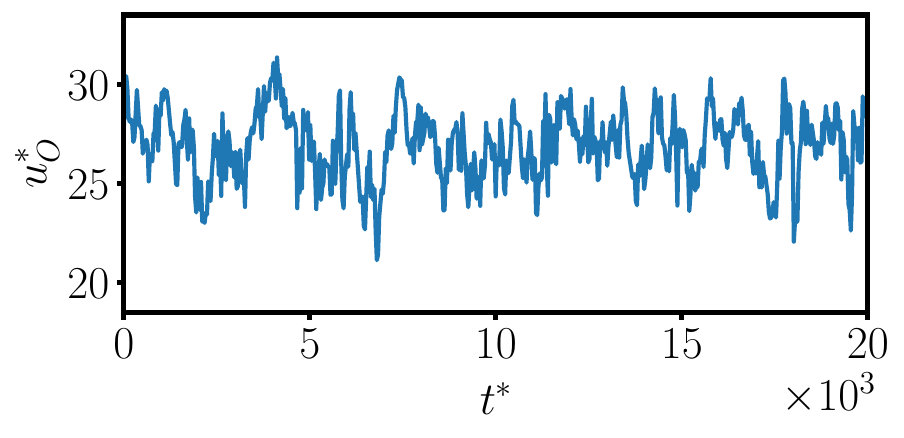}
        \includegraphics[width=\textwidth]{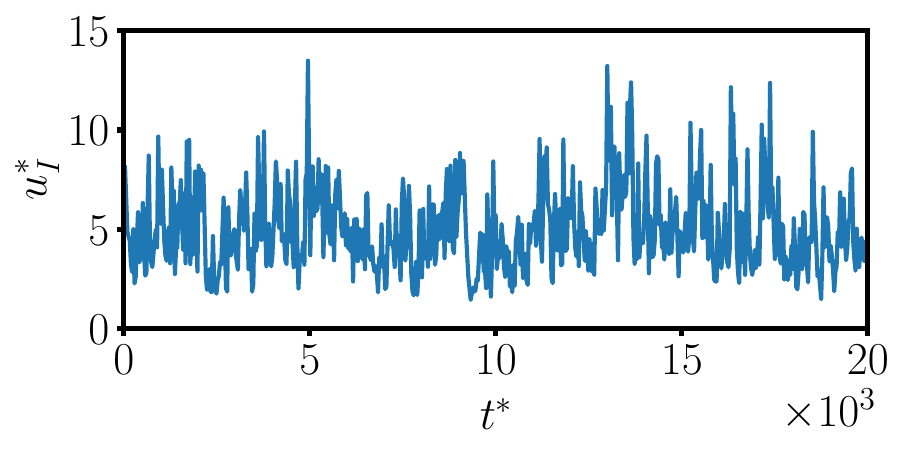}
    \end{minipage}
    \hspace{0.1cm}
    \includegraphics[height=0.305\textwidth]{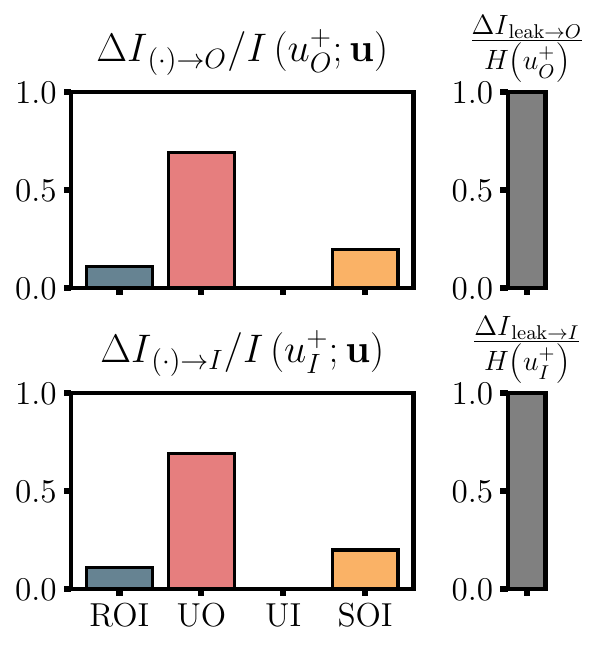}
    \includegraphics[height=0.29\textwidth]{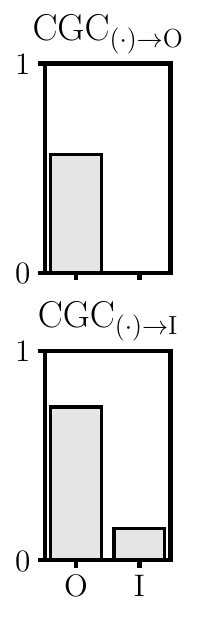}
    \includegraphics[height=0.29\textwidth]{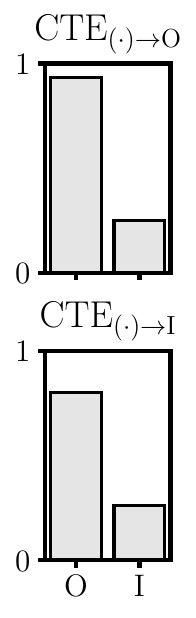}
    \includegraphics[height=0.29\textwidth]{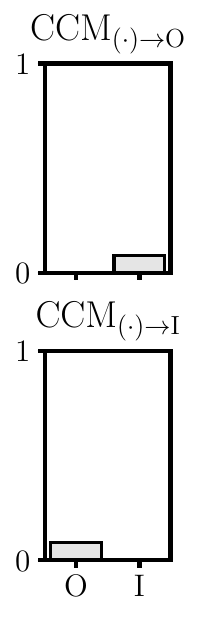}
    \includegraphics[height=0.29\textwidth]{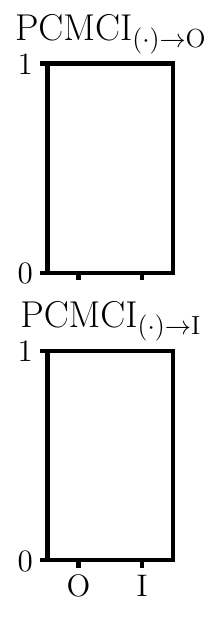}
    }
    \end{minipage}
    \caption{\textbf{Causality between streamwise velocity motions in
        a turbulent boundary layer}. (a) Schematic of outer-layer and
      inner-layer streamwise velocity motions in a turbulent boundary
      and their interactions via unique causality. The velocity
      signals $u_I(t)$ and $u_O(t)$ are experimentally measured at the
      wall-normal locations $y_I$ and $y_O$, respectively, and are
      shown in the panel below. The superscript $*$ denotes the inner
      scaling with friction velocity, $u_\tau$, and kinematic
      viscosity, $\nu$. (b) Redundant (R), unique (U), and synergistic
      (S) causalities among velocity signals in the inner (I) and
      outer (O) layer of a turbulent boundary layer. The gray bar is
      the causality leak. The results of CGC, CTE, CCM, and PCMCI are
      shown on the right. Details about data are provided in Methods.}
    \label{fig:inner-outer}
\end{figure}

\alvaro{The interaction of turbulent motions of different size within
  the thin fluid layers immediately adjacent to solid boundaries poses
  a significant challenge for both physical understanding and
  prediction. These layers are responsible for nearly 50\% of the
  aerodynamic drag on modern airliners and play a crucial role in the
  first hundred meters of the atmosphere, influencing broader
  meteorological phenomena\cite{marusic2010}.  Here, we leverage SURD
  to investigate the interaction between flow velocity motions in the
  outer layer (far from the wall) and inner layer (close to the wall)
  of a turbulent boundary layer. Figure \ref{fig:inner-outer}(a)
  illustrates the configuration used to examine the causal
  interactions between velocity motions. More specifically, the
  hypotheses under consideration are either i) a dominant influence of
  motions far from the wall on those closer, indicating top-down
  causality (a.k.a. Townsend's outer-layer similarity hypothesis
  \cite{townsend1976}), or ii) the opposite scenario, where influences
  move from areas closer to the wall outward, suggesting bottom-up
  causality. }

\alvaro{We use experimental data from a zero-pressure gradient
  turbulent boundary layer from the high Reynolds number wind tunnel
  at the University of Melbourne\cite{Baars2015, Baars2017,
    marusic2020}. The friction Reynolds number is $Re_{\tau} = u_\tau
  \delta / \nu = 14,750$, based on the thickness of the boundary layer
  $\delta$, the kinematic viscosity $\nu$, and the average friction
  velocity at the wall $u_\tau$. The time signals consists of the
  streamwise velocity at two wall-normal locations within the inner
  (I) and outer (O) layers, denoted by $u_I(t)$ and $u_O(t)$,
  respectively. }

\alvaro{Figure \ref{fig:inner-outer}(b) shows the redundant, unique,
  and synergistic causalities from SURD between the inner and outer
  layers. We use the subindices $I$ and $O$ to refer to causalities
  from/to $u_I(t)$ or $u_O(t)$, respectively. The primary observation
  is that the inner layer motions are predominantly influenced by the
  unique causality from the outer layer, $\Delta I^U_{O \rightarrow
    I}$. The redundant and synergistic causalities are lower, but they
  remain significant. Curiously, the unique causality $\Delta I^U_{I
    \rightarrow I}$ is zero, implying that, at the time scale
  considered, the inner layer motions are independent of their past
  history. For the outer-layer motions, most of the causality is
  self-induced $\Delta I^U_{O \rightarrow O}$ with no apparent
  influence from the inner layer. The results distinctly support the
  prevalence of top-down interactions: causality flows predominantly
  from the outer-layer large-scale motions to the inner-layer
  small-scale motions. The outcome is consistent with the modulation
  of near-wall scales by large-scale motions reported in previous
  investigations \cite{Hutchins2007, Mathis2009}. The lack of
  bottom-up causality from the inner to the outer layer also aligns
  with Townsend's outer-layer similarity hypothesis
  \cite{townsend1976} and previous observations in the literature
  \cite{flack2005, flores2006, busse2012, Mizuno2013, Chung2014,
    lozano2019x}.}

\alvaro{The causality leak, also shown in
  Fig. \ref{fig:inner-outer}(b), is 99\% for both $u_I$ and
  $u_O$. Such a high value implies that most of the causality
  determining the future of $u_I$ and $u_O$ is contained in other
  variables not considered in the analysis. This high value is
  unsurprising since most of the millions of degrees of freedom in the
  turbulent flow field have been neglected, and only two pointwise
  signals, $u_I$ and $u_O$, are retained to evaluate the causality.}

\alvaro{Finally, the results from SURD are contrasted with other
  methods. In this case, CCM and PCMCI do not support the hypothesis
  of top-down interactions between velocity motions. The reason behind
  the failure of these methods is unclear, but it might be related to
  the high causality leak. CGC and CTE are consistent with the flow of
  causality from the outer-layer large-scale motions to the
  inner-layer small-scale motions. However, as already highlighted in
  previous cases, none of these methods offer a detailed decomposition
  into redundant, unique, and synergistic causality, nor they account
  for the effect of unobserved variables as quantified by the
  causality leak in SURD.}

\section*{Discussion}


The cases presented in this study show that the faithful
quantification of causality remains elusive even in simple causal
networks.  The difficulties originate from a range of factors,
including complexities introduced by mediator, confounding and
collider effects; synergistic and duplicated variables; the influence
of stochastic noise; and the presence of unobserved variables. SURD
addresses these challenges by introducing several unique advances in
the field of causal inference that go beyond the insights provided by
other methods.

The first distinctive feature of SURD is its suitability for analyzing
causal networks involving mediator, confounder, and collider
effects---the building blocks of causal interactions between
variables. The success of SURD in capturing these fundamental
interactions stems from its ability to distinguish between redundant,
unique, and synergistic causalities, which is lacking in previous
methods. We have shown that the inability to disentangle redundant and
synergistic causalities can obscure the relationships among variables
within the system, leading to spurious causalities.  In the case of
PCMCI, incorporating a redundant variable into the set of conditioning
variables may lead to the identification of erroneous
links\cite{runge2019pcmci}.  The challenge posed by redundant
causality also extends to CTE\cite{verdes2005, lizier2008,
  barnett2009, lizier2010, cte2016}, as it evaluates the causality
between pairs of variables conditioned on the remaining set of
observed variables.  CGC encounters difficulties in the same
situations as CTE, since the former is a linear, parametric version of
the latter.
Several attempts have been made in the literature to account for
synergistic and redundant causality\cite{williams2010,griffith2014,
  griffith2015, ince2017, gutknecht2021, Lozano2022, kolchinsky2022},
for example, through the calculation of CTE in its multivariate
form\cite{Lozano2022}. However, these methods may yield negative
values of causality, which hinders the interpretability of the
results. In contrast, SURD ensures the non-negativity of all the
terms. This property significantly enhances the interpretability of
SURD terms, allowing for a clear distinction between redundant,
unique, and synergistic causality among variables.

SURD also introduces the concept of the causality leak, which
quantifies the extent of causality that remains unaccounted for due to
unobserved variables. The causality leak serves as a fundamental
metric to evaluate the significance of the causal links
identified. Low values of the causality leak imply that most of the
causality is accounted for by the observed variables. Conversely, high
values of the causality leak indicate that most of the causality can
be attributed to hidden, exogenous variables. In such situations, SURD
highlights the necessity of incorporating additional, currently
overlooked variables into the analysis. The capability to detect and
quantify missing causality is absent in other methods for causal
inference.

The foundational principle of SURD from
Equation~(\ref{eq:conservation_info}), along with the non-negativity
of causalities, also provides a natural normalization for causality
that is both intuitive and easily interpretable. Unique, redundant and
synergistic causalities are normalized using the mutual information
between all observed variables and the target, ensuring that their sum
equals 1. This normalization measures the relative importance of each
causality within the group of observed variables. Additionally, the
causality leak is naturally normalized by the information content of
the target variable, which bounds its value between 0 and 1. For
example, in a system with three variables $[Q_1, Q_2, Q_3]$ where we
are interested in the causes of $Q_2$, a normalized unique causality
from $Q_1 \rightarrow Q_2$ equal to 90\% implies that $Q_2$ and $Q_3$
play a minor role (i.e., 10\%) in influencing the future of
$Q_2$. However, the causality from $Q_1 \rightarrow Q_2$ would still
be deemed insignificant if the causality leak of $Q_2$ is 99\%,
indicating that most of the causality to $Q_2$ resides in other
variables not contained in the vector $[Q_1, Q_2, Q_3]$. Other
methods, such as CGC, do not offer bounded values nor a measure of
causality leak. Outcomes from PCMCI (when based on correlations) and
CCM can be normalized between 0 and 1; however, unlike SURD, the sum
of causalities in PCMCI and CCM does not add up to 1 or to any
conserved quantity. CTE shares many normalization properties with SURD
and allows for the concept of causality leak; however, the possibility
for negative values of causality in CTE complicates its
interpretation\cite{Lozano2022}.

Another essential aspect of SURD is its foundation in transitional
probability distributions, which ensures its invariance to
transformations such as shifting, rescaling, and other general
invertible transformation of the variables. This is applicable to
other methods such as CTE and PCMCI.
SURD is also robust across scenarios with varying amounts of samples
for causal inference, providing consistent causal links with fewer
than a thousand samples. This capability is realized through the
application of transport maps specifically designed for estimating
high-dimensional conditional probability
distributions\cite{baptista2023}. Even in situations where the sample
size is exceptionally small (e.g., of the order of hundreds) and the
number of variables is large, SURD is still capable of identifying
causal relationships up to a certain order of synergies [see Methods],
while higher-order synergistic interactions can be considered part of
the causality leak. Finally, methods such as CCM suffer from the
presence of increased noise, which complicates the reconstruction of
attractor manifolds and reduces the efficacy\cite{cobey2016,
  monster2017, runge2018chaos}. In contrast, our test cases have shown
that SURD is reliable even in the presence of noise. In summary, SURD stands as an effective tool in the field of causal
inference with the potential to drive progress across multiple
scientific and engineering domains, such as climate research,
neuroscience, economics, epidemiology, social sciences, and fluid
dynamics, among others.

\section*{Methods}

\subsection*{Assumptions for causal discovery}




SURD is an observational non-intrusive method that operates within a
probabilistic framework, where causal relationships emerge as a result
of transitional probabilities between states. The method adheres to
the principle of forward-in-time propagation of information, which
states that causation cannot occur backward in time. This formulation
is consistent with the identification of contemporaneous links, as
these can be interpreted as causal influences acting on a time scale
shorter than the measurement interval (e.g. inferring causality on an
eight-hour scale from daily measured data). Furthermore, the method
incorporates the concept of causality leak, which acts as a mechanism
of quality control by assessing the impact of unobserved variables.
This measure alleviates the need for the assumption of causal
sufficiency (i.e., all common causes of the variables must be
accurately measured), as it offers a quantifiable measure of the
extent to which information from unobserved variables remains
unaccounted for. Additionally, the method is model-free, i.e. no prior
knowledge about the system dynamics is required. This makes SURD
appealing for applications involving deterministic or stochastic
multivariate systems with linear and nonlinear dependencies. The
method also assumes that the time signals are stationary, which
ensures that their statistical properties do not vary over time.
Finally, the method can identify cyclic causal relationships, provided
that they adhere to the principle of forward-in-time propagation of
information.

\subsection*{Synergystic-Unique-Redundant Decomposition}

To perform the decomposition proposed in Equation
(\ref{eq:conservation_info}), we rely on the concept of mutual
information\cite{shannon1948, kullback1951, Kreer1957} between the
target variable $Q_j^+$ and the vector of observed variables
$\bQ$. This quantity can be mathematically described as:
\begin{equation}
  \label{eq:mutual_info}
  I(Q_j^+;\bQ) = \sum_{q_j^+,\bq} p(q_j^+,\bq)
  \log_2\left( \frac{p(q_j^+,\bq)}{p(q_j^+)p(\bq)}\right),
\end{equation}
where $q_j^+$ and $\bq$ represent all possible values or states of
$Q_j^+$ and $\bQ$, respectively. Mutual information measures how
different the joint probability distribution $p(q_j^+,\bq)$ is from
the hypothetical distribution $p(q_j^+)p(\bq)$, where $q_j^+$ and
$\bq$ are assumed to be independent. For instance, if $Q_j^+$ and
$\bQ$ are not independent, then $p(q_j^+,\bq)$ will differ
significantly from $p(q_j^+)p(\bq)$. Hence, we assess causality by
examining how the probability of $Q_j^+$ changes when accounting for
$\bQ$.

However, the source of causality might change depending on different
states $q_j^+$ of the target variable $Q_j^+$. For example, $Q_1$ can
only be causal to positive values of $Q_j^+$, whereas $Q_2$ can only
be causal to negative values of $Q_j^+$. Therefore, this decomposition
must be performed for all possible values of $Q_j^+$. To do that, we
define the \emph{specific} mutual information\cite{DeWeese1999} from
$\bQ$ to a particular event $Q_j^+=q_j^+$ as
\begin{equation}
  \label{eq:specific_mutual_info}
   \Is(q_j^+;\bQ) = \sum_{\bq} \frac{p(q_j^+,\bq)}{p(q_j^+)}
   \log_2\left( \frac{p(q_j^+,\bq)}{p(q_j^+)p(\bq)} \right) \geq 0.
\end{equation}
Note that the {specific} mutual information is a function of the
random variable $\bQ$ (which encompasses all its states) but only a
function of one particular state of the target variable (namely,
$q_j^+$). Similarly to Equation (\ref{eq:mutual_info}), the specific
mutual information quantifies the dissimilarity between $p(q_j^+)$ and
$p(q_j^+|\bq)$ but in this case for the particular state
$Q_j^+=q_j^+$. The mutual information between $Q_j^+$ and $\bQ$ is
recovered by $I(Q_j^+;\bQ) = \sum_{q_j^+} p(q_j^+)
\Is(q_j^+;\bQ)$. For simplicity, we will use $\Is_\bi (q_j^+) =
\Is(q_j^+;\bQ_\bi)$.

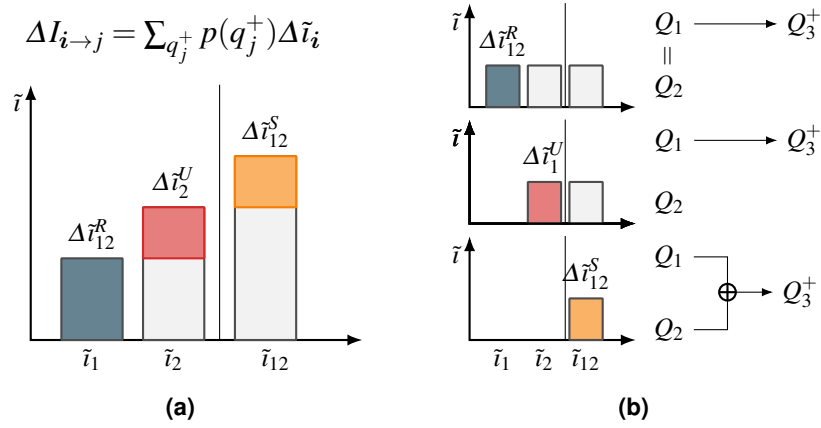
\begin{figure}[t!]
    \centering
    \subfloat[]{





\begin{tikzpicture}[
 			bar/.style={thick,black!70,fill=black!5,
            rounded corners=.1mm},
 			barline/.style={thick,black!70,rounded corners=.1mm},
 			myfill/.style={thick,rounded corners=.1mm},
            >={Latex[length=.2cm]},
            scale=1.35
    ]
    
    \begin{normalsize}


    \def\ths{{.8,1.3,1.8}}
    \def\lbs{{1,2,12}}


    \pgfmathsetmacro{\ypaneltwo}{0}


    \draw[thick,<->] (-.6,2.5) node[anchor=north east] {$\Is$} -- (-.6,0) --+ (3.25,0);

    \draw[barline,fill=myc1!60] (-.3,\ypaneltwo) --++ (.6,0) 
        --++ (0,\ths[0])  --++ (-.6,0) -- cycle; 
    \pgfmathsetmacro{\MyPgfMathResult}{{\lbs[0]}}
    \node[anchor=north] at (0,\ypaneltwo) {$\Is_{\MyPgfMathResult}$}; 

    \foreach \y in {1} {
        \draw[bar] (\y-.5,\ypaneltwo) --++ (.6,0) 
        --++ (0,\ths[\y])  --++ (-.6,0) 
        -- cycle; 
        \pgfmathsetmacro{\MyPgfMathResult}{{\lbs[\y]}}
        \node[anchor=north] at (\y-0.2,\ypaneltwo) {$\Is_{\MyPgfMathResult}$}; 
        }

    \foreach \y in {2} {
        \draw[bar] (\y-.6,\ypaneltwo) --++ (.6,0) 
        --++ (0,\ths[\y])  --++ (-.6,0) 
        -- cycle; 
        \pgfmathsetmacro{\MyPgfMathResult}{{\lbs[\y]}}
        \node[anchor=north] at (\y-0.2,\ypaneltwo) {$\Is_{\MyPgfMathResult}$}; 
        }



    \draw[myfill,myc2,fill=myc2!60] (1-.5,\ths[0]+\ypaneltwo) --++ (.6,0) --++ 
    (0,\ths[1]-\ths[0])  --++ (-.6,0) -- cycle;

    \foreach \y in {2} {
        \pgfmathparse{\ths[\y] - \ths[0] > 0.0}
        \ifnum \pgfmathresult=1
            \pgfmathsetmacro{\MyPgfMathResult}{{\lbs[\y]}}
            \node[anchor=south] at (\y-0.3,\ths[\y]+\ypaneltwo) {$\Delta \Is^S_{\MyPgfMathResult}$};     
            \pgfmathparse{\ths[\y-1] - \ths[0] > 0.0}
            \ifnum \pgfmathresult=1
                \draw[myfill,myc3,fill=myc3!60] (\y-.6,\ths[\y-1]+\ypaneltwo) --++ (.6,0) 
                --++ (0,\ths[\y]-\ths[\y-1])  --++ (-.6,0) 
                -- cycle;
            \else
                \draw[myfill,myc3,fill=myc3!60] (\y-.6,\ths[0]+\ypaneltwo) --++ (.6,0) 
                --++ (0,\ths[\y]-\ths[0])  --++ (-.6,0) 
                -- cycle;
            \fi
        \fi
        }

    \node[anchor=south] at (0,\ths[0]+\ypaneltwo) {$\Delta \Is^R_{12}$};
    \node[anchor=south] at (0.8,\ths[1]+\ypaneltwo) {$\Delta \Is^U_{2}$};

    \begin{large}
    \node[anchor=south] at (0.8,\ths[1]+\ypaneltwo+1.3) {$\Delta I_{\bi \rightarrow j} = \sum_{q_j^+} 
    p(q_j^+) \Delta \Is_{\bi}$};
    \end{large}
	
    \draw[thin] ( 1+.25,0+\ypaneltwo) --++ (0,2.5);
    
    \end{normalsize}

\end{tikzpicture}

}
    \hspace{0.05\tw} \subfloat[]{\begin{tikzpicture}[
 			bar/.style={thick,black!70,fill=black!5,
            rounded corners=.1mm},
 			barline/.style={thick,black!70,rounded corners=.1mm},
 			myfill/.style={thick,rounded corners=.1mm},
            >={Latex[length=.2cm]},
            scale=0.55
    ]
    


    \def\ths{{1.,1.,1.}}
    \def\lbs{{1,2,12}}


    \pgfmathsetmacro{\ypaneltwo}{+2.8}

    \draw[thick,<->] (-.8,2.5+\ypaneltwo) node[anchor=north east] {$\Is$} -- (-.8,\ypaneltwo) --+ (4,0);

    \draw[thick,<->] (-.8,2.5) node[anchor=north east] {$ $} -- (-.8,0) --+ (4,0);

    \draw[barline,fill=myc1!60] (-.4,\ypaneltwo) --++ (.8,0) 
        --++ (0,\ths[0])  --++ (-.8,0) -- cycle; 
    \pgfmathsetmacro{\MyPgfMathResult}{{\lbs[0]}}

    \foreach \y in {1,...,2} {
        \draw[bar] (\y-.4,\ypaneltwo) --++ (.8,0) 
        --++ (0,\ths[\y])  --++ (-.8,0) 
        -- cycle; 
        \pgfmathsetmacro{\MyPgfMathResult}{{\lbs[\y]}}
        }


    \node[anchor=south] at (0,\ths[0]+\ypaneltwo) {$\Delta \Is^R_{12}$};

    \draw[->,>={Latex[length=.15cm]}] (4.6,\ths[1]+\ypaneltwo+1) node[anchor=east] {$Q_1$} --++ (2,0) node[anchor=west] {$Q_3^+$};
    
    \node[anchor=south] at (4,\ths[1]+\ypaneltwo-0.25) {\rotatebox{90}{$=$}};
    \node[anchor=south] at (4,\ths[1]+\ypaneltwo-1) {$Q_2$};
    
    \draw[thin] ( 1+.5,0+\ypaneltwo) --++ (0,2.5);

    \pgfmathsetmacro{\ypaneltwo}{0}

    \draw[thick,<->] (-.8,2.5+\ypaneltwo) node[anchor=north east] {$\Is$} -- (-.8,\ypaneltwo) --+ (4,0);

    \draw[thick,<->] (-.8,2.5) node[anchor=north east] {$\Is$} -- (-.8,0) --+ (4,0);

    \draw[barline,fill=myc2!60] (1-.4,\ypaneltwo) --++ (.8,0) 
        --++ (0,\ths[1])  --++ (-.8,0) -- cycle; 
    \pgfmathsetmacro{\MyPgfMathResult}{{\lbs[0]}}

    \foreach \y in {2,...,2} {
        \draw[bar] (\y-.4,\ypaneltwo) --++ (.8,0) 
        --++ (0,\ths[\y])  --++ (-.8,0) 
        -- cycle; 
        \pgfmathsetmacro{\MyPgfMathResult}{{\lbs[\y]}}
        }

    \node[anchor=south] at (1,\ths[1]+\ypaneltwo) {$\Delta \Is^U_{1}$};

    \draw[->,>={Latex[length=.15cm]}] (4.6,\ths[1]+\ypaneltwo+1) node[anchor=east] {$Q_1$} --++ (2,0) node[anchor=west] {$Q_3^+$};
    \node[anchor=south] at (4,\ths[1]+\ypaneltwo-1) {$Q_2$};
    
	
    \draw[thin] ( 1+.5,0+\ypaneltwo) --++ (0,2.5);
    

    \pgfmathsetmacro{\ypaneltwo}{-2.8}

    \draw[thick,<->] (-.8,2.5+\ypaneltwo) node[anchor=north east] {$\Is$} -- (-.8,\ypaneltwo) --+ (4,0);

    \draw[thick,<->] (-.8,2.5) node[anchor=north east] {$\Is$} -- (-.8,0) --+ (4,0);

    \draw[barline,fill=myc3!60] (2-.4,\ypaneltwo) --++ (.8,0) 
        --++ (0,\ths[0])  --++ (-.8,0) -- cycle; 
    \pgfmathsetmacro{\MyPgfMathResult}{{\lbs[0]}}
    \node[anchor=north] at (0,\ypaneltwo) {$\Is_{\MyPgfMathResult}$}; 

    \foreach \y in {1,...,2} {
        \pgfmathsetmacro{\MyPgfMathResult}{{\lbs[\y]}}
        \node[anchor=north] at (\y,\ypaneltwo) {$\Is_{\MyPgfMathResult}$}; 
        }

    \node[anchor=south] at (2,\ths[0]+\ypaneltwo) {$\Delta \Is^S_{12}$};

    \node[anchor=south] (q1) at (4,\ths[1]+\ypaneltwo+0.5) {$Q_1$};
    \node[anchor=south] (q2) at (4,\ths[1]+\ypaneltwo-1.25) {$Q_2$};
    
    \coordinate (circ) at (5.4,\ths[1]+\ypaneltwo+0.15);
    \draw (q1.east) -| (circ) |- (q2) -- (q2.east);
    
    \draw[->,>={Latex[length=.15cm]}] (circ) --+ (1.1,0) node[anchor=west,pos=1] {$Q_3^+$};
            
    \draw[fill=white,anchor=south,thick] (circ) circle (0.2);
    \draw[thick] ($(circ)-(90:.2)$) --++ (90:.4);
    \draw[thick] ($(circ)-(0:.2)$) --++ (0:.4);
	
    \draw[thin] ( 1+.5,0+\ypaneltwo) --++ (0,2.5);
    

\end{tikzpicture}}
    \caption{\textbf{Schematic of the causal decomposition in SURD}.
    (a) For a given state $q_j^+$
    of the target variable $Q_j^+$, the specific causalities $\Is$
    are computed for all the possible combinations of past
    variables. These components are organized in ascending order,
    which allows to assign the redundant (blue), unique (red), and
    synergistic (yellow) causalities. This process is performed for
    all possible values $q_j^+$ of $Q_j^+$. (b) Schematic of simple
    examples and associated specific mutual information for (top
    panel) duplicated input, (middle panel) output equal to first
    input, and (bottom panel) exclusive-OR output.
    }
    \label{fig:method_increments}
\end{figure}

To perform the decomposition of the specific mutual information in its
redundant $\Delta \Is_{\bi}^R$, unique $\Delta \Is_{i}^U$, and
synergistic $\Delta \Is_{\bi}^S$ components, we quantify the
increments in specific information $\Delta \Is$ about $q_j^+$ obtained
by observing an individual or groups of components from $\bQ$. 
 For a given state $q_j^+$
    of the target variable $Q_j^+$, the specific causalities $\Is$
    are computed for all the possible combinations of past
    variables. These components are organized in ascending order,
    which allows to assign the redundant, unique, and
    synergistic  causalities.
Figure
\ref{fig:method_increments} shows an example of the decomposition of
$\Is(q_j^+;\bQ)$ for a particular state of the target variable and for
the three simple examples illustrated in Figure \ref{fig:method}. The
quantities in Equation (\ref{eq:conservation_info}) are then obtained
as the expectation of their corresponding values:
\begin{equation}
    \Delta I_{\bi \rightarrow j}^R = \sum_{q_j^+} p(q_j^+) \Delta \Is_{\bi}^R(q_j^+),
    \quad\quad\quad
    \Delta I_{i \rightarrow j}^U = \sum_{q_j^+} p(q_j^+) \Delta \Is_{i}^U(q_j^+),
    \quad\quad\quad
    \Delta I_{\bi \rightarrow j}^S = \sum_{q_j^+} p(q_j^+) \Delta \Is_{\bi}^S(q_j^+).
\end{equation}

\subsection*{Comparison with other causality methods}


We compared SURD with other established methods for causal inference
in time series: Conditional Granger Causality (CGC)\cite{geweke1984},
Conditional Transfer Entropy (CTE)\cite{barnett2009}, Convergent
Cross-Mapping (CCM)\cite{sugihara2012}, and Peter and Clark Momentary
Conditional Independence (PCMCI)\cite{runge2019pcmci}. The detailed
calculation of the previous methods has been documented in the
literature, and their corresponding source codes have been
employed\cite{ding2006, jidt, causalccm, runge2019pcmci}.
In this paper, an embedding dimension equivalent to the number of
variables was used for the application of the CCM method, which was
executed with a library size that ensured convergence of the
prediction skill for all cases. Therefore, the CCM value should be
close to 1 in order to consider a link significantly causal.
We used the version of PCMCI based on conditional mutual information
(CMI) independence test with the estimator $k$-nearest neighbor
($k$-NN)\cite{runge2018cmi}. All cases were evaluated at a
significance level of $1\%$. Furthermore, PCMCI was estimated with
$\alpha_{\rm{PC}} = 0.05$ and CMI-$k$NN parameters $k_{\rm{CMI}} =
0.1$, $k_{\rm{perm}} = 5$ and $B=200$ permutation surrogates. The same
time lag was used for all the methods. The reader is referred to the
Supplementary Materials for a more detailed discussion of the packages
used, the validation with test cases provided in each of the sources,
and the effect of convergence for the CCM method. A summary of the
results for PCMCI using different independence tests is also provided
in the Supplementary Materials, where the optimal confidence interval
$\alpha_{\rm{PC}}$ during the initial condition selection phase (PC
phase) was selected based on the Akaike Information
criterion~\cite{akaike2011} from a default list of values, i.e.,
$\alpha_{\rm{PC}} = [0.05, 0.1, 0.2, 0.3, 0.4, 0.5]$.

\subsection*{Validation data for mediator, confounder, and collider}

The mediator, confounder, and collider systems considered comprise
three variables $Q_1(t_n)$, $Q_2(t_n)$, and $Q_3(t_n)$ at discrete
times $t_n = n$. The system is initially set to $Q_1(1) = Q_2(1) =
Q_3(1) = 0$. A stochastic forcing, represented by $W_i(t_n)$, acts on
$Q_i(t_n)$ and follows a Gaussian distribution with a mean of zero and
a standard deviation of one. The computation of SURD is performed for
a time lag of $\Delta T=1$ using 100 uniform bins per variable. The
integration of the system is carried out over $10^8$ time steps, with
the first 10,000 steps excluded from the analysis to avoid transient
effects. CGC and CTE used the same samples as SURD, while CCM and
PCMCI used $5\times10^5$ samples due to computational constraints. The
latter methods were also evaluated using a smaller number of samples,
which were one order of magnitude lower, and no significant
differences were detected. An analysis of the impact of the number of
samples and the sensitivity to partition refinement for SURD is
provided in the Supplementary Materials.




\subsection*{Data for energy cascade in isotropic turbulence}

The case chosen to study the energy cascade is forced isotropic
turbulence in a triply periodic box.  The data were obtained from a
direct numerical simulation\cite{cardesa2015}, which is publicly
available in \url{https://torroja.dmt.upm.es/turbdata/}. In this
simulation, the Navier--Stokes equations are numerically integrated by
resolving the whole range of spatial and temporal scales of the
flow. The conservation of momentum and mass equations for an
incompressible flow are given by:
\begin{eqnarray}\label{eq:cau:NS}
\frac{\partial  u_i}{\partial t} + \frac{\partial  u_i  u_j  }{\partial x_j}
 = - \frac{\partial  \Pi}{\partial x_i} +
\nu \frac{\partial^2  u_i}{\partial x_j\partial x_j} + f_i,
\quad\frac{\partial  u_i}{\partial x_i}  = 0,
\end{eqnarray}
where repeated indices imply summation, $\bs{x}=[x_1, x_2, x_3]$ are
the spatial coordinates, $u_i$ for $i=1,2,3$ are the velocities
components, $\Pi$ is the pressure, $\nu$ is the kinematic viscosity,
and $f_i$ is a linear forcing sustaining the turbulent
flow\cite{rosales2005}. The simulation was conducted with $1024^3$
spatial Fourier modes, which is enough to accurately resolve all the
relevant length-scales of the flow\cite{cardesa2015}. To quantify the
transfer of kinetic energy among eddies at different length scales
over time, the $i$-th component of the instantaneous flow velocity in
Equation \ref{eq:cau:NS}, denoted as $u_i(\bs{x},t)$, is decomposed
into contributions from large and small scales according to
$u_i(\bs{x},t) = \bar{u}_i(\bs{x},t) + u'_i(\bs{x},t)$. The operator
$\bar{(\cdot)}$ signifies the low-pass Gaussian filter and is given
by:
\begin{equation}
    \bar{u}_i(\boldsymbol{x},t) = 
    \iiint_V \frac{\sqrt{\pi}}{\bar{\Delta}} \exp\left[-\pi^2( \boldsymbol{x} -
      \boldsymbol{x}')^2/\bar{\Delta}^2\right] u_i(\boldsymbol{x}',t) \mathrm{d}\boldsymbol{x}',
\end{equation}
where $\bar{\Delta}$ is the filter width and $V$ denotes integration
over the whole flow domain. Examples of filtered
velocity fields at four different filter widths are included in Figure \ref{fig:cascade}. The kinetic energy of the large-scale field evolves as
\begin{equation}
    \left( \frac{\partial}{\partial t} + \bar{u}_j \frac{\partial }{\partial x_j}\right)\frac{1}{2}\bar{u}_i\bar{u}_i 
    =-\frac{\partial }{\partial x_j}\left( \bar{u}_j \bar{\Pi} + \bar{u}_i\tau_{ij} -  2\nu \bar{u}_i \bar{S}_{ij} \right)
    + \tilde{\Sigma} - 2\nu \bar{S}_{ij}\bar{S}_{ij} + \bar{u}_i \bar{f}_i, 
\end{equation}
where $\tau_{ij} = (\overline{u_i u_j} - \bar{u}_i \bar{u}_j)$ is the
subgrid-scale stress tensor, which represents the effect of the
(filtered) small-scale eddies on the (resolved) large-scale
eddies and $\bar{S}_{ij} = (\partial \bar{u}_i/\partial x_j + \partial \bar{u}_j/\partial x_i)/2$ denotes the filtered strain-rate tensor. The interscale energy transfer $\tilde{\Sigma}_i(\bs{x},t ; \bar{\Delta}_i)$ between the filtered and
unfiltered scales is given by
\begin{equation}
\tilde{\Sigma}_i(\bs{x},t ; \bar{\Delta}_i) = \tau_{ij}(\bs{x},t ; \bar{\Delta}_i) \bar{S}_{ij}(\bs{x},t ; \bar{\Delta}_i),
\end{equation}
which is the quantity of interest. The velocity field was low-passed filtered at four filter widths:
$\bar{\Delta}_1=163 \eta$, $\bar{\Delta}_2=81\eta$,
$\bar{\Delta}_3=42\eta$, and $\bar{\Delta}_4=21\eta$. The filter
widths are selected to represent four different flow scales and are
located within the inertial range of the simulation: $L_\varepsilon >
\bar{\Delta}_i > \eta$, for $i=1,2,3$ and 4, where $L_\varepsilon$
represents the size of the largest scales and $\eta$ is the size of
the smallest scales. Finally, we volume-averaged $\tilde{\Sigma}_i$ over the
entire domain, denoted by $\Sigma_i$, which served as
a marker for the dynamics of the energy cascade. The generated data
are also resolved in time, with flow fields stored at intervals of
$\Delta t = 0.0076 T_\varepsilon$, where $T_\varepsilon$ is the
characteristic time of the largest flow scales. The simulation was
intentionally run for an extended period to ensure the accurate
computation of specific mutual information. The total simulated time
after transient effects was equal to $165 T_\varepsilon$. For a given
target variable, $\Sigma_j$, the time delay $\Delta
T_j$ used to evaluate causality was determined as the time required
for maximum $\Delta I^U_{i \rightarrow j}$ with $j \neq i$, where
$\Sigma_j^+$ is evaluated at $t + \Delta T_j$.

\subsection*{Data for turbulent boundary layer}

\alvaro{ The data used for analyzing inner/outer interactions in a
  turbulent boundary layer were obtained from a experimental campaign
  at the high Reynolds number wind tunnel at the University of
  Melbourne\cite{Baars2015,Baars2017,marusic2020}, which is publicly
  available in \url{https://fluids.eng.unimelb.edu.au/}. In this
  campaign, measurements were made at a streamwise distance of $x =
  21.65\,\rm{m}$ from the trip at the test section inlet ($x = 0$),
  with a free-stream velocity of nominally $20\,\rm{m/s}$. The
  boundary layer at this location has a thickness of $\delta =
  0.361\,\rm{m}$, and the friction velocity is $u_\tau =
  0.626\,\rm{m/s}$. Based on these values, the friction Reynolds
  number is $Re_{\tau} = u_\tau \delta / \nu = 14\,750$. The data used
  in this study includes measurements of the streamwise velocity
  obtained using two synchronous hot-wire anemometry probes with an
  acquisition rate $\Delta t ^* = 1.28$ at two different wall-normal
  locations: $y_I^* = 4.33$ (for the inner layer) and $y_O/\delta =
  0.31$ (for the outer layer). The superscript $*$ denotes the inner
  scaling with friction velocity, $u_\tau$, and kinematic viscosity,
  $\nu$.  At each location, the acquisition time consists of three
  cycles of approximately $T U_\infty / \delta = 20\,000$. Further
  details about the experimental setup can be found in Baars et
  al.\cite{Baars2015} and Marusic\cite{marusic2020}. The time lag
  utilized to evaluate causality is $\Delta T^*\approx 756$, which
  corresponds to the time lag for maximum cross-induced unique
  causality.}


\subsection*{Sensitivity of SURD to number of samples}
%

\begin{figure}
    \centering
    \begin{tikzpicture}
        \node[anchor=south west,inner sep=0] (image1) at (0,0) {
            \includegraphics[trim={2cm 2cm 1cm 1cm}, clip, width=0.3\textwidth]{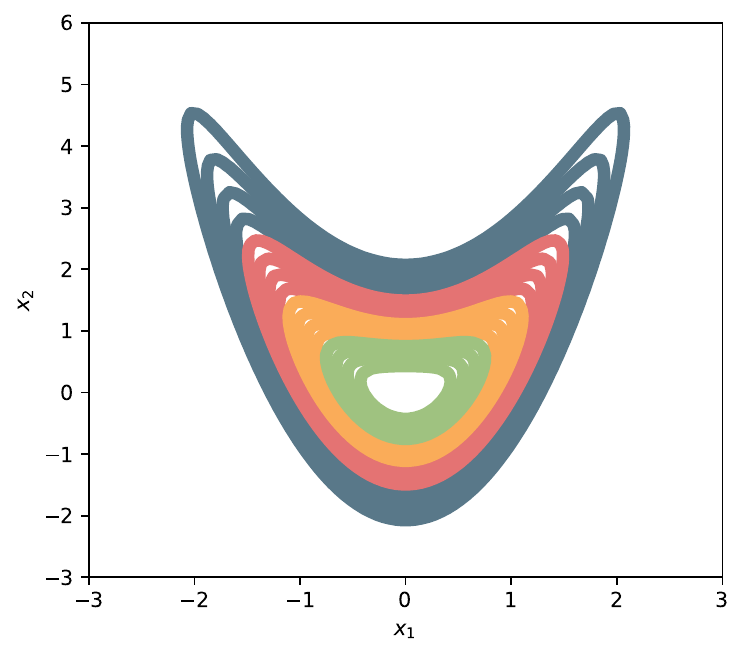}
        };
        \node[anchor=south] (label1) at ([yshift=-0.25cm]image1.north) {Prior distribution $\pi (\bs{x})$};

        \node[anchor=south west,inner sep=0] (image2) at (7,-0.15) { 
            \includegraphics[trim={2cm 2cm 1cm 1cm}, clip, width=0.24\textwidth]{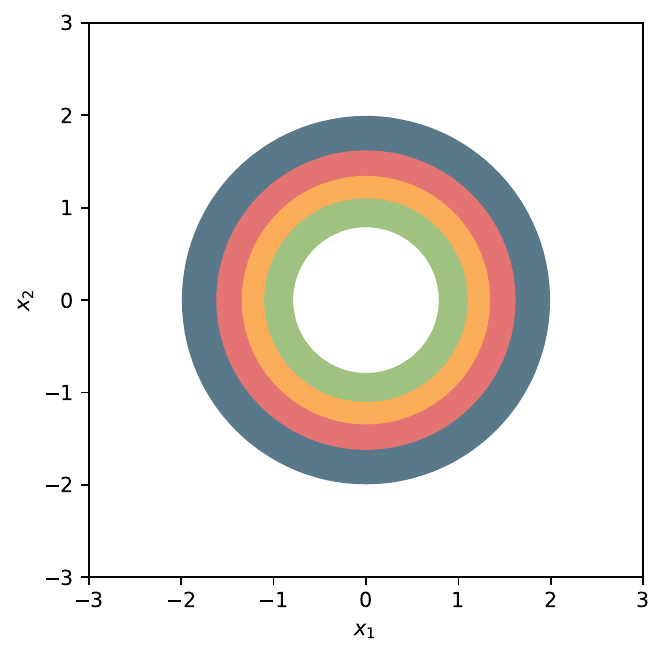}
        };
        \node[anchor=south] (label2) at ([yshift=-0.25cm, xshift=6.5cm]image1.north) {Posterior distribution $\eta (\bs{z})=\mathcal{N}(0,I)$};

        {\large
        \draw[->,>=latex,thick] ([xshift=-0.25cm, yshift=0cm]image1.east) to[bend left] node[midway,above] {$S(\bs{x})$} ([xshift=2cm, yshift=0cm]image1.east);
        \draw[<-,>=latex,thick] ([xshift=-0.25cm, yshift=-0.5cm]image1.east) to[bend right] node[midway,below] {$S^{-1}(\bs{z})$} ([xshift=2cm, yshift=-0.5cm]image1.east);
        }

        \node[anchor=south west,inner sep=0] (image3) at (12.1,-0.5) { 
            \includegraphics[width=0.31\tw]{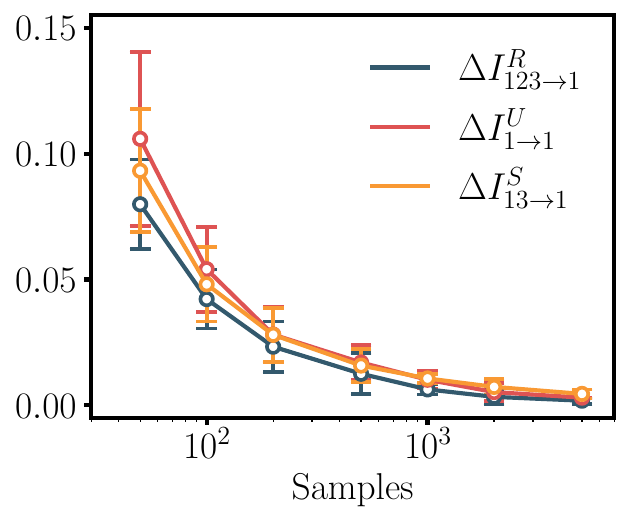}
        };
        \node[anchor=south] (label3) at ([yshift=-0.25cm, xshift=12.5cm]image1.north) {Error convergence with samples};
    \end{tikzpicture}
    \caption{(Left) Schematic representation of the construction
      process of the transport map $S$ that transforms an arbitrary
      distribution $\pi(\bs{x})$ to a Gaussian distribution
      $\eta(\bs{z}) = \mathcal{N}(0,I)$, where $\bs{x}$ and $\bs{z}$
      denote samples from each of the distributions. (Right) Evolution
      of the relative error with the number of samples for three
      different causalities from the confounder validation case. The
      results estimated using the transport map with $N=10^6$ samples
      and sixth-order polynomials are set as the ground truth.}
    \label{fig:transport-map}
\end{figure}
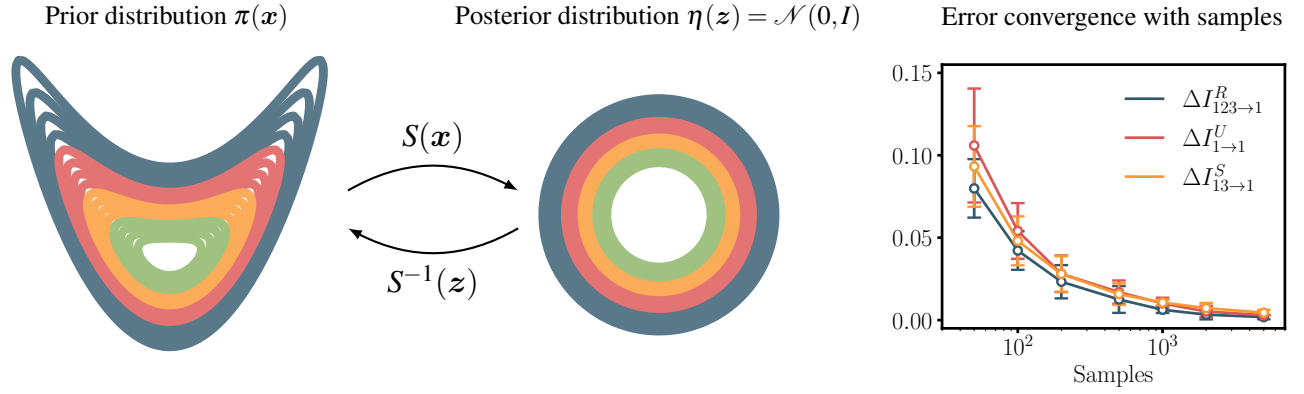

SURD relies on the estimation of probability distributions, which
becomes computationally intractable as the number of dimensions
increases. To address this limitation, we use the concept of transport
maps\cite{baptista2023} to estimate high-dimensional probability
distributions. The method relies on the estimation of a parsimonious
and interpretable nonlinear transformation from a complex distribution
$\pi(\bs{x})$ defined by the set of samples of the vector of observed
variables to a simpler reference distribution $\eta(\bs{z})$, e.g. a
Gaussian distribution $\mathcal{N}(0,I)$, as shown in Figure
\ref{fig:transport-map}. Although there are infinitely many
transformations that link these distributions, if $\pi$ is absolutely
continuous with respect to $\eta$, there exists a unique lower
triangular and monotone function $S:\mathbb{R}^d \rightarrow
\mathbb{R}^d $ that pushes forward $\pi$ to $\eta$. This type of
transformation is highly attractive since it provides a map that is
differentiable along with a differentiable inverse\cite{baptista2023}.

This transformation is a mere approximation of the real distribution,
which depends on the class of functions chosen. In this study, we used
sixth-order polynomials to estimate the transport map. Figure
\ref{fig:transport-map} shows the results for the evolution of the
relative error with the number of samples for redundant, unique, and
synergistic causalities from the confounder example in Figure
\ref{fig:confounder}. We set the results estimated using the transport
map with $N=10^6$ samples as a reference. Using this method, we can
obtain results with an error significantly lower than $5\%$ by using
only a number of samples in the order of a few hundred. Furthermore,
if we compare the reference results with those obtained using the
binning method with $N=10^8$ samples, we obtain differences lower than
$3\%$ for all causalities. Therefore, the approximation from the
transport map method allows us to estimate SURD causalities with
relatively high accuracy when the number of samples is low. A more
extensive analysis of the impact of sample size, partition refinement,
and order of the polynomials on the calculation of SURD is provided in
the Supplementary Materials for the binning and the transport map
methods.


\section*{Code availability}
The codes used for this work are available at:
\url{https://github.com/Computational-Turbulence-Group/SURD}.

\bibliography{references}

\begin{thebibliography}{100}
\urlstyle{rm}
\expandafter\ifx\csname url\endcsname\relax
  \def\url#1{\texttt{#1}}\fi
\expandafter\ifx\csname urlprefix\endcsname\relax\def\urlprefix{URL }\fi
\expandafter\ifx\csname doiprefix\endcsname\relax\def\doiprefix{DOI: }\fi
\providecommand{\bibinfo}[2]{#2}
\providecommand{\eprint}[2][]{\url{#2}}

\bibitem{pearl2000}
\bibinfo{author}{Pearl, J.}
\newblock \emph{\bibinfo{title}{Causality: Models, Reasoning, and Inference}} (\bibinfo{publisher}{Cambridge University Press}, \bibinfo{address}{New York, NY, USA}, \bibinfo{year}{2000}).

\bibitem{bunge1979}
\bibinfo{author}{Bunge, M.}
\newblock \emph{\bibinfo{title}{Causality and Modern Science}} (\bibinfo{publisher}{Dover Publications}, \bibinfo{year}{1979}).

\bibitem{Runge2023}
\bibinfo{author}{Runge, J.}, \bibinfo{author}{Gerhardus, A.}, \bibinfo{author}{Varando, G.}, \bibinfo{author}{Eyring, V.} \& \bibinfo{author}{Camps-Valls, G.}
\newblock \bibinfo{journal}{\bibinfo{title}{Causal inference for time series}}.
\newblock {\emph{\JournalTitle{Nature Reviews Earth {\&} Environment}}} \textbf{\bibinfo{volume}{4}}, \bibinfo{pages}{487--505}, \doiprefix\url{10.1038/s43017-023-00431-y} (\bibinfo{year}{2023}).

\bibitem{neuroscience2016}
\bibinfo{author}{Razi, A.} \& \bibinfo{author}{Friston, K.~J.}
\newblock \bibinfo{journal}{\bibinfo{title}{The connected brain: causality, models, and intrinsic dynamics}}.
\newblock {\emph{\JournalTitle{IEEE Signal Processing Magazine}}} \textbf{\bibinfo{volume}{33}}, \bibinfo{pages}{14--35} (\bibinfo{year}{2016}).

\bibitem{economic2008}
\bibinfo{author}{Chiou-Wei, S.~Z.}, \bibinfo{author}{Chen, C.-F.} \& \bibinfo{author}{Zhu, Z.}
\newblock \bibinfo{journal}{\bibinfo{title}{Economic growth and energy consumption revisited — evidence from linear and nonlinear granger causality}}.
\newblock {\emph{\JournalTitle{Energy Economics}}} \textbf{\bibinfo{volume}{30}}, \bibinfo{pages}{3063--3076}, \doiprefix\url{10.1016/j.eneco.2008.02.002} (\bibinfo{year}{2008}).
\newblock \bibinfo{note}{Technological Change and the Environment}.

\bibitem{epidemiology2005}
\bibinfo{author}{Rothman, K.~J.} \& \bibinfo{author}{Greenland, S.}
\newblock \bibinfo{journal}{\bibinfo{title}{Causation and causal inference in epidemiology}}.
\newblock {\emph{\JournalTitle{American journal of public health}}} \textbf{\bibinfo{volume}{95}}, \bibinfo{pages}{S144--S150} (\bibinfo{year}{2005}).

\bibitem{social2010}
\bibinfo{author}{Hedstr{\"o}m, P.} \& \bibinfo{author}{Ylikoski, P.}
\newblock \bibinfo{journal}{\bibinfo{title}{Causal mechanisms in the social sciences}}.
\newblock {\emph{\JournalTitle{Annual review of sociology}}} \textbf{\bibinfo{volume}{36}}, \bibinfo{pages}{49--67} (\bibinfo{year}{2010}).

\bibitem{lozano2020}
\bibinfo{author}{Lozano-Dur{\'a}n, A.}, \bibinfo{author}{Bae, H.~J.} \& \bibinfo{author}{Encinar, M.~P.}
\newblock \bibinfo{journal}{\bibinfo{title}{Causality of energy-containing eddies in wall turbulence}}.
\newblock {\emph{\JournalTitle{Journal of Fluid Mechanics}}} \textbf{\bibinfo{volume}{882}}, \bibinfo{pages}{A2} (\bibinfo{year}{2020}).

\bibitem{martinez2023}
\bibinfo{author}{Mart{\'\i}nez-S{\'a}nchez, {\'A}.} \emph{et~al.}
\newblock \bibinfo{journal}{\bibinfo{title}{Causality analysis of large-scale structures in the flow around a wall-mounted square cylinder}}.
\newblock {\emph{\JournalTitle{Journal of Fluid Mechanics}}} \textbf{\bibinfo{volume}{967}}, \bibinfo{pages}{A1} (\bibinfo{year}{2023}).

\bibitem{Eichler2013}
\bibinfo{author}{Eichler, M.}
\newblock \bibinfo{journal}{\bibinfo{title}{Causal inference with multiple time series: principles and problems}}.
\newblock {\emph{\JournalTitle{Philosophical Transactions of the Royal Society A: Mathematical, Physical and Engineering Sciences}}} \textbf{\bibinfo{volume}{371}}, \bibinfo{pages}{20110613}, \doiprefix\url{10.1098/rsta.2011.0613} (\bibinfo{year}{2013}).

\bibitem{barndorff2000}
\bibinfo{author}{Barndorff-Nielsen, O.~E.} \& \bibinfo{author}{Kluppelberg, C.}
\newblock \emph{\bibinfo{title}{Complex stochastic systems}} (\bibinfo{publisher}{Chapman and Hall, CRC, Boca Raton}, \bibinfo{year}{2001}).

\bibitem{Spirtes2001}
\bibinfo{author}{Spirtes, P.}, \bibinfo{author}{Glymour, C.} \& \bibinfo{author}{Scheines, R.}
\newblock \emph{\bibinfo{title}{{Causation, Prediction, and Search}}} (\bibinfo{publisher}{The MIT Press}, \bibinfo{year}{2001}).

\bibitem{Dawid2002}
\bibinfo{author}{Dawid, A.~P.}
\newblock \bibinfo{journal}{\bibinfo{title}{Influence diagrams for causal modelling and inference}}.
\newblock {\emph{\JournalTitle{International Statistical Review}}} \textbf{\bibinfo{volume}{70}}, \bibinfo{pages}{161--189}, \doiprefix\url{10.1111/j.1751-5823.2002.tb00354.x} (\bibinfo{year}{2002}).

\bibitem{kampa2008}
\bibinfo{author}{Kampa, M.} \& \bibinfo{author}{Castanas, E.}
\newblock \bibinfo{journal}{\bibinfo{title}{Human health effects of air pollution}}.
\newblock {\emph{\JournalTitle{Environmental pollution}}} \textbf{\bibinfo{volume}{151}}, \bibinfo{pages}{362--367} (\bibinfo{year}{2008}).

\bibitem{Altman2015}
\bibinfo{author}{Altman, N.} \& \bibinfo{author}{Krzywinski, M.}
\newblock \bibinfo{journal}{\bibinfo{title}{Association, correlation and causation}}.
\newblock {\emph{\JournalTitle{Nature Methods}}} \textbf{\bibinfo{volume}{12}}, \bibinfo{pages}{899--900}, \doiprefix\url{10.1038/nmeth.3587} (\bibinfo{year}{2015}).

\bibitem{pearson1895}
\bibinfo{author}{Pearson, K.} \& \bibinfo{author}{Galton, F.}
\newblock \bibinfo{journal}{\bibinfo{title}{{VII. Note on regression and inheritance in the case of two parents}}}.
\newblock {\emph{\JournalTitle{Proceedings of the Royal Society of London}}} \textbf{\bibinfo{volume}{58}}, \bibinfo{pages}{240--242}, \doiprefix\url{10.1098/rspl.1895.0041} (\bibinfo{year}{1895}).

\bibitem{Spearman1987}
\bibinfo{author}{Spearman, C.}
\newblock \bibinfo{journal}{\bibinfo{title}{The proof and measurement of association between two things}}.
\newblock {\emph{\JournalTitle{The American Journal of Psychology}}} \textbf{\bibinfo{volume}{100}}, \bibinfo{pages}{441--471} (\bibinfo{year}{1987}).

\bibitem{agresti2007art}
\bibinfo{author}{Agresti, A.} \& \bibinfo{author}{Franklin, C.}
\newblock \bibinfo{journal}{\bibinfo{title}{The art and science of learning from data}}.
\newblock {\emph{\JournalTitle{Upper Saddle River, New Jersey}}} \textbf{\bibinfo{volume}{88}} (\bibinfo{year}{2007}).

\bibitem{Horn2018}
\bibinfo{author}{Horn, K.~J.} \emph{et~al.}
\newblock \bibinfo{journal}{\bibinfo{title}{{Growth and survival relationships of 71 tree species with nitrogen and sulfur deposition across the conterminous U.S.}}}
\newblock {\emph{\JournalTitle{PLOS ONE}}} \textbf{\bibinfo{volume}{13}}, \bibinfo{pages}{1--19}, \doiprefix\url{10.1371/journal.pone.0205296} (\bibinfo{year}{2018}).

\bibitem{duan2020}
\bibinfo{author}{Duan, R.-R.}, \bibinfo{author}{Hao, K.} \& \bibinfo{author}{Yang, T.}
\newblock \bibinfo{journal}{\bibinfo{title}{Air pollution and chronic obstructive pulmonary disease}}.
\newblock {\emph{\JournalTitle{Chronic Diseases and Translational Medicine}}} \textbf{\bibinfo{volume}{6}}, \bibinfo{pages}{260--269}, \doiprefix\url{https://doi.org/10.1016/j.cdtm.2020.05.004} (\bibinfo{year}{2020}).

\bibitem{camps2023}
\bibinfo{author}{Camps-Valls, G.} \emph{et~al.}
\newblock \bibinfo{journal}{\bibinfo{title}{Discovering causal relations and equations from data}}.
\newblock {\emph{\JournalTitle{Physics Reports}}} \textbf{\bibinfo{volume}{1044}}, \bibinfo{pages}{1--68}, \doiprefix\url{10.1016/j.physrep.2023.10.005} (\bibinfo{year}{2023}).

\bibitem{eberhardt2007}
\bibinfo{author}{Eberhardt, F.} \& \bibinfo{author}{Scheines, R.}
\newblock \bibinfo{journal}{\bibinfo{title}{Interventions and causal inference}}.
\newblock {\emph{\JournalTitle{Philosophy of Science}}} \textbf{\bibinfo{volume}{74}}, \bibinfo{pages}{981--995} (\bibinfo{year}{2007}).

\bibitem{runge2019pcmci}
\bibinfo{author}{Runge, J.}, \bibinfo{author}{Nowack, P.}, \bibinfo{author}{Kretschmer, M.}, \bibinfo{author}{Flaxman, S.} \& \bibinfo{author}{Sejdinovic, D.}
\newblock \bibinfo{journal}{\bibinfo{title}{Detecting and quantifying causal associations in large nonlinear time series datasets}}.
\newblock {\emph{\JournalTitle{Science Advances}}} \textbf{\bibinfo{volume}{5}}, \bibinfo{pages}{eaau4996}, \doiprefix\url{10.1126/sciadv.aau4996} (\bibinfo{year}{2019}).

\bibitem{wiener1956}
\bibinfo{author}{Wiener, N.}
\newblock \emph{\bibinfo{title}{{The theory of prediction. Modern mathematics for engineers}}}, vol. \bibinfo{volume}{165} (\bibinfo{publisher}{New York}, \bibinfo{year}{1956}).

\bibitem{granger1969}
\bibinfo{author}{Granger, C. W.~J.}
\newblock \bibinfo{journal}{\bibinfo{title}{Investigating causal relations by econometric models and cross-spectral methods}}.
\newblock {\emph{\JournalTitle{Econometrica}}} \textbf{\bibinfo{volume}{37}}, \bibinfo{pages}{424--438} (\bibinfo{year}{1969}).

\bibitem{geweke1984}
\bibinfo{author}{Geweke, J.~F.}
\newblock \bibinfo{journal}{\bibinfo{title}{Measures of conditional linear dependence and feedback between time series}}.
\newblock {\emph{\JournalTitle{Journal of the American Statistical Association}}} \textbf{\bibinfo{volume}{79}}, \bibinfo{pages}{907--915} (\bibinfo{year}{1984}).

\bibitem{barnett2009}
\bibinfo{author}{Barnett, L.}, \bibinfo{author}{Barrett, A.~B.} \& \bibinfo{author}{Seth, A.~K.}
\newblock \bibinfo{journal}{\bibinfo{title}{Granger causality and transfer entropy are equivalent for gaussian variables}}.
\newblock {\emph{\JournalTitle{Physical Review Letters}}} \textbf{\bibinfo{volume}{103}}, \bibinfo{pages}{238701} (\bibinfo{year}{2009}).

\bibitem{barnett2014}
\bibinfo{author}{Barnett, L.} \& \bibinfo{author}{Seth, A.~K.}
\newblock \bibinfo{journal}{\bibinfo{title}{{The MVGC multivariate Granger causality toolbox: A new approach to Granger-causal inference}}}.
\newblock {\emph{\JournalTitle{Journal of Neuroscience Methods}}} \textbf{\bibinfo{volume}{223}}, \bibinfo{pages}{50--68}, \doiprefix\url{10.1016/j.jneumeth.2013.10.018} (\bibinfo{year}{2014}).

\bibitem{Barnett2015}
\bibinfo{author}{Barnett, L.} \& \bibinfo{author}{Seth, A.~K.}
\newblock \bibinfo{journal}{\bibinfo{title}{Granger causality for state-space models}}.
\newblock {\emph{\JournalTitle{Phys. Rev. E}}} \textbf{\bibinfo{volume}{91}}, \bibinfo{pages}{040101}, \doiprefix\url{10.1103/PhysRevE.91.040101} (\bibinfo{year}{2015}).

\bibitem{Hiemstra1994}
\bibinfo{author}{Hiemstra, C.} \& \bibinfo{author}{Jones, J.~D.}
\newblock \bibinfo{journal}{\bibinfo{title}{{Testing for Linear and Nonlinear Granger Causality in the Stock Price-Volume Relation}}}.
\newblock {\emph{\JournalTitle{The Journal of Finance}}} \textbf{\bibinfo{volume}{49}}, \bibinfo{pages}{1639--1664}, \doiprefix\url{10.1111/j.1540-6261.1994.tb04776.x} (\bibinfo{year}{1994}).

\bibitem{Bell9967}
\bibinfo{author}{Bell, D.}, \bibinfo{author}{Kay, J.} \& \bibinfo{author}{Malley, J.}
\newblock \bibinfo{journal}{\bibinfo{title}{A non-parametric approach to non-linear causality testing}}.
\newblock {\emph{\JournalTitle{Economics Letters}}} \textbf{\bibinfo{volume}{51}}, \bibinfo{pages}{7--18}, \doiprefix\url{10.1016/0165-1765(95)00791-1} (\bibinfo{year}{1996}).

\bibitem{abhyankar1998}
\bibinfo{author}{Abhyankar, A.}
\newblock \bibinfo{journal}{\bibinfo{title}{Linear and nonlinear granger causality: Evidence from the uk stock index futures market}}.
\newblock {\emph{\JournalTitle{The Journal of Futures Markets (1986-1998)}}} \textbf{\bibinfo{volume}{18}}, \bibinfo{pages}{519} (\bibinfo{year}{1998}).

\bibitem{tissot2014}
\bibinfo{author}{Tissot, G.}, \bibinfo{author}{Lozano-Dur{\'a}n, A.}, \bibinfo{author}{Jim{\'e}nez, J.}, \bibinfo{author}{Cordier, L.} \& \bibinfo{author}{Noack, B.~R.}
\newblock \bibinfo{journal}{\bibinfo{title}{Granger causality in wall-bounded turbulence}}.
\newblock {\emph{\JournalTitle{J. Phys. Conf. Ser}}} \textbf{\bibinfo{volume}{506}}, \bibinfo{pages}{012006}, \doiprefix\url{10.1088/1742-6596/506/1/012006} (\bibinfo{year}{2014}).

\bibitem{ancona2004}
\bibinfo{author}{Ancona, N.}, \bibinfo{author}{Marinazzo, D.} \& \bibinfo{author}{Stramaglia, S.}
\newblock \bibinfo{journal}{\bibinfo{title}{Radial basis function approach to nonlinear granger causality of time series}}.
\newblock {\emph{\JournalTitle{Physical Review E}}} \textbf{\bibinfo{volume}{70}}, \bibinfo{pages}{056221} (\bibinfo{year}{2004}).

\bibitem{bueso2020}
\bibinfo{author}{Bueso, D.}, \bibinfo{author}{Piles, M.} \& \bibinfo{author}{Camps-Valls, G.}
\newblock \bibinfo{journal}{\bibinfo{title}{Explicit granger causality in kernel hilbert spaces}}.
\newblock {\emph{\JournalTitle{Physical Review E}}} \textbf{\bibinfo{volume}{102}}, \bibinfo{pages}{062201} (\bibinfo{year}{2020}).

\bibitem{sugihara2012}
\bibinfo{author}{Sugihara, G.} \emph{et~al.}
\newblock \bibinfo{journal}{\bibinfo{title}{Detecting causality in complex ecosystems}}.
\newblock {\emph{\JournalTitle{Science}}} \textbf{\bibinfo{volume}{338}}, \bibinfo{pages}{496--500}, \doiprefix\url{10.1126/science.1227079} (\bibinfo{year}{2012}).

\bibitem{pai2014}
\bibinfo{author}{McCracken, J.~M.} \& \bibinfo{author}{Weigel, R.~S.}
\newblock \bibinfo{journal}{\bibinfo{title}{Convergent cross-mapping and pairwise asymmetric inference}}.
\newblock {\emph{\JournalTitle{Phys. Rev. E}}} \textbf{\bibinfo{volume}{90}}, \bibinfo{pages}{062903}, \doiprefix\url{10.1103/PhysRevE.90.062903} (\bibinfo{year}{2014}).

\bibitem{mccm2015}
\bibinfo{author}{Clark, A.~T.} \emph{et~al.}
\newblock \bibinfo{journal}{\bibinfo{title}{Spatial convergent cross mapping to detect causal relationships from short time series}}.
\newblock {\emph{\JournalTitle{Ecology}}} \textbf{\bibinfo{volume}{96}}, \bibinfo{pages}{1174--1181}, \doiprefix\url{10.1890/14-1479.1} (\bibinfo{year}{2015}).

\bibitem{extendedccm2015}
\bibinfo{author}{Ye, H.}, \bibinfo{author}{Deyle, E.~R.}, \bibinfo{author}{Gilarranz, L.~J.} \& \bibinfo{author}{Sugihara, G.}
\newblock \bibinfo{journal}{\bibinfo{title}{Distinguishing time-delayed causal interactions using convergent cross mapping}}.
\newblock {\emph{\JournalTitle{Scientific Reports}}} \textbf{\bibinfo{volume}{5}}, \bibinfo{pages}{14750}, \doiprefix\url{10.1038/srep14750} (\bibinfo{year}{2015}).

\bibitem{pcm2020}
\bibinfo{author}{Leng, S.} \emph{et~al.}
\newblock \bibinfo{journal}{\bibinfo{title}{Partial cross mapping eliminates indirect causal influences}}.
\newblock {\emph{\JournalTitle{Nature Communications}}} \textbf{\bibinfo{volume}{11}}, \bibinfo{pages}{2632}, \doiprefix\url{10.1038/s41467-020-16238-0} (\bibinfo{year}{2020}).

\bibitem{lccm2021}
\bibinfo{author}{Brouwer, E.~D.}, \bibinfo{author}{Arany, A.}, \bibinfo{author}{Simm, J.} \& \bibinfo{author}{Moreau, Y.}
\newblock \bibinfo{title}{Latent convergent cross mapping}.
\newblock In \emph{\bibinfo{booktitle}{International Conference on Learning Representations}} (\bibinfo{year}{2021}).

\bibitem{Takens1981}
\bibinfo{author}{Takens, F.}
\newblock \bibinfo{title}{Detecting strange attractors in turbulence}.
\newblock In \bibinfo{editor}{Rand, D.} \& \bibinfo{editor}{Young, L.-S.} (eds.) \emph{\bibinfo{booktitle}{Dynamical Systems and Turbulence, Warwick 1980}}, \bibinfo{pages}{366--381} (\bibinfo{publisher}{Springer Berlin Heidelberg}, \bibinfo{address}{Berlin, Heidelberg}, \bibinfo{year}{1981}).

\bibitem{cont_scaling_2022}
\bibinfo{author}{Ying, X.} \emph{et~al.}
\newblock \bibinfo{journal}{\bibinfo{title}{Continuity scaling: A rigorous framework for detecting and quantifying causality accurately}}.
\newblock {\emph{\JournalTitle{Research}}} \textbf{\bibinfo{volume}{2022}}, \doiprefix\url{10.34133/2022/9870149} (\bibinfo{year}{2022}).

\bibitem{shannon1948}
\bibinfo{author}{Shannon, C.~E.}
\newblock \bibinfo{journal}{\bibinfo{title}{A mathematical theory of communication}}.
\newblock {\emph{\JournalTitle{The Bell System Technical Journal}}} \textbf{\bibinfo{volume}{27}}, \bibinfo{pages}{379--423}, \doiprefix\url{10.1002/j.1538-7305.1948.tb01338.x} (\bibinfo{year}{1948}).

\bibitem{Lozano2022}
\bibinfo{author}{Lozano-Dur\'an, A.} \& \bibinfo{author}{Arranz, G.}
\newblock \bibinfo{journal}{\bibinfo{title}{Information-theoretic formulation of dynamical systems: Causality, modeling, and control}}.
\newblock {\emph{\JournalTitle{Phys. Rev. Res.}}} \textbf{\bibinfo{volume}{4}}, \bibinfo{pages}{023195}, \doiprefix\url{10.1103/PhysRevResearch.4.023195} (\bibinfo{year}{2022}).

\bibitem{yuan2024}
\bibinfo{author}{Yuan, Y.} \& \bibinfo{author}{Lozano-Dur\'an, A.}
\newblock \bibinfo{journal}{\bibinfo{title}{Limits to extreme event forecasting in chaotic systems}}.
\newblock {\emph{\JournalTitle{Physica D: Nonlinear Phenomena}}} \textbf{\bibinfo{volume}{467}}, \bibinfo{pages}{134246}, \doiprefix\url{https://doi.org/10.1016/j.physd.2024.134246} (\bibinfo{year}{2024}).

\bibitem{massey1990}
\bibinfo{author}{Massey, J.}
\newblock \bibinfo{title}{Causality, feedback and directed information}.
\newblock In \emph{\bibinfo{booktitle}{Proc. 1990 Int. Symp. on Infom. Theory and its Applications}}, \bibinfo{pages}{27--30} (\bibinfo{year}{1990}).

\bibitem{kramer1998}
\bibinfo{author}{Kramer, G.}
\newblock \emph{\bibinfo{title}{Directed information for channels with feedback}}.
\newblock \bibinfo{type}{{PhD Thesis}}, \bibinfo{school}{ETH Z\"urich}, \bibinfo{address}{Z\"urich} (\bibinfo{year}{1998}).
\newblock \doiprefix\url{10.3929/ethz-a-001988524}.

\bibitem{schreiber2000}
\bibinfo{author}{Schreiber, T.}
\newblock \bibinfo{journal}{\bibinfo{title}{Measuring information transfer}}.
\newblock {\emph{\JournalTitle{Phys. Rev. Lett.}}} \textbf{\bibinfo{volume}{85}}, \bibinfo{pages}{461} (\bibinfo{year}{2000}).

\bibitem{verdes2005}
\bibinfo{author}{Verdes, P.}
\newblock \bibinfo{journal}{\bibinfo{title}{Assessing causality from multivariate time series}}.
\newblock {\emph{\JournalTitle{Physical Review E}}} \textbf{\bibinfo{volume}{72}}, \bibinfo{pages}{026222} (\bibinfo{year}{2005}).

\bibitem{lizier2008}
\bibinfo{author}{Lizier, J.~T.}, \bibinfo{author}{Prokopenko, M.} \& \bibinfo{author}{Zomaya, A.~Y.}
\newblock \bibinfo{journal}{\bibinfo{title}{Local information transfer as a spatiotemporal filter for complex systems}}.
\newblock {\emph{\JournalTitle{Physical Review E}}} \textbf{\bibinfo{volume}{77}}, \bibinfo{pages}{026110} (\bibinfo{year}{2008}).

\bibitem{lizier2010}
\bibinfo{author}{Lizier, J.~T.}, \bibinfo{author}{Prokopenko, M.} \& \bibinfo{author}{Zomaya, A.~Y.}
\newblock \bibinfo{journal}{\bibinfo{title}{Information modification and particle collisions in distributed computation}}.
\newblock {\emph{\JournalTitle{Chaos: An Interdisciplinary Journal of Nonlinear Science}}} \textbf{\bibinfo{volume}{20}} (\bibinfo{year}{2010}).

\bibitem{cte2016}
\bibinfo{author}{Bossomaier, T.}, \bibinfo{author}{Barnett, L.}, \bibinfo{author}{Harré, M.} \& \bibinfo{author}{Lizier, J.~T.}
\newblock \emph{\bibinfo{title}{An Introduction to Transfer Entropy: Information Flow in Complex Systems}} (\bibinfo{publisher}{Springer International Publishing}, \bibinfo{address}{Cham}, \bibinfo{year}{2016}), \bibinfo{edition}{1st ed. 2016.} edn.

\bibitem{pompe2011}
\bibinfo{author}{Pompe, B.} \& \bibinfo{author}{Runge, J.}
\newblock \bibinfo{journal}{\bibinfo{title}{Momentary information transfer as a coupling measure of time series}}.
\newblock {\emph{\JournalTitle{Phys. Rev. E}}} \textbf{\bibinfo{volume}{83}}, \bibinfo{pages}{051122}, \doiprefix\url{10.1103/PhysRevE.83.051122} (\bibinfo{year}{2011}).

\bibitem{liang2006}
\bibinfo{author}{Liang, X.~S.} \& \bibinfo{author}{Kleeman, R.}
\newblock \bibinfo{journal}{\bibinfo{title}{Information transfer between dynamical system components}}.
\newblock {\emph{\JournalTitle{Phys. Rev. Lett.}}} \textbf{\bibinfo{volume}{95}}, \bibinfo{pages}{244101}, \doiprefix\url{10.1103/PhysRevLett.95.244101} (\bibinfo{year}{2006}).

\bibitem{liang2016if}
\bibinfo{author}{Liang, X.~S.}
\newblock \bibinfo{journal}{\bibinfo{title}{Information flow and causality as rigorous notions ab initio}}.
\newblock {\emph{\JournalTitle{Physical Review E}}} \textbf{\bibinfo{volume}{94}}, \bibinfo{pages}{052201} (\bibinfo{year}{2016}).

\bibitem{liang2008if}
\bibinfo{author}{Liang, X.~S.}
\newblock \bibinfo{journal}{\bibinfo{title}{Information flow within stochastic dynamical systems}}.
\newblock {\emph{\JournalTitle{Phys. Rev. E}}} \textbf{\bibinfo{volume}{78}}, \bibinfo{pages}{031113} (\bibinfo{year}{2008}).

\bibitem{liang2013if}
\bibinfo{author}{Liang, X.~S.}
\newblock \bibinfo{journal}{\bibinfo{title}{The liang-kleeman information flow: Theory and applications}}.
\newblock {\emph{\JournalTitle{Entropy}}} \textbf{\bibinfo{volume}{15}}, \bibinfo{pages}{327--360} (\bibinfo{year}{2013}).

\bibitem{spirtes1991}
\bibinfo{author}{Spirtes, P.} \& \bibinfo{author}{Glymour, C.}
\newblock \bibinfo{journal}{\bibinfo{title}{An algorithm for fast recovery of sparse causal graphs}}.
\newblock {\emph{\JournalTitle{Social Science Computer Review}}} \textbf{\bibinfo{volume}{9}}, \bibinfo{pages}{62--72} (\bibinfo{year}{1991}).

\bibitem{runge2018cmi}
\bibinfo{author}{Runge, J.}
\newblock \bibinfo{title}{Conditional independence testing based on a nearest-neighbor estimator of conditional mutual information}.
\newblock In \bibinfo{editor}{Storkey, A.} \& \bibinfo{editor}{Perez-Cruz, F.} (eds.) \emph{\bibinfo{booktitle}{Proceedings of the Twenty-First International Conference on Artificial Intelligence and Statistics}}, vol.~\bibinfo{volume}{84} of \emph{\bibinfo{series}{Proceedings of Machine Learning Research}}, \bibinfo{pages}{938--947} (\bibinfo{publisher}{PMLR}, \bibinfo{year}{2018}).

\bibitem{runge2023comment}
\bibinfo{author}{Runge, J.}
\newblock \bibinfo{journal}{\bibinfo{title}{Modern causal inference approaches to investigate biodiversity-ecosystem functioning relationships}}.
\newblock {\emph{\JournalTitle{Nature Communications}}} \textbf{\bibinfo{volume}{14}}, \bibinfo{pages}{1917}, \doiprefix\url{10.1038/s41467-023-37546-1} (\bibinfo{year}{2023}).

\bibitem{runge2020pcmci+}
\bibinfo{author}{Runge, J.}
\newblock \bibinfo{title}{Discovering contemporaneous and lagged causal relations in autocorrelated nonlinear time series datasets}.
\newblock In \emph{\bibinfo{booktitle}{Conference on Uncertainty in Artificial Intelligence}}, \bibinfo{pages}{1388--1397} (\bibinfo{organization}{PMLR}, \bibinfo{year}{2020}).

\bibitem{Gerhardus2020LPCMCI}
\bibinfo{author}{Gerhardus, A.} \& \bibinfo{author}{Runge, J.}
\newblock \bibinfo{title}{High-recall causal discovery for autocorrelated time series with latent confounders}.
\newblock In \bibinfo{editor}{Larochelle, H.}, \bibinfo{editor}{Ranzato, M.}, \bibinfo{editor}{Hadsell, R.}, \bibinfo{editor}{Balcan, M.} \& \bibinfo{editor}{Lin, H.} (eds.) \emph{\bibinfo{booktitle}{Advances in Neural Information Processing Systems}}, vol.~\bibinfo{volume}{33}, \bibinfo{pages}{12615--12625} (\bibinfo{publisher}{Curran Associates, Inc.}, \bibinfo{year}{2020}).

\bibitem{saggioro2020RPCMCI}
\bibinfo{author}{Saggioro, E.}, \bibinfo{author}{de~Wiljes, J.}, \bibinfo{author}{Kretschmer, M.} \& \bibinfo{author}{Runge, J.}
\newblock \bibinfo{journal}{\bibinfo{title}{{Reconstructing regime-dependent causal relationships from observational time series}}}.
\newblock {\emph{\JournalTitle{Chaos: An Interdisciplinary Journal of Nonlinear Science}}} \textbf{\bibinfo{volume}{30}}, \bibinfo{pages}{113115}, \doiprefix\url{10.1063/5.0020538} (\bibinfo{year}{2020}).

\bibitem{kullback1951}
\bibinfo{author}{Kullback, S.} \& \bibinfo{author}{Leibler, R.~A.}
\newblock \bibinfo{journal}{\bibinfo{title}{On information and sufficiency}}.
\newblock {\emph{\JournalTitle{Ann. Math. Stat.}}} \textbf{\bibinfo{volume}{22}}, \bibinfo{pages}{79--86} (\bibinfo{year}{1951}).

\bibitem{Kreer1957}
\bibinfo{author}{Kreer, J.}
\newblock \bibinfo{journal}{\bibinfo{title}{A question of terminology}}.
\newblock {\emph{\JournalTitle{IRE Transactions on Information Theory}}} \textbf{\bibinfo{volume}{3}}, \bibinfo{pages}{208--208}, \doiprefix\url{10.1109/TIT.1957.1057418} (\bibinfo{year}{1957}).

\bibitem{Robin2017}
\bibinfo{author}{Ince, R. A.~A.}
\newblock \bibinfo{journal}{\bibinfo{title}{Measuring multivariate redundant information with pointwise common change in surprisal}}.
\newblock {\emph{\JournalTitle{Entropy}}} \textbf{\bibinfo{volume}{19}}, \doiprefix\url{10.3390/e19070318} (\bibinfo{year}{2017}).

\bibitem{lotka1925}
\bibinfo{author}{Lotka, A.~J.}
\newblock \emph{\bibinfo{title}{Elements of physical biology}} (\bibinfo{publisher}{Williams \& Wilkins}, \bibinfo{year}{1925}).

\bibitem{volterra1926}
\bibinfo{author}{Volterra, V.}
\newblock \bibinfo{journal}{\bibinfo{title}{Fluctuations in the abundance of a species considered mathematically}}.
\newblock {\emph{\JournalTitle{Nature}}} \textbf{\bibinfo{volume}{118}}, \bibinfo{pages}{558--560} (\bibinfo{year}{1926}).

\bibitem{moran1953}
\bibinfo{author}{Moran, P.~A.}
\newblock \bibinfo{journal}{\bibinfo{title}{The statistical analysis of the canadian lynx cycle.}}
\newblock {\emph{\JournalTitle{Australian Journal of Zoology}}} \textbf{\bibinfo{volume}{1}}, \bibinfo{pages}{291--298} (\bibinfo{year}{1953}).

\bibitem{ding2006}
\bibinfo{author}{Ding, M.}, \bibinfo{author}{Chen, Y.} \& \bibinfo{author}{Bressler, S.}
\newblock \bibinfo{title}{Granger causality: Basic theory and application to neuroscience}.
\newblock In \bibinfo{editor}{Schelter, B.}, \bibinfo{editor}{Winterhalder, M.} \& \bibinfo{editor}{Timmer, J.} (eds.) \emph{\bibinfo{booktitle}{Handbook of Time Series Analysis: Recent Theoretical Developments and Applications}}, \bibinfo{pages}{2437--2459} (\bibinfo{publisher}{Wiley-VCH}, \bibinfo{address}{Berlin}, \bibinfo{year}{2006}).

\bibitem{may1976}
\bibinfo{author}{May, R.~M.}
\newblock \bibinfo{journal}{\bibinfo{title}{Simple mathematical models with very complicated dynamics}}.
\newblock {\emph{\JournalTitle{Nature}}} \textbf{\bibinfo{volume}{261}}, \bibinfo{pages}{459--467} (\bibinfo{year}{1976}).

\bibitem{lorenz1963}
\bibinfo{author}{Lorenz, E.~N.}
\newblock \bibinfo{journal}{\bibinfo{title}{Deterministic nonperiodic flow}}.
\newblock {\emph{\JournalTitle{J. Atmos. Sci.}}} \textbf{\bibinfo{volume}{20}}, \bibinfo{pages}{130--141} (\bibinfo{year}{1963}).

\bibitem{rossler1977}
\bibinfo{author}{R{\"o}ssler, O.~E.}
\newblock \bibinfo{title}{Continuous chaos}.
\newblock In \emph{\bibinfo{booktitle}{Synergetics: A Workshop Proceedings of the International Workshop on Synergetics at Schloss Elmau, Bavaria, May 2--7, 1977}}, \bibinfo{pages}{184--197} (\bibinfo{organization}{Springer}, \bibinfo{year}{1977}).

\bibitem{richardson1922}
\bibinfo{author}{Richardson, L.~F.}
\newblock \emph{\bibinfo{title}{Weather Prediction by Numerical Process}} (\bibinfo{publisher}{Cambridge University Press}, \bibinfo{year}{1922}).

\bibitem{obukhov1941}
\bibinfo{author}{Obukhov, A.~M.}
\newblock \bibinfo{journal}{\bibinfo{title}{On the distribution of energy in the spectrum of turbulent flow}}.
\newblock {\emph{\JournalTitle{Izv. Akad. Nauk USSR, Ser. Geogr. Geofiz.}}} \textbf{\bibinfo{volume}{5}}, \bibinfo{pages}{453--466} (\bibinfo{year}{1941}).

\bibitem{kolmogorov1941}
\bibinfo{author}{Kolmogorov, A.~N.}
\newblock \bibinfo{title}{{The Local Structure of Turbulence in Incompressible Viscous Fluid for Very Large {R}eynolds' Numbers}}.
\newblock In \emph{\bibinfo{booktitle}{Dokl. Akad. Nauk SSSR}}, vol.~\bibinfo{volume}{30}, \bibinfo{pages}{301--305} (\bibinfo{year}{1941}).

\bibitem{Baars2015}
\bibinfo{author}{Baars, W.~J.}, \bibinfo{author}{Talluru, K.~M.}, \bibinfo{author}{Hutchins, N.} \& \bibinfo{author}{Marusic, I.}
\newblock \bibinfo{journal}{\bibinfo{title}{Wavelet analysis of wall turbulence to study large-scale modulation of small scales}}.
\newblock {\emph{\JournalTitle{Experiments in Fluids}}} \textbf{\bibinfo{volume}{56}}, \bibinfo{pages}{188}, \doiprefix\url{10.1007/s00348-015-2058-8} (\bibinfo{year}{2015}).

\bibitem{Baars2017}
\bibinfo{author}{Baars, W.~J.}, \bibinfo{author}{Hutchins, N.} \& \bibinfo{author}{Marusic, I.}
\newblock \bibinfo{journal}{\bibinfo{title}{Reynolds number trend of hierarchies and scale interactions in turbulent boundary layers}}.
\newblock {\emph{\JournalTitle{Philosophical Transactions of the Royal Society A: Mathematical, Physical and Engineering Sciences}}} \textbf{\bibinfo{volume}{375}}, \bibinfo{pages}{20160077}, \doiprefix\url{10.1098/rsta.2016.0077} (\bibinfo{year}{2017}).

\bibitem{marusic2020}
\bibinfo{author}{Marusic, I.}
\newblock \bibinfo{journal}{\bibinfo{title}{{Two-point high Reynolds number zero-pressure gradient turbulent boundary layer dataset}}}.
\newblock {\emph{\JournalTitle{University of Melbourne}}} \doiprefix\url{10.26188/5e919e62e0dac} (\bibinfo{year}{2020}).

\bibitem{Quiroga2000}
\bibinfo{author}{Quiroga, R.~Q.}, \bibinfo{author}{Arnhold, J.} \& \bibinfo{author}{Grassberger, P.}
\newblock \bibinfo{journal}{\bibinfo{title}{Learning driver-response relationships from synchronization patterns}}.
\newblock {\emph{\JournalTitle{Phys. Rev. E}}} \textbf{\bibinfo{volume}{61}}, \bibinfo{pages}{5142--5148}, \doiprefix\url{10.1103/PhysRevE.61.5142} (\bibinfo{year}{2000}).

\bibitem{Krakovska2018}
\bibinfo{author}{Krakovsk\'a, A.} \emph{et~al.}
\newblock \bibinfo{journal}{\bibinfo{title}{Comparison of six methods for the detection of causality in a bivariate time series}}.
\newblock {\emph{\JournalTitle{Phys. Rev. E}}} \textbf{\bibinfo{volume}{97}}, \bibinfo{pages}{042207}, \doiprefix\url{10.1103/PhysRevE.97.042207} (\bibinfo{year}{2018}).

\bibitem{causalccm}
\bibinfo{author}{Javier, P. J.~E.}
\newblock \bibinfo{title}{{causal-ccm: a Python implementation of Convergent Cross Mapping}} (\bibinfo{year}{2021}).

\bibitem{kolmogorov1962}
\bibinfo{author}{Kolmogorov, A.~N.}
\newblock \bibinfo{journal}{\bibinfo{title}{A refinement of previous hypotheses concerning the local structure of turbulence in a viscous incompressible fluid at high {Reynolds} number}}.
\newblock {\emph{\JournalTitle{J. Fluid Mech.}}} \textbf{\bibinfo{volume}{13}}, \bibinfo{pages}{82--85}, \doiprefix\url{10.1017/S0022112062000518} (\bibinfo{year}{1962}).

\bibitem{aoyama2005}
\bibinfo{author}{Aoyama, T.} \emph{et~al.}
\newblock \bibinfo{journal}{\bibinfo{title}{Statistics of energy transfer in high-resolution direct numerical simulation of turbulence in a periodic box}}.
\newblock {\emph{\JournalTitle{J. Phys. Soc. Jpn.}}} \textbf{\bibinfo{volume}{74}}, \bibinfo{pages}{3202--3212} (\bibinfo{year}{2005}).

\bibitem{falkovich2009}
\bibinfo{author}{Falkovich, G.}
\newblock \bibinfo{journal}{\bibinfo{title}{Symmetries of the turbulent state}}.
\newblock {\emph{\JournalTitle{J. Phys. A}}} \textbf{\bibinfo{volume}{42}}, \bibinfo{pages}{123001}, \doiprefix\url{10.1088/1751-8113/42/12/123001} (\bibinfo{year}{2009}).

\bibitem{cardesa2017}
\bibinfo{author}{Cardesa, J.~I.}, \bibinfo{author}{Vela-Mart{\'\i}n, A.} \& \bibinfo{author}{Jim{\'e}nez, J.}
\newblock \bibinfo{journal}{\bibinfo{title}{The turbulent cascade in five dimensions}}.
\newblock {\emph{\JournalTitle{Science}}} \textbf{\bibinfo{volume}{357}}, \bibinfo{pages}{782--784} (\bibinfo{year}{2017}).

\bibitem{Yamada2008}
\bibinfo{author}{Yamada, T.} \emph{et~al.}
\newblock \bibinfo{journal}{\bibinfo{title}{Anatomy of plasma turbulence}}.
\newblock {\emph{\JournalTitle{Nature Physics}}} \textbf{\bibinfo{volume}{4}}, \bibinfo{pages}{721--725}, \doiprefix\url{10.1038/nphys1029} (\bibinfo{year}{2008}).

\bibitem{veynante2002}
\bibinfo{author}{Veynante, D.} \& \bibinfo{author}{Vervisch, L.}
\newblock \bibinfo{journal}{\bibinfo{title}{Turbulent combustion modeling}}.
\newblock {\emph{\JournalTitle{Prog. Energy Combust. Sci.}}} \textbf{\bibinfo{volume}{28}}, \bibinfo{pages}{193--266} (\bibinfo{year}{2002}).

\bibitem{bodenschatz2015}
\bibinfo{author}{Bodenschatz, E.}
\newblock \bibinfo{journal}{\bibinfo{title}{Clouds resolved}}.
\newblock {\emph{\JournalTitle{Science}}} \textbf{\bibinfo{volume}{350}}, \bibinfo{pages}{40--41}, \doiprefix\url{10.1126/science.aad1386} (\bibinfo{year}{2015}).

\bibitem{young2017}
\bibinfo{author}{Young, R. M.~B.} \& \bibinfo{author}{Read, P.~L.}
\newblock \bibinfo{journal}{\bibinfo{title}{Forward and inverse kinetic energy cascades in {J}upiter's turbulent weather layer}}.
\newblock {\emph{\JournalTitle{Nat. Phys.}}} \textbf{\bibinfo{volume}{13}}, \bibinfo{pages}{1135--1140} (\bibinfo{year}{2017}).

\bibitem{sirovich1997}
\bibinfo{author}{Sirovich, L.} \& \bibinfo{author}{Karlsson, S.}
\newblock \bibinfo{journal}{\bibinfo{title}{Turbulent drag reduction by passive mechanisms}}.
\newblock {\emph{\JournalTitle{Nature}}} \textbf{\bibinfo{volume}{388}}, \bibinfo{pages}{753--755} (\bibinfo{year}{1997}).

\bibitem{hof2010}
\bibinfo{author}{Hof, B.}, \bibinfo{author}{De~Lozar, A.}, \bibinfo{author}{Avila, M.}, \bibinfo{author}{Tu, X.} \& \bibinfo{author}{Schneider, T.~M.}
\newblock \bibinfo{journal}{\bibinfo{title}{Eliminating turbulence in spatially intermittent flows}}.
\newblock {\emph{\JournalTitle{Science}}} \textbf{\bibinfo{volume}{327}}, \bibinfo{pages}{1491--1494}, \doiprefix\url{10.1126/science.1186091} (\bibinfo{year}{2010}).

\bibitem{marusic2010}
\bibinfo{author}{Marusic, I.}, \bibinfo{author}{Mathis, R.} \& \bibinfo{author}{Hutchins, N.}
\newblock \bibinfo{journal}{\bibinfo{title}{Predictive model for wall-bounded turbulent flow}}.
\newblock {\emph{\JournalTitle{Science}}} \textbf{\bibinfo{volume}{329}}, \bibinfo{pages}{193--196}, \doiprefix\url{10.1126/science.1188765} (\bibinfo{year}{2010}).

\bibitem{kuhnen2018}
\bibinfo{author}{K{\"u}hnen, J.} \emph{et~al.}
\newblock \bibinfo{journal}{\bibinfo{title}{Destabilizing turbulence in pipe flow}}.
\newblock {\emph{\JournalTitle{Nat. Phys.}}} \textbf{\bibinfo{volume}{14}}, \bibinfo{pages}{386--390}, \doiprefix\url{10.1038/s41567-017-0018-3} (\bibinfo{year}{2018}).

\bibitem{Vela2021}
\bibinfo{author}{Vela-Mart{\'\i}n, A.} \& \bibinfo{author}{Jim{\'e}nez, J.}
\newblock \bibinfo{journal}{\bibinfo{title}{Entropy, irreversibility and cascades in the inertial range of isotropic turbulence}}.
\newblock {\emph{\JournalTitle{Journal of Fluid Mechanics}}} \textbf{\bibinfo{volume}{915}}, \bibinfo{pages}{A36} (\bibinfo{year}{2021}).

\bibitem{Vela2022}
\bibinfo{author}{Vela-Mart{\'\i}n, A.}
\newblock \bibinfo{journal}{\bibinfo{title}{Subgrid-scale models of isotropic turbulence need not produce energy backscatter}}.
\newblock {\emph{\JournalTitle{Journal of Fluid Mechanics}}} \textbf{\bibinfo{volume}{937}}, \bibinfo{pages}{A14} (\bibinfo{year}{2022}).

\bibitem{taylor1935}
\bibinfo{author}{Taylor, G.~I.}
\newblock \bibinfo{journal}{\bibinfo{title}{Statistical theory of turbulence}}.
\newblock {\emph{\JournalTitle{Proceedings of the Royal Society of London. Series A-Mathematical and Physical Sciences}}} \textbf{\bibinfo{volume}{151}}, \bibinfo{pages}{444--454}, \doiprefix\url{10.1098/rspa.1935.0158} (\bibinfo{year}{1935}).

\bibitem{frisch1995}
\bibinfo{author}{Frisch, U.}
\newblock \emph{\bibinfo{title}{Turbulence: The Legacy of A. N. Kolmogorov}} (\bibinfo{publisher}{Cambridge University Press}, \bibinfo{year}{1995}).

\bibitem{zhou1993a}
\bibinfo{author}{Zhou, Y.}
\newblock \bibinfo{journal}{\bibinfo{title}{Degrees of locality of energy transfer in the inertial range}}.
\newblock {\emph{\JournalTitle{Phys. Fluids}}} \textbf{\bibinfo{volume}{5}}, \bibinfo{pages}{1092--1094} (\bibinfo{year}{1993}).

\bibitem{eyink2005}
\bibinfo{author}{Eyink, G.~L.}
\newblock \bibinfo{journal}{\bibinfo{title}{Locality of turbulent cascades}}.
\newblock {\emph{\JournalTitle{Phys. D: Nonlinear Phenomena}}} \textbf{\bibinfo{volume}{207}}, \bibinfo{pages}{91--116} (\bibinfo{year}{2005}).

\bibitem{mininni2006}
\bibinfo{author}{Mininni, P.}, \bibinfo{author}{Alexakis, A.} \& \bibinfo{author}{Pouquet, A.}
\newblock \bibinfo{journal}{\bibinfo{title}{Large-scale flow effects, energy transfer, and self-similarity on turbulence}}.
\newblock {\emph{\JournalTitle{Phys. Rev. E}}} \textbf{\bibinfo{volume}{74}}, \bibinfo{pages}{016303} (\bibinfo{year}{2006}).

\bibitem{aluie2009}
\bibinfo{author}{Aluie, H.} \& \bibinfo{author}{Eyink, G.~L.}
\newblock \bibinfo{journal}{\bibinfo{title}{Localness of energy cascade in hydrodynamic turbulence. ii. sharp spectral filter}}.
\newblock {\emph{\JournalTitle{Phys. Fluids}}} \textbf{\bibinfo{volume}{21}}, \bibinfo{pages}{115108} (\bibinfo{year}{2009}).

\bibitem{domaradzki2009}
\bibinfo{author}{Domaradzki, J.~A.}, \bibinfo{author}{Teaca, B.} \& \bibinfo{author}{Carati, D.}
\newblock \bibinfo{journal}{\bibinfo{title}{Locality properties of the energy flux in turbulence}}.
\newblock {\emph{\JournalTitle{Phys. Fluids}}} \textbf{\bibinfo{volume}{21}}, \bibinfo{pages}{025106} (\bibinfo{year}{2009}).

\bibitem{townsend1976}
\bibinfo{author}{Townsend, A.~A.}
\newblock \emph{\bibinfo{title}{The structure of turbulent shear flow}} (\bibinfo{publisher}{Cambridge University Press}, \bibinfo{year}{1976}).

\bibitem{Hutchins2007}
\bibinfo{author}{Hutchins, N.} \& \bibinfo{author}{Marusic, I.}
\newblock \bibinfo{journal}{\bibinfo{title}{Evidence of very long meandering features in the logarithmic region of turbulent boundary layers}}.
\newblock {\emph{\JournalTitle{J. Fluid Mech.}}} \textbf{\bibinfo{volume}{579}}, \bibinfo{pages}{1--28} (\bibinfo{year}{2007}).

\bibitem{Mathis2009}
\bibinfo{author}{Mathis, R.}, \bibinfo{author}{Hutchins, N.} \& \bibinfo{author}{Marusic, I.}
\newblock \bibinfo{journal}{\bibinfo{title}{Large-scale amplitude modulation of the small-scale structures in turbulent boundary layers}}.
\newblock {\emph{\JournalTitle{J. Fluid Mech.}}} \textbf{\bibinfo{volume}{628}}, \bibinfo{pages}{311--337} (\bibinfo{year}{2009}).

\bibitem{flack2005}
\bibinfo{author}{Flack, K.~A.}, \bibinfo{author}{Schultz, M.~P.} \& \bibinfo{author}{Shapiro, T.~A.}
\newblock \bibinfo{journal}{\bibinfo{title}{Experimental support for townsend’s reynolds number similarity hypothesis on rough walls}}.
\newblock {\emph{\JournalTitle{Phys. Fluids}}} \textbf{\bibinfo{volume}{17}}, \bibinfo{pages}{035102} (\bibinfo{year}{2005}).

\bibitem{flores2006}
\bibinfo{author}{Flores, O.} \& \bibinfo{author}{Jim\'enez, J.}
\newblock \bibinfo{journal}{\bibinfo{title}{Effect of wall-boundary disturbances on turbulent channel flows}}.
\newblock {\emph{\JournalTitle{J. Fluid Mech.}}} \textbf{\bibinfo{volume}{566}}, \bibinfo{pages}{357--376} (\bibinfo{year}{2006}).

\bibitem{busse2012}
\bibinfo{author}{Busse, B.} \& \bibinfo{author}{Sandham, A.}
\newblock \bibinfo{journal}{\bibinfo{title}{Parametric forcing approach to rough-wall turbulent channel flow}}.
\newblock {\emph{\JournalTitle{J. Fluid Mech.}}} \textbf{\bibinfo{volume}{712}}, \bibinfo{pages}{169--202} (\bibinfo{year}{2012}).

\bibitem{Mizuno2013}
\bibinfo{author}{Mizuno, Y.} \& \bibinfo{author}{Jim{\'e}nez, J.}
\newblock \bibinfo{journal}{\bibinfo{title}{Wall turbulence without walls}}.
\newblock {\emph{\JournalTitle{J. Fluid Mech.}}} \textbf{\bibinfo{volume}{723}}, \bibinfo{pages}{429--455} (\bibinfo{year}{2013}).

\bibitem{Chung2014}
\bibinfo{author}{Chung, D.}, \bibinfo{author}{Monty, J.~P.} \& \bibinfo{author}{Ooi, A.}
\newblock \bibinfo{journal}{\bibinfo{title}{An idealised assessment of townsend's outer-layer similarity hypothesis for wall turbulence}}.
\newblock {\emph{\JournalTitle{J. Fluid Mech.}}} \textbf{\bibinfo{volume}{742}}, \doiprefix\url{10.1017/jfm.2014.17} (\bibinfo{year}{2014}).

\bibitem{lozano2019x}
\bibinfo{author}{Lozano-Dur{\'a}n, A.} \& \bibinfo{author}{Bae, H.~J.}
\newblock \bibinfo{journal}{\bibinfo{title}{{Characteristic scales of Townsend's wall-attached eddies}}}.
\newblock {\emph{\JournalTitle{J. Fluid Mech.}}} \textbf{\bibinfo{volume}{868}}, \bibinfo{pages}{698--725}, \doiprefix\url{10.1017/jfm.2019.209} (\bibinfo{year}{2019}).

\bibitem{williams2010}
\bibinfo{author}{Williams, P.~L.} \& \bibinfo{author}{Beer, R.~D.}
\newblock \bibinfo{journal}{\bibinfo{title}{Nonnegative decomposition of multivariate information}}.
\newblock {\emph{\JournalTitle{arXiv preprint arXiv:1004.2515}}}  (\bibinfo{year}{2010}).

\bibitem{griffith2014}
\bibinfo{author}{Griffith, V.} \& \bibinfo{author}{Koch, C.}
\newblock \bibinfo{title}{Quantifying synergistic mutual information}.
\newblock In \emph{\bibinfo{booktitle}{Guided Self-Organization: Inception}}, \bibinfo{pages}{159--190}, \doiprefix\url{10.1007/978-3-642-53734-9_6} (\bibinfo{publisher}{Springer}, \bibinfo{address}{Berlin, Heidelberg}, \bibinfo{year}{2014}).

\bibitem{griffith2015}
\bibinfo{author}{Griffith, V.} \& \bibinfo{author}{Ho, T.}
\newblock \bibinfo{journal}{\bibinfo{title}{Quantifying redundant information in predicting a target random variable}}.
\newblock {\emph{\JournalTitle{Entropy}}} \textbf{\bibinfo{volume}{17}}, \bibinfo{pages}{4644--4653} (\bibinfo{year}{2015}).

\bibitem{ince2017}
\bibinfo{author}{Ince, R.~A.}
\newblock \bibinfo{journal}{\bibinfo{title}{Measuring multivariate redundant information with pointwise common change in surprisal}}.
\newblock {\emph{\JournalTitle{Entropy}}} \textbf{\bibinfo{volume}{19}}, \bibinfo{pages}{318} (\bibinfo{year}{2017}).

\bibitem{gutknecht2021}
\bibinfo{author}{Gutknecht, A.~J.}, \bibinfo{author}{Wibral, M.} \& \bibinfo{author}{Makkeh, A.}
\newblock \bibinfo{journal}{\bibinfo{title}{Bits and pieces: understanding information decomposition from part-whole relationships and formal logic}}.
\newblock {\emph{\JournalTitle{Proc. R. Soc. A}}} \textbf{\bibinfo{volume}{477}}, \bibinfo{pages}{20210110}, \doiprefix\url{10.1098/rspa.2021.0110} (\bibinfo{year}{2021}).

\bibitem{kolchinsky2022}
\bibinfo{author}{Kolchinsky, A.}
\newblock \bibinfo{journal}{\bibinfo{title}{A novel approach to the partial information decomposition}}.
\newblock {\emph{\JournalTitle{Entropy}}} \textbf{\bibinfo{volume}{24}}, \bibinfo{pages}{403} (\bibinfo{year}{2022}).

\bibitem{baptista2023}
\bibinfo{author}{Baptista, R.}, \bibinfo{author}{Marzouk, Y.} \& \bibinfo{author}{Zahm, O.}
\newblock \bibinfo{journal}{\bibinfo{title}{On the representation and learning of monotone triangular transport maps}}.
\newblock {\emph{\JournalTitle{Foundations of Computational Mathematics}}} \bibinfo{pages}{1--46} (\bibinfo{year}{2023}).

\bibitem{cobey2016}
\bibinfo{author}{Cobey, S.} \& \bibinfo{author}{Baskerville, E.~B.}
\newblock \bibinfo{journal}{\bibinfo{title}{Limits to causal inference with state-space reconstruction for infectious disease}}.
\newblock {\emph{\JournalTitle{PLOS ONE}}} \textbf{\bibinfo{volume}{11}}, \bibinfo{pages}{1--22}, \doiprefix\url{10.1371/journal.pone.0169050} (\bibinfo{year}{2016}).

\bibitem{monster2017}
\bibinfo{author}{Mønster, D.}, \bibinfo{author}{Fusaroli, R.}, \bibinfo{author}{Tylén, K.}, \bibinfo{author}{Roepstorff, A.} \& \bibinfo{author}{Sherson, J.~F.}
\newblock \bibinfo{journal}{\bibinfo{title}{Causal inference from noisy time-series data — testing the convergent cross-mapping algorithm in the presence of noise and external influence}}.
\newblock {\emph{\JournalTitle{Future Generation Computer Systems}}} \textbf{\bibinfo{volume}{73}}, \bibinfo{pages}{52--62}, \doiprefix\url{10.1016/j.future.2016.12.009} (\bibinfo{year}{2017}).

\bibitem{runge2018chaos}
\bibinfo{author}{Runge, J.}
\newblock \bibinfo{journal}{\bibinfo{title}{{Causal network reconstruction from time series: From theoretical assumptions to practical estimation}}}.
\newblock {\emph{\JournalTitle{Chaos: An Interdisciplinary Journal of Nonlinear Science}}} \textbf{\bibinfo{volume}{28}}, \bibinfo{pages}{075310}, \doiprefix\url{10.1063/1.5025050} (\bibinfo{year}{2018}).

\bibitem{DeWeese1999}
\bibinfo{author}{DeWeese, M.~R.} \& \bibinfo{author}{Meister, M.}
\newblock \bibinfo{journal}{\bibinfo{title}{How to measure the information gained from one symbol}}.
\newblock {\emph{\JournalTitle{Network: Computation in Neural Systems}}} \textbf{\bibinfo{volume}{10}}, \bibinfo{pages}{325}, \doiprefix\url{10.1088/0954-898X/10/4/303} (\bibinfo{year}{1999}).

\bibitem{jidt}
\bibinfo{author}{Lizier, J.~T.}
\newblock \bibinfo{journal}{\bibinfo{title}{{JIDT: An Information-Theoretic Toolkit for Studying the Dynamics of Complex Systems}}}.
\newblock {\emph{\JournalTitle{Frontiers in Robotics and AI}}} \textbf{\bibinfo{volume}{1}}, \doiprefix\url{10.3389/frobt.2014.00011} (\bibinfo{year}{2014}).

\bibitem{akaike2011}
\bibinfo{author}{Akaike, H.}
\newblock \bibinfo{title}{Akaike’s information criterion}.
\newblock In \bibinfo{editor}{Lovric, M.} (ed.) \emph{\bibinfo{booktitle}{International Encyclopedia of Statistical Science}}, \doiprefix\url{10.1007/978-3-642-04898-2_110} (\bibinfo{publisher}{Springer}, \bibinfo{address}{Berlin, Heidelberg}, \bibinfo{year}{2011}).

\bibitem{cardesa2015}
\bibinfo{author}{Cardesa, J.~I.}, \bibinfo{author}{Vela-Mart\'in, A.}, \bibinfo{author}{Dong, S.} \& \bibinfo{author}{Jim\'enez, J.}
\newblock \bibinfo{journal}{\bibinfo{title}{The temporal evolution of the energy flux across scales in homogeneous turbulence}}.
\newblock {\emph{\JournalTitle{Phys. Fluids}}} \textbf{\bibinfo{volume}{27}}, \bibinfo{pages}{111702}, \doiprefix\url{10.1063/1.4935812} (\bibinfo{year}{2015}).

\bibitem{rosales2005}
\bibinfo{author}{Rosales, C.} \& \bibinfo{author}{Meneveau, C.}
\newblock \bibinfo{journal}{\bibinfo{title}{Linear forcing in numerical simulations of isotropic turbulence: Physical space implementations and convergence properties}}.
\newblock {\emph{\JournalTitle{Phys. Fluids}}} \textbf{\bibinfo{volume}{17}}, \bibinfo{pages}{095106} (\bibinfo{year}{2005}).

\end{thebibliography}

\section*{Acknowledgements}
The project that gave rise to these results received the support of a
fellowship from the "la Caixa" Foundation (ID 100010434). The
fellowship code is LCF/BQ/EU22/11930094. This work was supported by
the National Science Foundation under Grant No. 2140775 and MISTI
Global Seed Funds and UPM. G.~A. was partially supported by the
Predictive Science Academic Alliance Program (PSAAP; grant
DE-NA0003993) managed by the NNSA (National Nuclear Security
Administration) Office of Advanced Simulation and Computing and the
STTR N68335-21-C-0270 with Cascade Technologies, Inc. and the Naval
Air Systems Command.  The authors acknowledge the MIT SuperCloud and
Lincoln Laboratory Supercomputing Center for providing HPC resources
that have contributed to the research results reported within this
paper. The authors would like to thank Mathieu Le Provost for his
assistance with the implementation of the transport map method.


\section*{Author contributions statement}
A.~M.-S.: Methodology, Software, Validation, Investigation, Data
Curation, Writing – Original Draft, Writing – Review \& Editing,
Visualization. G.~A.: Methodology, Software, Investigation, Writing –
Review \& Editing. A.~L.-D.: Ideation, Methodology, Writing – Review
\& Editing, Supervision, Resources, Funding acquisition.

\section*{Competing Interests}
Authors declare no competing interests.

\end{document}


\maketitle
\tableofcontents
\renewcommand{\thesection}{S\arabic{section}}
\renewcommand{\thesubsection}{\thesection.\arabic{subsection}}
\renewcommand{\thesubsubsection}{\thesubsection.\arabic{subsubsection}}
\newpage

\section{Method formulation}

\subsection{Fundamentals of information theory}
\label{sec:fundamentals}

Consider $N$ quantities of interest at time $t$ represented by the
vector of observable variables $\bQ = [Q_1(t),Q_2(t), \ldots,Q_N(t)]$.
We treat $\bQ$ as a random variable and consider a finite partition of
the observable phase space $D =\{D_1, D_2, \dots, D_{N_D}\}$, where
$N_D$ is the number of partitions, such that $D = \cup_{i=1}^{N_D}
D_i$ and $D_i \cap D_j = \emptyset$ for all $i \neq j$ (i.e.,
non-overlapping partitions that cover all the space $D$).  We use
upper case $Q$ to denote the random variable itself; and lower case
$q$ to denote a particular state contained in one $D_i$ (also referred
to as a value or an event) of $Q$.  The probability of finding the
system at state $D_i$ at time $t$ is $p( \bQ(t) \in D_i )$, that in
general depends on the partition $D$.  For simplicity, we refer to the
latter probability as $p(\bq)$.

The information contained in the variable $\bQ$ is given
by~\cite{shannon1948}:
\begin{equation}\label{eq:entropy}
  H(\bQ) = \sum_{\bq} -p(\bq) \log_2[p(\bq)]\geq 0,
\end{equation}
where the summation is over all the states of $\bQ$. The quantity $H$
is referred to as the Shannon information or
entropy~\cite{shannon1948}. The units of $H$ are set by the base
chosen, in this case `bits' for base 2. For example, consider a fair
coin with $Q\in\{\mathrm{heads},\mathrm{tails}\}$ such that
$p(\mathrm{heads})=p(\mathrm{tails})=0.5$. The information of the
system `tossing a fair coin $n$ times' is $H = -\sum 0.5^n
\log_2(0.5^n) = n$ bits, where the summation is carried out across all
possible outcomes (namely, $2^n$).  If the coin is completely biased
towards heads, $p(\mathrm{heads})=1$, then $H = 0$ bits (taking $0\log
0 = 0$), i.e., no information is gained as the outcome was already
known before tossing the coin.  The Shannon information can also be
interpreted in terms of uncertainty: $H(\bQ)$ is the average number of
bits required to unambiguously determine $\bQ$.  $H$ is maximum when
all the possible outcomes are equiprobable (indicating a high level of
uncertainty in the state of the system) and zero when the process is
completely deterministic (indicating no uncertainty in the outcome).

\begin{figure}[t]
\centering
  \begin{tikzpicture}[scale=2.4,>={Latex[length=.2cm]}]
    \colorlet{c1}{myc1}
    \colorlet{c2}{myc2}

    \colorlet{MI}{myc3}
    \pgfmathsetmacro{\dis}{.6}
    \begin{normalsize}
        \draw[c1!80!black,fill=c1!40] (-\dis,0) circle (1);
        \draw[c2!80!black,fill=c2!40] (\dis,0) circle (1);
        
        \begin{scope}
            \clip (-\dis,0) circle (1);
            \draw[draw=MI,fill=MI!40] (\dis,0) circle (1);
        \end{scope}
        \begin{scope}
            \clip (\dis,0) circle (1);
            \draw[MI!80!black] (-\dis,0) circle (1);
        \end{scope}

        \node[anchor=east] at (-\dis,0) {$H(\bQ |\bQ')$};
        \node[anchor=west] at (+\dis,0) {$H(\bQ'|\bQ)$};

        \node[anchor=center] at (0,0) {$I(\bQ;\bQ')$};

        \draw[<-] (+\dis,0) ++ (30:1)  --+ (30:.6)  node[fill=white,pos=1] {$H(\bQ')$};
        \draw[<-] (-\dis,0) ++ (150:1) --+ (150:.6) node[fill=white,pos=1] {$H(\bQ)$};

    \end{normalsize}
\end{tikzpicture} 
    \caption{Venn diagram of the Shannon information, conditional Shannon
      information and mutual information between two random variables
      $\bQ$ and $\bQ'$.}\label{fig:basics}
\end{figure}
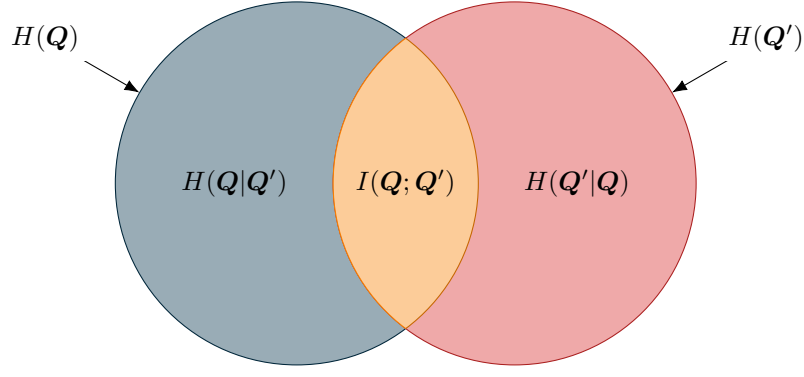

The Shannon information of $\bQ$ conditioned on another variable
$\bQ'$ is defined as:
\begin{equation}\label{eq:entropy conditional}
H( \bQ | \bQ' ) = \sum_{\bq,\bq'} -p(\bq,\bq') \log_2[p(\bq|\bq')].
\end{equation}
where $p(\bq|\bq') = p(\bq,\bq')/p(\bq')$ with $p(\bq')\neq 0$ is the
conditional probability distribution, and $p(\bq') = \sum_{\bq}
p(\bq,\bq')$ is the marginal probability distribution of $\bq'$. It is
useful to interpret $H(\bQ|\bQ')$ as the uncertainty in the variable
$\bQ$ after conducting the `measurement' of $\bQ'$.  If $\bQ$ and
$\bQ'$ are independent random variables, then $H(\bQ|\bQ')=H(\bQ)$,
i.e., knowing $\bQ'$ does not reduce the uncertainty in
$\bQ$. Conversely, $H(\bQ|\bQ')=0$ if knowing $\bQ'$ implies that
$\bQ$ is completely determined. Finally, the mutual information between
the random variables $\bQ$ and $\bQ'$ is
\begin{equation}\label{eq:optimal}
   I(\bQ;\bQ') =  H(\bQ) - H(\bQ|\bQ') = H(\bQ') - H(\bQ'|\bQ),
\end{equation}
which is a symmetric measure $I(\bQ;\bQ')=I(\bQ';\bQ)$ representing
the information shared among the variables $\bQ$ and $\bQ'$. Figure
\ref{fig:basics} depicts the relationship between the Shannon
information, conditional Shannon information, and mutual information.

The definitions above can be extended to continuous random variables
by replacing summation by integration and the probability mass
functions by probability density functions:
\begin{subequations}
  \label{eq:Hc}
  \begin{align}    
    H_c(\bQ) &= \int_{\bQ} -\rho(\bq) \log_2[\rho(\bq)] \dd \bq, \\
    H_c(\bQ|\bQ') &= \int_{\bQ,\bQ'} -\rho(\bq,\bq') \log_2[\rho(\bq|\bq')] \dd \bq\dd \bq', \\
    I_c(\bQ;\bQ') &= H_c(\bQ) - H_c(\bQ|\bQ') = H_c(\bQ') - H_c(\bQ'|\bQ),
  \end{align}
\end{subequations}  
where $H_c$ is referred to as the differential entropy, $\bQ$ and
$\bQ'$ are now continuous random variables, $\rho$ denotes probability
density function, and the integrals are performed over the support set
of $\bQ$ and $\bQ'$.  The differential entropy shares many of the
properties of the discrete entropy. However, it can be infinitely
large, positive or negative.  SURD relies on the use of mutual
information, which is non-negative in the continuous
case. Additionally, it can be shown that if
$\rho(\bq,\bq')\log_2[\rho(\bq,\bq')]$ is Riemann integrable, then
$I(\bQ^\Delta;\bQ'^\Delta) \rightarrow I_c(\bQ;\bQ')$ for $\Delta
\rightarrow 0$, where $\bQ^\Delta$ and $\bQ'^\Delta$ are the quantized
versions of $\bQ$ and $\bQ'$, respectively, defined over a finite
partition $D = \cup_{i=1}^{N_D} D_i$ with a characteristic partition
size for $D_i$ equal to $\Delta$. In the following
section, SURD is presented using discrete mutual information;
nevertheless, a similar formulation is applicable to the continuous
case.

\subsection{Synergistic-Unique-Redundant Decomposition of causality (SURD)}
\label{subsec:infocau}

\begin{figure}
  \begin{center}
    \begin{minipage}{0.4\textwidth}
        {
\pgfmathdeclarefunction{gauss}{3}{%
  \pgfmathparse{1/(#3*sqrt(2*pi))*exp(-((#1-#2)^2)/(2*#3^2))}%
}
 
\begin{tikzpicture}[scale=.8]

    \def\B{12};   
    \def\S{14};   
    \def\T{11.5}; 

    \def\Bs{4.50};  
    \def\Ss{3.20};  
    \def\Ts{4.40};  

    \def\xmax{25};
    \def\ymin{{-0.15*gauss(\B,\B,\Bs)}};
 
    \begin{axis}[every axis plot post/.append style={
                 mark=none,domain={-0.5*(\xmax)}:{1.08*\xmax},samples=80,smooth},
                 xmin={-0.1*(\xmax)}, xmax=\xmax,
                 ymin=\ymin, ymax={1.3*gauss(\B,\B,\Bs)},
                 axis lines=middle,
                 axis line style=thick,
                 enlargelimits=upper, 
                 ticks=none,
                 xlabel=$q_j^+$,
                 y axis line style={opacity=0},
                 x label style={at={(axis description cs:0.9,0.05)},anchor=north},
                 width=9cm, height=5cm,
                ]

      \addplot[name path=B,thick,black!10!myc1,fill=myc1!10] {gauss(x,\B,\Bs)};
      \addplot[name path=T,thick,black!10!myc3,dashed ] {gauss(x,\T,\Ts)};
 
      \node[above,black!20!myc1 ] at (\B,{gauss(.1*\B,\B,\Bs)}) {$p(q_j^+)$};
      \node[left,black!20!myc3 ] at 
        (.9*\T,{gauss(.9*\T,\T,\Ts)}) {$p(q_j^+| \bq )$};

    \end{axis}

\end{tikzpicture}}
    \end{minipage}
    \begin{minipage}{0.4\textwidth}
        {
\pgfmathdeclarefunction{gauss}{3}{%
  \pgfmathparse{1/(#3*sqrt(2*pi))*exp(-((#1-#2)^2)/(2*#3^2))}%
}
 
\begin{tikzpicture}[scale=.8]

    \def\B{12};   
    \def\S{14};   
    \def\T{11.5}; 

    \def\Bs{4.50};  
    \def\Ss{3.20};  
    \def\Ts{4.40};  

    \def\xmax{25};
    \def\ymin{{-0.15*gauss(\B,\B,\Bs)}};
 
    \begin{axis}[every axis plot post/.append style={
                 mark=none,domain={-0.5*(\xmax)}:{1.08*\xmax},samples=80,smooth},
                 xmin={-0.1*(\xmax)}, xmax=\xmax,
                 ymin=\ymin, ymax={1.3*gauss(\B,\B,\Bs)},
                 axis lines=middle,
                 axis line style=thick,
                 enlargelimits=upper, 
                 ticks=none,
                 xlabel=$q_j^+$,
                 y axis line style={opacity=0},
                 x label style={at={(axis description cs:0.9,0.05)},anchor=north},
                 width=9cm, height=5cm,
                ]

      \addplot[name path=B,thick,black!10!myc1,fill=myc1!10] {gauss(x,\B,\Bs)};
      \addplot[name path=S,thick,black!10!myc2 ] {gauss(x,\S,\Ss)};
 
      \node[above,black!20!myc1 ] at (\B,{gauss(.1*\B,\B,\Bs)}) {$p(q_j^+)$};
      \node[right,black!20!myc2 ] at 
        (1.1*\S,{gauss(1.1*\S,\S,\Ss)}) {$p(q_j^+| \bq )$};

    \end{axis}

\end{tikzpicture}}
    \end{minipage}
  \end{center}
\caption{Dissimilarity between $p(q_j^+)$ and $p(q_j^+|\bq)$
  contributing to $I(Q_j^+;\bQ)$.  Examples of (a) $p(q_j^+|\bq)$
  resembling $p(q_j^+)$, which barely contributes to $I(Q_j^+;\bQ)$;
  and (b) $p(q_j^+|\bq)$ different from $p(q_j^+)$, which increases
  the value of $I(Q_j^+;\bQ)$. Causality from $\bQ$ to $Q_j^+$ is
  quantified by the expectation of $\log_2[p(q_j^+|\bq)/p(q_j^+)]$.}
  \label{fig:interpretation_mutual}
\end{figure}
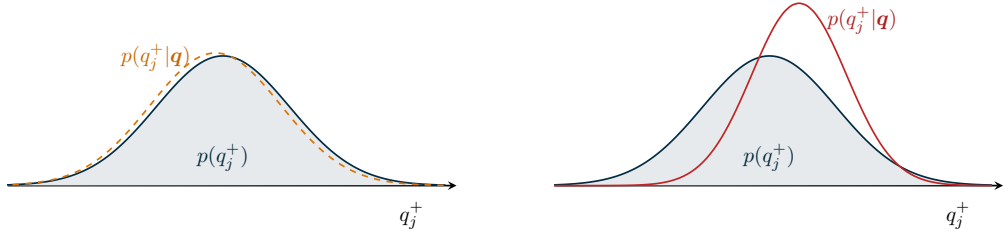

Our objective is to quantify the causality from the components of
$\bQ(t)$ to the future of the variable $Q_j^+ = Q_j(t+\Delta T)$,
where $Q_j$ could be one of the components of $\bQ$ and $\Delta T>0$
represents an arbitrary time lag. Moreover, for each component of
$\bQ$, the causality is decomposed into \emph{redundant},
\emph{unique}, and \emph{synergistic} contributions to $Q_j^+$. The
theoretical foundation of the method is rooted in the forward
propagation of information in dynamical systems: information can only
flow toward the future. Let us consider the information in the
variable $Q_j^+$, given by $H(Q_j^+)$. Assuming that all the
information in $Q_j^+$ is determined by the past states of the system,
we can write the equation for the forward propagation of
information~\cite{Lozano2022}
\begin{equation}
  \label{eq:conservation_info_sup}
 H(Q_j^+) = \Delta I(Q_j^+; \bQ) + \Delta I_{\text{leak}\rightarrow j},
\end{equation}
where $\Delta I(Q_j^+; \bQ)$ is the information flow from $\bQ$ to
$Q_j^+$, and $\Delta I_{\text{leak}\rightarrow j}$ is the causality
\emph{leak}, representing the causality from unobserved variables that
influence the dynamics of $Q_j^+$ but are not part of $\bQ$. The
causality leak can be expressed in closed form as a function of the
observed variables:
\begin{equation}
  \Delta I_{\text{leak}\rightarrow j} = H(Q_j^+|\bQ),
\end{equation}  
that is the uncertainty in $Q_j^+$ given the information in $\bQ$. The
amount of available information about $Q_j^+$ given $\bQ$ is
\begin{equation}
H(Q_j^+)-\Delta I_{\text{leak}\rightarrow j}= \Delta I(Q_j^+; \bQ)= H(Q_j^+)-H(Q_j^+|\bQ)=I(Q_j^+;\bQ),
\end{equation}
which is the mutual information between $Q_j^+$ and $\bQ$,
\begin{equation}
  \label{eq:mutual_info_decom}
  I(Q_j^+;\bQ) = \sum_{q_j^+,\bq} p(q_j^+,\bq) \log_2\left(
  \frac{p(q_j^+|\bq)}{p(q_j^+)}\right) = \sum_{q_j^+,\bq} p(q_j^+,\bq)
  \log_2\left( \frac{p(q_j^+,\bq)}{p(q_j^+)p(\bq)}\right),
\end{equation}
or in continuous form
\begin{equation}
  \label{eq:mutual_info_decom_cont}
  I(Q_j^+;\bQ) = \int_{Q_j^+,\bQ} \rho(q_j^+,\bq) \log_2\left(
  \frac{\rho(q_j^+|\bq)}{\rho(q_j^+)}\right) \dd q_j^+\dd \bq = \int_{Q_j^+,\bQ} \rho(q_j^+,\bq)
  \log_2\left( \frac{\rho(q_j^+,\bq)}{\rho(q_j^+)p(\bq)}\right) \dd q_j^+\dd \bq.
\end{equation}
%
Equation~\eqref{eq:mutual_info_decom}, quantifies the average
dissimilarity between $p(q_j^+)$ and $p(q_j^+|\bq)$. In terms of the
Kullback-Leibler divergence~\cite{kullback1951},
Equation~\eqref{eq:mutual_info_decom} measures the dissimilarity between
$p(q_j^+,\bq)$ and the distribution obtained under the assumption of
independence between $Q_j^+$ and $\bQ$, viz. $p(q_j^+)p(\bq)$. Hence,
SURD quantifies the causality from all the components of $\bQ$ to
$Q_j^+$ by examining how the probability of $Q_j^+$ changes when
accounting for $\bQ$.  Figure~\ref{fig:interpretation_mutual} provides
an interpretation of the quantification of causality based on
Equation~\eqref{eq:mutual_info_decom}.
%
%

The next step involves decomposing $I(Q_j^+;\bQ)$ into its unique,
redundant, and synergistic components as
\begin{equation}
  \label{eq:conservation_info_2}
I(Q_j^+;\bQ) \equiv \sum_{i=1}^N \Delta I_{i\rightarrow j}^U + \sum_{\bi
  \in \mathcal{C}} \Delta I_{\bi \rightarrow j}^R + \sum_{\bi\in
  \mathcal{C}} \Delta I_{\bi \rightarrow j}^S,
\end{equation}
where $\Delta I_{i\rightarrow j}^U$ is the unique causality from $Q_i$
to $Q_j^+$, $\Delta I_{\bi \rightarrow j}^R$ is the redundant
causality among the variables in $\bQ_{\bi}$ with $\bi = [i_1, i_2,
  \ldots]$ being a collection of indices, $\Delta I_{\bi \rightarrow
  j}^S$ is the synergistic causality from the variables in
$\bQ_{\bi}$, and $\mathcal{C}$ is the set of all the combinations of
numbers from 1 to $N$ with more than one element and less than or
equal to $N$ elements. For example,
Equation~\eqref{eq:conservation_info_2} can be expanded for $N=4$ as
\begin{subequations}
\begin{align}    
  I(Q_j^+;\bQ) &\equiv  \Delta I_{1\rightarrow j}^U +  \Delta I_{2\rightarrow j}^U  +  \Delta I_{3\rightarrow j}^U +  \Delta I_{4\rightarrow j}^U \\  
  &+  \Delta I_{12 \rightarrow j}^R + \Delta I_{13 \rightarrow j}^R + \Delta I_{14 \rightarrow j}^R  +
  \Delta I_{23 \rightarrow j}^R + \Delta I_{24 \rightarrow j}^R + \Delta I_{34 \rightarrow j}^R  + \\
  &+  \Delta I_{12 \rightarrow j}^S + \Delta I_{13 \rightarrow j}^S + \Delta I_{14 \rightarrow j}^S  +
  \Delta I_{23 \rightarrow j}^S + \Delta I_{24 \rightarrow j}^S + \Delta I_{34 \rightarrow j}^S  + \\  
  &+ \Delta I_{123 \rightarrow j}^R + \Delta I_{124 \rightarrow j}^R + 
  \Delta I_{134 \rightarrow j}^R + \Delta I_{234 \rightarrow j}^R + \\
    &+ \Delta I_{123 \rightarrow j}^S + \Delta I_{124 \rightarrow j}^S + 
  \Delta I_{134 \rightarrow j}^S + \Delta I_{234 \rightarrow j}^S \\
  &+ \Delta I_{1234 \rightarrow j}^R + \Delta I_{1234 \rightarrow j}^S.
\end{align}    
\end{subequations}

An important insight used to define SURD is the realization that the
source of causality might change depending on the value of $Q_j^+$.
For example, $Q_1$ can be only causal to positive values of $Q_j^+$,
whereas $Q_2$ can be only causal to negative values of $Q_j^+$. For
that reason, we define the specific mutual
information~\cite{DeWeese1999} from $\bQ_{\bi}$ to a particular event
$Q_j^+=q_j^+$ as
\begin{equation}
  \label{eq:specific_mutual_2}
   \Is(Q_j^+ = q_j^+;\bQ_{\bi}) = \sum_{\bq_{\bi}} p(\bq_{\bi} | q_j^+)
  \log_2\left( \frac{p(q_j^+|\bq_{\bi})}{p(q_j^+)} \right) \geq 0.
\end{equation}
Note that the specific mutual information is a function of the random
variable $\bQ_{\bi}$ (which encompasses all its states) but only a
function of one particular state of the target variable (namely,
$q_j^+$). For the sake of simplicity, we will use the notation
$\Is_{\bi}(q_j^+) = \Is(Q_j^+ = q_j^+;\bQ_{\bi})$.
Similarly to Equation~\eqref{eq:mutual_info_decom}, the specific mutual
information quantifies the dissimilarity between $p(q_j^+)$ and
$p(q_j^+|\bq)$ but in this case for the particular state
$Q_j^+=q_j^+$. The mutual information between $Q_j^+$and $\bQ_{\bi}$
is recovered by $I(Q_j^+;\bQ_{\bi}) = \sum_{q_j^+} p(q_j^+)
\Is_{\bi}(q_j^+)$.

\begin{figure}
    \centering
    \begin{tikzpicture}
        \node[anchor=north west] (fig) at (0,0) {\includegraphics[width=.75\textwidth]{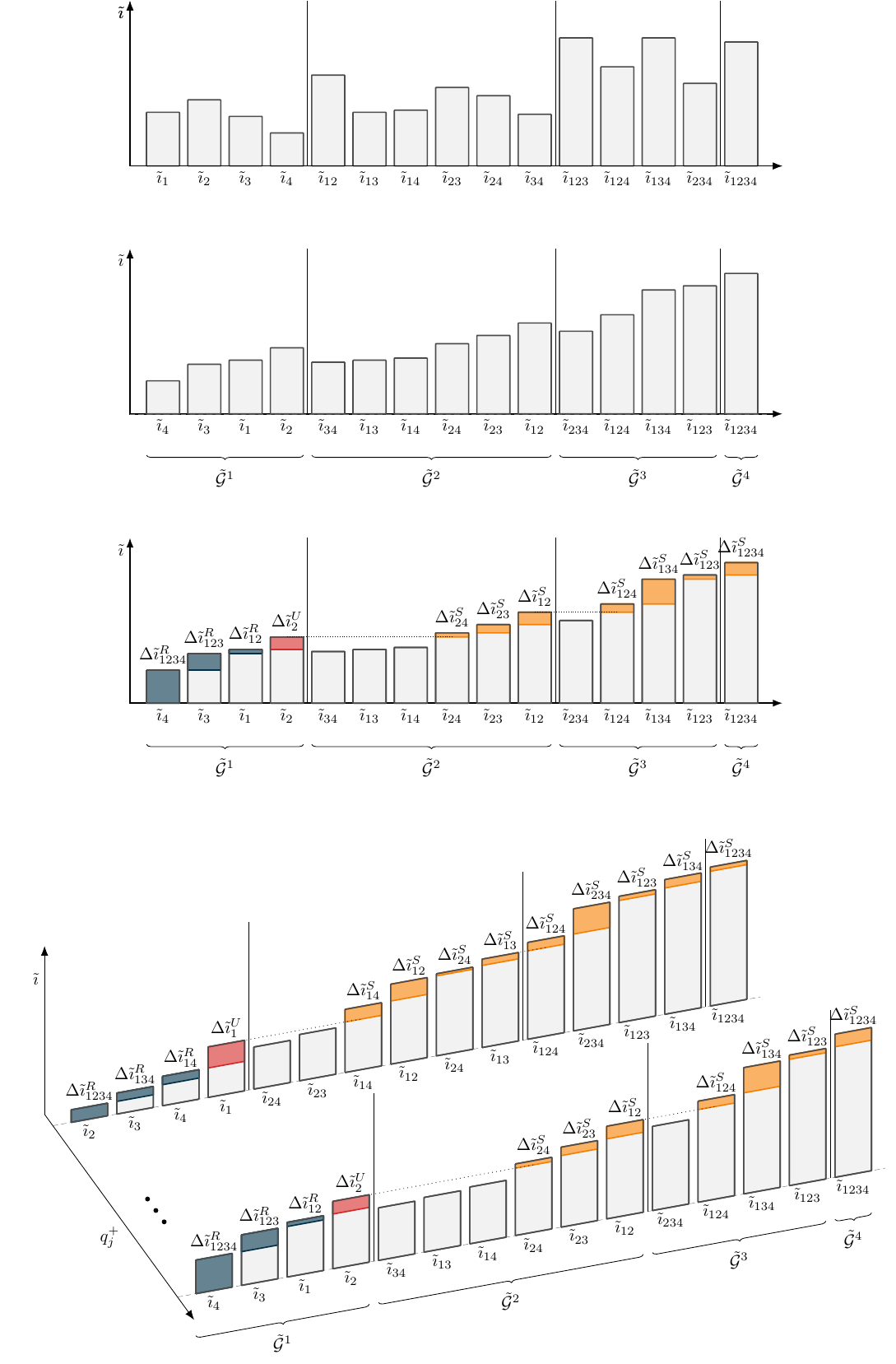}};
        \node at (0,0) {(a)};
        \node at (0,-3.5) {(b)};
        \node at (0,-7.5) {(c)};
        \node at (0,-12) {(d)};
    \end{tikzpicture}
\caption{ Schematic of the steps involved in the calculation of
  specific causalities. For a given state $Q_j^+=q_j^+$, the panels
  illustrate: (a) all possible specific mutual information values for
  a collection of four variables; (b) tuples of specific mutual
  information with the components organized in ascending order; (c)
  the increments corresponding to specific redundant (blue), unique
  (red), and synergistic (yellow) causalities; and (d) examples of
  specific causalities for different states of $Q_j^+$.}
  \label{fig:sup_method}
\end{figure}

We are now in the position of introducing the steps involved in the
calculation of redundant, unique, and synergistic causalities
(Figure~\ref{fig:sup_method}).  Our definitions are motivated by the
following intuition:
\begin{itemize}
\item Redundant causality from $\bQ_{\bi}=[Q_{i_1},Q_{i_2},\ldots]$ to
  $Q_j^+$ is the common causality shared among all the components of
  $\bQ_{\bi}$, where $\bQ_{\bi}$ is a subset of $\bQ$ with two or more
  components.
\item Unique causality from $Q_i$ to $Q_j^+$ is the causality from
  $Q_i$ that cannot be obtained from any other individual variable
  $Q_k$ with $k\neq i$.
\item Synergistic causality from $\bQ_{\bi}=[Q_{i_1},Q_{i_2},\ldots]$
  to $Q^+_j$ is the causality arising from the joint effect of the
  variables in $\bQ_{\bi}$.

\item Redundant and unique causalities must depend only on probability
  distributions based on $Q_i$ and $Q_j^+$, that is,
  $p(q_i,q_j^+)$. On the other hand, synergistic causality must depend
  on the joint probability distribution of $\bQ_{\bi}$ and $Q^+_j$,
  i.e., $p(\bq_{\bi},q_j^+)$.
\end{itemize}

For a given value $Q_j^+=q_j^+$, the specific redundant, unique, and
synergistic causalities are calculated as follows:
\begin{enumerate}
\item The specific mutual information are computed for all possible
  combinations of variables in $\bQ$. This includes specific mutual
  information of order one ($\Is_1, \Is_2, \ldots$), order two
  ($\Is_{12}, \Is_{13}, \ldots$), order three ($\Is_{123}, \Is_{124},
  \ldots$), and so forth. One example is shown in
  Figure~\ref{fig:sup_method}(a).
\item The tuples containing the specific mutual information of order
  $M$, denoted by $\GIs^M$, are constructed for $M=1,\ldots,N$.  The
  components of each $\GIs^M$ are organized in ascending order as
  shown in Figure~\ref{fig:sup_method}(b).
\item
  The specific redundant causality is the increment in information
  gained about $q_j^+$ that is common to all the components of
  $\bQ_{\bj_k}$ (blue contributions in Figure~\ref{fig:sup_method}c):
  \begin{equation}
    \Delta \Is^R_{\bj_k} =
    \begin{cases}
      \Is_{i_k}-\Is_{i_{k-1}},& \text{for} \ \Is_{i_k},\Is_{i_{k-1}} \in \GIs^1 \ \text{and} \ k\neq n_1\\
      0,              & \text{otherwise},
    \end{cases}
  \end{equation}
  where we take $\Is_{i_0}=0$, $\bj_k = [j_{k1}, j_{k2}, \ldots]$ is
  the vector of indices satisfying $\Is_{j_{kl}} \geq \Is_{i_k}$ for
  $\Is_{j_{kl}}, \Is_{i_k} \in \GIs^1$, and $n_1$ is the number of
  elements in $\GIs^1$.
\item
  The specific unique causality is the increment in information gained
  by $Q_{i_k}$ about $q_j^+$ that cannot be obtained by any other
  individual variable (red contribution in
  Figure~\ref{fig:sup_method}c):
    \begin{equation}
      \Delta \Is^U_{i_k} = 
      \begin{cases}
        \Is_{i_k}-\Is_{i_{k-1}}, & \text{for}\ i_k=n_1, \ \Is_{i_{k}},\Is_{i_{k-1}} \in \GIs^1\\
        0,                 & \text{otherwise}.
      \end{cases}
    \end{equation}
    %
  \item
  The specific synergistic causality is the increment in information
  gained by the combined effect of all the variables in $\bQ_{\bi_k}$
  that cannot be gained by other combination of variables
  $\bQ_{\bj_k}$ (yellow contributions in Figure~\ref{fig:sup_method}c)
  such that $\Is_{\bj_k} \leq \Is_{\bi_k}$ for $\Is_{\bi_k} \in
  \GIs^M$ and $\Is_{\bj_k} \in \{\GIs^1,\ldots,\GIs^{M}\}$ with $M>1$
  (dotted line in Figure~\ref{fig:sup_method}c):
    \begin{equation}
      \Delta \Is^S_{\bi_k} = 
      \begin{cases}
        \Is_{\bi_k} - \Is_{\bi_{k-1}}, & \text{for} \ \Is_{\bi_{k-1}}\geq \max\{\GIs^{M-1}\}, \ \text{and} \ \Is_{\bi_{k}}, \Is_{\bi_{k-1}} \in \GIs^M \\
        \Is_{\bi_k} - \max\{\GIs^{M-1}\}, & \text{for} \ \Is_{\bi_{k}}>\max\{\GIs^{M-1}\}>\Is_{\bi_{k-1}}, \ \text{and} \ \Is_{\bi_{k}},  \Is_{\bi_{k-1}} \in \GIs^M\\
        0,              & \text{otherwise}.
      \end{cases}
    \end{equation}
    %
\item The specific redundant, unique and synergistic causalities that
  do not appear in the steps above are set to zero.
\item The steps (1) to (6) are repeated for all the states of $Q_j^+$
  (Figure~\ref{fig:sup_method}d).
\item Redundant, unique, and synergistic causalities are obtained as
  the expectation of their corresponding specific values with respect
  to $Q_j^+$,
  \begin{subequations}
    \begin{align}  
      \Delta I^R_{\bi \rightarrow j} &= \sum_{q_j^+} p(q_j^+) \Delta \Is^R_{\bi}(q_j^+), \\
      \Delta I^U_{i\rightarrow j}   &= \sum_{q_j^+} p(q_j^+) \Delta \Is^U_{i}(q_j^+), \\
      \Delta I^S_{\bi\rightarrow j} &= \sum_{q_j^+} p(q_j^+) \Delta \Is^S_{\bi}(q_j^+). 
    \end{align}  
  \end{subequations}
\item 
  Finally, we define the average order of the specific causalities
  with respect to $Q_j^+$ as
  \begin{equation}
    N^{\alpha}_{\bi \rightarrow j} = \sum_{q_j^+} p(q_j^+)
    n^{\alpha}_{\bi \rightarrow j}(q_j^+),
  \end{equation}
  where $\alpha$ denotes R, U, or S, $n^{\alpha}_{\bi \rightarrow
    j}(q_j^+)$ is the order of appearance of $\Delta
  \Is^{\alpha}_{\bi}(q_j^+)$ from left to right as in the example
  shown in Figure~\ref{fig:sup_method}.  The values of
  $N^{\alpha}_{\bi \rightarrow j}$ are used to plot $\Delta
  I^{\alpha}_{\bi \rightarrow j}$ following the expected order of
  appearance of $\Delta \Is^{\alpha}_{\bi \rightarrow j}$. All the
  causalities from SURD presented in this work are plotted in order
  from left to right, following $N^{\alpha}_{\bi \rightarrow j}$.
\end{enumerate}
%
%

{It is worth noting that the problem of defining
  redundant, unique, and synergistic causalities can be generally
  framed as the task of decomposing the mutual information
  $I(Q_j^+;\bQ)$ into multiple components. The definitions proposed
  above are motivated by their consistency with the properties
  presented in the following sections along with the ease of
  interpretability.  Alternative definitions are possible and other
  decompositions have been suggested in the
  literature~\cite{williams2010, griffith2014, griffith2015, ince2017,
    gutknecht2021, Lozano2022, kolchinsky2022}; however, these do not
  comply with the properties discussed next or result in an
  unmanageable number of terms.  For instance, one of the most
  referenced decompositions by Williams \& Beer \cite{williams2010}
  results in a number of terms that grows as the Dedekind numbers. In
  the case of 9 variables, this decomposition yields over $10^{23}$
  terms, whereas SURD produces only 512 terms.}

{In scenarios with a large number of terms,
  synergistic and redundant causalities in SURD can be grouped by
  different orders to facilitate interpretation. For example, for
  three variables, we can define $\Delta I^S_{\text{2nd}} = \Delta
  I^S_{12\rightarrow 1}+\Delta I^S_{23\rightarrow 1} + \Delta
  I^S_{13\rightarrow 1}$, which represents the second-order causal
  synergy to the target variable 1 (similarly for other orders and
  redundancies). This is possible in SURD due to the additivity of its
  causal components. Additionally, if synergistic causalities above a
  given order are not computed, they are accounted for by the
  causality leak. In \S\,\ref{sec:eight}, we apply SURD to a system of
  eight interacting species and calculate synergistic causalities up
  to the fourth order. The remaining synergistic causalities are
  considered as causality leaks. This approach effectively manages the
  challenge of dimensionality by focusing on lower-order synergistic
  interactions, which are often sufficient for practical analyses.}

\subsection{Application of SURD to multiple time lags}
\label{subsec:multiple lags}

The vector of observables can also include variables at multiple time
lags:
\begin{equation}
\bQ = \left[ Q_1(t), Q_1(t -\Delta T_1), \dots, Q_1(t-\Delta
  T_p), \dots, Q_2(t), Q_2(t -\Delta T_1), \dots, Q_2(t-\Delta T_p),
  \dots \right],
\end{equation}
where $\Delta T_i>0$ for $i=1,\dots,p$. For instance, for $N=2$ and
$p=1$, the vector is
\begin{equation}
  \label{eq:Qlagged}
  \bQ = \left[ Q_1(t), Q_1(t- \Delta T), Q_2(t),
    Q_2(t- \Delta T) \right].
\end{equation}
For simplicity, we will use the notation $\bQ = \left[ Q_{1_1},
  Q_{1_2}, Q_{2_1}, Q_{2_2}\right]$, where the first subindex denotes
the variable number and the second subindex denotes the present time
for $i=1$ and past times for $i>1$, e.g., $Q_{1_1} = Q_1(t)$, $Q_{1_2}
= Q_1(t-\Delta T)$, $Q_{2_1} = Q_2(t)$, $Q_{2_2} = Q_2(t- \Delta T)$
and so on.  The formulation of SURD presented above is equally
applicable to the observable $\bQ$ from Equation~\eqref{eq:Qlagged}.
Following the example above for $N=2$ and $p=1$, the mutual
information between the target variable $Q_j^+$ and $\bQ$ is
decomposed as:
\begin{subequations}
\begin{align}
I(Q_j^+;\bQ) &= \Delta I_{{1_1}\rightarrow j}^U + \Delta I_{{2_1}\rightarrow j}^U + \Delta I_{{1_2}\rightarrow j}^U + \Delta I_{{2_2}\rightarrow j}^U \\
&+ \Delta I_{{1_1}{2_1} \rightarrow j}^R + \Delta I_{{1_1}{1_2} \rightarrow j}^R + \Delta I_{{1_1}{2_2} \rightarrow j}^R +
\Delta I_{{2_1}{1_2} \rightarrow j}^R + \Delta I_{{2_1}{2_2} \rightarrow j}^R + \Delta I_{{1_2}{2_2} \rightarrow j}^R + \\
&+ \Delta I_{{1_1}{2_1} \rightarrow j}^S + \Delta I_{{1_1}{1_2} \rightarrow j}^S + \Delta I_{{1_1}{2_2} \rightarrow j}^S +
\Delta I_{{2_1}{1_2} \rightarrow j}^S + \Delta I_{{2_1}{2_2} \rightarrow j}^S + \Delta I_{{1_2}{2_2} \rightarrow j}^S + \\
&+ \Delta I_{{1_1}{2_1}{1_2} \rightarrow j}^R + \Delta I_{{1_1}{2_1}{2_2} \rightarrow j}^R +
\Delta I_{{1_1}{1_2}{2_2} \rightarrow j}^R + \Delta I_{{2_1}{1_2}{2_2} \rightarrow j}^R + \\
&+ \Delta I_{{1_1}{2_1}{1_2} \rightarrow j}^S + \Delta I_{{1_1}{2_1}{2_2} \rightarrow j}^S +
\Delta I_{{1_1}{1_2}{2_2} \rightarrow j}^S + \Delta I_{{2_1}{1_2}{2_2} \rightarrow j}^S \\
&+ \Delta I_{{1_1}{2_1}{1_2}{2_2} \rightarrow j}^R + \Delta I_{{1_1}{2_1}{1_2}{2_2} \rightarrow j}^S.
\end{align}
\end{subequations}

\subsection{Properties of SURD}
\label{subsec:properties}

\begin{itemize}
\item \emph{Non-negativity}. All the terms in
  Equation~\eqref{eq:conservation_info_2} are non-negative by the
  definition of the redundant, unique and synergistic causalities, and
  the non-negativity of the specific mutual
  information~\cite{DeWeese1999}.
  \begin{equation}
    \Delta I_{i\rightarrow j}^U \geq 0, \ \text{for all} \ i=1,\ldots,N, \ \Delta I_{\boldsymbol{i} \rightarrow j}^R\geq 0, \ \Delta I_{\boldsymbol{i} \rightarrow j}^S \geq 0, \ \text{for all} \ \bi \in \mathcal{C}. 
  \end{equation}
  %
  %
\item \emph{Reconstruction of individual mutual information}. The
  mutual information between $Q_i$ and $Q_j^+$ is equal to the unique
  and redundant causalities containing $Q_i$
  \begin{equation}\label{eq:individual_causality}
    I(Q_i; Q_j^+ ) = \Delta I_{i\rightarrow j}^U +
    \sum_{\boldsymbol{i}\in \mathcal{C}_i} \Delta I_{\boldsymbol{i}
      \rightarrow j}^R,
  \end{equation}
  where $\mathcal{C}_i$ is the set of the combinations in
  $\mathcal{C}$ containing the variable $Q_i$. This condition is
  consistent with the notion that the information shared between $Q_i$
  and $Q_j^+$ should comprise contributions from unique and redundant
  causalities only, whereas synergistic causalities should arise
  through the combined effects of variables. This property, along with
  the non-negativity and forward propagation of information, enables
  the construction of the causality diagrams as the example depicted
  in Figure~\ref{fig:diagram_mutuals} for three variables. The
  properties mentioned above are also responsible for avoiding the
  duplication of causalities within the system.
  %
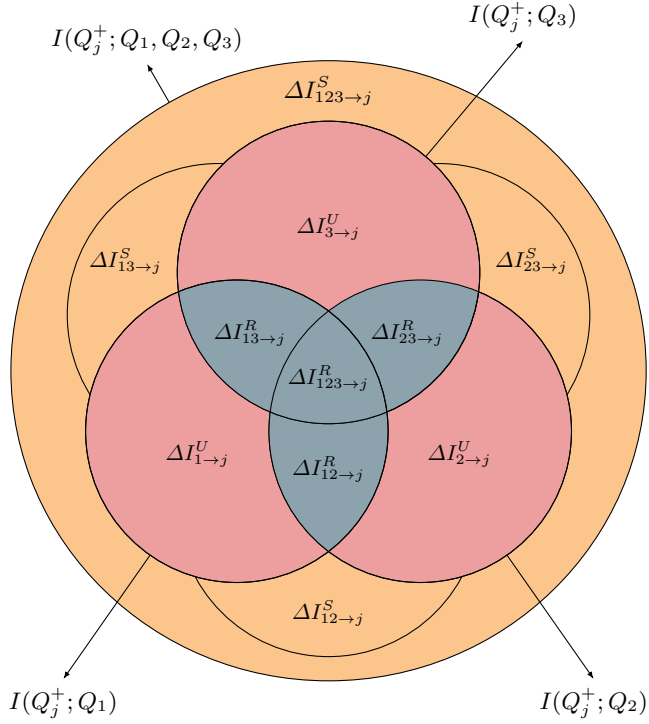
\begin{figure}
  \begin{center}
    \begin{minipage}{1\textwidth}
    \centering
        {

\colorlet{c1}{myc1}
\colorlet{c2}{myc2}
\colorlet{c3}{myc3}

\colorlet{c1s}{c1!50}
\colorlet{c2s}{c2!50}
\colorlet{c3s}{c3!50}

\begin{tikzpicture}[scale=2,thin,>={Latex[length=.1cm]}
]
    
    \begin{small}
    \pgfmathsetmacro{\rdi}{1}
    \pgfmathsetmacro{\rdiO}{.9*\rdi}
    \pgfmathsetmacro{\dis}{.7}

    \coordinate (O3) at (90:\dis);
    \coordinate (O2) at (-30:\dis);
    \coordinate (O1) at (-150:\dis);

    \coordinate (O4) at (270:1.2*\dis);
    \coordinate (O5) at ( 30:1.2*\dis);
    \coordinate (O6) at (150:1.2*\dis);
   
    \draw[thin,->] (0,0) --++ (120:2.4) node[anchor=south]{$I(Q_j^+;Q_1,Q_2,Q_3)$};
    \draw[fill=c3!45] (0,.05) ellipse (2.1 and 2.05 );

    \draw[fill=c3!45] (O4) circle (1);
    \draw[fill=c3!45] (O5) circle (1);
    \draw[fill=c3!45] (O6) circle (1);


    \draw[fill=c2!45] (O1) circle (1);
    \draw[fill=c2!45] (O2) circle (1);
    \draw[fill=c2!45] (O3) circle (1);
    
    \begin{scope}
        \clip (O2) circle (1);
        \draw[fill=c1!45] (O1) circle (1);
    \end{scope}

    \begin{scope}
        \clip (O2) circle (1);
        \draw[fill=c1!45] (O3) circle (1);
    \end{scope}

    \begin{scope}
        \clip (O3) circle (1);
        \draw[fill=c1!45] (O1) circle (1);
    \end{scope}

    \draw[name path=C1] (O1) circle (1);
    \draw[name path=C2] (O2) circle (1);
    \draw[name path=C3] (O3) circle (1);



    \draw[thin,->] (O2)+(-55:\rdi) --++ (-55:2) node[anchor=north]{$I(Q_j^+;Q_2)$};
    \draw[thin,->] (O1)+(235:\rdi) --++ (235:2) node[anchor=north]{$I(Q_j^+;Q_1)$};
    \draw[thin,->] (O3)+( 50:\rdi) --++ ( 50:2) node[anchor=south]{$I(Q_j^+;Q_3)$};

   
    \node[scale=.9] at (90:1*\rdi)   {$\Delta I^U_{3\to j}$};
    \node[scale=.9] at (-30:1*\rdi)  {$\Delta I^U_{2\to j}$};
    \node[scale=.9] at (-150:1*\rdi) {$\Delta I^U_{1\to j}$};

    \node[scale=.9] at ( 30:.6*\rdi) {$\Delta I^R_{23\to j}$};
    \node[scale=.9] at (150:.6*\rdi) {$\Delta I^R_{13\to j}$};
    \node[scale=.9] at (270:.6*\rdi) {$\Delta I^R_{12\to j}$};

    \node[scale=.9] at (0:0) {$\Delta I^R_{123\to j}$};

    \node[scale=.9] at ( 30:1.55*\rdi) {$\Delta I^S_{23\to j}$};
    \node[scale=.9] at (150:1.55*\rdi) {$\Delta I^S_{13\to j}$};
    \node[scale=.9] at (270:1.55*\rdi) {$\Delta I^S_{12\to j}$};

    \node at (90:1.9*\rdi) {$\Delta I^S_{123\to j}$};


    \end{small}
\end{tikzpicture}}
    \end{minipage}
 \end{center}
\caption{Diagram of the decomposition into redundant, unique, and
  synergistic causalities and contributions to total and individual
  mutual information for three observable variables $\bQ = [Q_1, Q_2,
    Q_3]$ and the target $Q_j^+$.}\label{fig:diagram_mutuals}
\end{figure}
  \item \emph{Zero-causality property}. If $Q_j^+$ is independent of
    $Q_i$, then $\Delta I^R_{\bi \rightarrow j} = 0$ for $\bi \in
    \mathcal{C}_i$ and $\Delta I^U_{i \rightarrow j}=0$ as long as
    $Q_i$ is observable.
\item \emph{Invariance under invertible transformations}. The
  redundant, unique, and synergistic causalities are invariant under
  invertible transformations of $\bQ$. This property follows from the
  invariance of the mutual information.
%
\end{itemize}

\subsection{Example of SURD in logic gates}
\label{subsec:simple}

\begin{figure}[h]
  \begin{center}
    \includegraphics[width=\textwidth, trim={0 2cm 0 0}, clip]{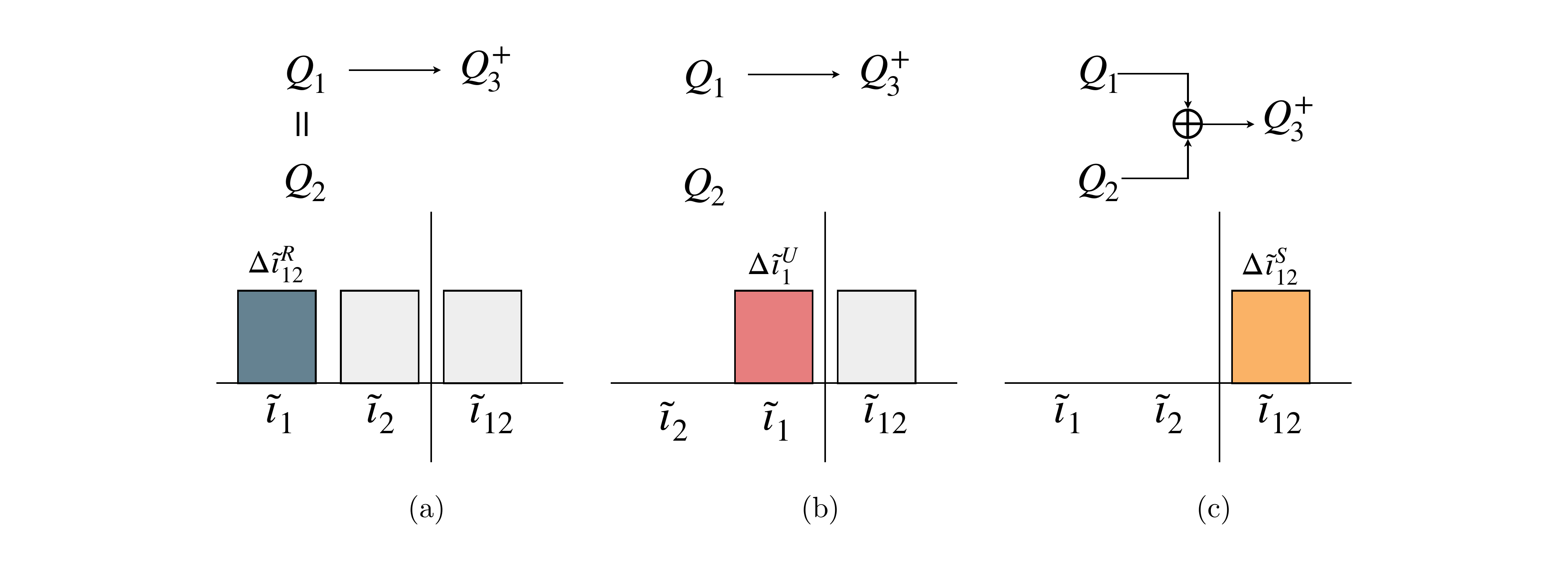}
 \end{center}
\caption{ Schematic of logic gates (top panels) and associated
  specific mutual information (bottom panels) for (a) duplicated input
  (pure redundant causality), (b) output equal to first input (pure
  unique causality), and (c) exclusive-OR output (pure synergistic
  causality). The schematics of the specific mutual information apply
  to both states $Q_3^+=0$ and $Q_3^+=1$.} \label{fig:simple}
\end{figure}

We illustrate the concept of redundant, unique, and synergistic
causality in three simple examples. The examples represent a system
with two inputs $Q_1$ and $Q_2$ and one output $Q_3^+ =
f(Q_1,Q_2)$. Both input and output are binary variables distributed
randomly and independently. The causal description of the system is
characterized by the four components:
\begin{equation}
  \label{eq:simple}
  H(Q_3^+) = \Delta I^U_{1 \rightarrow 3} + \Delta I^U_{2 \rightarrow 3} +
  \Delta I^R_{12 \rightarrow 3} + \Delta I^S_{12 \rightarrow 3},
\end{equation}  
where $\Delta I_{\text{leak} \rightarrow 3} = 0$ as $H(Q_3^+ | Q_1,
Q_2) =0$.  The results for the three cases are summarized in
Figure~\ref{fig:simple}.

The first example represents a system in which $Q_2\equiv Q_1$
(duplicated input) and the output is given by $Q_3^+=Q_1$. In this
case, both $Q_1$ and $Q_2$ provide the same information about the
output and the only non-zero term in Equation~\eqref{eq:simple} is the
redundant causality $\Delta I^R_{12 \rightarrow 3} = 1$ bit.  In the
second example, the output is given by $Q_3^+ = Q_1$ with no
dependence on $Q_2$, which only results in the unique causality
$\Delta I^U_{1\rightarrow 3} = 1$ bit.  In the last example, the
output is given by the exclusive-OR operator: $Q_3^+=Q_1 \oplus Q_2$
such that $Q_3^+ = 1$ if $Q_1\neq Q_2$ and $Q_3^+ = 0$ otherwise. In
this case, the output behaves randomly when observing $Q_1$ or $Q_2$
independently.  However, the outcome is completely determined when the
joint variable $[Q_1,Q_2]$ is considered. Hence, $[Q_1,Q_2]$ contains
more information than their individual components and all the
causality is synergistic $\Delta I^S_{12 \rightarrow 3} = 1$ bit.

\newpage
\section{Other methods for causal inference: description and implementation}

In this section, we discuss the methods for causal inference
implemented in our study. For each method, we provide a description of
the approach, details of the implementation, and its verification in
documented cases.

\subsection{Conditional Granger causality (CGC)}

Granger causality (GC)~\cite{granger1969} is a statistical technique
for assessing causality between two time series. This approach relies
on the ability of past values of one time series, $Q_2(t)$, to enhance
the predictability of another series, $Q_1(t)$. The time signal
$Q_2(t)$ is considered to Granger-cause $Q_1(t)$ if the historical
data of $Q_2(t)$ significantly improves the forecast of $Q_1(t)$. A
foundational principle of Granger causality rests on the assumptions
that the causative factor must provide unique information not
available in the past values of the effect itself, and that this
unique information can be detected via a forecasting model.

GC analysis is typically implemented by means of vector autoregressive
(VAR) modeling of the time series. In the bivariate case, Granger
causality between $Q_1$ and $Q_2$ is assessed by computing the change
in the error between two VAR models:
\begin{align}
    Q_1(t) &= \hat{a}_0 + \sum_{j=1}^{p} {\hat{a}}_{j} {Q_1}({t-\Delta T_j}) + \hat{\varepsilon}(t),\label{eq:gc1}\\
    Q_1(t) &= a_0 + \sum_{j=1}^{p} {a}_{j} {Q_1}({t-\Delta T_j}) + \sum_{j=1}^{p} b_{j} {Q_2} ({t-\Delta T_j}) + \varepsilon(t),\label{eq:gc2}
\end{align}
where ${\hat{a}}_{j}$, ${a}_{j}$ and ${b}_{j}$ are regression
coefficients that represent the influences of the past values of the
time series of $Q_1(t)$ and $Q_2(t)$, $\hat{\varepsilon}(t)$ and
$\varepsilon(t)$ denote the errors of the models at time $t$, $j$ is
the lag into the past used to predict the values of $Q_1(t)$ and $p$
represents the maximum time lag used. Granger causality from $Q_2$ to
$Q_1$ is defined as a measure of the extent to which inclusion of
$Q_2$ in the second model (\ref{eq:gc2}) reduces the prediction error
of the first model (\ref{eq:gc1}). The standard measure of Granger
causality is given by the natural logarithm of the ratio of the
residual variance between both models:
\begin{equation}
\label{eq:gcf}
    \mathrm{GC}_{2\rightarrow 1} = \log_2{
    \left( 
    \frac{
    {\mathrm{var}} \left(\hat{\varepsilon}\right)
    }
    {
    {\mathrm{var}} \left({\varepsilon}
    \right)
    } 
    \right) \geq 0
    }.
\end{equation}
Therefore, if the past values of $Q_2(t)$ improve the prediction of
$Q_1(t)$, the residual variance of the second model will be smaller
than that of the first model, i.e., ${\mathrm{var}} \left({\varepsilon}
\right) < {\mathrm{var}} \left(\hat{\varepsilon}\right) $, and the Granger
causality measure will be greater than zero,
$\mathrm{GC}_{2\rightarrow 1} > 0$. Conversely, if the residual
variance between both models is exactly the same after introducing
$Q_2$ in the model, i.e., ${\mathrm{var}} \left({\varepsilon} \right) =
{\mathrm{var}} \left(\hat{\varepsilon}\right) $, the past values of
$Q_2(t)$ do not improve the prediction of $Q_1(t)$ and the Granger
causality measure will be zero, $\mathrm{GC}_{2\rightarrow 1} = 0$.

In this work, we use the extension first proposed by
Geweke~\cite{geweke1984}, in which both models use an additional
vector of variables, $\bQ'(t)=[Q_3(t), Q_4(t),\dots,Q_N(t)]$, which do
not contain $Q_1$ and $Q_2$. The method is referred to as conditional
Granger causality (CGC) and allows us to compute the Granger causality
value from $Q_2$ to $Q_1$ conditioned on $\bQ'$ by comparing the errors
from:
\begin{align}
    Q_1(t) &= \hat{a}_0 +\sum_{j=1}^{p} {\hat{a}}_{j} {Q_1}({t-\Delta T_j}) + \sum_{j=1}^{p}{\hat{\boldsymbol{c}}}_{j} {\bQ'}({t-\Delta T_j}) + \hat{\varepsilon}(t) ,\label{eq:cgc1}\\
    Q_1(t) &= a_0 + \sum_{j=1}^{p} {a}_{j} {Q_1}({t-\Delta T_j}) + \sum_{j=1}^{p} b_{j} {Q_2} ({t-\Delta T_j}) + \sum_{j=1}^{p}{\boldsymbol{c}}_{j} {\bQ'}({t-\Delta T_j}) +\varepsilon(t),\label{eq:cgc}
\end{align}
where ${\hat{\boldsymbol{c}}}_{j}$ and ${\boldsymbol{c}}_{j}$ are the
regression coefficient matrices associated with $\bQ'(t-\Delta T_j)$.
Causality is then computed as:
\begin{equation}
\label{eq:cgcf}
    \mathrm{CGC}_{2\rightarrow 1} = \log_2{
    \left( 
    \frac{
    {\mathrm{var}} \left(\hat{\varepsilon}\right)
    }
    {
    {\mathrm{var}} \left({\varepsilon}
    \right)
    } 
    \right) \geq 0
    }.
\end{equation}
The CGC formulation has been extensively investigated and further
developed in a multivariate setting--the Multivariate Granger
causality (MVGC)~\cite{Barnett2015}.

\begin{figure}
    \centering
    \begin{tikzpicture}
    
      \node[anchor=south west,inner sep=0] (image1) at (0,0) {\includegraphics[width=0.4\textwidth]{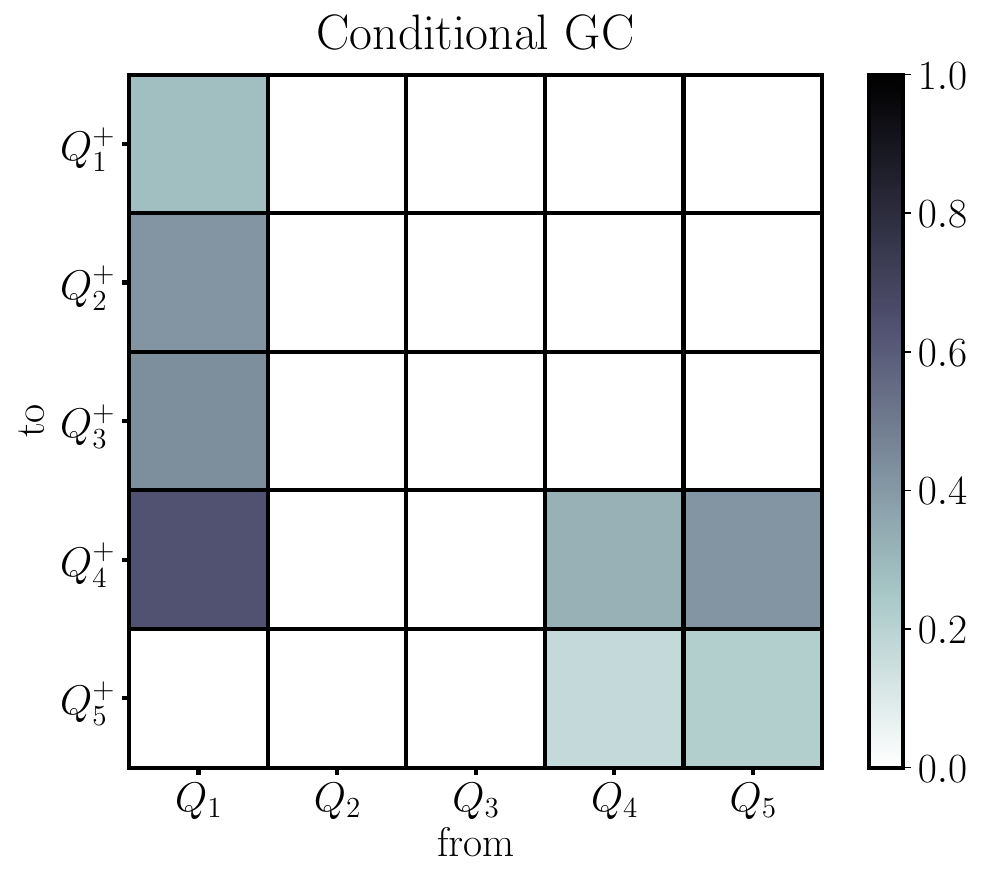}};
      \begin{scope}[x={(image1.south east)},y={(image1.north west)}]
        \fill[white] (0,0.925) rectangle (0.8,1);
        \node[fill=white, anchor=north, yshift=-0cm] at (0.48,1) {CGC};
      \end{scope}
    
      \node[inner sep=0] (space) at (0.45\textwidth, 0.5) {};
    
      \node[anchor=south west,inner sep=0] (image2) at (0.5\textwidth,0) {\includegraphics[width=0.4\textwidth]{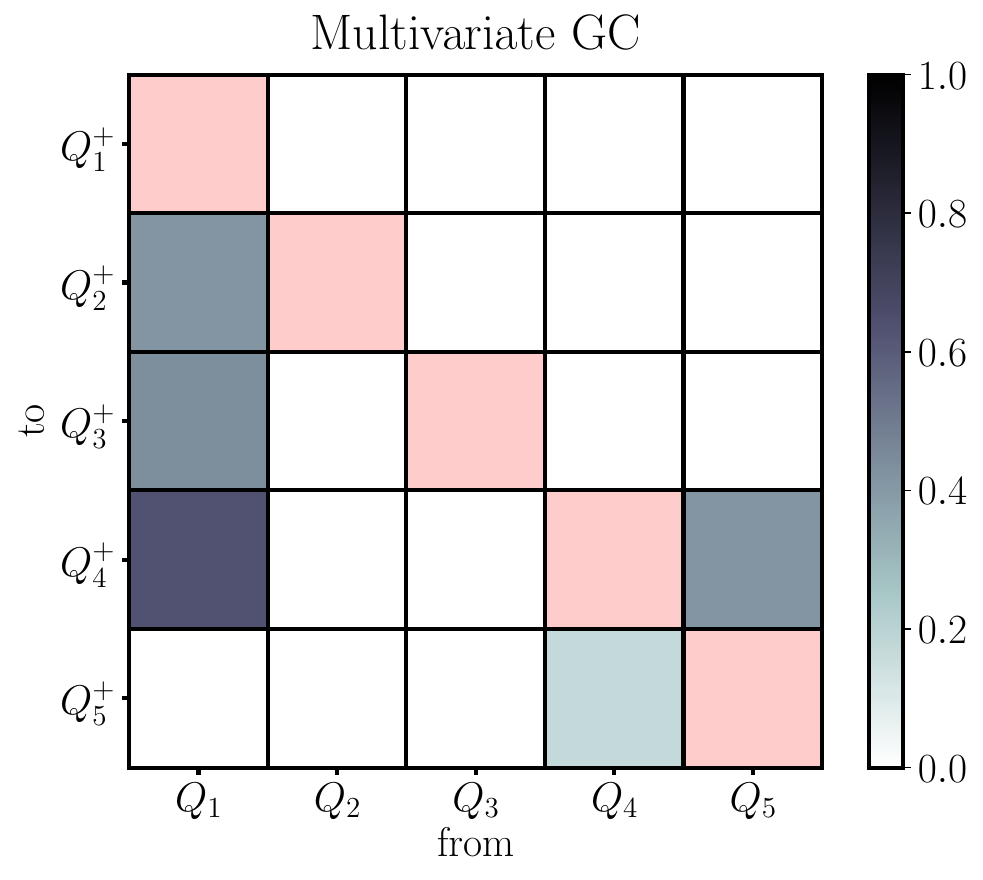}};
      \begin{scope}[x={(image2.south east)},y={(image2.north west)}]
        \fill[white] (0,0.925) rectangle (0.4,1);
        \node[fill=white, anchor=north, yshift=-0.cm] at (0.216,1) {MVGC};
      \end{scope}
    \end{tikzpicture}
    \caption{Comparison of current implementation of conditional
      Granger causality (CGC) and multivariate Granger causality
      toolbox (MVGC)~\cite{barnett2014} for a five-node oscillatory
      network. The MVGC does not provide self-causality values and the
      diagonal is colored in light pink.  Note that in this case the results
      from CGC and MVGC have not been normalised, as compared with the results 
      presented in the main text. 
        }
    \label{fig:cgc vs mvgc}
\end{figure}

We have validated our implementation of CGC with the MVGC
toolbox~\cite{barnett2014}. Here, we show as an example the five-node
oscillatory network given by the following
equations~\cite{barnett2014}: 
%
\begin{align}
  \label{eq:five-nodes}
  \begin{split}
Q_1(n+1) &= \left(0.95\sqrt{2} - 0.9025\right) Q_1(n) + \eta_1(n)
\\ Q_2(n+1) &= 0.5 Q_1(n) + \eta_2(n) \\ Q_3(n+1) &= -0.4 Q_1(n) +
\eta_3(n) \\ Q_4(n+1) &= -0.5 Q_1(n) + 0.25\sqrt{2} Q_4(n) +
0.25\sqrt{2} Q_5(n) + \eta_4(n) \\ Q_5(n+1) &= -0.25\sqrt{2} Q_4(n) +
0.25\sqrt{2} Q_5(n) + \eta_5(n),
\end{split}
\end{align}
where $\eta_i(t)$ for $i=1,2,3,4,5$ represents Gaussian white noise
processes with variances $\sigma^2_i = \left[0.6, 0.5, 0.3, 0.3,
  0.6\right]$. The results, shown in Figure~\ref{fig:cgc vs mvgc},
indicate that both CGC and the MVGC toolkit yield identical values
that correctly identify the cross-induced causalities. Our
implementation of CGC also computes the self-induced causalities,
although MVGC does not provide this information.

It is known that CGC is subject to some important limitations
\cite{camps2023}. First, CGC cannot account for hidden confounding
effects or capture non-linear causal relationships. This limitation
stems from the assumed linear relationship between variables, which
may not be appropriate for complex nonlinear systems. Several
nonlinear extensions of Granger causality have been proposed in the
literature \cite{Hiemstra1994, Bell9967, abhyankar1998,ancona2004,
  tissot2014,tissot2014, bueso2020}, although their adoption is much
more limited than that of the linear counterpart.

\subsection{Convergent cross-mapping (CCM)}

Convergent cross-mapping (CCM)~\cite{sugihara2012} is a statistical
method for causal inference grounded in the theory of dynamical
systems. The approach relies on Takens' embedding
theorem~\cite{Takens1981}, which states the conditions under which the
dynamics of chaotic nonlinear systems can be captured by observing the
trajectory of a single variable over time.

Consider two time series \(Q_1(t)\) and \(Q_2(t)\). We can define the
vector of delay coordinates for $Q_1$ (similarly for $Q_2$) as \(\bQ_1
= [Q_1(t), Q_1(t - \Delta T),..., Q_1(t - (E-1)\Delta T)]\), where
\(\Delta T\) is the time lag, and \(E\) is the embedding dimension
that determines the number of time lagged observations.  The
lagged-coordinate embeddings $\bQ_1$ and $\bQ_2$ lie within the shadow
manifolds \(M_{1}\) and \(M_{2}\), respectively.  A consequence of
Takens' embedding theorem is that if \(Q_1\) and \(Q_2\) belong to the
same manifold, then \(M_{1}\) and \(M_{2}\) are topologically
equivalent.  CCM leverages this property and states that if \(Q_1\)
causally influences \(Q_2\), then local neighborhoods on $M_1$ should
correspond to local neighborhoods on $M_2$.

The reconstruction of \(Q_2(t)\) from the manifold of \(Q_1\), denoted as
\(\hat{Q}_2(t)|_{M_{1}}\), is calculated as a weighted average:
\begin{equation}
    \hat{Q}_2(t)|_{M_{1}} = \sum_{i=1}^{E+1} w_i Q_2(t_i),
\end{equation}
where \(t_i\) represent the time indices of the nearest neighbors in
\(M_{1}\), and \(w_i\) are the weights.  The process of estimating
$Q_2(t)$ from $\hat{Q}_2(t)|_{M_{1}}$ is referred to as cross mapping.
The weights are computed as
\begin{equation}
  w_i = \frac{u_i}{\sum_{j=1}^{E+1} u_j}, \quad u_i =
  \exp\left(-\frac{d(\bQ_1(t), \bQ_1(t_i))}{d(\bQ_1(t),
    \bQ_1(t_1))}\right),
\end{equation}
where \(d(\bQ_1(t), \bQ_1(t_i))\) is the Euclidean distance between
the lagged-coordinate vectors in \(M_{1}\), and \(d(\bQ_1(t),
\bQ_1(t_1))\) is the distance to the nearest neighbor, serving as a
normalization factor to scale the weights. This ensures that the
prediction is most influenced by the states closest to the current
state, whereas the effect of more distant states decays
exponentially. The causality from \(Q_2\) to \(Q_1\) is evaluated by
examining the correlation coefficient between the actual and estimated
values of \(Q_2\) across the timeline:
\begin{equation}
  \mathrm{CCM}_{2\rightarrow 1} = \mathrm{corr}( Q_2, \hat{Q}_2(t)|_{M_{1}} ).
\end{equation}
where a causal link is detected for \(\mathrm{CCM}_{2\rightarrow 1}
\rightarrow 1\) as the length of the time series increases. An analogous definition applies to
\(\mathrm{CCM}_{1\rightarrow 2}\) using \(Q_1(t)\) and
\(\hat{Q}_1(t)|_{M_{2}}\).

The assessment of causality in CCM is predicated upon the convergence
of \(\mathrm{CCM}_{1\rightarrow 2}\) as the length of the time series
$N$ increases. As \(N\) grows, the shadow manifolds become more
densely populated, leading to a reduction in the distances among the
\(E+1\) nearest neighbors. This increased density enables more
accurate predictions as \(\hat{Q}_2(t)|_{M_1}\) converges to
\(Q_2(t)\). Observing the convergence of the nearest neighbors and the
resulting improvement in prediction accuracy is crucial for
substantiating claims about the influence of one variable on another
within the framework of CCM. Generally, the stronger the causal link
between variables, the faster the convergence with \(N\).

In this work, we employ the Python implementation of CCM by
Erneszer~\cite{causalccm}. The method was used with an embedding
dimension equivalent to the number of variables of the system and executed
with a library size that ensured convergence of the prediction
capabilities for all cases. In this section, we illustrate the
assessment of causality from CCM for the two test cases used by
Sugihara \textit{et al.}~\cite{sugihara2012}.  The system used
consists of a nonlinear logistic difference system with constant
coefficients that exhibits chaotic behaviour and represents the
phenomenon of mirage correlation. The dynamical system is given by:
\begin{subequations}
\label{eq: sugihara1}
\begin{align}
    Q_1(n+1) &= Q_1(n) \left[r_1 - r_1 Q_1(n) - \beta_{2\rightarrow 1} Q_2(n) \right], \\
    Q_2(n+1) &= Q_2(n) \left[r_2 - r_2 Q_2(n) - \beta_{1\rightarrow 2} Q_1(n) \right].
\end{align}
\end{subequations}
For certain combinations of the coefficients, the variables of this
system can be positively coupled for long periods of time and can
spontaneously become anticorrelated or decoupled. This can become
challenging when fitting models for the variables as in the Granger
causality framework. 

\begin{figure}
    \centering
    \begin{tikzpicture}
        \node[anchor=south west,inner sep=0] (image1) at (0,0) {\includegraphics[width=0.425\textwidth]{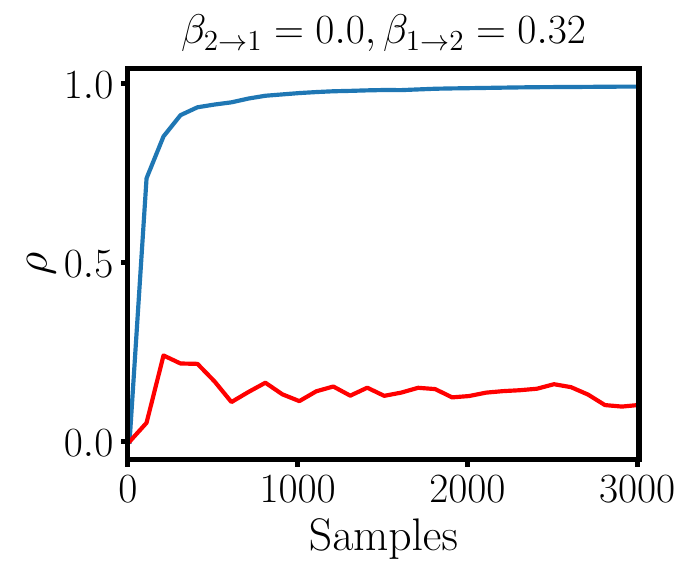}};
        \node[anchor=south west,inner sep=0] (image2) at (0.5\textwidth,0) {\includegraphics[width=0.425\textwidth]{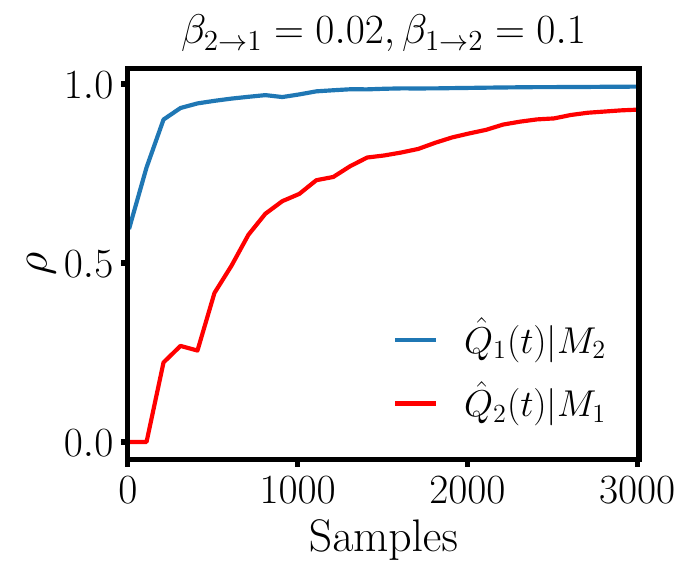}};

        \node[rotate=90, anchor=south, fill=white] at ([yshift=-27mm,xshift=7mm]image1.north west) {\large\(\mathrm{CCM}_{i\rightarrow j}\)};
        \node[rotate=90, anchor=south, fill=white] at ([yshift=-27mm,xshift=7mm]image2.north west) {\large\(\mathrm{CCM}_{i\rightarrow j}\)};

        \node[anchor=south, fill=white] at ([yshift=-40.1mm,xshift=58mm]image2.north west) {\large \(\mathrm{CCM}_{1\rightarrow 2}\)};
        \node[anchor=south, fill=white] at ([yshift=-47.25mm,xshift=58mm]image2.north west) {\large \(\mathrm{CCM}_{2\rightarrow 1}\)};
    \end{tikzpicture}
    \caption{Convergent cross-mapping for two cases in the nonlinear
      logistic difference system from Equation~\eqref{eq:
        sugihara1}. \(\mathrm{CCM}_{i\rightarrow j}\) is the
      prediction of cross-map estimates represented by the correlation
      coefficient as a function of the number of samples.  (Left)
      \(\mathrm{CCM}_{i\rightarrow j}\) for the system with
      \(\beta_{2\rightarrow 1} = 0\) and \(\beta_{1\rightarrow 2} =
      0.32\) (i.e., \(Q_2\) has no effect on \(Q_1\)) and (right)
      \(\beta_{2\rightarrow 1} = 0.02\) and \(\beta_{1\rightarrow 2} =
      0.1\) (i.e., the effect of \(Q_1\) on \(Q_2\) is stronger than
      in the previous case). }
    \label{fig:ccm convergence}
\end{figure}

We analyze two cases proposed by Sugihara et
al.~\cite{sugihara2012}. The results, shown in Figure \ref{fig:ccm
  convergence}, aim to illustrate the role of convergence in CCM. For
the first case, the variable $Q_1$ is decoupled from $Q_2$, while
$Q_2$ is driven by $Q_1$, namely $\beta_{2\rightarrow 1} = 0$ and
$\beta_{1\rightarrow 2} \neq 0$, respectively. The results for this
case are displayed in Figure~\ref{fig:ccm convergence}(left), where
the aforementioned dependencies are clearly captured. For the second
model, both variables are coupled, but $\beta_{1\rightarrow 2} >
\beta_{2\rightarrow 1}$. In this scenario, observing the evolution of
prediction accuracy can aid in understanding the influence of each
variable on the other. Figure~\ref{fig:ccm convergence}(right)
demonstrates how the lower coupling coefficient $\beta_{2\rightarrow
  1}$ leads to a lesser prediction accuracy as the sample size
increases: cross mapping $Q_1$ using $M_2$ converges faster than cross
mapping $Q_2$ using $M_1$. Consequently, the causality from variable
$Q_2$ to $Q_1$ is weaker compared to that from $Q_1$ to $Q_2$.

CCM is particularly suited for data measured from a deterministic
nonlinear attractor. For time series of stochastic nature, CCM is
known to underperform~\cite{runge2018chaos}. Additionally, the
presence of noise in the signals complicates the reconstruction
process of the attractor manifold and could reduce the reliability of
the CCM, as observed in this work and previously reported in the
literature~\cite{cobey2016, monster2017, runge2018chaos}. Moreover,
CCM fails to accurately predict the causality direction in cases where
the coupling is strong enough to lead to the synchronization of
variables.

Finally, this study focused on the original CCM algorithm introduced
by Sugihara \textit{et al.}~\cite{sugihara2012}. However, more recent
algorithms have been developed to examine the impact of noise,
external signals, and synchronisation of variables such as pairwise
asymmetric inference (PAI)~\cite{pai2014}, multispatial CCM
(MCCM)~\cite{mccm2015}, partial cross-mapping (PCM)~\cite{pcm2020},
and latent CCM \cite{lccm2021}.

\subsection{Peter–Clark algorithm with momentary conditional independence test (PCMCI)}


Conditional independence-based methods uncover the causal structure of
interactions, often represented as directed acyclic graphs, by
examining the conditional dependencies among
variables~\cite{runge2023comment}. The core idea is that if two
variables are conditionally independent given a set of other
variables, then there is no direct causal link between them.
%
The approach was popularized by the Peter–Clark (PC)
algorithm~\cite{spirtes1991}, with subsequent extensions incorporating
tests for momentary conditional independence
(PCMCI)~\cite{runge2019pcmci}. PCMCI algorithms are formulated under
two main assumptions: (i) the causal Markov condition, which assumes
that a variable is independent of its non-descendants, given its
parents in a causal graph, and (ii) the faithfulness assumption, which
posits that if two variables are statistically independent, then they
must be conditionally independent given some set of variables in the
causal network. Here, the parents of a particular variable are all
those variables in the causal graph that have a direct arrow pointing
to it, whereas non-descendant refers to any variable in the graph that
is not a direct or indirect outcome of the given variable.

Contrary to methods that condition on the entire past of all
processes, PCMCI seeks to identify a reduced set of conditioning
variables that includes, at a minimum, the parents of the target
variable. These parents are associated with the particular time lag at
which the causal relationship occurs in the system. The algorithm
unfolds in two phases:
\begin{itemize}
\item The initial phase, based on the PC algorithm, is a selection
  stage aiming to infer a superset of the parents of each variable
  $Q_{j}$ at time $t$, denoted as $\hat{\mathcal{P}}[Q_{j}(t)]$. This phase
  starts with a fully connected graph and tests the independence of
  $Q_{i}(t-\Delta T)$ and $Q_{j}(t)$, given conditioning sets of
  increasing size. The goal of this phase is the removal of irrelevant
  links and it is designed to have an initial estimate of the parents
  of each variable $Q_j$, namely $\hat{\mathcal{P}}[Q_{j}(t)]$. 
  
    %
  \item The second phase of the algorithm conducts the momentary
    conditional independence (MCI) test for each pair of variables
    $Q_i (t - \Delta T)$ and $Q_j(t)$ using the estimated parents $\hat{\mathcal{P}}[Q_{i}(t - \Delta T )]$ and $\hat{\mathcal{P}}[Q_{j}(t)]$ as conditions. The test examines the null
    hypothesis at a significance threshold $\alpha_{\rm{MCI}}$:
    \begin{equation} \label{eq: pcmci}
        Q_i (t - \Delta T) \!\perp\!\!\!\perp Q_j(t) |
        \hat{\mathcal{P}}\left[Q_j(t)\right] \setminus Q_i (t - \Delta T),
        \hat{\mathcal{P}}\left[Q_i (t - \Delta T)\right],
    \end{equation}
    which denotes the conditional independence of $ Q_i(t - \Delta T)$
    and $ Q_j(t) $, given the causes (or parents) of $Q_j(t)$
    excluding $Q_i(t - \Delta T)$, and the causes of $ Q_i(t - \Delta
    T)$. If this hypothesis is not rejected at a significance threshold $\alpha_{\rm{MCI}}$, the causal link between
    $Q_{i}(t-\Delta T)$ and $Q_j(t)$ is removed. This phase
    effectively eliminates autodependencies and controls false
    positives, while improving detection power compared to other
    adaptations of the PC algorithm. 
\end{itemize}

In our study, we use the Python implementation of PCMCI provided by
the package \emph{Tigramite}. It is worth noting that PCMCI
accommodates different independence tests, such as partial
correlation, nonlinear two-step conditional independence test, and a
fully non-parametric test based on conditional mutual information
(CMI). For comparison purposes with SURD, we selected the CMI with the
$k$-nearest neighbor ($k$-NN) estimator. The causal strength of the causal link is determined by the statistic value of the test in Equation \ref{eq: pcmci}:
\begin{equation}
  \text{PCMCI}_{i_\rightarrow j} = I\left(
  Q_i (t - \Delta T) ; Q_j(t) | C \right),
\end{equation}
where $C = \hat{\mathcal{P}}\left[Q_j(t)\right] \setminus Q_i (t - \Delta T),
        \hat{\mathcal{P}}\left[Q_i (t - \Delta T)\right]$ and $I(\cdot,\cdot |
\cdot)$ is the conditional mutual information. $\text{PCMCI}_{i_\rightarrow j}$ is non-zero only if the null hypothesis is not rejected at a certain significance threshold. All cases examined in
this study were evaluated at a significance level of 1\%. Furthermore,
PCMCI was estimated with $\alpha_{\text{PC}} = 0.05$ and CMI-$k$NN
parameters $k_{\text{CMI}} = 0.1$, $k_{\text{sn}} = 5$, and $B=200$
permutation surrogates. The parameter $\alpha_{\text{PC}}$ denotes the
signficance level for the parent selection phase of the algorithm,
$k_{\text{CMI}}$ determines the size of hypercubes, i.e., the
data-adaptive local length-scale used in the $k$-NN estimator,
$k_{\text{sn}}$ denotes the number of neighbours to which each sample
is mapped randomly, and $B$ the number of surrogates to approximate the null distribution. More details about these parameters and their
effect on the PCMCI algorithm are provided in
Ref.~\cite{runge2018chaos}.
 
\begin{figure}
    \centering
    \subfloat[]{
    \begin{minipage}{0.3\textwidth}
      \vspace{-6.8cm}
      \includegraphics[trim={0 1cm 0 0},clip,width=\textwidth]{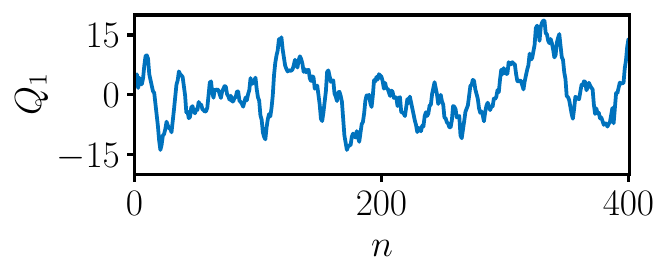}
      \includegraphics[trim={0 1cm 0 0},clip,width=\textwidth]{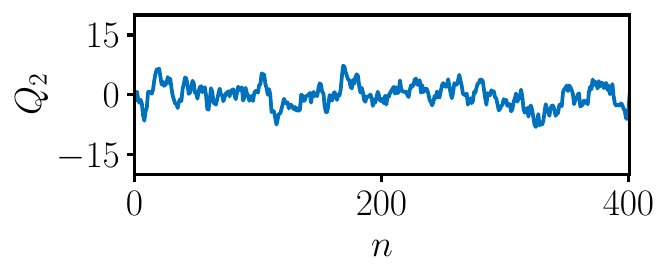}
      \includegraphics[trim={0 1cm 0 0},clip,width=\textwidth]{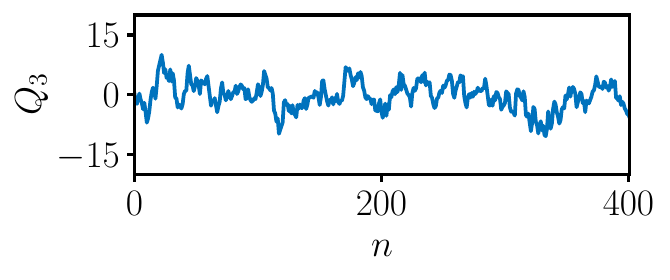}
      \includegraphics[trim={0 0 0 0},clip,width=\textwidth]{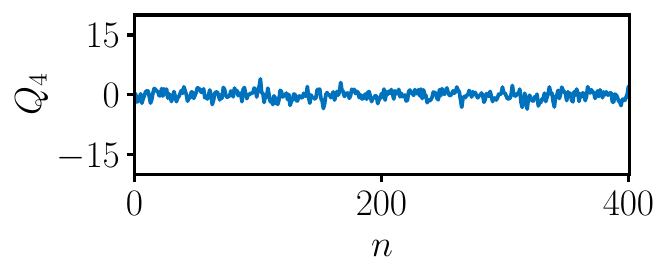}
    \end{minipage}
    }
    \hspace{0.005\textwidth}
    \subfloat[]{
    \begin{minipage}{0.305\textwidth}
      \vspace{-7.5cm}      
      \includegraphics[width=\textwidth]{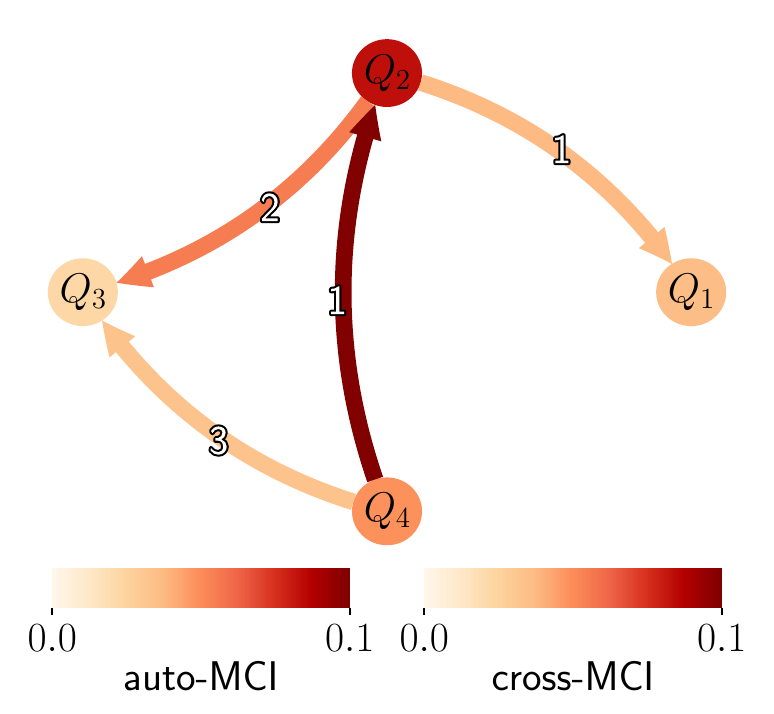}      
    \end{minipage}
    }
    \hspace{0.005\textwidth}
    \subfloat[]{
    \includegraphics[width=0.33\textwidth]{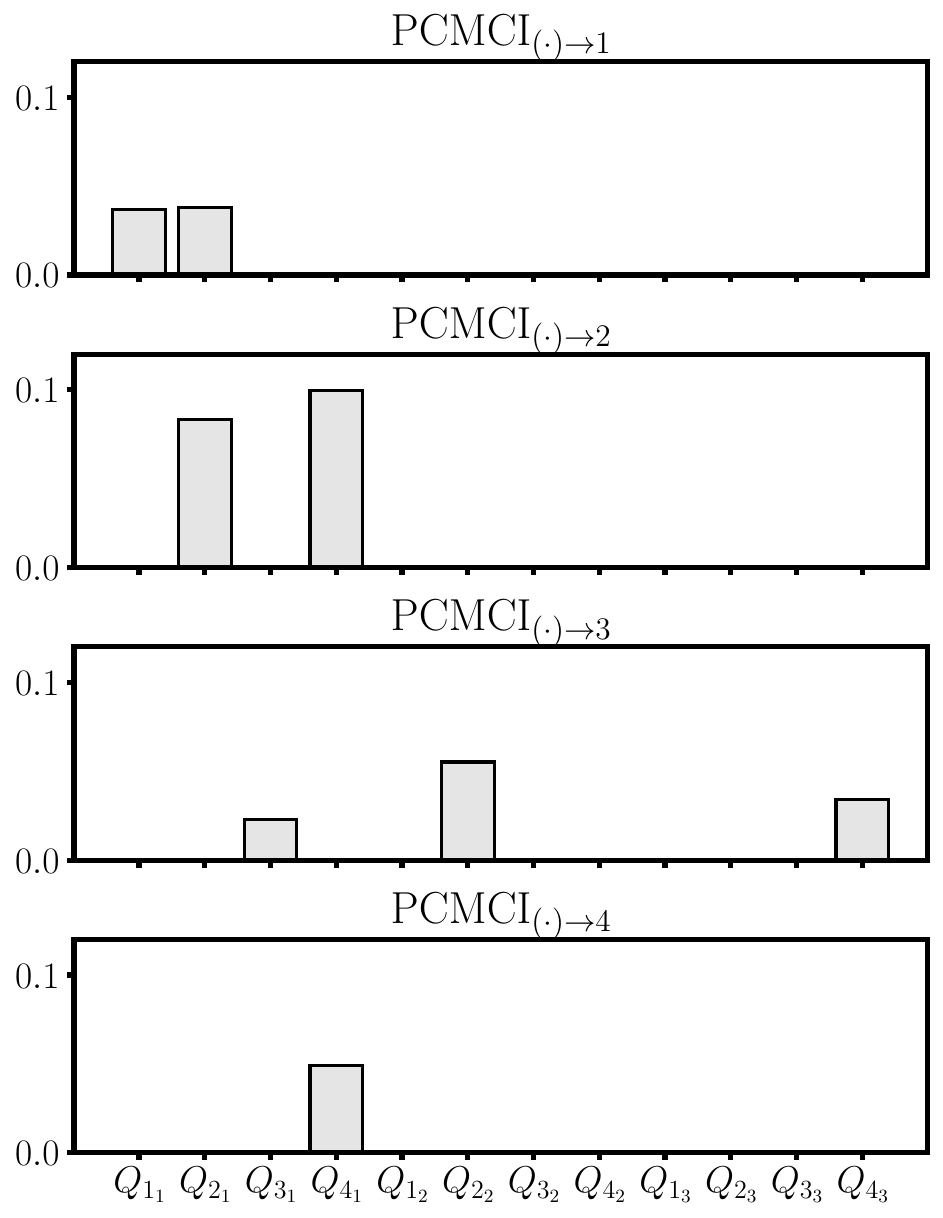}
    }
    \caption{Validation for PCMCI. (a) Time series evolution for the
      signals of the system. (b) Graphical result of the causal
      connections identified by PCMCI. The indices on top of the
      arrows represent the time lag at which the causal links are
      identified.  (c) Representation of PCMCI results in bar
      format. The first subindex represents the variable, and the
      second subindex represents the time lag. The height of the bar in (c) represents the color of the links in (b).}
    \label{fig:validation pcmci}
\end{figure}

The graphical results for PCMCI are now reported for the test case
provided by the package \emph{Tigramite}. The system is a set of time
series given by:
\begin{align*}
Q_1(n) &= 0.7Q_1(n-1) - 0.8Q_2(n-1) + \eta_1(n), \\
Q_2(n) &= 0.8Q_2(n-1) + 0.8Q_4(n-1) + \eta_2(n), \\
Q_3(n) &= 0.5Q_3(n-1) + 0.5Q_2(n-2) + 0.6Q_4(n-3) + \eta_3(n), \\
Q_4(n) &= 0.7Q_4(n-1) + \eta_4(n),
\end{align*}
where $\eta$ are independent Gaussian
variables. PCMCI results for this system are provided in
Figure~\ref{fig:validation pcmci}, where we also introduce the
notation and organization of the results from PCMCI used in our study.

Finally, we discuss some of the limitations of the PCMCI
algorithm. First, the method relies on the selection of parameters
during the parent-selection and link-selection steps. The presence or
absence of a link can be affected by the confidence level used during
these steps. Certain parameters used in the $k$-NN estimator, such as
permutation surrogates, might also affect the results: lower values
alleviate computational cost; however, they could lead to reduced
confidence in the results.  The choice of independence test may also
impact the results depending on the type of relationships in the
data. The CMI is conceptually the most reliable approach but also
computationally expensive and data demanding, particularly for a large
number of neighbors and permutations. A summary of the results presented in this work for different independence tests is reported in Table \ref{tab:summary pcmci}. Lastly, the presence of
redundant variables in the conditioning set can yield to unidentified
links, even when the conditioning set encompasses the entire history
of all other processes~\cite{runge2019pcmci}, as demonstrated in the
examples from this work.

\subsection{Conditional transfer entropy (CTE)}

Transfer entropy (TE) is a method for assessing the directional
information transfer between two time series in a non-parametric
manner~\cite{schreiber2000}. The TE between two time signals $Q_1$ and
$Q_2$ is defined as follows:
\begin{equation} \label{eq: te}
      \mathrm{TE}_{2\rightarrow 1} = H \left(Q_{1} (t) | \bQ_{1}^{(k)}\right) - H \left(Q_{1}(t) | \bQ_{1}^{(k)}, \bQ_{2}^{(l)}\right),
\end{equation}
where $H(\cdot \mid \cdot)$ is the conditional Shannon entropy defined
in Equation~\eqref{eq:entropy conditional}, $k$ and $l$ are constants
denoting the time lag of the variables, $\bQ_{1}^{(k)} = Q_1(t-\Delta
T), \dots, Q_1(t-k \Delta T)$ and $\bQ_{2}^{(l)} = Q_2(t-\Delta T),
\dots, Q_2(t-\Delta T)$.  TE measures the amount of information about
the future state of $Q_2$ that is exclusively provided by the current
state of $Q_1$, beyond what is already known from the past of $Q_2$
itself. The method is particularly effective for analyzing complex
systems where traditional linear methods may not be suitable, such as
in the study of nonlinear dynamics.

The multivariate extension of TE is given by the conditional transfer
entropy (CTE)~\cite{verdes2005, lizier2008, lizier2010, Lozano2022}.
The approach is usually formulated for one single time lag as:
\begin{equation}
    {\mathrm{CTE}}_{\bi\rightarrow j}(\Delta T) = H(Q_j^+ |
    \bQ_{\cancel{\bi}}) - H(Q_j^+ | \bQ),
\end{equation}
where $\bQ=[Q_1(t-\Delta T),Q_2(t-\Delta T),\dots,Q_N(t-\Delta T)]$
represents the vector of observed variables and $\bQ_{\cancel{\bi}}$
denotes the same vector excluding the components given by the indices
in $\bi$. In this work, we have restricted our analysis to
applications of CTE with one single component in $\bi$. The results of
CTE using combinations of variables is discussed below in Figure~\ref{fig:info flux bblocks}.

Barnett \textit{et al.}~\cite{barnett2009} showed that conditional
Granger causality and CTE are equivalent up to a factor of 2 when the
variables in $\bQ$ follow a joint multivariate Gaussian distribution:
\begin{equation}
  \label{eq:CGC_factor}
    \mathrm{CTE}_{i\rightarrow j} = \frac{1}{2} \mathrm{CGC}_{i\rightarrow j}.
\end{equation}
Consequently, both measurements of causality share the same upper
bound, given by the maximum value of the mutual information between
the target and the vector of observed variables $\bQ$. Throughout our
study, $\mathrm{CGC}_{i\rightarrow j}$ is divided by 2 to allow for
direct comparisons with CTE.

\begin{figure}
    \centering
    \subfloat[]{
    \includegraphics[width=0.95\textwidth]{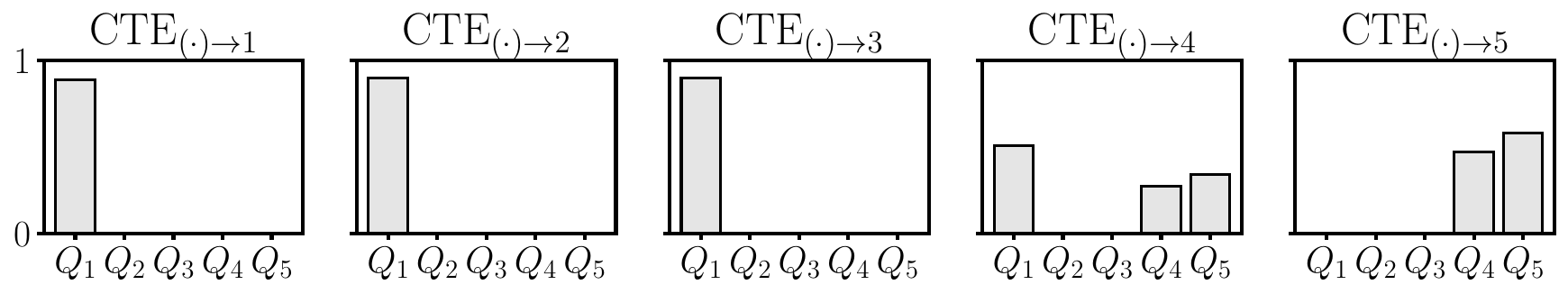}
    }\\
    \subfloat[]{
    \includegraphics[width=0.95\textwidth]{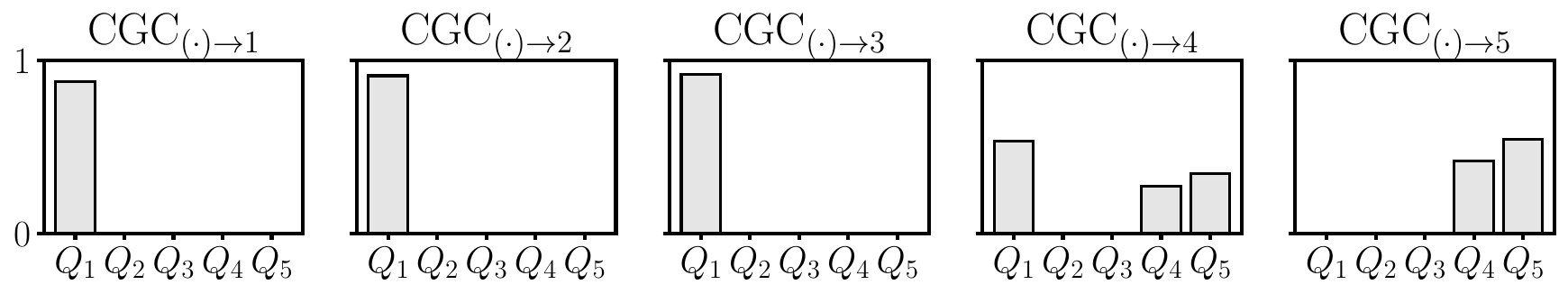}
    }
    \caption{Comparison of current implementation of conditional
      transfer entropy (CTE) and conditional Granger causality (CGC)
      for a five-node oscillatory
      network~\cite{barnett2014}. $\mathrm{CGC}_{i\rightarrow j}$ is
      divided by 2 for consistency with $\mathrm{CTE}_{i\rightarrow
        j}$. Both quantities are then normalized with the mutual
      information between the target and the vector of observed
      variables. The system is integrated for $N=10^8$ steps and the
      phase space was partitioned using 50 bins for each variable to
      calculate CTE.}
    \label{fig:cte vs mvgc}
\end{figure}

We validate the current implementation of CTE in the five-node
oscillatory network~\cite{barnett2014} from
Equation~\eqref{eq:five-nodes}. In this case, CTE and CGC should yield
similar results, as the variables are Gaussian distributed. The
results for CTE and CGC are shown in Figure~\ref{fig:cte vs mvgc},
where both methods provide same causal links with almost identical
strengths.
%

While CTE offers a robust framework for causal inference in time
series data, its practical application requires careful consideration
of sample size and estimation methods. Various approaches, such as
binning, nearest neighbors, and non-uniform embedding, are employed to
effectively estimate the probabilities in entropy
calculations. Regarding the effect of multiple variables, some
attempts have been made in the literature to account for causality
from combinations of variables, for example, through the calculation
of CTE in its multivariate form~\cite{Lozano2022}. However, these
methods may yield negative values of causality, which can limit the
interpretability of the results. Examples are discuss below in Figure
\ref{fig:info flux bblocks}. 
 
\section{Additional validation cases}

We discuss additional validation cases for SURD, CGC, CTE, CCM, and
PCMCI. The metric for success is based on whether the results are
consistent with the functional dependencies of the system, rather than
on the concrete value of the causal strength provided by each method.

\subsection{Lotka--Volterra prey-predator model}

\begin{figure}[t!]
    \centering
    \begin{minipage}{0.31\textwidth}
        \includegraphics[width=\textwidth]{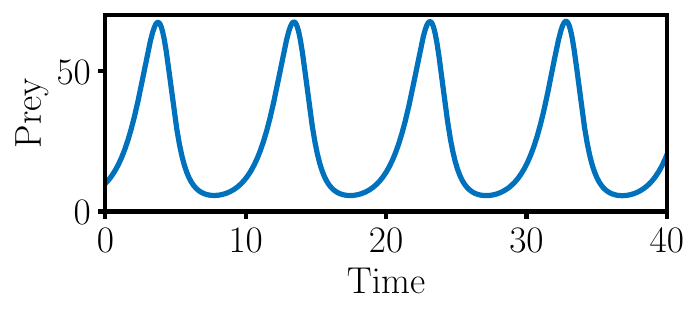}
        \includegraphics[width=\textwidth]{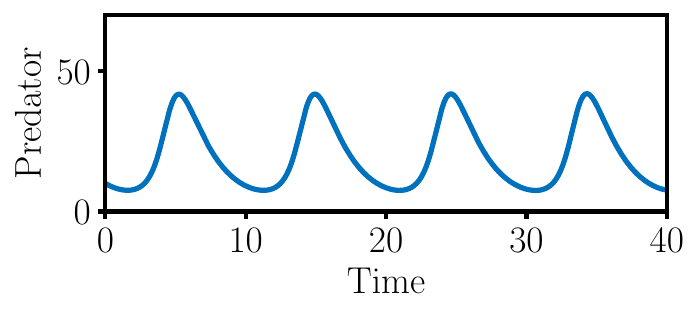}
    \end{minipage}
    \hspace*{0.001\textwidth}
    \begin{minipage}{0.215\textwidth}
    \vspace{-0.5cm}
        \includegraphics[width=\textwidth]{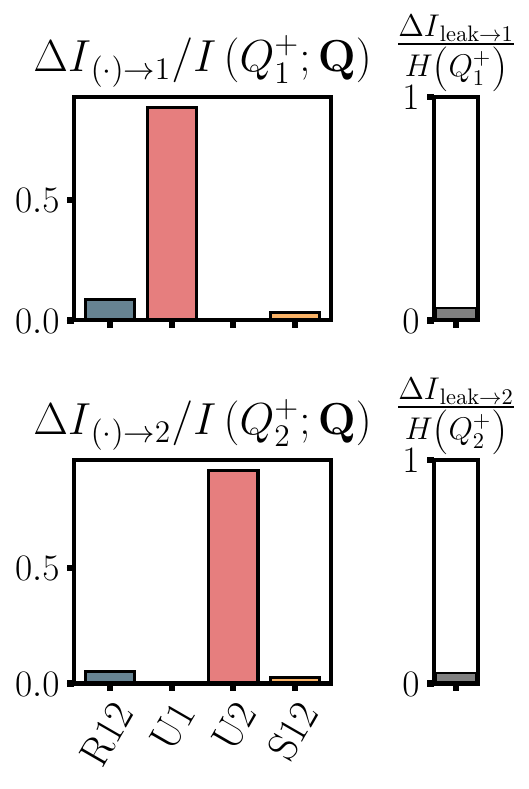}
    \end{minipage}
    \begin{minipage}{0.1\textwidth}
    \vspace{-0.5cm}
        \includegraphics[width=\textwidth]{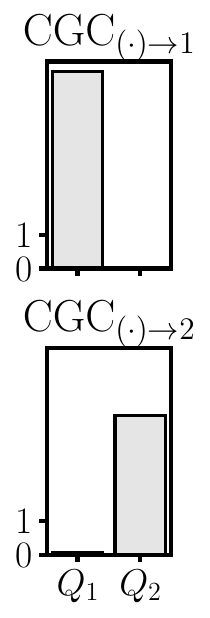}
    \end{minipage}
    \begin{minipage}{0.1\textwidth}
    \vspace{-0.5cm}
        \includegraphics[width=\textwidth]{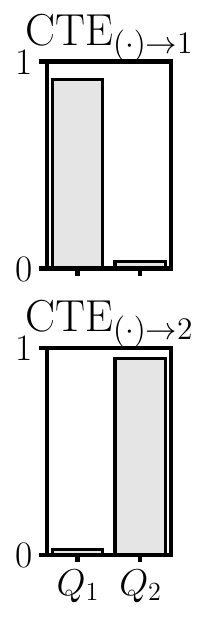}
    \end{minipage}
    \begin{minipage}{0.1\textwidth}
    \vspace{-0.5cm}
        \includegraphics[width=\textwidth]{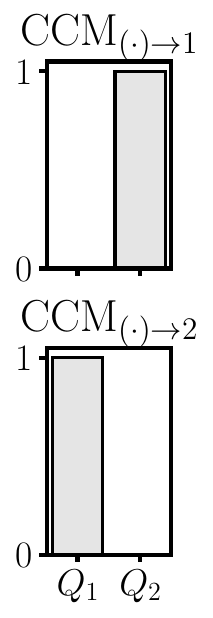}
    \end{minipage}
    \begin{minipage}{0.1175\textwidth}
    \vspace{-0.5cm}
        \includegraphics[width=\textwidth]{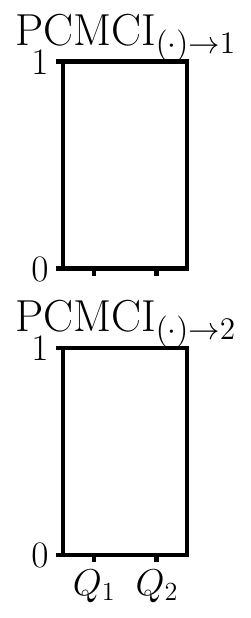}
    \end{minipage}
    \caption{Lotka--Volterra prey-predator model
      \cite{lotka1925,volterra1926}. Redundant (R), unique (U), and
      synergistic (S) causalities. The gray bar is the causality
      leak. The results from CGC, CTE, CCM, and PCMCI are depicted on
      the right. CGC and CTE use the same normalization as SURD. The methods
      are applied at a time lag $\Delta t=1.5$
      corresponding to the delay between the maximum in the predator
      and prey population numbers.}
    \label{fig:lotka-volterra}
\end{figure}

The Lotka--Volterra predator-prey model~\cite{lotka1925, volterra1926}
was envisioned to describe the dynamics of biological systems in which
two species interact. The model can be expressed as a pair of
first-order nonlinear differential equations:
\begin{align}
\label{eq:lotka-volterra}
    \frac{{\mathrm{d}}Q_1}{{\mathrm{d}}t} &= \alpha Q_1 - \beta Q_1 Q_2, \\
    \frac{{\mathrm{d}}Q_2}{{\mathrm{d}}t} &= \Gamma Q_1 Q_2 - \gamma Q_2,
\end{align}
where $Q_1$ and $Q_2$ denote the prey and the predator population
number, respectively, and $\alpha = 1$, $\beta = 0.05$, $\Gamma =
0.02$, and $\gamma = 0.5$ represent the prey reproduction rate,
predation rate, predator reproduction rate and predator death rate,
respectively. Figure \ref{fig:lotka-volterra} shows a visualization of
the time signals of the model together with the results from SURD,
CGC, CTE, CCM and PCMCI. SURD identifies synergistic causality as the
most significant contribution for both variables, i.e., $\Delta
I^U_{1\rightarrow 1}$ and $\Delta I^U_{2\rightarrow 2}$, together with
some smaller redundant and synergistic causalities for both
variables. This is in agreement with the low values of the nonlinear
coupling between variables in Equation~\eqref{eq:lotka-volterra}.  In
relation to the other approaches, we observe relationships consistent
with the functional dependencies in Equation~\eqref{eq:lotka-volterra}
for CCM, CGC and CTE; however, PCMCI does not detect any significant
causal link.

\subsection{Moran effect model}

\begin{figure}[t!]
    \centering
    \begin{minipage}{0.345\textwidth}
    \vspace{0.25cm}
        \includegraphics[width=\textwidth]{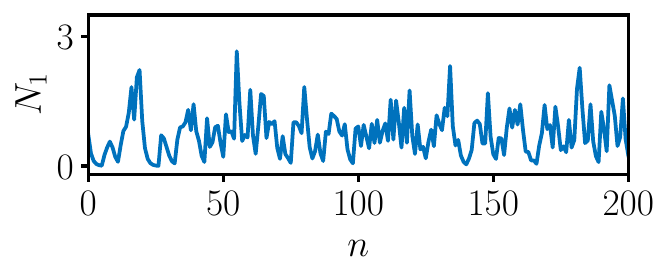}
        \includegraphics[width=\textwidth]{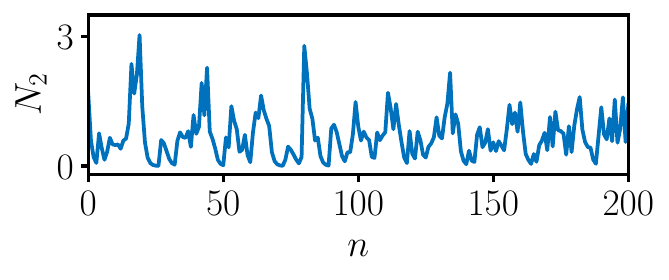}
    \end{minipage}
    \begin{minipage}{0.21\textwidth}
    \vspace{-0.45cm}
        \includegraphics[width=\textwidth]{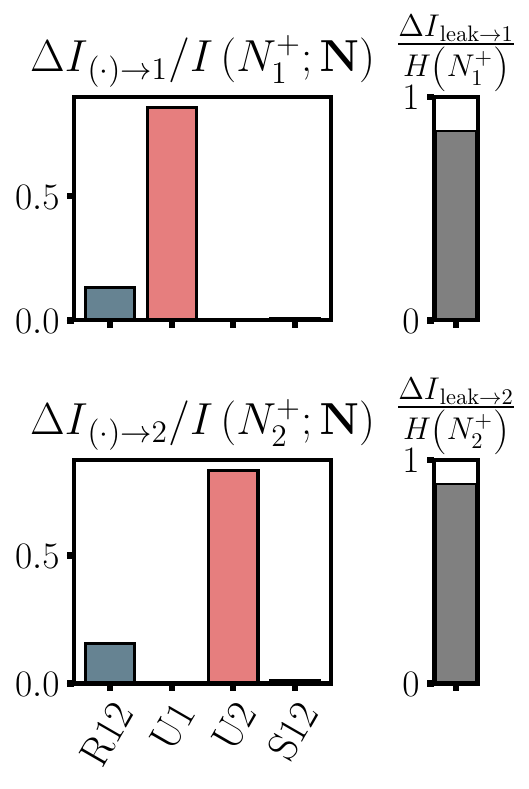}
    \end{minipage}
    \hspace*{-0.01\textwidth}
    \begin{minipage}{0.099\textwidth}
    \vspace{-0.52cm}
        \includegraphics[width=\textwidth]{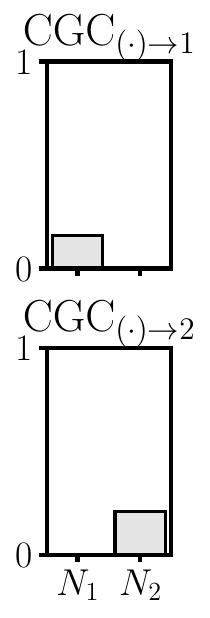}
    \end{minipage}
    \begin{minipage}{0.0975\textwidth}
    \vspace{-0.5cm}
        \includegraphics[width=\textwidth]{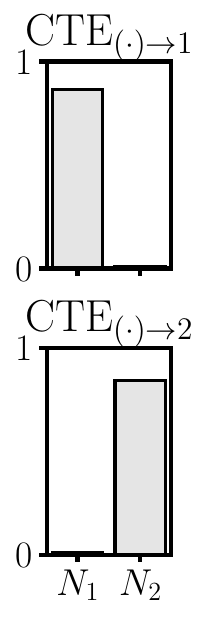}
    \end{minipage}
    \begin{minipage}{0.1\textwidth}
    \vspace{-0.5cm}
        \includegraphics[width=\textwidth]{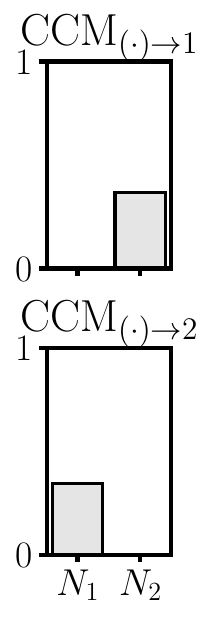}
    \end{minipage}
    \hspace*{-0.02\textwidth}
    \begin{minipage}{0.1225\textwidth}
    \vspace{-0.5cm}
        \includegraphics[width=\textwidth]{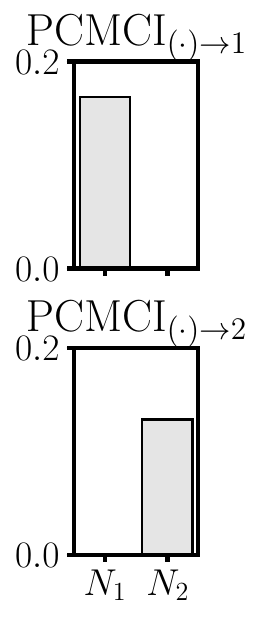}
    \end{minipage}
    \caption{Moran effect model~\cite{moran1953}. 
    (Left panel)  Time evolution of independent variables $N_1$ and $N_2$. 
    (Right panel) Redundant
      (R), unique (U), and synergistic (S) causalities with the vector
      of observed variables $\bN = [N_1, N_2]$. The gray bar is the
      causality leak. The results from CGC, CTE, CCM, and PCMCI are
      depicted on the right. CGC and CTE use the same normalization as
      SURD. The model was run for $10^8$ time steps with the set of
      parameters: $r_1 = 3.4, r_2 = 2.9, \Psi_1 = 0.5, \Psi_2 = 0.6,
      s_1 = 0.4, s_2 = 0.35, D_1 = 3, D_2 = 3, R_1(0)=R_2(0)=1,
      N_1(0)=N_2(0)=0.5$.}
    \label{fig:moran_effect}
\end{figure}

The Moran effect~\cite{moran1953} refers to a phenomenon in population
ecology that describes how spatially separated populations can exhibit
synchronous fluctuations in their sizes due to a common environmental
factor affecting them all simultaneously. The system comprises two
variables, $N_1$ and $N_2$, that do not exhibit causal relationships
but are significantly correlated in their time series due to shared
external forcing, $V$. The latter follows a Gaussian distribution with
a mean of zero and a standard deviation of one, acting on $N_1$ and
$N_2$ through the mediator variables $R_1$ and $R_2$,
respectively. The equations of the model are given by:
\begin{align*}
    R_i(n+1) &= r_i N_i(n)[1-N_i(n)]e^{-\Psi_i V(n)}\\
    N_i(n+1) &= s_i N_i(n) + \max{[R_i(n-D_i),0]}.
\end{align*}
The vector of observed variables is $\boldsymbol{N} = [N_1, N_2]$, and
the aim is to assess whether different methods can discern the causal
independence between $N_1$ and $N_2$ despite their significant
correlation due to a shared confounder. The results are provided in
Figure~\ref{fig:moran_effect}. The most significant causal
interactions detected by SURD are self-unique causalities, while the
remaining components are redundant causalities. These outcomes align
with the functional dependencies between the variables in
Figure~\ref{fig:moran_effect}, and SURD accurately captures the
confounding effect. The other approaches also identify the causal
independence among variables. CCM yields a non-zero value for
cross-induced causal relationships. However, the inspection of this
value for increasing number of samples revealed that it does not
converge to 1, suggesting no causal link between the variables.
%

\subsection{Coupled logistic difference system}

\begin{figure}[t!]
    \centering
    \subfloat[$\beta_{2\rightarrow 1}=0.1$, $\beta_{1\rightarrow 2}=0$]{
    \begin{minipage}{0.34\textwidth}
    \vspace{0.25cm}
        \includegraphics[width=\textwidth]{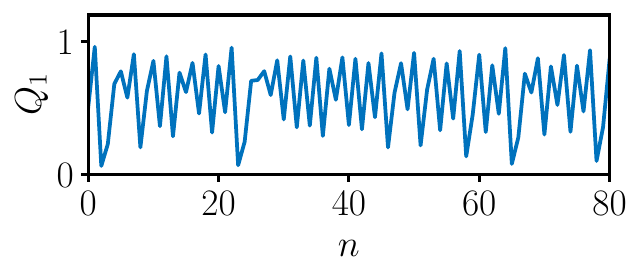}
        \includegraphics[width=\textwidth]{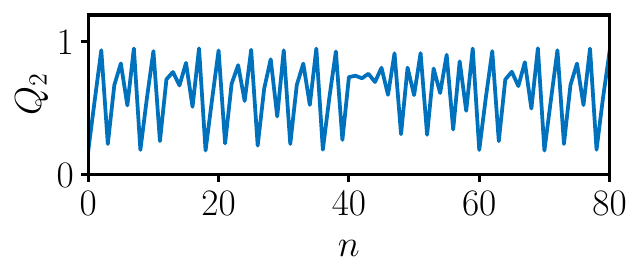}
    \end{minipage}
    \hspace*{0.01\textwidth}
    \begin{minipage}{0.215\textwidth}
    \vspace{-0.45cm}
        \includegraphics[width=\textwidth]{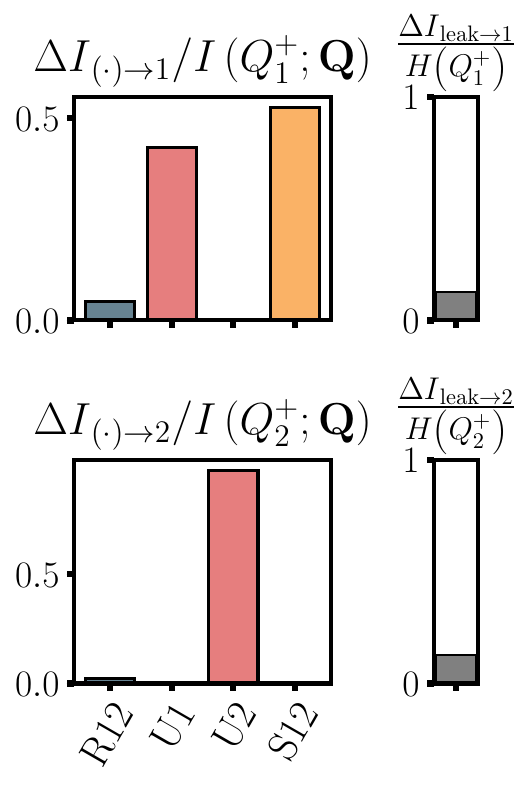}
    \end{minipage}
    \hspace*{-0.01\textwidth}
    \begin{minipage}{0.101\textwidth}
    \vspace{-0.51cm}
        \includegraphics[width=\textwidth]{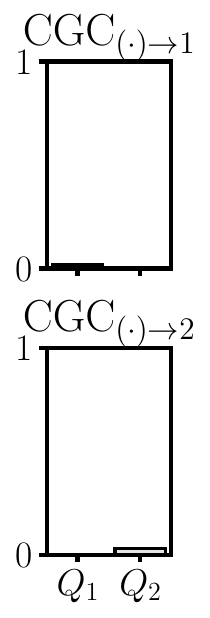}
    \end{minipage}
    \begin{minipage}{0.1\textwidth}
    \vspace{-0.5cm}
        \includegraphics[width=\textwidth]{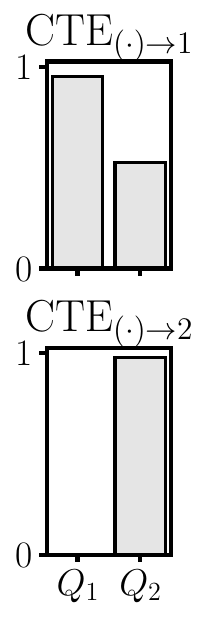}
    \end{minipage}
    \begin{minipage}{0.104\textwidth}
    \vspace{-0.5cm}
        \includegraphics[width=\textwidth]{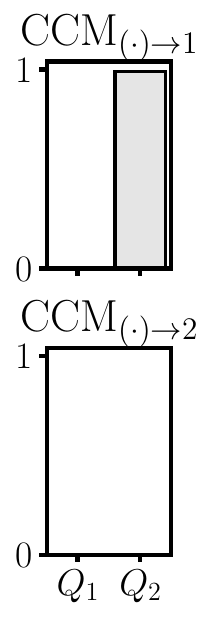}
    \end{minipage}
    \hspace*{-0.02\textwidth}
    \begin{minipage}{0.125\textwidth}
    \vspace{-0.5cm}
        \includegraphics[width=\textwidth]{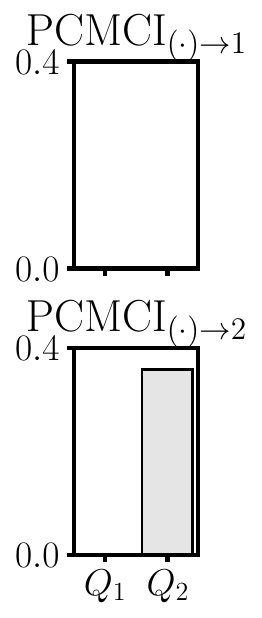}
    \end{minipage}
    }
    \vspace{0.7cm}

    \subfloat[$\beta_{2\rightarrow 1}=0$, $\beta_{1\rightarrow 2}=0.32$]{
    \begin{minipage}{0.34\textwidth}
    \vspace{0.25cm}
        \includegraphics[width=\textwidth]{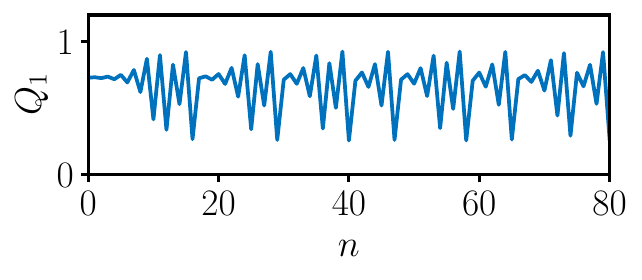}
        \includegraphics[width=\textwidth]{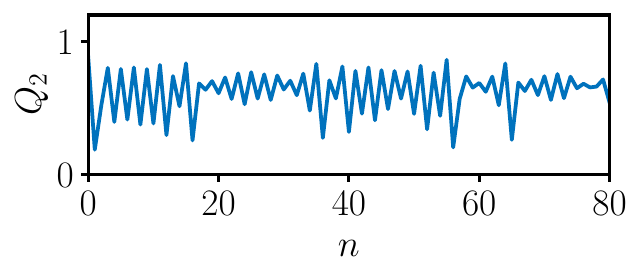}
    \end{minipage}
    \hspace*{0.01\textwidth}
    \begin{minipage}{0.215\textwidth}
    \vspace{-0.45cm}
        \includegraphics[width=\textwidth]{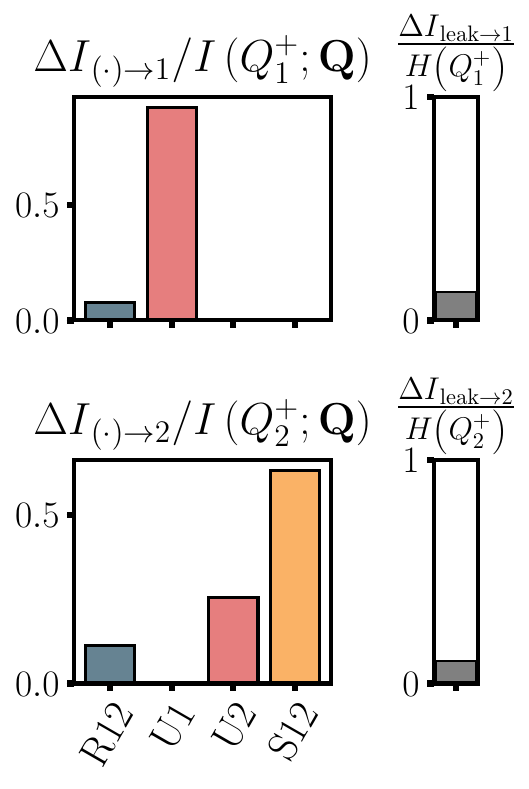}
    \end{minipage}
    \hspace*{-0.01\textwidth}
    \begin{minipage}{0.101\textwidth}
    \vspace{-0.51cm}
        \includegraphics[width=\textwidth]{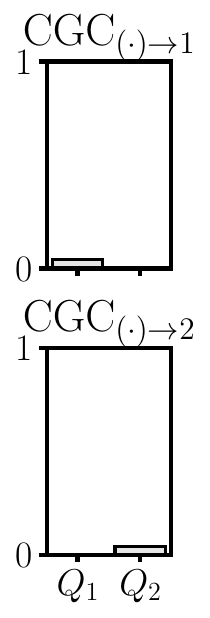}
    \end{minipage}
    \begin{minipage}{0.1\textwidth}
    \vspace{-0.5cm}
        \includegraphics[width=\textwidth]{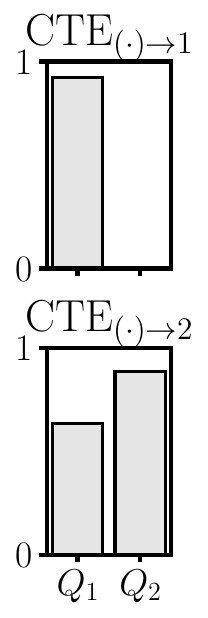}
    \end{minipage}
    \begin{minipage}{0.104\textwidth}
    \vspace{-0.5cm}
        \includegraphics[width=\textwidth]{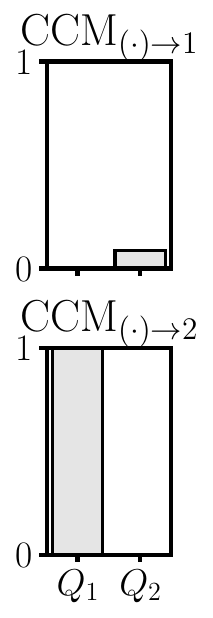}
    \end{minipage}
    \hspace*{-0.02\textwidth}
    \begin{minipage}{0.125\textwidth}
    \vspace{-0.5cm}
        \includegraphics[width=\textwidth]{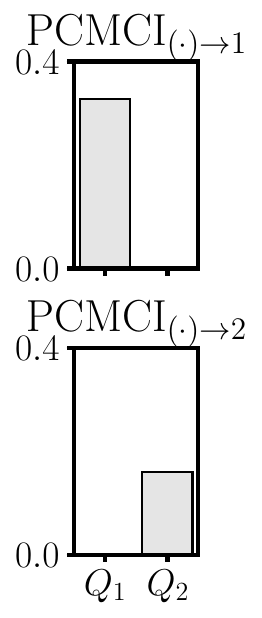}
    \end{minipage}
    }    
    \caption{Nonlinear logistic difference system with one-way coupled
      components~\cite{sugihara2012}. Redundant (R), unique (U), and
      synergistic (S) causalities. The gray bar is the causality
      leak. The results of CGC, CTE, PCMCI and CCM are depicted on the
      right. CGC and CTE use the same normalization as SURD.}
    \label{fig:sugihara oneway}
\end{figure}

\begin{figure}[t!]
    \centering

    \subfloat[$\beta_{2\rightarrow 1}=0.2$, $\beta_{1\rightarrow 2}=0.01$]{
    \begin{minipage}{0.34\textwidth}
    \vspace{0.25cm}
        \includegraphics[width=\textwidth]{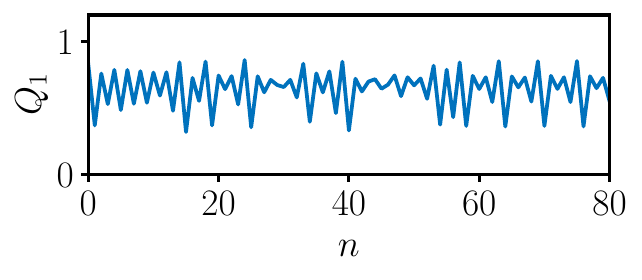}
        \includegraphics[width=\textwidth]{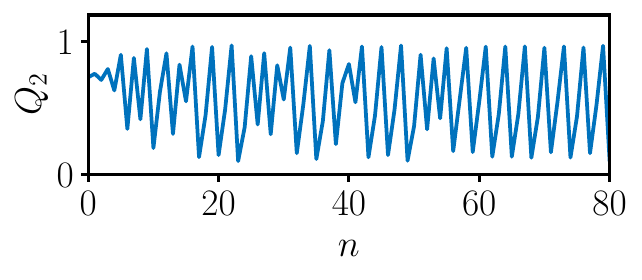}
    \end{minipage}
    \hspace*{0.01\textwidth}
    \begin{minipage}{0.215\textwidth}
    \vspace{-0.45cm}
        \includegraphics[width=\textwidth]{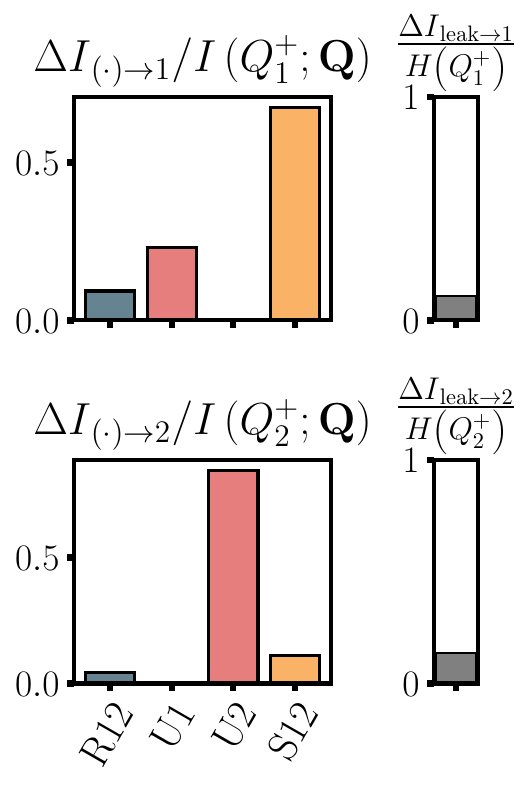}
    \end{minipage}
    \hspace*{-0.01\textwidth}
    \begin{minipage}{0.101\textwidth}
    \vspace{-0.51cm}
        \includegraphics[width=\textwidth]{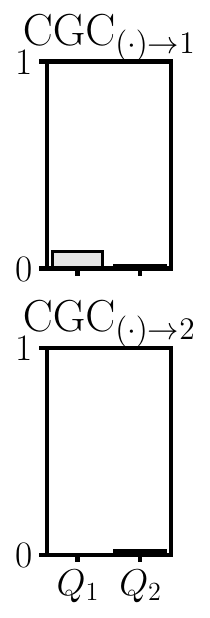}
    \end{minipage}
    \begin{minipage}{0.1\textwidth}
    \vspace{-0.5cm}
        \includegraphics[width=\textwidth]{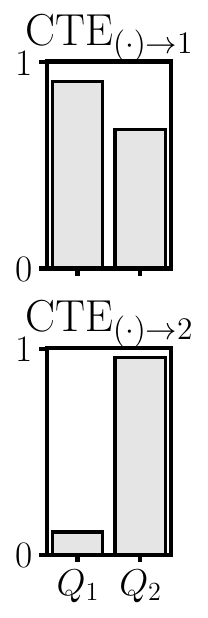}
    \end{minipage}
    \begin{minipage}{0.104\textwidth}
    \vspace{-0.5cm}
        \includegraphics[width=\textwidth]{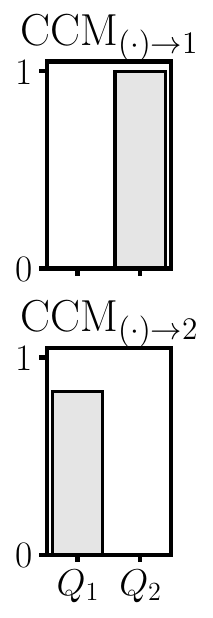}
    \end{minipage}
    \hspace*{-0.02\textwidth}
    \begin{minipage}{0.125\textwidth}
    \vspace{-0.5cm}
        \includegraphics[width=\textwidth]{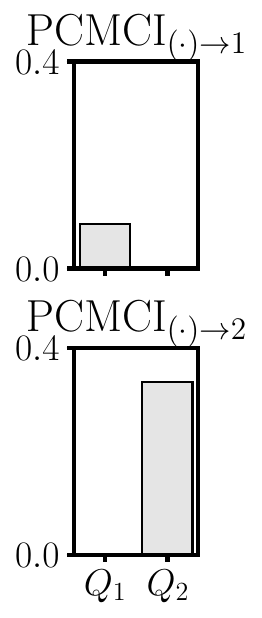}
    \end{minipage}
    }
    \vspace{0.7cm}

    \subfloat[$\beta_{2\rightarrow 1}=0.02$, $\beta_{1\rightarrow 2}=0.1$]{
    \begin{minipage}{0.34\textwidth}
    \vspace{0.25cm}
        \includegraphics[width=\textwidth]{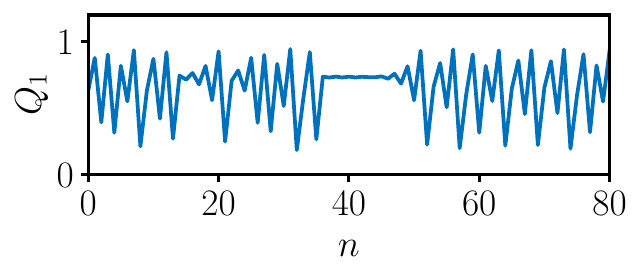}
        \includegraphics[width=\textwidth]{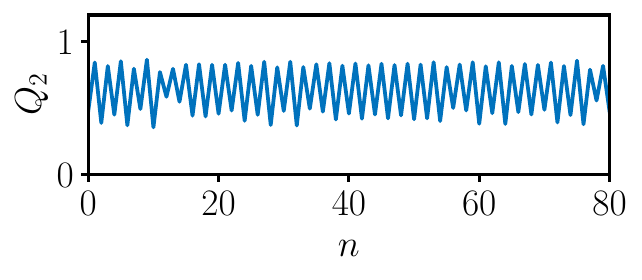}
    \end{minipage}
    \hspace*{0.01\textwidth}
    \begin{minipage}{0.215\textwidth}
    \vspace{-0.45cm}
        \includegraphics[width=\textwidth]{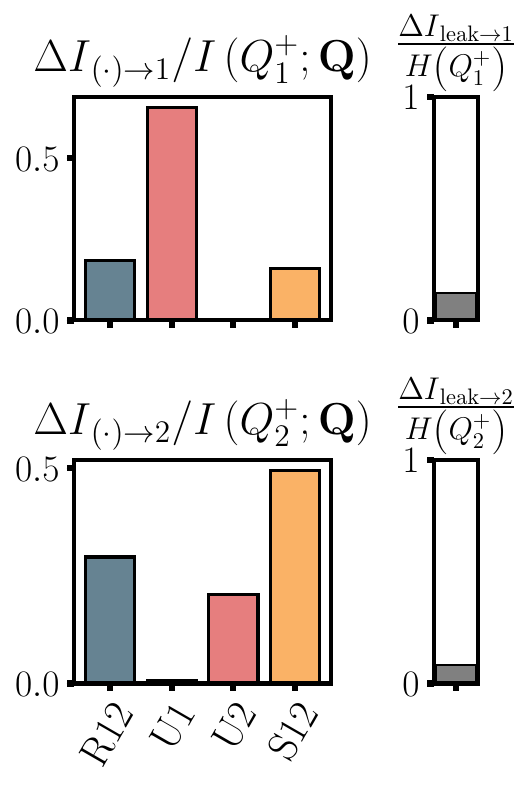}
    \end{minipage}
    \hspace*{-0.01\textwidth}
    \begin{minipage}{0.101\textwidth}
    \vspace{-0.51cm}
        \includegraphics[width=\textwidth]{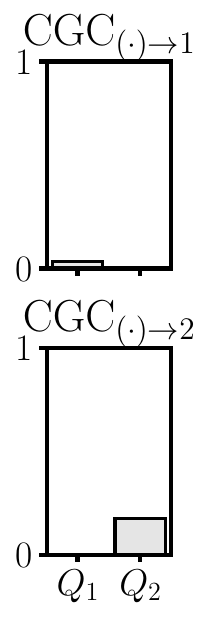}
    \end{minipage}
    \begin{minipage}{0.1\textwidth}
    \vspace{-0.5cm}
        \includegraphics[width=\textwidth]{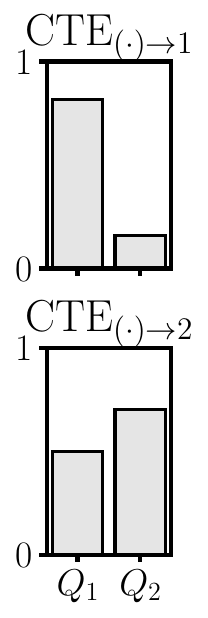}
    \end{minipage}
    \begin{minipage}{0.104\textwidth}
    \vspace{-0.5cm}
        \includegraphics[width=\textwidth]{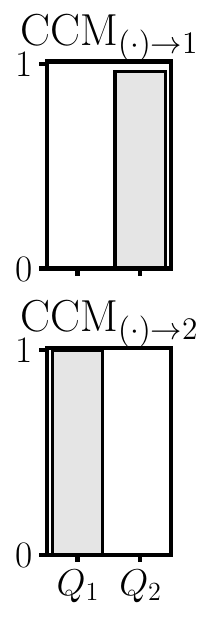}
    \end{minipage}
    \hspace*{-0.02\textwidth}
    \begin{minipage}{0.125\textwidth}
    \vspace{-0.5cm}
        \includegraphics[width=\textwidth]{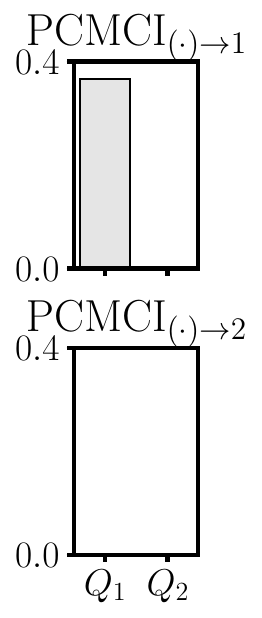}
    \end{minipage}
    }    
    \caption{Nonlinear logistic difference system with two-way coupled
      components~\cite{sugihara2012}. Redundant (R), unique (U), and
      synergistic (S) causalities. The gray bar is the causality
      leak. The results of CGC, CTE, PCMCI and CCM are depicted on the
      right. CGC and CTE use the same normalization as SURD.}
    \label{fig:sugihara two way}
\end{figure}

We consider the deterministic, nonlinear, logistic difference system
proposed by Sugihara et al.~\cite{sugihara2012}.  The system, given in
Equation~\eqref{eq: sugihara1}, illustrates the concept of mirage
correlation, namely, a perceived but spurious correlation between two
variables. To illustrate this phenomenon, we use four versions of the
models for different values of the coupling parameters
$\beta_{2\rightarrow 1}$ and $\beta_{1\rightarrow 2}$. A comparison of
the results for all methods is shown in Figure~\ref{fig:sugihara oneway} for one-way coupling between variables and in Figure
\ref{fig:sugihara two way} for the two-way coupling case.
\begin{itemize}
    \item One-way coupling $Q_2 \rightarrow Q_1$: SURD identifies the
      causalities $\Delta I_{12\rightarrow 1} ^ S$ and $\Delta
      I_{1\rightarrow 1} ^ U$ as the most important causalities
      driving $Q_1$, whereas only $\Delta I_{2\rightarrow 2 }^ U$ is
      identified for $Q_2$. This is consistent with the non-zero
      coupling constant $\beta_{2 \rightarrow 1}$.
    \item One-way coupling $Q_1 \rightarrow Q_2$: similar conclusions
      can be drawn from this case. SURD identifies the causalities
      $\Delta I_{12\rightarrow 2} ^ S$ and $\Delta I_{2\rightarrow 2}
      ^ U$ as the most important variables driving $Q_2$, whereas only
      $\Delta I_{1\rightarrow 1}^ U$ is identified for $Q_1$. This is
      consistent with the non-zero coupling constant $\beta_{1
        \rightarrow 2}$. 
    \item Two-way coupling $\beta_{2\rightarrow 1} >
      \beta_{1\rightarrow 2}$: SURD identifies the causality $\Delta
      I^U_{2\rightarrow 2}$ for variable $Q_2$, which implies that
      $Q_2$ is mostly self-causal. This is consistent with the low
      value of the coupling constant $\beta_{1\rightarrow 2}$. There
      are also redundant and synergistic contributions given by
      $\Delta I^R_{12\rightarrow 2}$ and $\Delta I^S_{12\rightarrow
        2}$, respectively, although these are smaller than the
      self-induced causality. In the case of $Q_1$, the synergistic
      contribution dominates over the unique causality $\Delta
      I^U_{1\rightarrow 1}$. This is agrees with the fact that the
      coupling parameter is significantly larger from $Q_2\rightarrow
      Q_1$ (i.e., $\beta_{2 \rightarrow 1}$) than from $Q_1\rightarrow
      Q_2$ ($\beta_{1\rightarrow 2}$). 
    \item Two-way coupling $\beta_{2\rightarrow 1} <
      \beta_{1\rightarrow 2}$: SURD identifies the causality $\Delta
      I^U_{1\rightarrow 1}$ for variable $Q_1$, which implies that
      $Q_1$ is mostly self-causal. This is consistent with the low
      value of the coupling constant $\beta_{1\rightarrow 2}$ from
      $Q_2\rightarrow Q_1$. There are also redundant and synergistic
      contributions given by $\Delta I^R_{12\rightarrow 1}$ and
      $\Delta I^S_{12\rightarrow 1}$, respectively, although these are
      smaller than the self-induced causality. In the case of $Q_2$,
      the redundant and synergistic contributions dominate over the
      unique causality from $Q_2$ to $Q_2$. This is consistent with
      the fact that the coupling parameter is significantly larger
      from $Q_1\rightarrow Q_2$ than from $Q_2\rightarrow Q_1$ (i.e.,
      $\beta_{1\rightarrow 2} > \beta_{2\rightarrow 1}$).
\end{itemize}

Among the other methods, only CTE and CCM can identify all the
functional dependencies between variables in all cases. PCMCI
correctly identifies self-causal links in all cases; however, it
detects cross-induced causalities only in the two-way coupling case,
which are very low in intensity. It is also important to mention that
these methods fail to provide information on whether the coupling in
the system is introduced by individual variables acting alone, or
whether it is the combined effect of multiple variables that drives
the coupling. In these scenarios, SURD offers a more comprehensive
understanding of the system through the synergistic contribution
$\Delta I^S_{12\rightarrow j}$.

%

\subsection{Stochastic system with time-lagged dependencies}

We examine bivariate stochastic systems with linear and nonlinear
time-lagged dependencies previously studied in \cite{ding2006} and
\cite{bueso2020}, respectively.  These cases demonstrate the
performance of causal methods when the relationships are introduced
with different time lags. The notation for the observable vector is
$\bQ =[Q_{1_1}, Q_{1_2}, Q_{2_1}, Q_{2_2}]$, where the second subindex
denotes the time lag, e.g., $Q_{1_1}=Q_1(n-1)$ and $Q_{1_2}=Q_1(n-2)$.

For the first case, the relationships between variables are linear and
they are described as:
\begin{align*}
Q_1(n+1) &= 0.95 \sqrt{2} Q_1(n) - 0.9025 Q_1(n-1) + W_1(n) \\
Q_2(n+1) &= 0.5 \cdot Q_1(n-1) + W_2(n)
\end{align*}
where $W_i \sim \mathcal{N}(0,1)$ represents a stochastic forcing following 
a Gaussian distribution with a mean of zero and a variance of one. In this case,
$Q_1(n)$ is self-caused at two different time lags, i.e., $Q_1(n-1)$
and $Q_1(n-2)$, while $Q_1(n-2)$ drives $Q_2(n)$. These relationships
are well captured by the SURD, where the most significant causal
relationships are $\Delta I^S_{1_1 1_2\rightarrow 1}$ and $\Delta
I^U_{1_2\rightarrow 2}$, with the subscript indicating the time lag of
the variable. Note that since the relationships
between the variables are linear, the functional dependencies of $Q_1$
can also be expressed as a function of $Q_2(n-1)$. This is why SURD
identifies synergistic causalities $\Delta I^S_{1_1 2_1\rightarrow 1}$
and $\Delta I^S_{1_1 2_2\rightarrow 1}$.

Among the other methods, we observe that only CGC and CTE can identify
all causal relationships between variables across all time lags. PCMCI
clearly detects the links $Q_1(n-1) \rightarrow Q_1(n)$ and $Q_1(n-2)
\rightarrow Q_2(n)$, but depending on the threshold used the remaining identified connections exhibit
lower intensity values compared to these. The identification of these
latter links depends on the confidence threshold used. For a strict threshold of $\alpha = 0.01$, only $Q_1(n-1) \rightarrow Q_1(n)$ and $Q_1(n-2)
\rightarrow Q_2(n)$ are identified. For CCM, causality needs to be assessed upon the convergence of the prediction skill to 1 as the length of the time series increases. 
In this system, the prediction
skill for $Q_2$ using variable $Q_1(n-2)$ is slightly higher than that
for $Q_1(n-1)$, implying that the manifold associated with $Q_1(n-2)$
enables a better reconstruction of the states of $Q_2(n)$ than the
states of $Q_1(n-1)$. A similar conclusion can be drawn for $Q_2$,
where the manifold associated with $Q_2(n-1)$ provides better
reconstruction of the states of $Q_1(n)$.
\begin{figure}[t!]
    \includegraphics[width=0.45\textwidth]{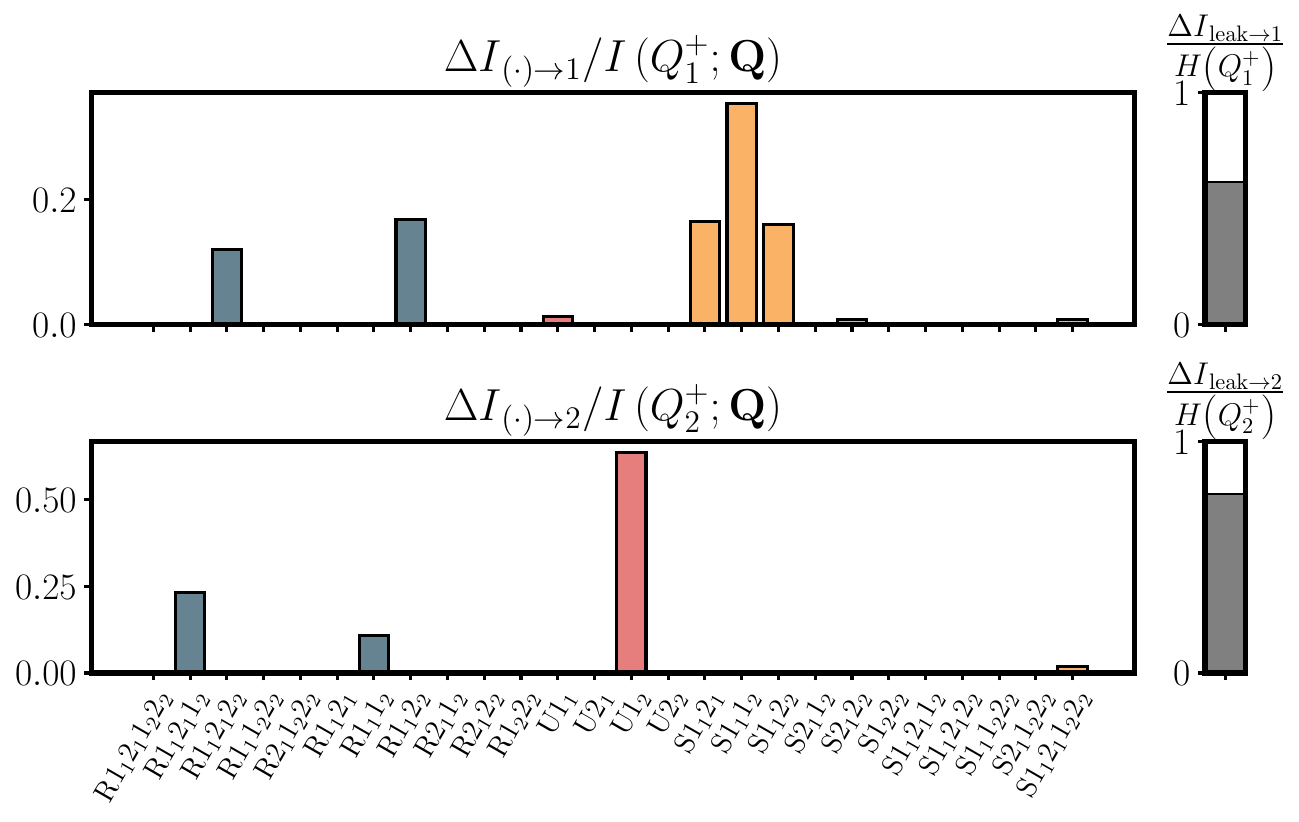}
    \begin{minipage}{0.1275\textwidth}
        \vspace{-5.3cm}
        \includegraphics[width=\textwidth]{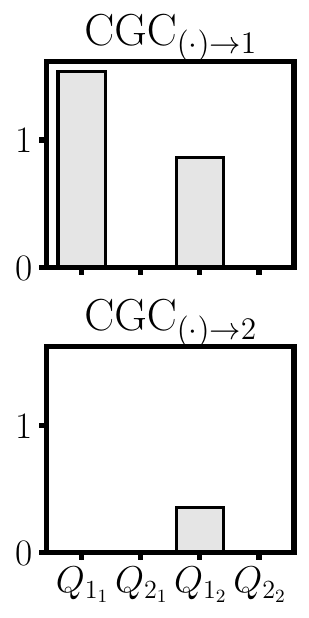}
    \end{minipage}
    \begin{minipage}{0.1275\textwidth}
        \vspace{-5.3cm}
        \includegraphics[width=\textwidth]{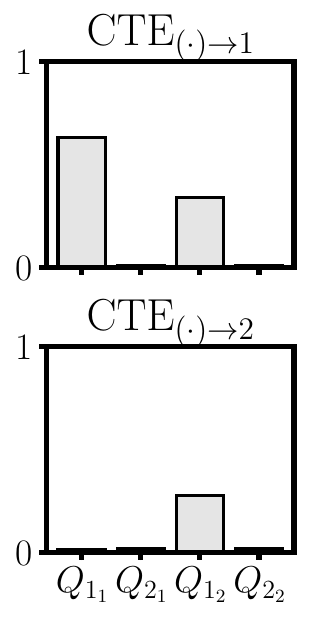}
    \end{minipage}
    \begin{minipage}{0.1275\textwidth}
        \vspace{-5.3cm}
        \includegraphics[width=\textwidth]{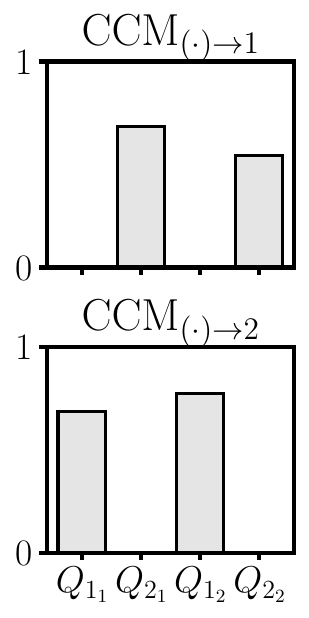}
    \end{minipage}
    \begin{minipage}{0.14\textwidth}
        \vspace{-5.3cm}
        \includegraphics[width=\textwidth]{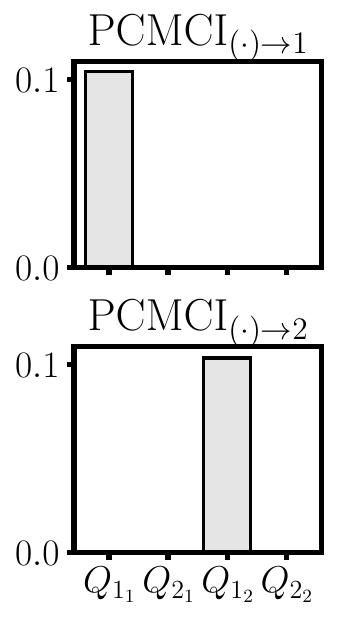}
    \end{minipage}
    \caption{Stochastic systems with linear time-lagged dependencies
      \cite{ding2006}. Redundant (R), unique (U), and synergistic (S)
      causalities in blue, orange and yellow, respectively. The
      subindex of the labels represents the time delay associated with
      the variable. The gray bar is the causality leak. The results from CGC, CTE, CCM, and PCMCI are depicted on
      the right. CGC and CTE use the same normalization as SURD.}
    \label{fig:ding}
\end{figure}

The second case is a system with non-linear relationships between
variables. This system was previously studied by Bueso \textit{et al.}
\cite{bueso2020} for a single-lag dependency. In this work, we employ
an extension of the system for multiple time lags to demonstrate how
SURD can also be applied to identify nonlinear relationships at
multiple lags. The equations describing the system are given by:
\begin{align*}
Q_1(n+1) &= 3.4 Q_1(n) (1 - Q_1(n)^2) \exp\left(-Q_1(n-1)^2\right) + W_1(n) \\
Q_2(n+1) &= 3.4 Q_2(n) (1 - Q_2(n)^2) \exp\left(-Q_2(n)^2\right) + \frac{Q_1(n-1) Q_2(n)}{2} + W_2(n)
\end{align*}
where the time-varying stochastic forcing $W_i$ that affects each variable
follows a Gaussian distribution with a mean of zero and a variance of
0.4. For this model, $Q_1(n-1)$ and $Q_1(n-2)$ are common drivers of
$Q_1(n)$, and $Q_2(n-1)$ and $Q_1(n-2)$ for $Q_2(n)$. 

\begin{figure}[t!]
    \includegraphics[width=0.45\textwidth]{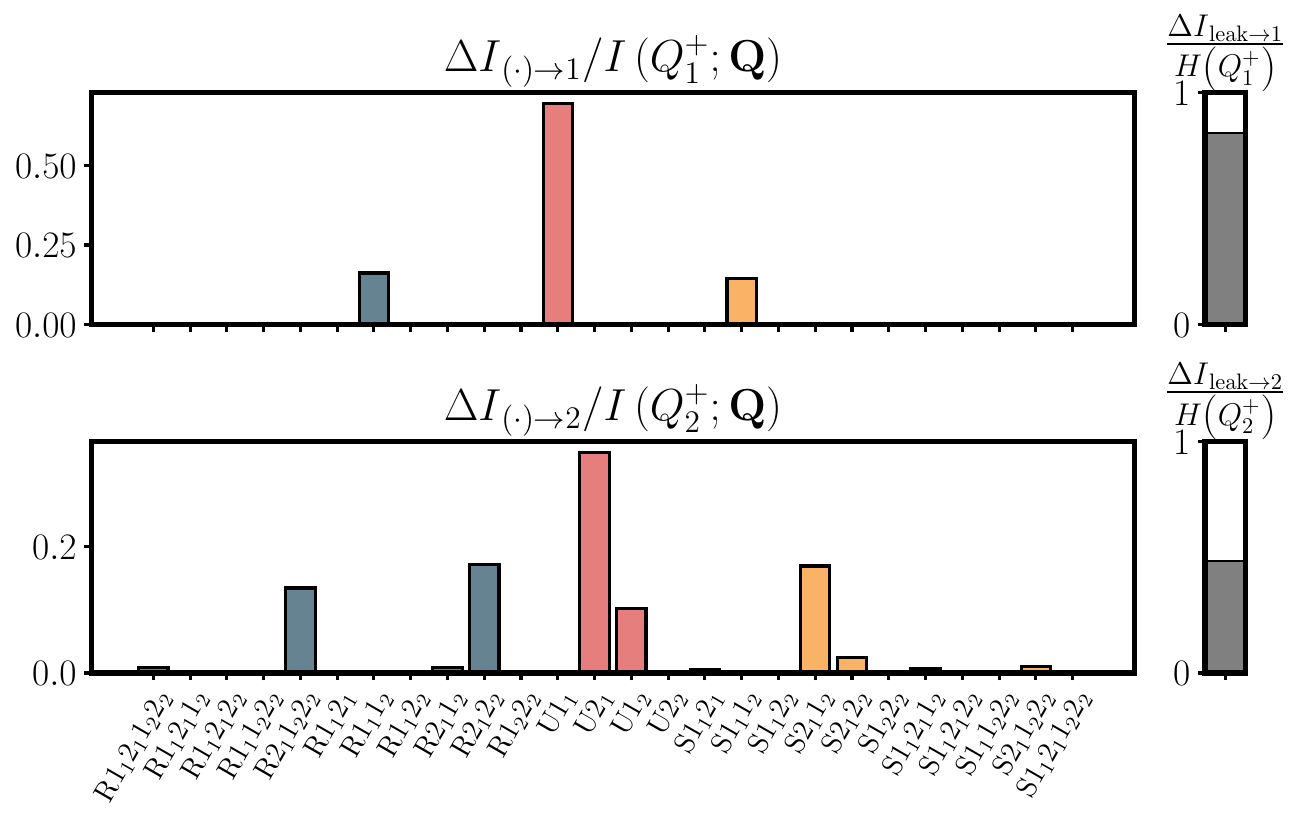}
    \begin{minipage}{0.1275\textwidth}
        \vspace{-5.3cm}
        \includegraphics[width=\textwidth]{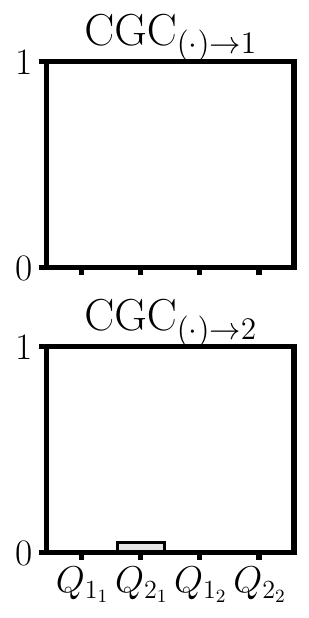}
    \end{minipage}
    \begin{minipage}{0.1275\textwidth}
        \vspace{-5.3cm}
        \includegraphics[width=\textwidth]{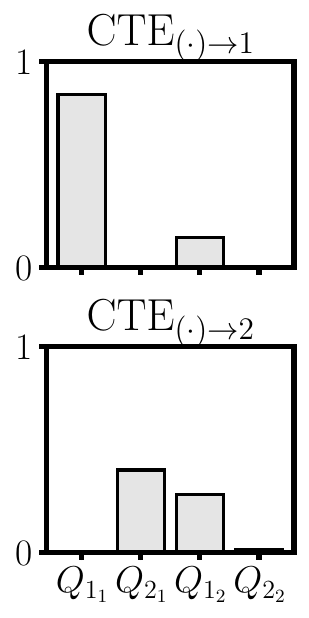}
    \end{minipage}
    \begin{minipage}{0.1275\textwidth}
        \vspace{-5.3cm}
        \includegraphics[width=\textwidth]{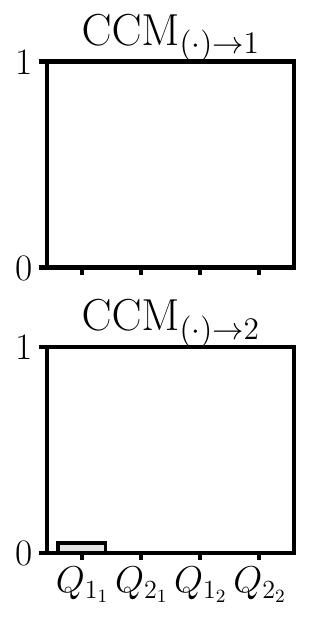}
    \end{minipage}
    \begin{minipage}{0.145\textwidth}
        \vspace{-5.3cm}
        \includegraphics[width=\textwidth]{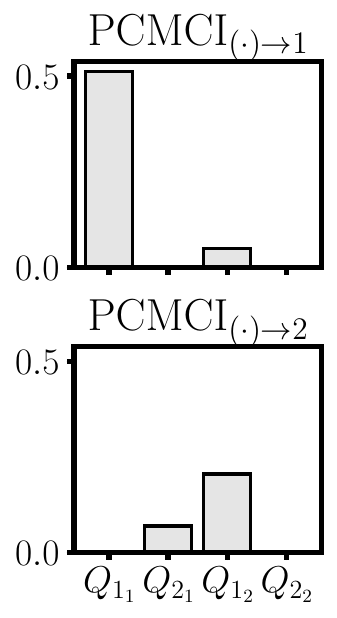}
    \end{minipage}
    \caption{Stochastic systems with nonlinear time-lagged
      dependencies \cite{bueso2020}. Redundant (R), unique (U), and
      synergistic (S) causalities in blue, orange and yellow,
      respectively. The subindex of the labels represents the time
      delay associated with the variable. The gray bar is the
      causality leak. The results from CGC, CTE, CCM, and PCMCI are
      depicted on the right. CGC and CTE use the same normalization as
      SURD.}
    \label{fig:bueso}
\end{figure}

The results are shown in Figure~\ref{fig:bueso}.  The causal
connections identified by SURD $\Delta I^U_{1_1\rightarrow 1}$,
$\Delta I^S_{1_1 1_2\rightarrow 1}$, $\Delta I^U_{2_1\rightarrow 2}$,
and $\Delta I^U_{1_2\rightarrow 2}$ correctly capture the functional
dependencies of the system. Since the relationships between variables
are highly nonlinear, CGC and CCM are no longer able to identify the causal
connections between variables.  PCMCI and CTE are still capable of
discerning the relationships $[Q_1(n-1),Q_1(n-2)]\rightarrow Q_1(n)$
and $[Q_2(n-1),Q_1(n-2)]\rightarrow Q_2(n)$ in a consistent manner
with the functional dependencies between
variables.

\subsection{Synchronization in logistic maps}
The one-dimensional logistic map is a recurrence given by the
relationship,
\begin{equation}
  \label{eq:logistic_1}
  Q_1(n+1) = \alpha_1 Q_1(n)[1-Q_1(n)],  
\end{equation}
where $n$ is the time step and $\alpha_1$ is a
constant. Equation~(\ref{eq:logistic_1}) exhibits a chaotic behavior
for $\alpha_1\approx 3.57-4$ \cite{may1976}.  We consider the three
logistic maps:
\begin{subequations}
\begin{align}
  Q_1(n+1) &= \alpha_1 Q_1(n)[1-Q_1(n)],\\
  Q_2(n+1) &= \alpha_2 f_{12}[1-f_{12}],\\
  Q_3(n+1) &= \alpha_3 f_{123}[1-f_{123}],
  \end{align}
\end{subequations}
which are coupled by
\begin{subequations}
\begin{align}
  f_{12}   &= \frac{Q_2(n)+c_{1\rightarrow2}Q_1(n)}{1+c_{1\rightarrow2}},\\
  f_{123}  &= \frac{Q_3(n)+c_{12\rightarrow3}Q_1(n)+c_{12\rightarrow3}Q_2(n)}{1+2c_{12\rightarrow3}},
\end{align}
\end{subequations}
where $\alpha_1 = 3.68$, $\alpha_2 = 3.67$, and $\alpha_3 = 3.78$ are
constants, $c_{1\rightarrow2}$ is the parameter coupling $Q_2$ with
$Q_1$, and $c_{12\rightarrow3}$ is the parameter coupling $Q_3$ with
$Q_2$ and $Q_1$. The clear directionality of the variables in this
system for different values of $c_{1\rightarrow2}$ and
$c_{12\rightarrow3}$ offers a simple testbed to illustrate the
behavior of the causality. The causal analysis for SURD is performed
for one time-step lag after integrating the system for $10^8$
steps. The phase-space was partitioned using 100 bins for each
variables.
%
\begin{figure}[p!]
\subfloat[]{
    \begin{minipage}{0.175\textwidth}
        \vspace{-2\textwidth}
        ${\hbox{


\begin{tikzpicture}[cir/.style={circle,draw=black!70!white,,thick,inner sep=.5em},
    >={Latex[length=.2cm]},scale=.8]
    

    \colorlet{cA}{Q1!30!white}
    \colorlet{cB}{Q2!30!white}
    \colorlet{cC}{Q3!30!white}


    \node [cir,fill=cB] (q1) at (-1,+1)   {$Q_1$}; 
    \node [cir,fill=cA] (q2) at (-1,-1)   {$Q_2$}; 
    \node [cir,fill=cC] (q3) at (+0.73,0)    {$Q_3$}; 
    


    \path[thick,<-] (q3) edge[loop right] node {} (q3);
    \path[thick,<-] (q2) edge[loop below] node {} (q2);
    \path[thick,<-] (q1) edge[loop above] node {} (q1);
    
\end{tikzpicture}

}}$
    \end{minipage}
    \hspace{0.01\textwidth} 
    \includegraphics[width=0.37\textwidth]{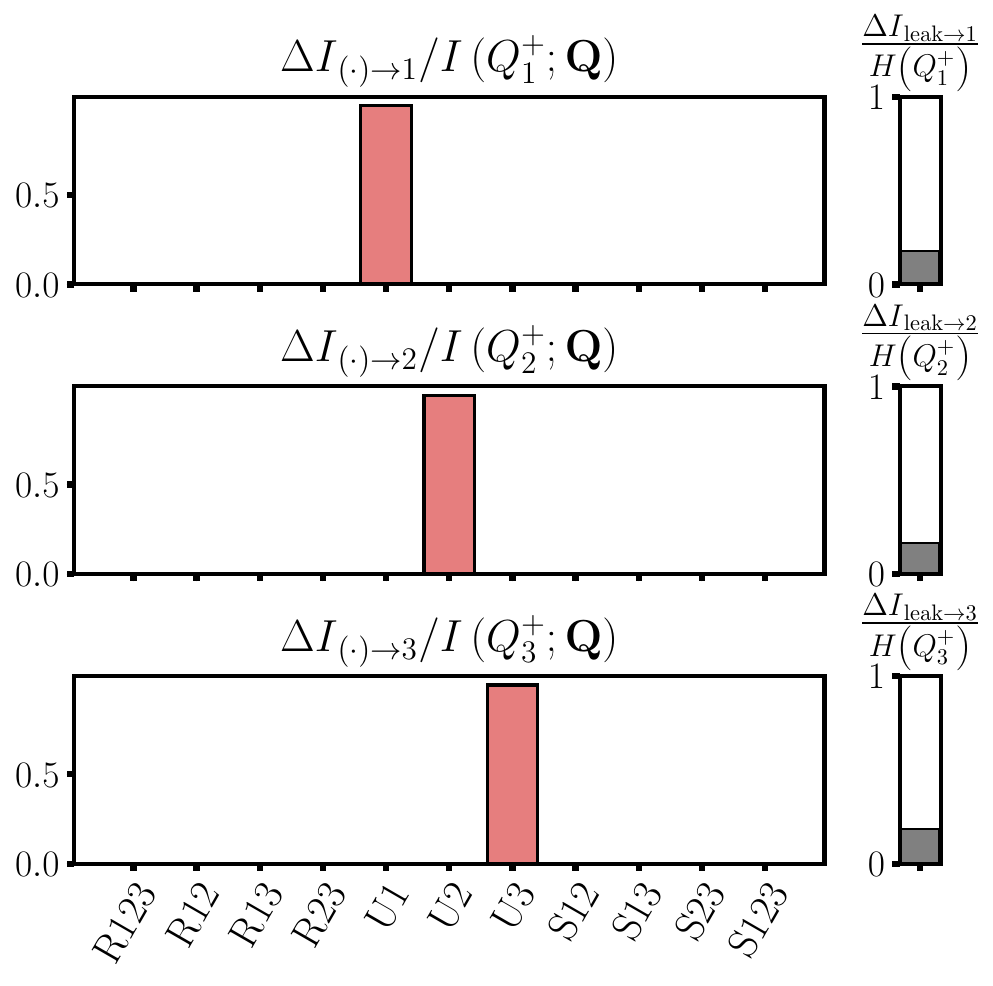}
    \begin{minipage}{0.1\textwidth}
        \vspace{-6.5cm}
        \includegraphics[width=\textwidth]{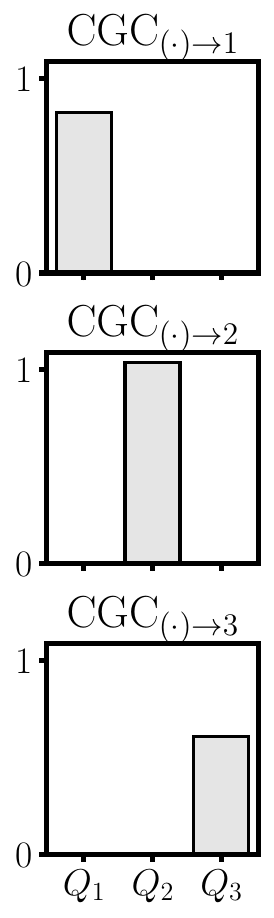}
    \end{minipage}
    \begin{minipage}{0.1\textwidth}
        \vspace{-6.5cm}
        \includegraphics[width=\textwidth]{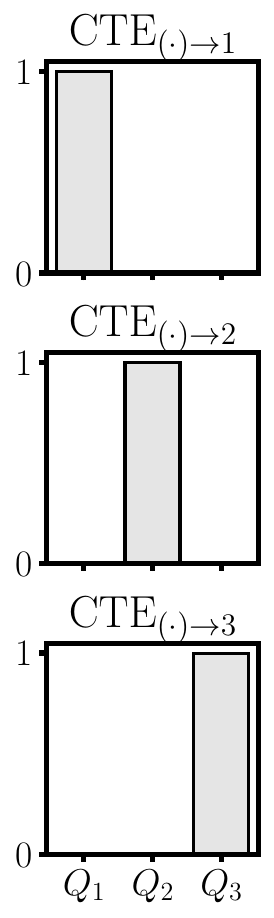}
    \end{minipage}
    \begin{minipage}{0.1\textwidth}
        \vspace{-6.5cm}
        \includegraphics[width=\textwidth]{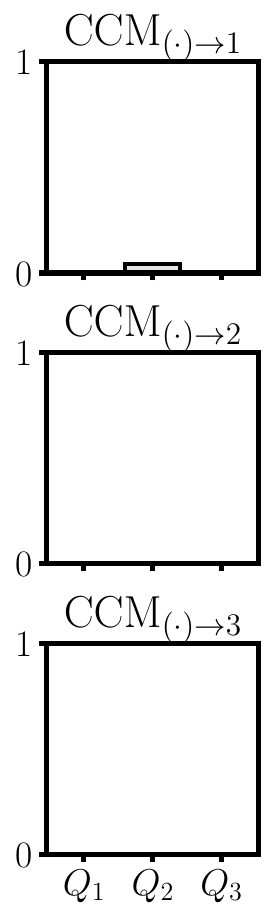}
    \end{minipage}
    \begin{minipage}{0.1125\textwidth}
        \vspace{-6.5cm}
        \includegraphics[width=\textwidth]{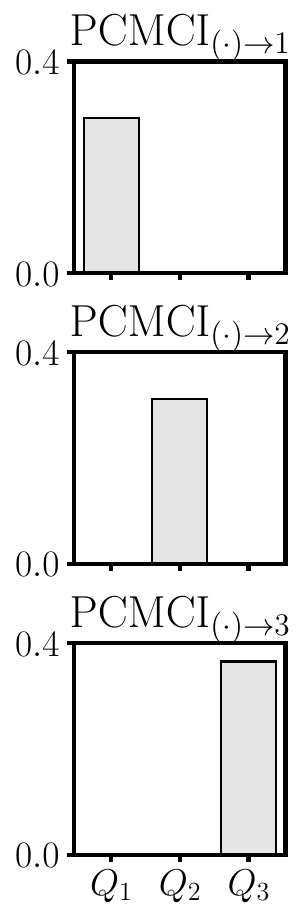}
    \end{minipage}
}

\subfloat[]{    
\begin{minipage}{0.175\textwidth}
        \vspace{-2\textwidth}
    ${\hbox{


\begin{tikzpicture}[cir/.style={circle,draw=black!70!white,,thick,inner sep=.5em},
    >={Latex[length=.2cm]},scale=.8]
    

    \colorlet{cA}{Q1!30!white}
    \colorlet{cB}{Q2!30!white}
    \colorlet{cC}{Q3!30!white}


    \node [cir,fill=cB] (q1) at (-1,+1)   {$Q_1$}; 
    \node [cir,fill=cA] (q2) at (-1,-1)   {$Q_2$}; 
    \node [cir,fill=cC] (q3) at (+0.73,0)    {$Q_3$}; 
    


    \path[thick,->]
        (q1) edge[bend right] node [left] {} (q2);
    \path[thick,<-] (q3) edge[loop right] node {} (q3);
    \path[thick,<-] (q2) edge[loop below] node {} (q2);
    \path[thick,<-] (q1) edge[loop above] node {} (q1);
    
\end{tikzpicture}

}}$
    \end{minipage}
    \hspace{0.01\textwidth} 
    \includegraphics[width=0.37\textwidth]{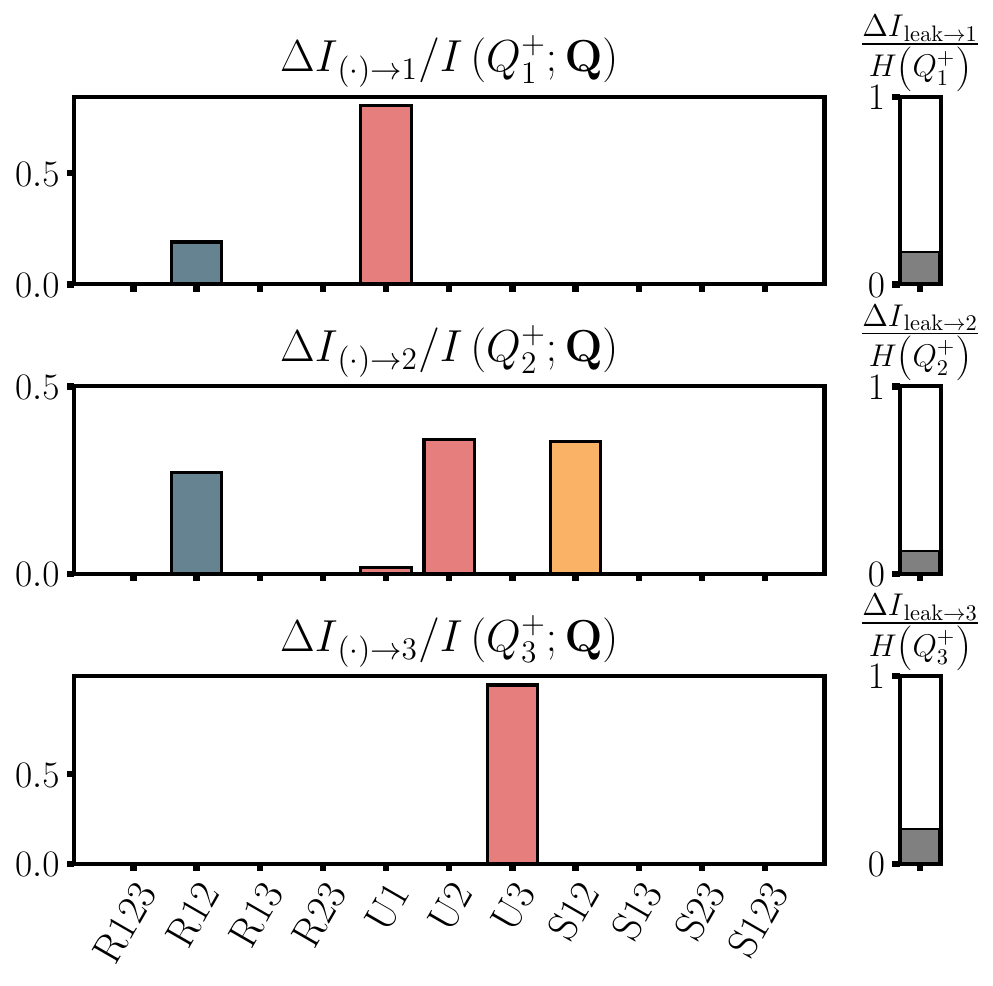}
    \begin{minipage}{0.1\textwidth}
        \vspace{-6.5cm}
        \includegraphics[width=\textwidth]{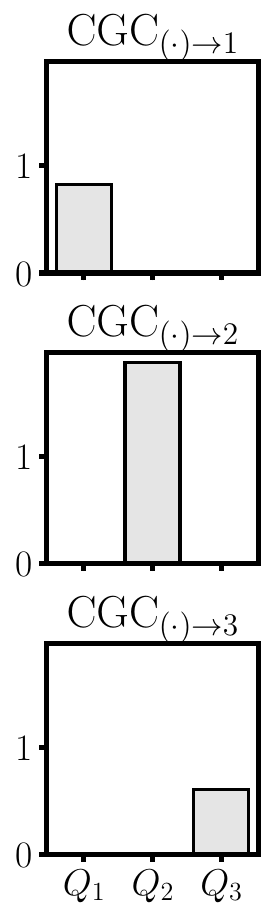}
    \end{minipage}
    \begin{minipage}{0.1\textwidth}
        \vspace{-6.5cm}
        \includegraphics[width=\textwidth]{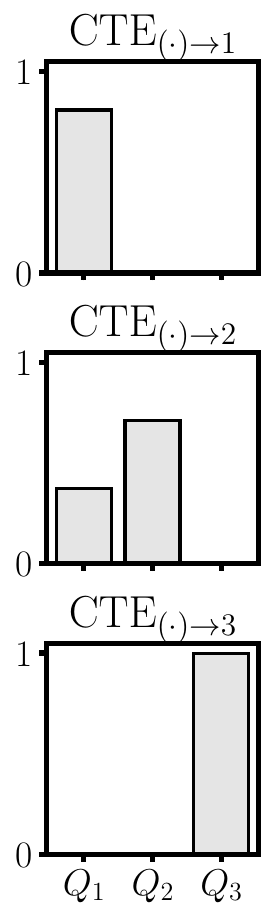}
    \end{minipage}
    \begin{minipage}{0.1\textwidth}
        \vspace{-6.5cm}
        \includegraphics[width=\textwidth]{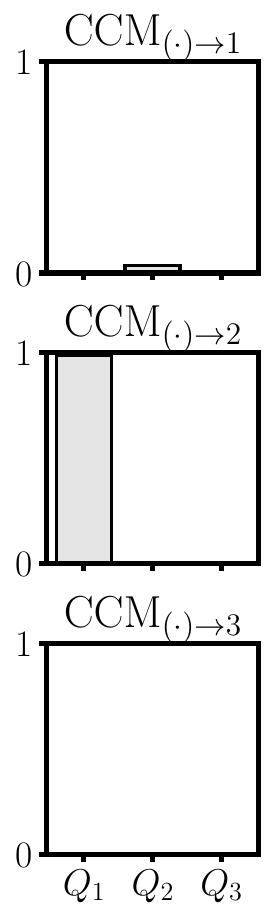}
    \end{minipage}
    \begin{minipage}{0.1125\textwidth}
        \vspace{-6.5cm}
        \includegraphics[width=\textwidth]{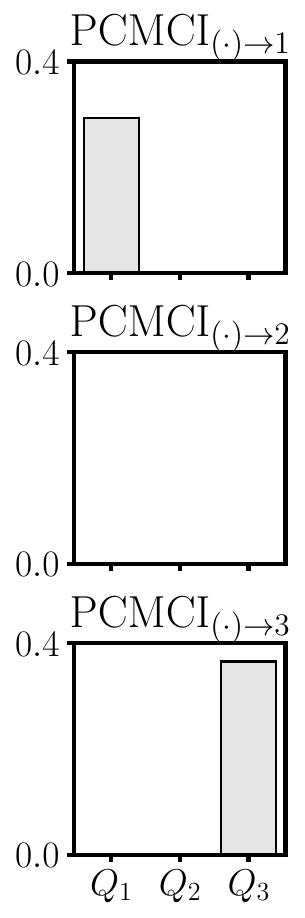}
    \end{minipage}}
    
\subfloat[]{    
\begin{minipage}{0.175\textwidth}
        \vspace{-2\textwidth}
    ${\hbox{


\begin{tikzpicture}[cir/.style={circle,draw=black!70!white,,thick,inner sep=.5em},
    >={Latex[length=.2cm]},scale=.8]
    

    \colorlet{cA}{Q1!30!white}
    \colorlet{cB}{Q2!30!white}
    \colorlet{cC}{Q3!30!white}


    \node [cir,fill=cB] (q1) at (-1,+1)   {$Q_1$}; 
    \node [cir,fill=cA] (q2) at (-1,-1)   {$Q_2$}; 
    \node [cir,fill=cC] (q3) at (+0.73,0)    {$Q_3$}; 
    


    \path[line width=2pt,->]
        (q1) edge[bend right] node [left] {} (q2);
    \path[thick,<-] (q3) edge[loop right] node {} (q3);
    \path[thick,<-] (q2) edge[loop below] node {} (q2);
    \path[thick,<-] (q1) edge[loop above] node {} (q1);

    
\end{tikzpicture}

}}$
    \end{minipage}
\hspace{0.01\textwidth} 
\includegraphics[width=0.37\textwidth]{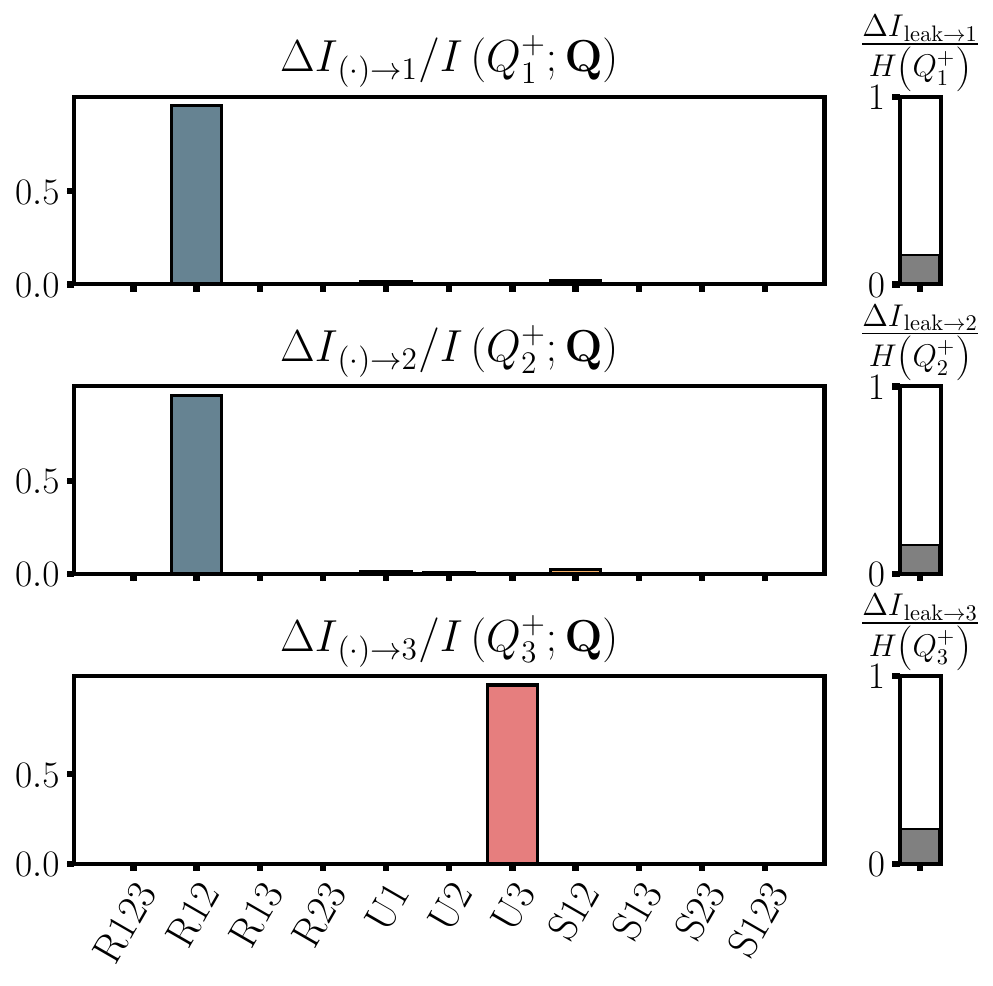}
    \hspace{-0.01\textwidth}
    \begin{minipage}{0.11\textwidth}
        \vspace{-6.5cm}
        \includegraphics[width=\textwidth]{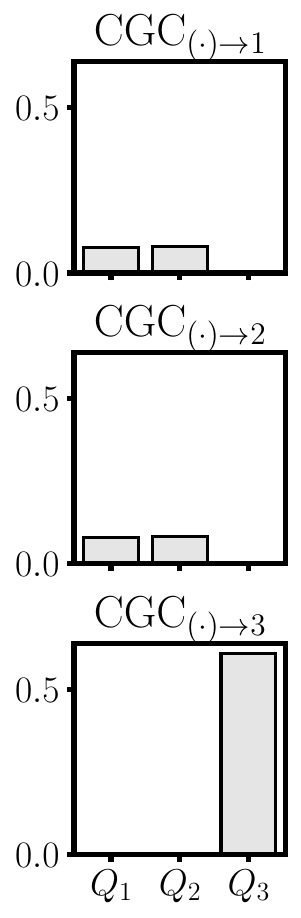}
    \end{minipage}
    \begin{minipage}{0.1\textwidth}
        \vspace{-6.5cm}
        \includegraphics[width=\textwidth]{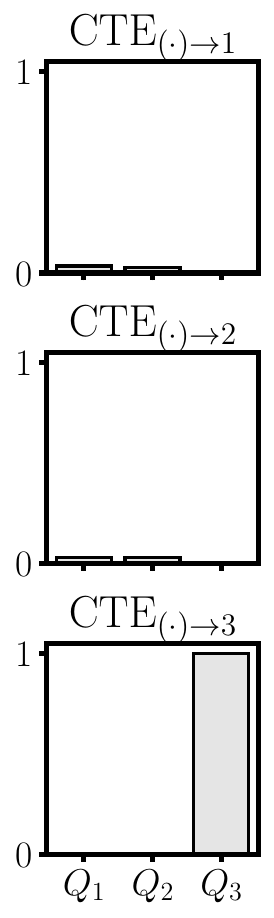}
    \end{minipage}
    \begin{minipage}{0.1\textwidth}
        \vspace{-6.5cm}
        \includegraphics[width=\textwidth]{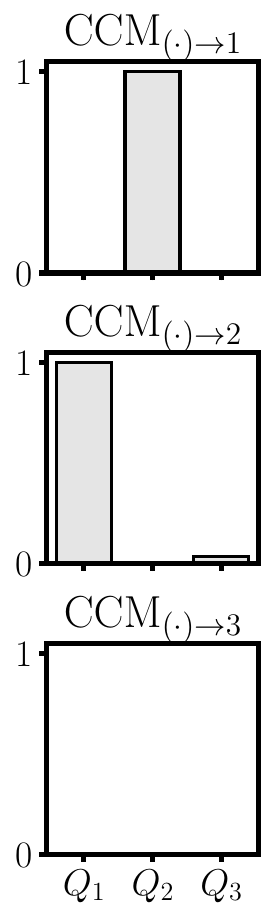}
    \end{minipage}
    \hspace{-0.01\textwidth}
    \begin{minipage}{0.1125\textwidth}
        \vspace{-6.5cm}
        \includegraphics[width=\textwidth]{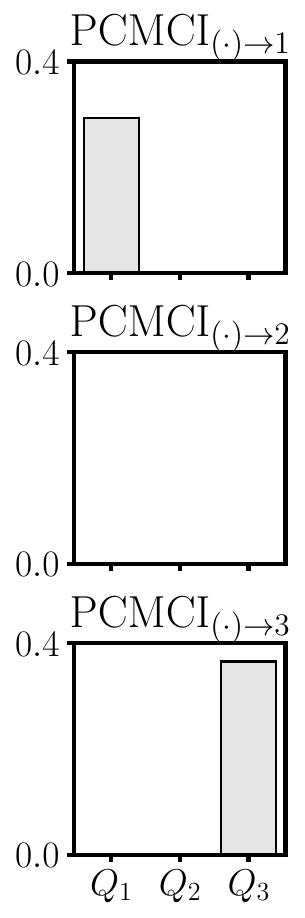}
    \end{minipage}}
\caption{Logistic maps with none or one variables coupled. Redundant (R), unique (U) and synergistic (S) causalities for
  $Q_1$, $Q_2$, $Q_3$ in coupled logistic maps for (a) uncoupled variables, $c_{1\rightarrow 2}=0$ and $c_{12\rightarrow 3}=0$, (b) intermediate coupling $Q_1 \rightarrow Q_2$,
  $c_{1\rightarrow 2}=0.1$ and $c_{12\rightarrow 3}=0$ and (c) strong
  coupling $Q_1 \rightarrow Q_2$ $c_{1\rightarrow2}=1$ and
  $c_{12\rightarrow3}=0$. The results from CGC, CTE, CCM, and PCMCI are depicted on
      the right. CGC and CTE use the same normalization as SURD.} \label{fig:logistic_12}
\end{figure}

First, we consider three cases with different degrees of coupling
between $Q_1$ and $Q_2$ while maintaining $Q_3$ uncoupled. The results
are shown in Figure~\ref{fig:logistic_12}.
\begin{itemize}
\item Uncoupled systems ($c_{1\rightarrow2}=c_{12\rightarrow3}=0$). In
  this case, $Q_1$, $Q_2$, and $Q_3$ are completely uncoupled and the
  only non-zero causalities are the self-induced unique components
  $\Delta I^U_{1\rightarrow 1}$, $\Delta I^U_{2\rightarrow 2}$, and
  $\Delta I^U_{3\rightarrow 3}$, as shown by the left panels in
  Figure~\ref{fig:logistic_12}(a).
\item Intermediate coupling $Q_1 \rightarrow Q_2$
  ($c_{1\rightarrow2}=0.1$ and $c_{12\rightarrow3}=0$). In this case,
  the dynamics of $Q_2$ are affected by $Q_1$. This is shown in
  Figure~\ref{fig:logistic_12}(b) by the non-zero terms $\Delta
  I^R_{12\rightarrow 2}\neq 0$, $\Delta I^U_{1\rightarrow 1}\neq 0$
  and $\Delta I^S_{12\rightarrow 2}\neq 0$. The latter is the
  synergistic causality due to the combined effect of $Q_1$ and $Q_2$,
  which is a manifestation of the coupling term $f_{1\rightarrow
    2}$. We can also observe that $\Delta I^R_{12\rightarrow 1}\neq
  0$. The latter redundant causality appears due to
    the emerging synchronization between $Q_1$ and $Q_2$. However,
    note that there is no other contribution (unique or synergistic)
    from $Q_2$ to $Q_1$. Hence, the redundant causality $\Delta I^R_{12\rightarrow 1}$ does not
    necessarily imply that conducting an intervention on 
    $Q_2$ (e.g., a perturbation) will alter the value of
    $Q_1$. Instead, it should be interpreted as $Q_2$ being able to
    inform about the future of $Q_1$, which is expected since $Q_1$ is
    contained in the right-hand side of the equation for $Q_2$.  The
  only non-zero causality for $Q_3$ is again $\Delta I^U_{3\rightarrow
    3}$, as it is uncoupled from $Q_1$ and $Q_2$. This result also
  demonstrates that the detection power of SURD is not affected by the
  inclusion of new independent variables in the analysis.
\item Strong coupling $Q_1 \rightarrow Q_2$ ($c_{1\rightarrow2}=1$ and
  $c_{12\rightarrow3}=0$). Taking the limit
  $c_{1\rightarrow2}\rightarrow \infty$, it can be seen that $Q_2
  \equiv Q_1$. It is also known that even for lower values of
  $c_{12\rightarrow3}\sim1$, $Q_1$ and $Q_2$ synchronize and both
  variables exhibit identical dynamics. This is revealed in
  Figure~\ref{fig:logistic_12}(c), where the only non-zero causalities
  are $\Delta I^R_{12\rightarrow 1}=\Delta I^R_{12\rightarrow 2}\neq
  0$. The identical redundant causalities along with the absence of
  any unique or synergistic causality between $Q_1$ and $Q_2$, imply
  that both variables are fully synchronized. In this situation, the
  directionality of the causality cannot be established, as $Q_1$ and
  $Q_2$ behave as a single variable, {but SURD still
    effectively identifies the state of synchronization}. Similar to
  the two previous cases, $Q_3$ remains unaffected ($\Delta
  I^U_{3\rightarrow 3} \neq 0$).
\end{itemize}
%
Compared to other methods, SURD provides consistent results across the
three coupling scenarios. While all methods correctly identified links
in the uncoupled case, introducing intermediate and strong coupling
between $Q_1$ and $Q_2$ hindered the identification of causal
relationships. In such cases, the synchronization of $Q_1$ and $Q_2$
can be observed in SURD through redundant causality, a capability that
other methods lack.
%
\begin{figure}[t!]
\subfloat[]{
    \begin{minipage}{0.175\textwidth}
        \vspace{-2\textwidth}
        ${\hbox{


\begin{tikzpicture}[cir/.style={circle,draw=black!70!white,,thick,inner sep=.5em},
    >={Latex[length=.2cm]},scale=.8]
    

    \colorlet{cA}{Q1!30!white}
    \colorlet{cB}{Q2!30!white}
    \colorlet{cC}{Q3!30!white}


    \node [cir,fill=cB] (q1) at (-1,+1)   {$Q_1$}; 
    \node [cir,fill=cA] (q2) at (-1,-1)   {$Q_2$}; 
    \node [cir,fill=cC] (q3) at (+0.73,0)    {$Q_3$}; 
    


    \path[line width=2pt,->]
        (q1) edge[bend left] node [left] {} (q3)
        (q2) edge[bend right] node [left] {} (q3);
    \path[thick,<-] (q3) edge[loop right] node {} (q3);
    \path[thick,<-] (q2) edge[loop below] node {} (q2);
    \path[thick,<-] (q1) edge[loop above] node {} (q1);
    
\end{tikzpicture}

}}$
    \end{minipage}
    \hspace{0.01\textwidth} 
    \includegraphics[width=0.37\textwidth]{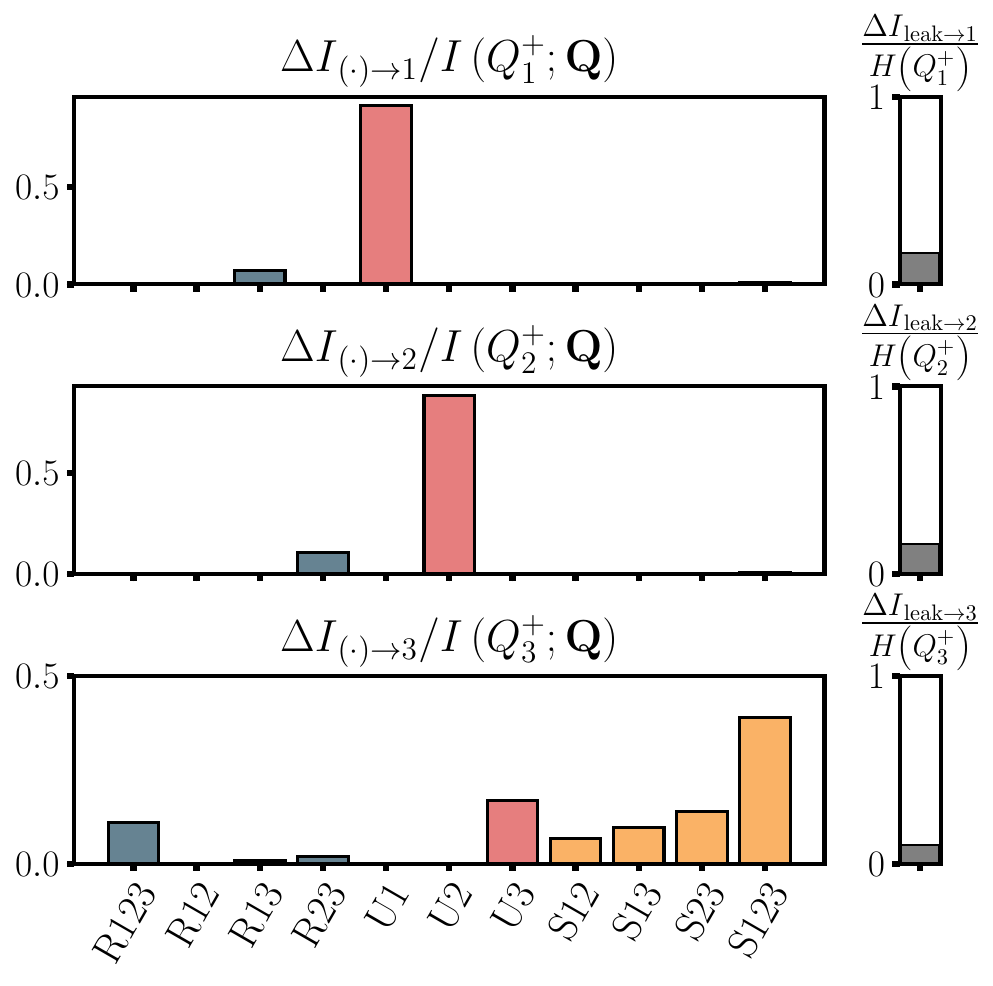}
    \begin{minipage}{0.11\textwidth}
        \vspace{-6.5cm}
        \includegraphics[width=\textwidth]{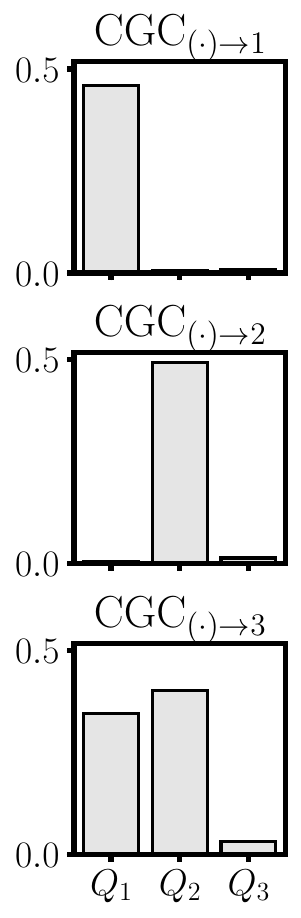}
    \end{minipage}
    \begin{minipage}{0.1\textwidth}
        \vspace{-6.5cm}
        \includegraphics[width=\textwidth]{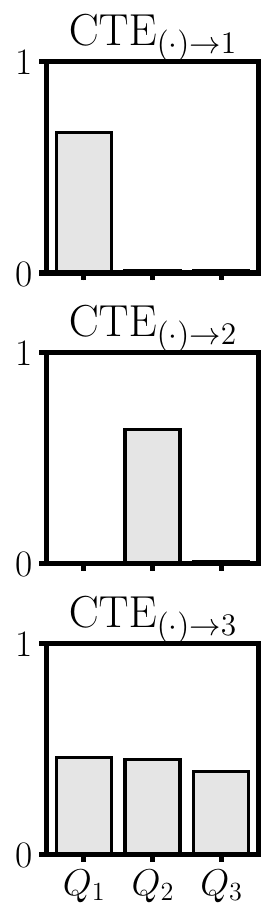}
    \end{minipage}
    \begin{minipage}{0.1\textwidth}
        \vspace{-6.5cm}
        \includegraphics[width=\textwidth]{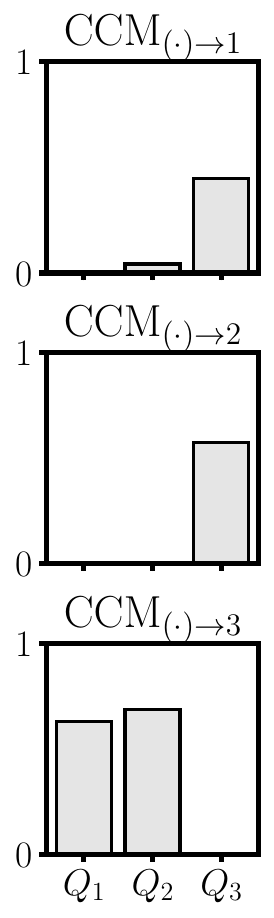}
    \end{minipage}
    \begin{minipage}{0.1125\textwidth}
        \vspace{-6.5cm}
        \includegraphics[width=\textwidth]{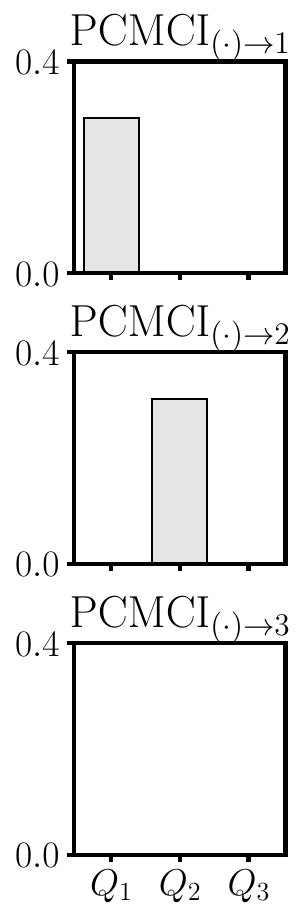}
    \end{minipage}
}

    
\subfloat[]{    
\begin{minipage}{0.175\textwidth}
        \vspace{-2\textwidth}
    ${\hbox{


\begin{tikzpicture}[cir/.style={circle,draw=black!70!white,,thick,inner sep=.5em},
    >={Latex[length=.2cm]},scale=.8]
    

    \colorlet{cA}{Q1!30!white}
    \colorlet{cB}{Q2!30!white}
    \colorlet{cC}{Q3!30!white}


    \node [cir,fill=cB] (q1) at (-1,+1)   {$Q_1$}; 
    \node [cir,fill=cA] (q2) at (-1,-1)   {$Q_2$}; 
    \node [cir,fill=cC] (q3) at (+0.73,0)    {$Q_3$}; 
    


    \path[line width=2pt,->]
        (q1) edge[bend right] node [left] {} (q2)
        (q1) edge[bend left] node [left] {} (q3)
        (q2) edge[bend right] node [left] {} (q3);
    \path[thick,<-] (q3) edge[loop right] node {} (q3);
    \path[thick,<-] (q2) edge[loop below] node {} (q2);
    \path[thick,<-] (q1) edge[loop above] node {} (q1);
    
\end{tikzpicture}

}}$
    \end{minipage}
\hspace{0.01\textwidth} 
\includegraphics[width=0.37\textwidth]{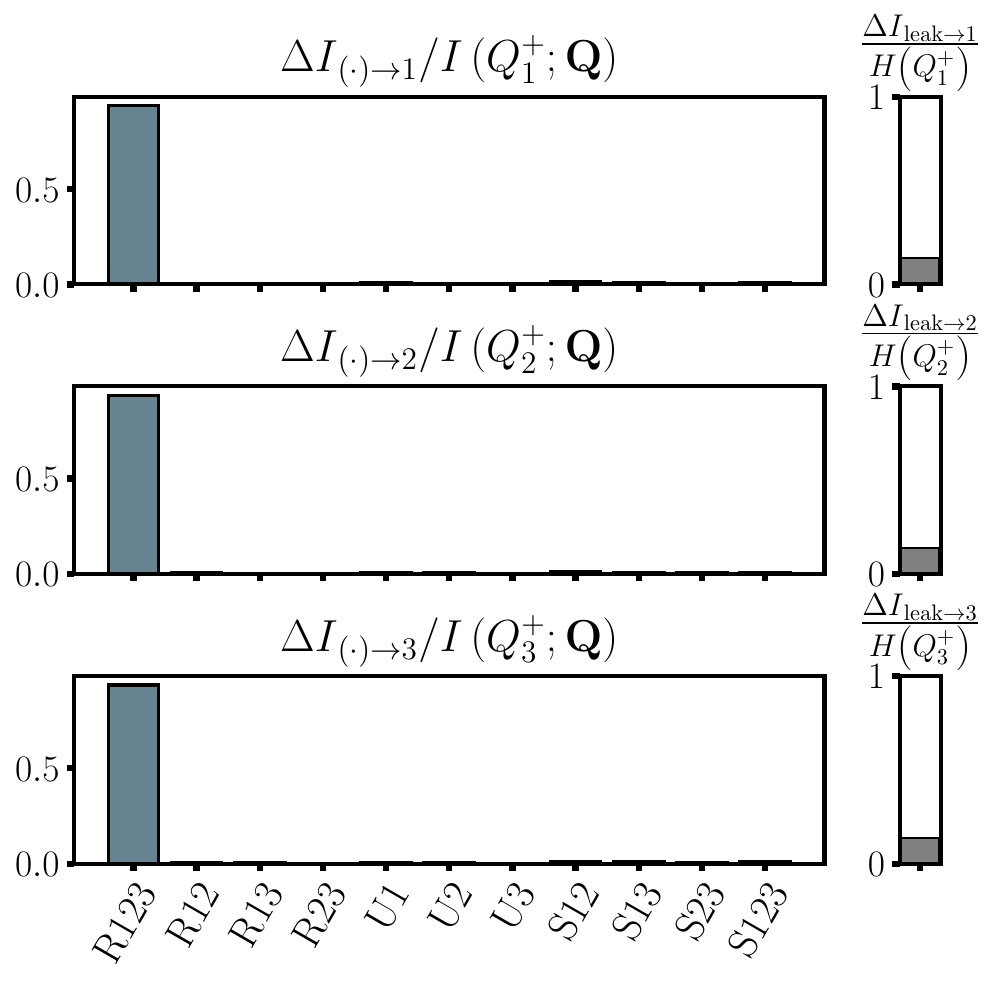}
    \hspace{-0.01\textwidth}
    \begin{minipage}{0.1125\textwidth}
        \vspace{-6.5cm}
        \includegraphics[width=\textwidth]{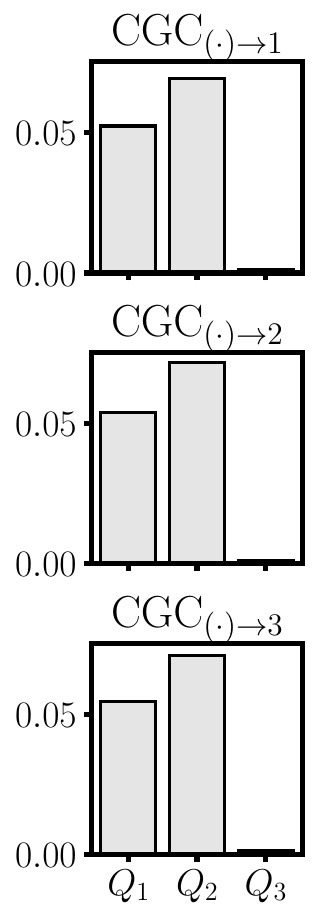}
    \end{minipage}
    \begin{minipage}{0.1\textwidth}
        \vspace{-6.5cm}
        \includegraphics[width=\textwidth]{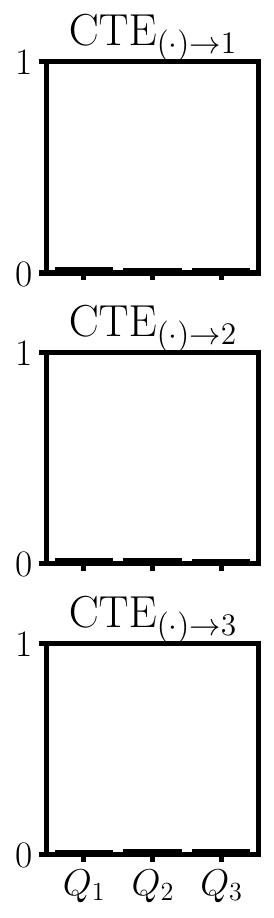}
    \end{minipage}
    \begin{minipage}{0.1\textwidth}
        \vspace{-6.5cm}
        \includegraphics[width=\textwidth]{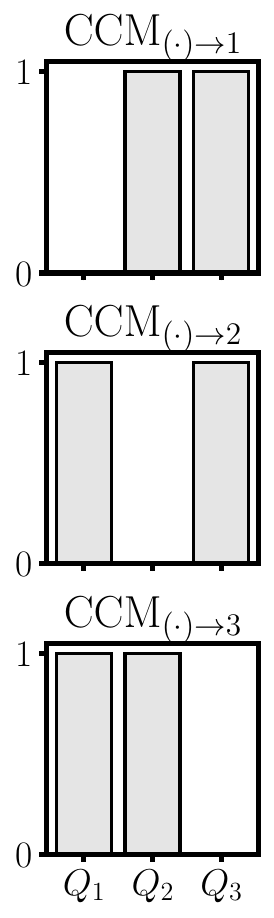}
    \end{minipage}
    \begin{minipage}{0.1125\textwidth}
        \vspace{-6.5cm}
        \includegraphics[width=\textwidth]{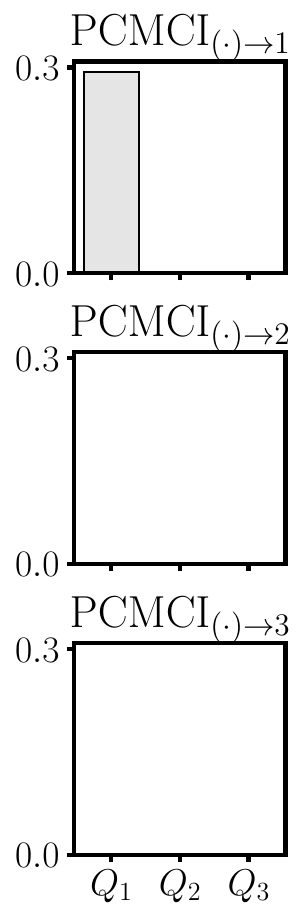}
    \end{minipage}}
\caption{Logistic maps with two or three variables coupled. Redundant
  (R), unique (U) and synergistic (S) causalities for $Q_1$, $Q_2$,
  $Q_3$ for (a) uncoupled variables, $c_{1\rightarrow 2}=0$ and
  $c_{12\rightarrow 3}=1$, (b) strong coupling $Q_1 \rightarrow
  Q_2$ and $[Q_1,Q_2] \rightarrow Q_3$, $c_{12\rightarrow3}=1$ and
  $c_{12\rightarrow3}=1$. The results from CGC, CTE, CCM, and PCMCI are depicted on
      the right. CGC and CTE use the same normalization as SURD.}\label{fig:logistic_123}
\end{figure}

Next, we consider two additional cases in which $Q_3$ is coupled with
$Q_1$ and $Q_2$. The results are shown in
Figure~\ref{fig:logistic_123}.
\begin{itemize}
\item Strong coupling $[Q_2,Q_1]\rightarrow Q_3$ and no coupling
  $Q_1\rightarrow Q_2$ ($c_{1\rightarrow2}=0$ and
  $c_{12\rightarrow3}=1$). The results, included in
  Figure~\ref{fig:logistic_123}(a), show that most of the causality to
  $Q_1$ and $Q_2$ is self-induced and unique ($\Delta
  I^U_{1\rightarrow 1}\neq0$ and $\Delta I^U_{2\rightarrow 2}\neq0$,
  respectively).  There is also a strong causality from $Q_1$ and
  $Q_2$ to $Q_3$ in the form of synergistic causality, being $\Delta
  I^S_{123 \rightarrow 3}$ the dominant component consistent with the
  coupling term $f_{123}$.
\item Strong coupling $[Q_2,Q_1]\rightarrow Q_3$ and $Q_1\rightarrow
  Q_2$ ($c_{1\rightarrow2}=1$ and $c_{12\rightarrow3}=1$).  In this
  case, the three variables synchronize such that $\Delta
  I^R_{123\rightarrow 1} = \Delta I^R_{123\rightarrow 2} = \Delta
  I^R_{123\rightarrow 3}\neq 0$ (i.e., they can be interpreted as
  exact copies of each other). The results are shown in
  Figure~\ref{fig:logistic_123}(b).
\end{itemize}
%

A summary of the results provided by other methods is provided in Table
\ref{tab:summary logisitc maps}. We note that CCM is the only method that can
offer insights into the dynamics of the system, since the logistic maps analyzed in this section are given by a deterministic dynamical system where the coupled variables are part of the same manifold. The only case in which this method failed is the one of strong coupling between variables $\left[ Q_1, Q_2 \right] \rightarrow Q_3$, where the CCM did not converge to a value of one for a relatively high number of samples. This is the case where synergistic effects are important according to SURD and this might be playing a role in the identification of causalities from CCM. 

CTE also provides consistent results for those cases in which the coupling between variables is inexistent or intermediate. However, when the coupling is strong the method completely cases. For these cases, SURD identifies significant redundant causalities, which implies those variables are fully synchronized and act as the same variables. Therefore, introducing those variables into the conditioning set of CTE for a given variable hinders the identification of causalities. A similar conclusion can be drawn for CGC and PCMCI, which only identified consistent causal relationships for the case in which there is no coupling between variables and for $\left[ Q_1, Q_2 \right] \rightarrow Q_3$ in the CGC case.


\begin{table}[t!]
\centering
\begin{tabular}{|l|c|c|c|c|c|c|c|}
\hline
\textbf{Case}  & \textbf{CGC} & \textbf{CTE} & \textbf{CCM} & \textbf{PCMCI} & \textbf{SURD} \\
\hline
No coupling             & \cmark & \cmark & \cmark & \cmark  & \cmark \\
One-way intermediate coupling             & \xmark & \cmark & \cmark & \xmark  & \cmark \\
One-way strong coupling             & \xmark & \xmark & \cmark & \xmark  & \cmark \\
Two-way strong coupling             & \cmark & \cmark & \xmark$^\dag$ & \xmark  & \cmark \\
Three-way strong coupling               & \xmark & \xmark & \cmark & \xmark  & \cmark \\
\hline
\end{tabular}
\caption{{{Summary of the performance of the different methods for logistic maps with none to 
    three coupled variables.}} $^\dag$The value of the prediction skill from CCM does not converge to a value of one.}
\label{tab:summary logisitc maps}
\end{table}

\subsection{Coupled R{\"o}ssler--Lorenz system}
\label{sec:Rossler}

We study a coupled version of the Lorenz system~\cite{lorenz1963} and
the R\"ossler system~\cite{rossler1977}. The former was developed by
Lorenz as a simplified model of a viscous fluid flow. R\"ossler proposed
a simpler version of the Lorenz's equations in order to facilitate the
study of its chaotic properties. The governing equations are
\begin{subequations}
\begin{align}  
  \label{eq:RL}
 \frac{\dd Q_1}{\dd t} &= -6[ Q_2 + Q_3 ], \\
 \frac{\dd Q_2}{\dd t} &= 6[Q_1 + 0.2Q_2 ], \\
 \frac{\dd Q_3}{\dd t} &= 6\left[ 0.2+Q_3[Q_1-5.7] \right], \\
 \frac{\dd Q_4}{\dd t} &= 10[ Q_5 - Q_4 ], \\
 \frac{\dd Q_5}{\dd t} &= Q_4 [28 - Q_6 ] - Q_5 + c Q_2^2, \\
 \frac{\dd Q_6}{\dd t} &= Q_4 Q_5 - \frac{8}{3}Q_6,
\end{align}
\end{subequations}
where $[Q_1,Q_2,Q_3]$ correspond to the R\"ossler system and
$[Q_4,Q_5,Q_6]$ to the Lorenz system. The coupling is unidirectional
from the R\"ossler system to the Lorenz system from $ Q_2 \rightarrow
Q_5$ via the parameter $c$. This coupled system has previously been
studied by \cite{Quiroga2000} and \cite{Krakovska2018}. 
We use this case to study the behavior of SURD in a continuous
dynamical system when some of the variables are hidden. The observable
variables are $\bQ = [Q_1, Q_2, Q_5, Q_6]$. The system was integrated
for $10^6 t_{\text{ref}}$ where $t_{\text{ref}}$ is the time for which
$I(Q_1^+ ; Q_1)/I(Q_1 ; Q_1) = 0.5$. The time-lag selected for causal
inference is $\Delta T \approx t_{\text{ref}}$ and 50 bins per
variable were used to partition the observed phase space.
%
\begin{figure}[t!]
\begin{minipage}{\textwidth}
    \centering
    \hspace{-0.025\textwidth}
    \begin{minipage}{0.35\textwidth}
        \includegraphics[trim={0 6cm 0 0},clip,width=\textwidth]{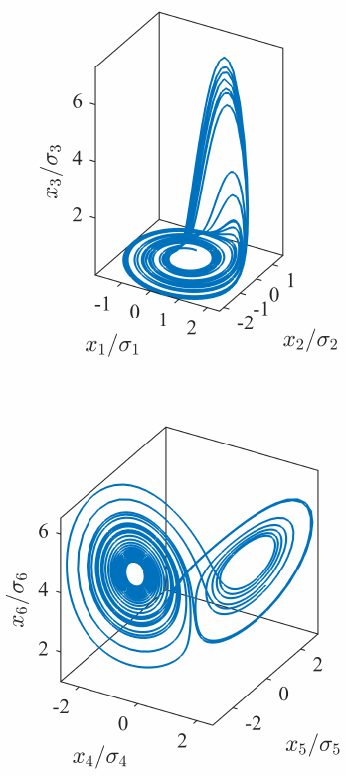}
    \end{minipage}
    \hspace{0.025\textwidth}
    \begin{minipage}{0.35\textwidth}
    \vspace{-2cm}
        \includegraphics[trim={0 0 0 6cm},clip,width=\textwidth]{supplementary/Rossler_Lorenz_trajectory_c0.pdf}
    \end{minipage}
    \vspace{0.1cm}
  \end{minipage}
  \begin{minipage}{0.5\textwidth}
  \vspace{-1cm}
      \includegraphics[width=\textwidth]{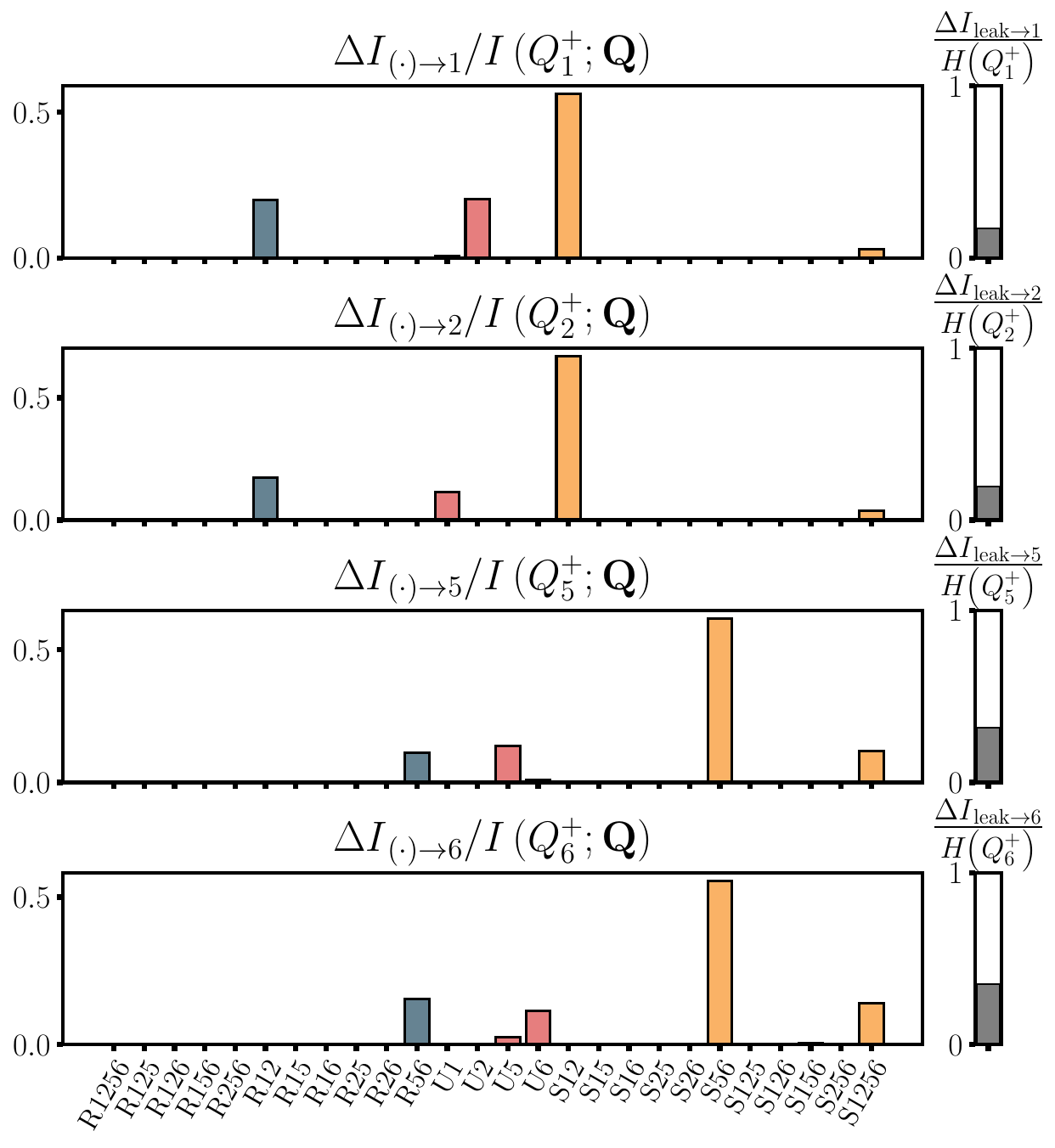}
  \end{minipage}
  \hspace{-0.01\textwidth}
  \begin{minipage}{0.115\textwidth}
  \vspace{-1cm}
      \includegraphics[width=\textwidth]{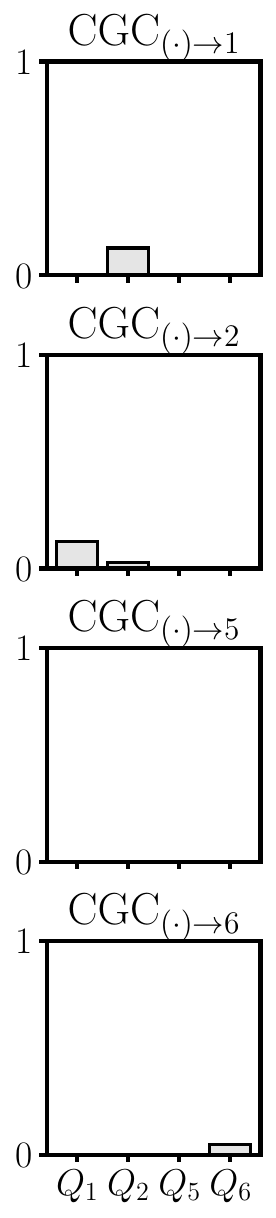}
  \end{minipage}
  \begin{minipage}{0.115\textwidth}
  \vspace{-1cm}
      \includegraphics[width=\textwidth]{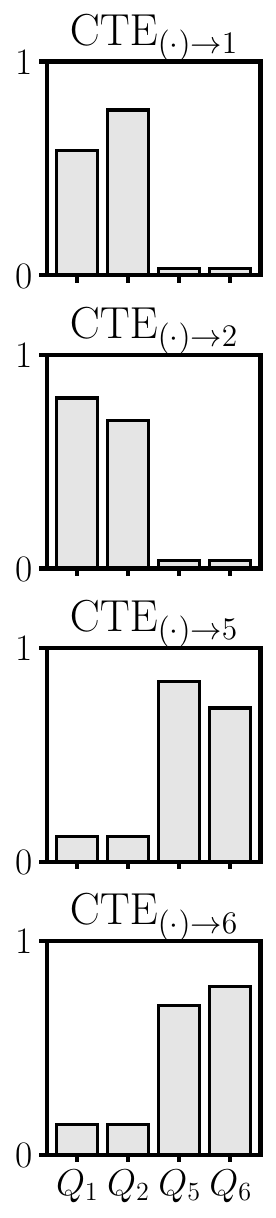}
  \end{minipage}
  \begin{minipage}{0.115\textwidth}
  \vspace{-1cm}
      \includegraphics[width=\textwidth]{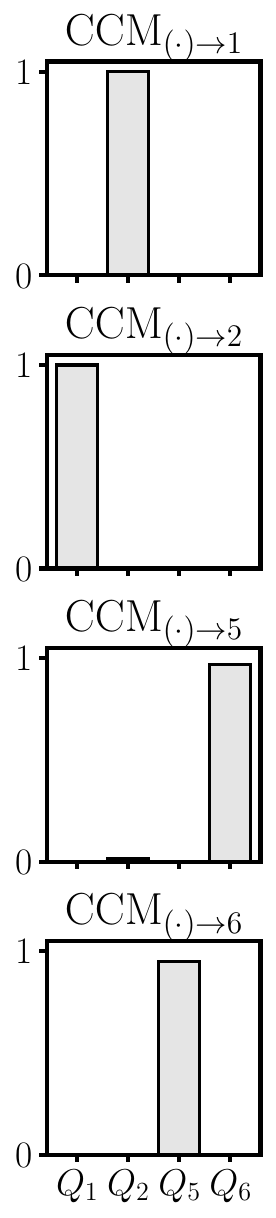}
  \end{minipage}
  \begin{minipage}{0.1275\textwidth}
  \vspace{-1cm}
      \includegraphics[width=\textwidth]{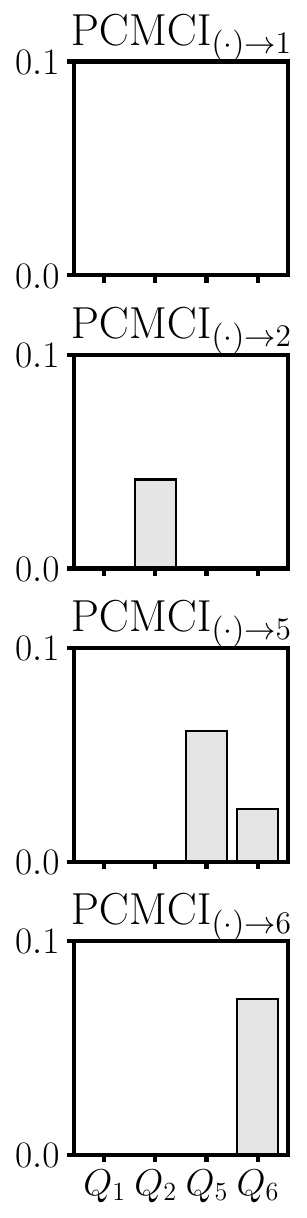}
  \end{minipage}
\caption{ Uncoupled R\"ossler-Lorenz system ($c=0$). The left panels
  show excerpts of the trajectories pertaining to R\"ossler systems
  $[Q_1,Q_2,Q_3]$ (left) and Lorenz system $[Q_4,Q_5,Q_6]$
  (right). The left panels show the redundant (R), unique (U), and
  synergistic (S) causalities among $[Q_1, Q_2, Q_5, Q_6]$. The
  causality leak for each variable is also shown in the right-hand
  side bar. The
  results from CGC, CTE, CCM, and PCMCI are depicted on the right. CGC
  and CTE use the same normalization as
  SURD.}\label{fig:Rossler_c0}
\end{figure}

The results for uncoupled systems ($c=0$) are shown in Figure
\ref{fig:Rossler_c0}. The upper panel portrays typical trajectories
of the systems.  Unsurprisingly, SURD shows that both systems are
uncoupled. Moreover, the unique, redundant and synergistic causal
structure identified in the R\"ossler and Lorenz systems are
consistent with structure of Equation~\eqref{eq:RL}. The causality leak is
roughly 25\% due to the unobserved variables.
%

\begin{figure}[t!]
\begin{minipage}{\textwidth}
    \centering
    \begin{minipage}{0.35\textwidth}
        \includegraphics[trim={0 6cm 0 0},clip,width=\textwidth]{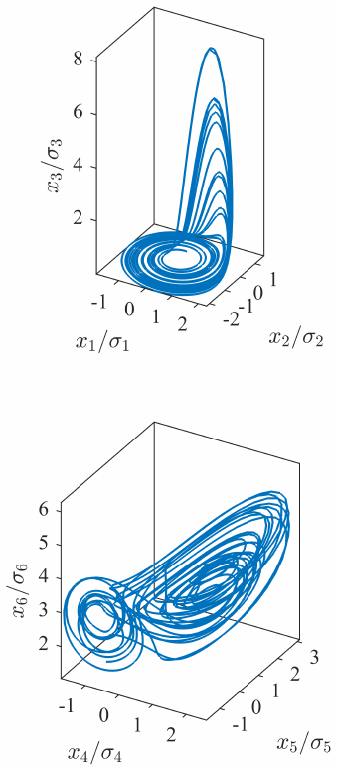}
    \end{minipage}
    \hspace{0.025\textwidth}
    \begin{minipage}{0.35\textwidth}
    \vspace{-2cm}
        \includegraphics[trim={0 0 0 6cm},clip,width=\textwidth]{supplementary/Rossler_Lorenz_trajectory_c2.pdf}
    \end{minipage}
    \vspace{0.1cm}
  \end{minipage}
  \begin{minipage}{0.5\textwidth}
  \vspace{-1cm}
      \includegraphics[width=\textwidth]{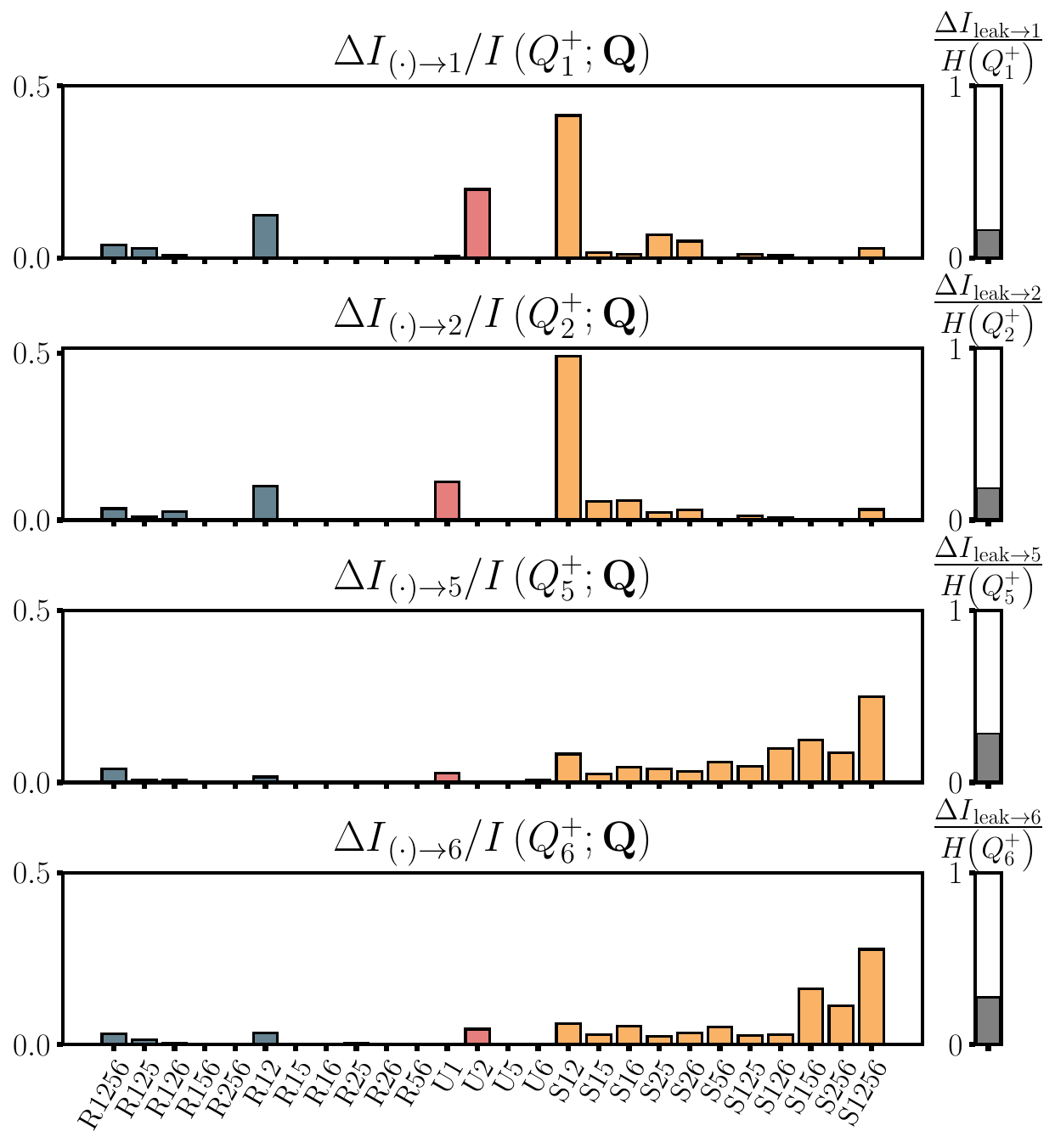}
  \end{minipage}
  \hspace{-0.01\textwidth}
  \begin{minipage}{0.115\textwidth}
  \vspace{-1cm}
      \includegraphics[width=\textwidth]{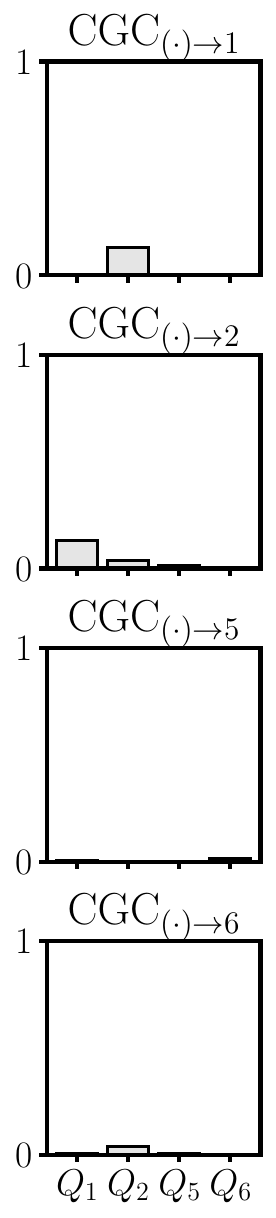}
  \end{minipage}
  \begin{minipage}{0.115\textwidth}
  \vspace{-1cm}
      \includegraphics[width=\textwidth]{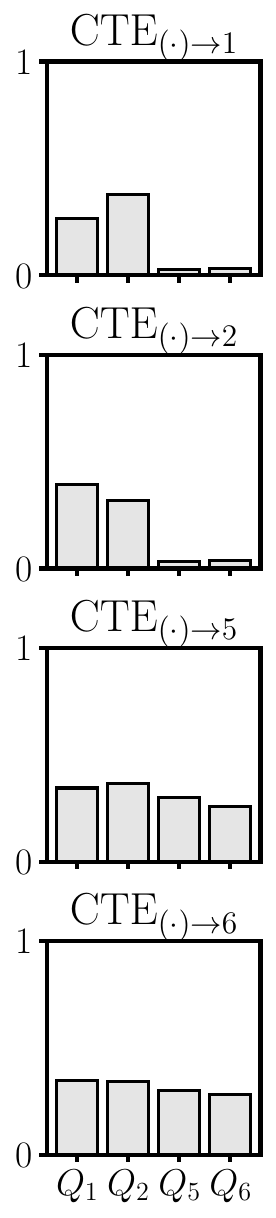}
  \end{minipage}
  \begin{minipage}{0.115\textwidth}
  \vspace{-1cm}
      \includegraphics[width=\textwidth]{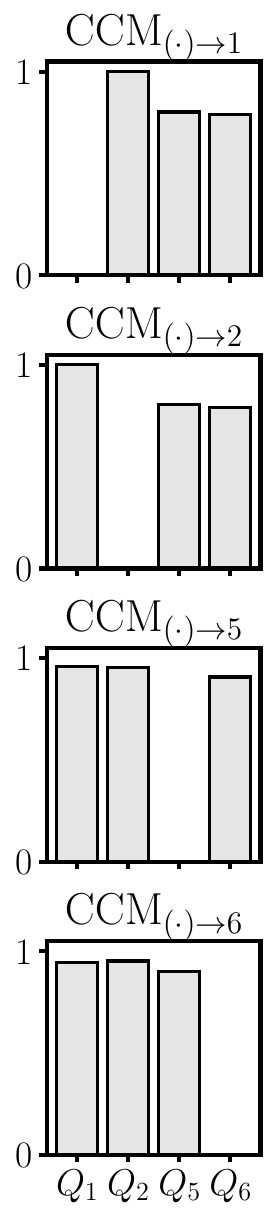}
  \end{minipage}
  \begin{minipage}{0.1275\textwidth}
  \vspace{-1cm}
      \includegraphics[width=\textwidth]{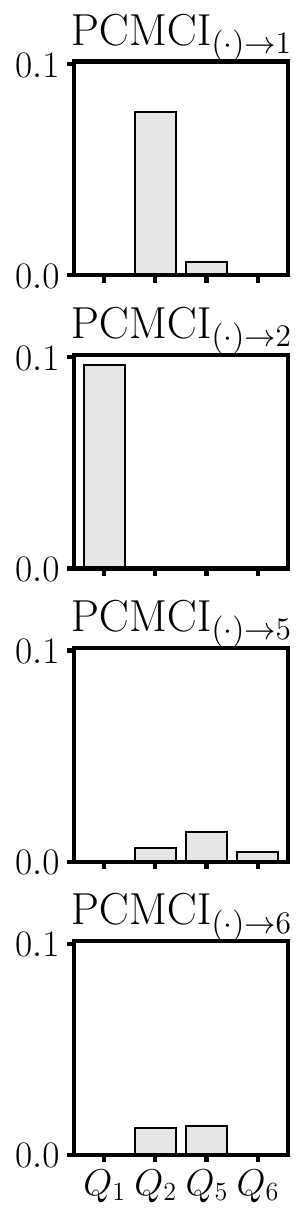}
  \end{minipage}
\caption{ Coupled R\"ossler-Lorenz system ($c=2$). The left panels
  show excerpts of the trajectories pertaining to R\"ossler systems
  $[Q_1,Q_2,Q_3]$ (left) and Lorenz system $[Q_4,Q_5,Q_6]$
  (right). The left panels show the redundant (R), unique (U), and
  synergistic (S) causalities among $[Q_1, Q_2, Q_5, Q_6]$. The
  causality leak for each variable is also shown in the right-hand
  side bar. The results from CGC, CTE, CCM, and PCMCI are depicted on
  the right. CGC and CTE use the same normalization as
  SURD.} \label{fig:Rossler_c2}
\end{figure}
The results for the coupled system ($c=2$) are shown in Figure
\ref{fig:Rossler_c2}.  The left panel shows how the trajectory of the
Lorenz system is severely impacted by the coupling.  The causalities
in the R\"ossler system remain comparable to the uncoupled case besides some
small redundancies and synergies due to the effect of unobserved
variables. On the contrary, the causalities in the Lorenz system
undergo deeper changes. This is evidenced by the emergence of multiple
synergistic causalities involving $Q_1$ and $Q_2$. This effect is
consistent with the coupling of both systems. 


Among the other methods, only the CTE and CCM provide insight into the
dynamics of the system, given that the system is given by a
deterministic dynamical system where the group of variables
$\left[Q_1, Q_2\right]$ and $\left[Q_5, Q_6\right]$ are part of the
same manifold, respectively, in the uncoupled R{\"o}ssler--Lorenz
system and $\left[Q_1, Q_2, Q_5, Q_6\right]$ in the coupled
R{\"o}ssler--Lorenz case. However, CCM cannot clearly identify that
the coupling is from $Q_2$ to $Q_5$, since the prediction skill of
variables $Q_1$ and $Q_2$ using $Q_5$ and $Q_6$ is also very high,
although lower than in the reverse case. For CTE, the direction of
this coupling is properly identified. Finally, the methods CGC and
PCMCI completely fail in both cases.

\subsection{Three interacting species}
\begin{figure}[p!]
\subfloat[]{
    \begin{minipage}{0.175\textwidth}
        \vspace{-2\textwidth}
        ${\hbox{


\begin{tikzpicture}[cir/.style={circle,draw=black!70!white,,thick,inner sep=.5em},
    >={Latex[length=.2cm]},scale=.8]
    

    \colorlet{cA}{Q1!30!white}
    \colorlet{cB}{Q2!30!white}
    \colorlet{cC}{Q3!30!white}


    \node [cir,fill=cB] (q1) at (-1,+1)   {$Q_1$}; 
    \node [cir,fill=cA] (q2) at (-1,-1)   {$Q_2$}; 
    \node [cir,fill=cC] (q3) at (+0.73,0)    {$Q_3$}; 
    


    \path[thick,->]
        (q1) edge[bend right] node [left] {} (q2);
    \path[thick,<-] (q3) edge[loop right] node {} (q3);
    \path[thick,<-] (q2) edge[loop below] node {} (q2);
    \path[thick,<-] (q1) edge[loop above] node {} (q1);
    
\end{tikzpicture}

}}$
    \end{minipage}
    \hspace{0.01\textwidth} 
    \includegraphics[width=0.37\textwidth]{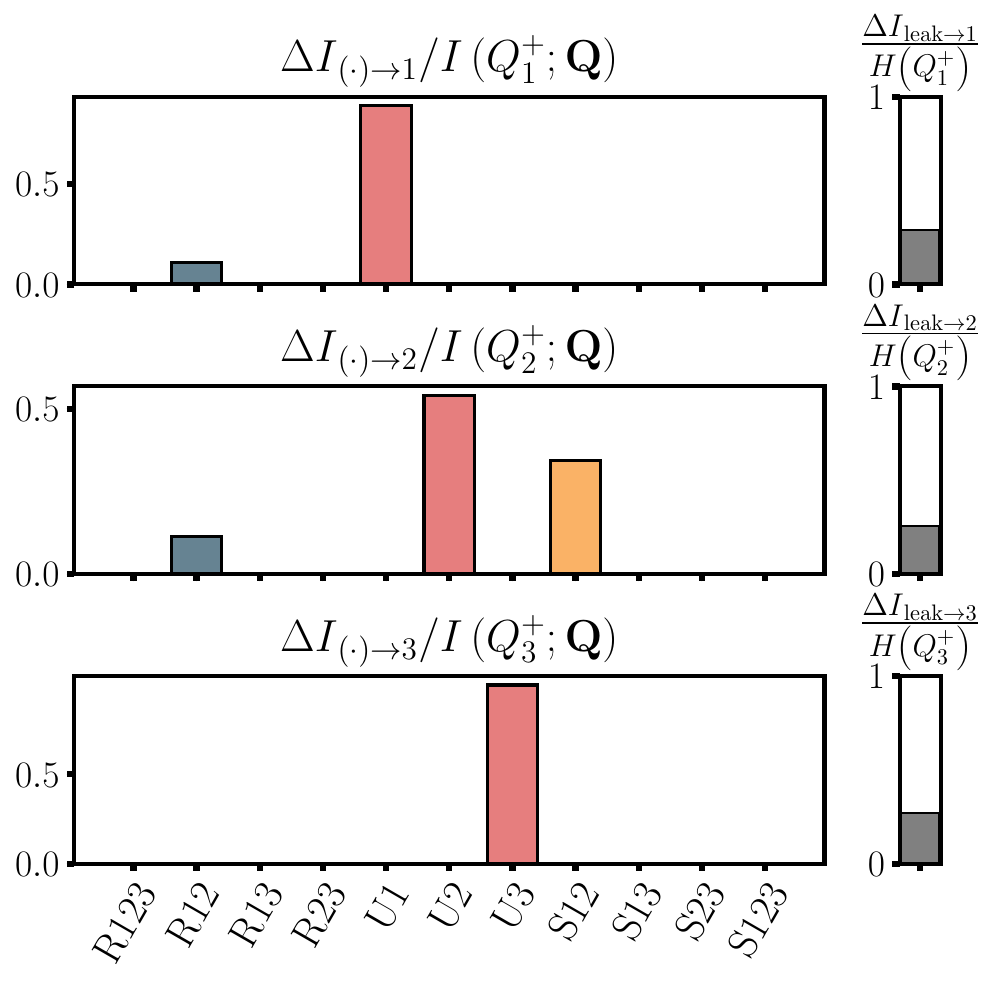}
    \begin{minipage}{0.1\textwidth}
        \vspace{-6.5cm}
        \includegraphics[width=\textwidth]{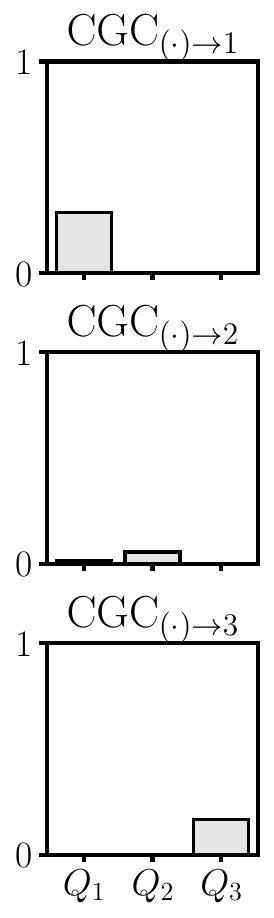}
    \end{minipage}
    \begin{minipage}{0.1\textwidth}
        \vspace{-6.5cm}
        \includegraphics[width=\textwidth]{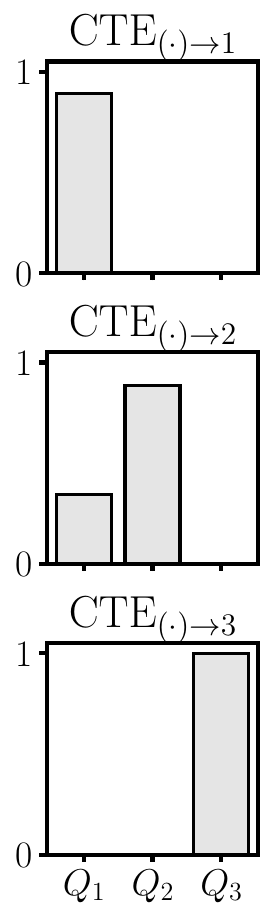}
    \end{minipage}
    \begin{minipage}{0.1\textwidth}
        \vspace{-6.5cm}
        \includegraphics[width=\textwidth]{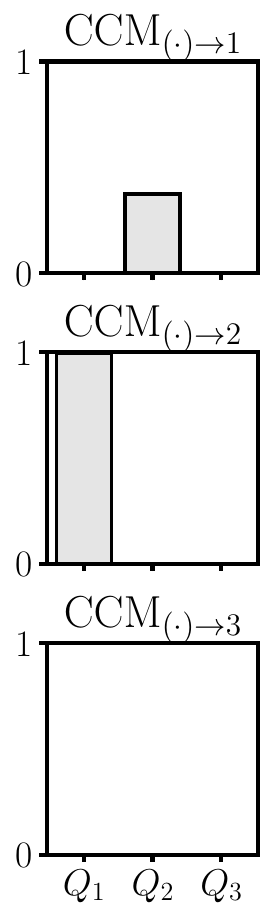}
    \end{minipage}
    \begin{minipage}{0.1125\textwidth}
        \vspace{-6.5cm}
        \includegraphics[width=\textwidth]{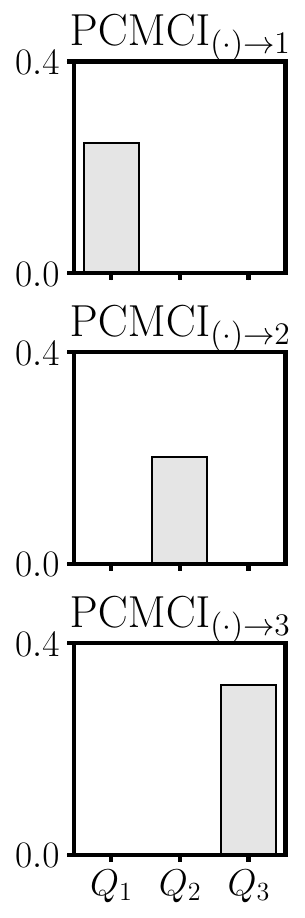}
    \end{minipage}
}
    
\subfloat[]{    
\begin{minipage}{0.175\textwidth}
        \vspace{-2\textwidth}
    ${\hbox{


\begin{tikzpicture}[cir/.style={circle,draw=black!70!white,,thick,inner sep=.5em},
    >={Latex[length=.2cm]},scale=.8]
    

    \colorlet{cA}{Q1!30!white}
    \colorlet{cB}{Q2!30!white}
    \colorlet{cC}{Q3!30!white}


    \node [cir,fill=cB] (q1) at (-1,+1)   {$Q_1$}; 
    \node [cir,fill=cA] (q2) at (-1,-1)   {$Q_2$}; 
    \node [cir,fill=cC] (q3) at (+0.73,0)    {$Q_3$}; 
    


    \path[thick,->]
        (q1) edge[bend left] node [left] {} (q3)
        (q3) edge[bend left] node [right] {} (q2);
    \path[thick,<-] (q3) edge[loop right] node {} (q3);
    \path[thick,<-] (q2) edge[loop below] node {} (q2);
    \path[thick,<-] (q1) edge[loop above] node {} (q1);
    
\end{tikzpicture}

}}$
    \end{minipage}
\hspace{0.01\textwidth} 
\includegraphics[width=0.37\textwidth]{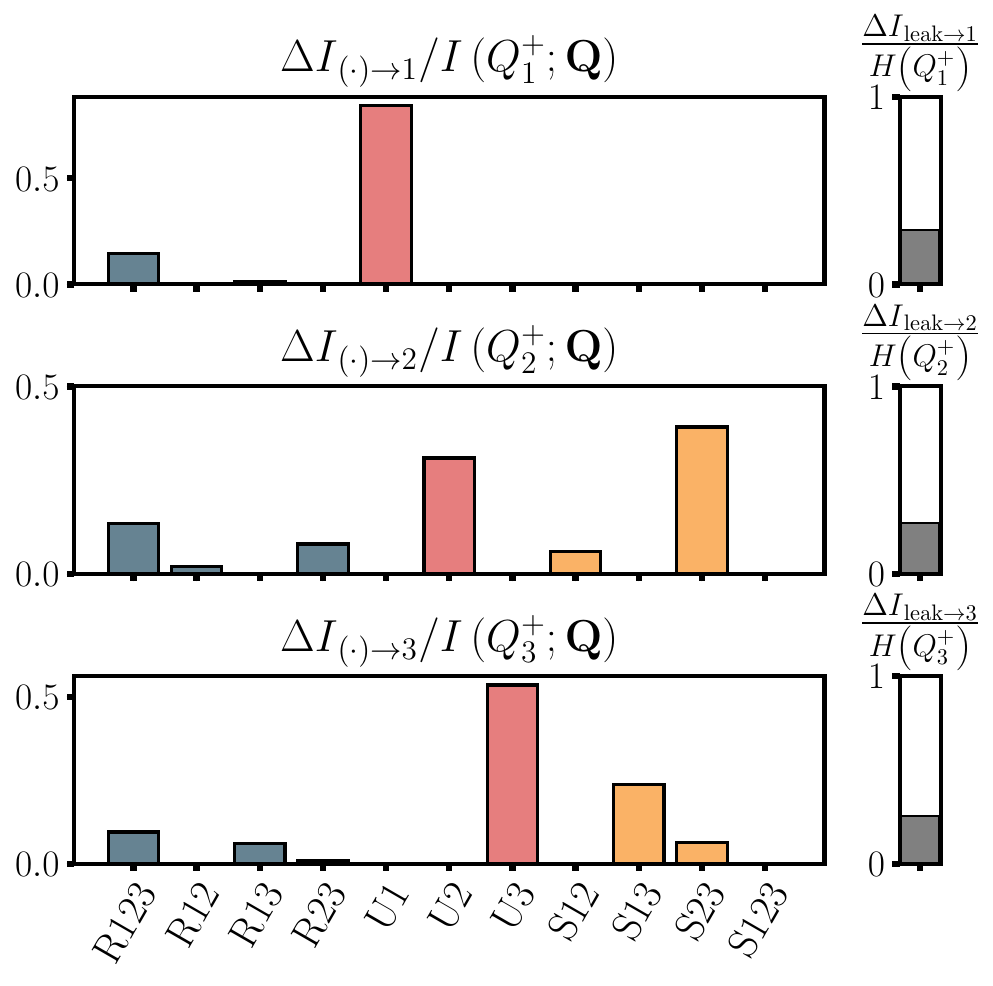}
    \hspace{-0.01\textwidth}
    \begin{minipage}{0.1\textwidth}
        \vspace{-6.5cm}
        \includegraphics[width=\textwidth]{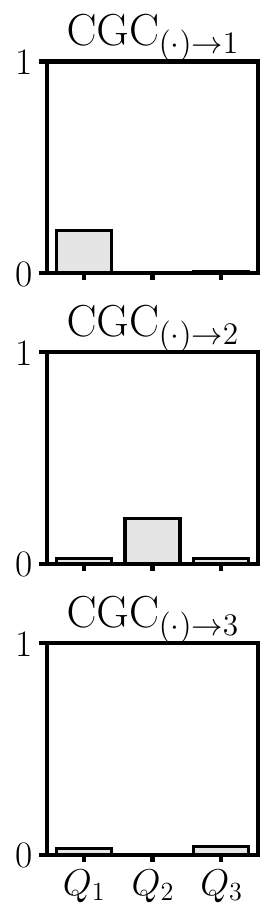}
    \end{minipage}
    \begin{minipage}{0.1\textwidth}
        \vspace{-6.5cm}
        \includegraphics[width=\textwidth]{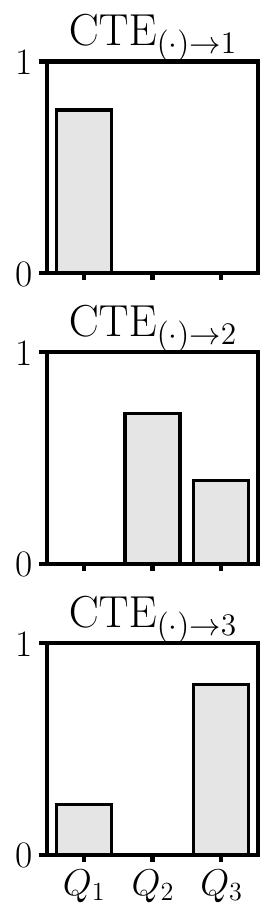}
    \end{minipage}
    \begin{minipage}{0.1\textwidth}
        \vspace{-6.5cm}
        \includegraphics[width=\textwidth]{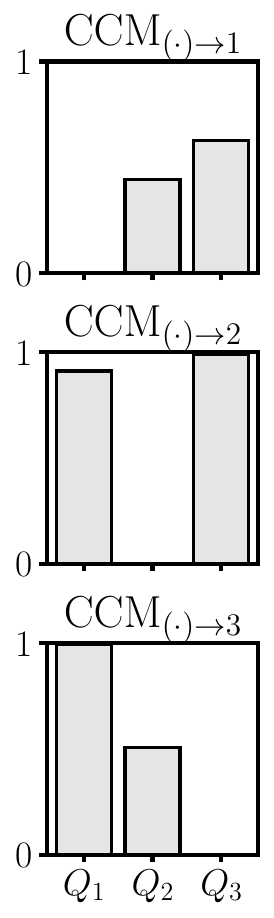}
    \end{minipage}
    \begin{minipage}{0.1125\textwidth}
        \vspace{-6.5cm}
        \includegraphics[width=\textwidth]{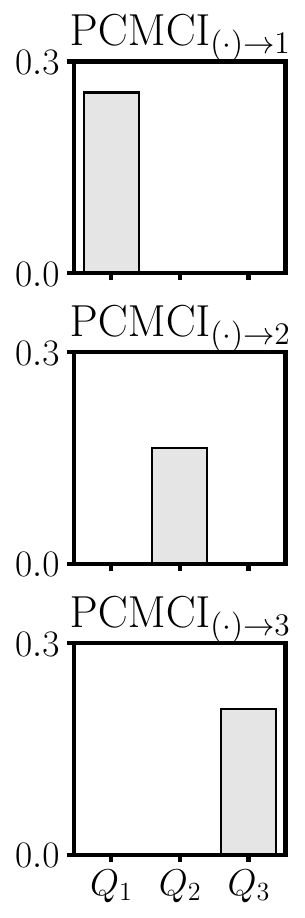}
    \end{minipage}}

\subfloat[]{    
\begin{minipage}{0.175\textwidth}
        \vspace{-2\textwidth}
    ${\hbox{


\begin{tikzpicture}[cir/.style={circle,draw=black!70!white,,thick,inner sep=.5em},
    >={Latex[length=.2cm]},scale=.8]
    

    \colorlet{cA}{Q1!30!white}
    \colorlet{cB}{Q2!30!white}
    \colorlet{cC}{Q3!30!white}


    \node [cir,fill=cB] (q1) at (-1,+1)   {$Q_1$}; 
    \node [cir,fill=cA] (q2) at (-1,-1)   {$Q_2$}; 
    \node [cir,fill=cC] (q3) at (+0.73,0)    {$Q_3$}; 
    


    \path[thick,->]
        (q1) edge[bend left] node [left] {} (q3)
        (q3) edge[bend left] node [right] {} (q2)
        (q2) edge[bend left] node [right] {} (q1);
    \path[thick,<-] (q3) edge[loop right] node {} (q3);
    \path[thick,<-] (q2) edge[loop below] node {} (q2);
    \path[thick,<-] (q1) edge[loop above] node {} (q1);
    
\end{tikzpicture}

}}$
    \end{minipage}
\hspace{0.01\textwidth} 
\includegraphics[width=0.37\textwidth]{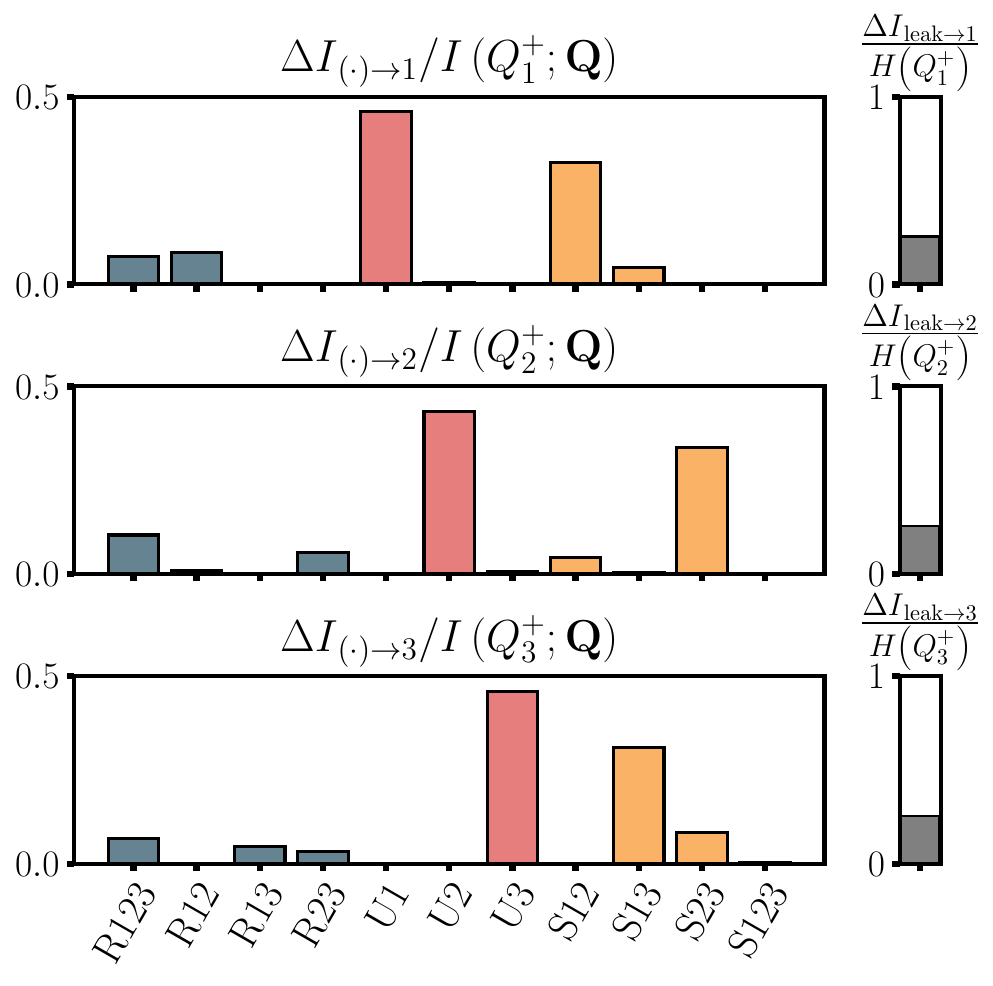}
    \hspace{-0.01\textwidth}
    \begin{minipage}{0.1\textwidth}
        \vspace{-6.5cm}
        \includegraphics[width=\textwidth]{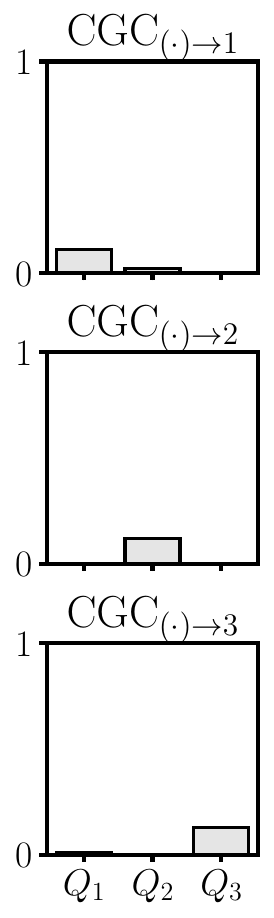}
    \end{minipage}
    \begin{minipage}{0.1\textwidth}
        \vspace{-6.5cm}
        \includegraphics[width=\textwidth]{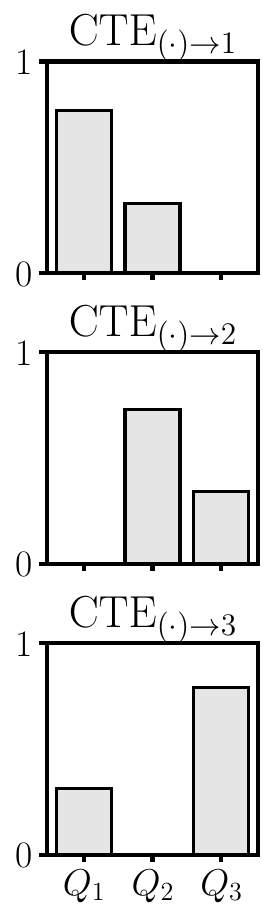}
    \end{minipage}
    \begin{minipage}{0.1\textwidth}
        \vspace{-6.5cm}
        \includegraphics[width=\textwidth]{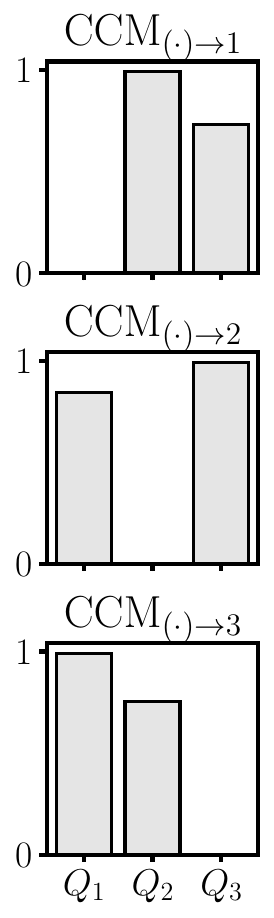}
    \end{minipage}
    \begin{minipage}{0.1125\textwidth}
        \vspace{-6.5cm}
        \includegraphics[width=\textwidth]{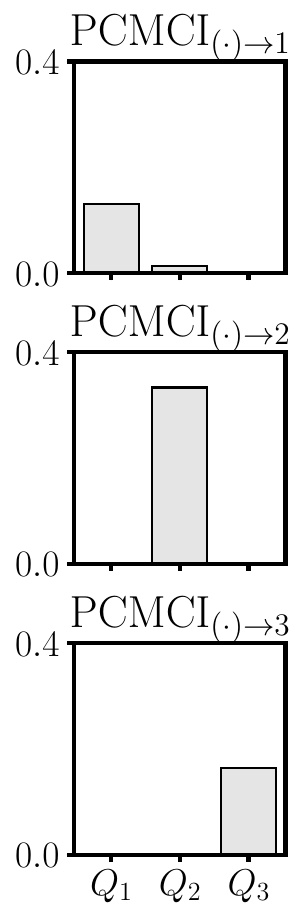}
    \end{minipage}}
\caption{Three interacting species~\cite{pcm2020}. Redundant (R),
  unique (U) and synergistic (S) causalities for $Q_1$, $Q_2$, $Q_3$
  in coupled logistic maps for (a) fan-in with $\beta_{1\rightarrow 2}
  = 0.35$, (b) fan-out with $\beta_{1\rightarrow 3} =
  \beta_{3\rightarrow 2} = 0.35$, and (c) cascading structures with
  $\beta_{1\rightarrow 3} = \beta_{3\rightarrow 2} =
  \beta_{2\rightarrow 1} = 0.35$. The results from CGC, CTE, CCM, and PCMCI are depicted on
      the right. CGC and CTE use the same normalization as SURD.}\label{fig:three interacting
  systems}
\end{figure}

We analyze causality in a system of three interacting species. The
case, proposed by Leng \textit{et al.}~\cite{pcm2020}, validates the
Partial Cross Mapping (PCM) method --a variation of CCM that
eliminates indirect causal influences. This benchmark served as an
example where other methods, including CGC, CTE, and CCM, failed to
detect causal links. The equations of the system are given by:
\begin{align}
    Q_1(n+1) &= Q_1(n) \left[\alpha_1 - \alpha_1 Q_1(n) - \beta_{2\rightarrow 1} Q_2(n) \right] + \eta_1 (n)\\
    Q_2(n+1) &= Q_2(n) \left[\alpha_2 - \alpha_2 Q_2(n) - \beta_{1\rightarrow 2} Q_1(n) - \beta_{3\rightarrow 2} Q_3(n)\right] + \eta_2 (n)\\
    Q_3(n+1) &= Q_3(n) \left[\alpha_3 - \alpha_3 Q_3(n) - \beta_{1\rightarrow 3} Q_3(n) \right] + \eta_3 (n)
\end{align}
where $\alpha_1 = 3.6$, $\alpha_2 = 3.72$, $\alpha_3 = 3.68$,
$\beta_{i\rightarrow j}$ denotes the coupling constants between
variables, and $\eta_i$ represents white noise with zero mean and a
standard deviation of 0.005. Different choices of coupling parameters
$\beta_{i\rightarrow j}$ can lead to various modes of interaction. We
analyze the same combinations reported by Leng \textit{et
  al.}~\cite{pcm2020}, which represent three possible interaction
structures among three species: fan-in ($Q_1\rightarrow Q_2$), fan-out
($Q_1\rightarrow Q_3 \rightarrow Q_2$), and cascading structures
($Q_1\rightarrow Q_3 \rightarrow Q_2 \rightarrow Q_1$).

The results, shown in Figure~\ref{fig:three interacting systems},
demonstrate that SURD identifies links consistent with the governing
equations across all scenarios. CCM fails to identify correct causal
links consistently as already reported by \cite{pcm2020}. PCM (not
shown) operates successfully; however, it does not provide information
about the strength of self-causal links, a feature that SURD
offers. CGC underperforms, while CTE can offer a good insight into the
causal relationships between variables. Nonetheless, CTE cannot
distinguish between unique and synergistic causalities. This
distinction is useful in these scenarios, as variables may be coupled
through both cross-induced causal relationships and self-causal
effects.

\subsection{Eight interacting species}
\label{sec:eight}

\begin{figure}[t]
    \centering
    \begin{tikzpicture}
        \node[anchor=south west,inner sep=0] (image) at (0,0) {
            \includegraphics[trim={0 23.cm 0 0},clip,width=0.485\textwidth]{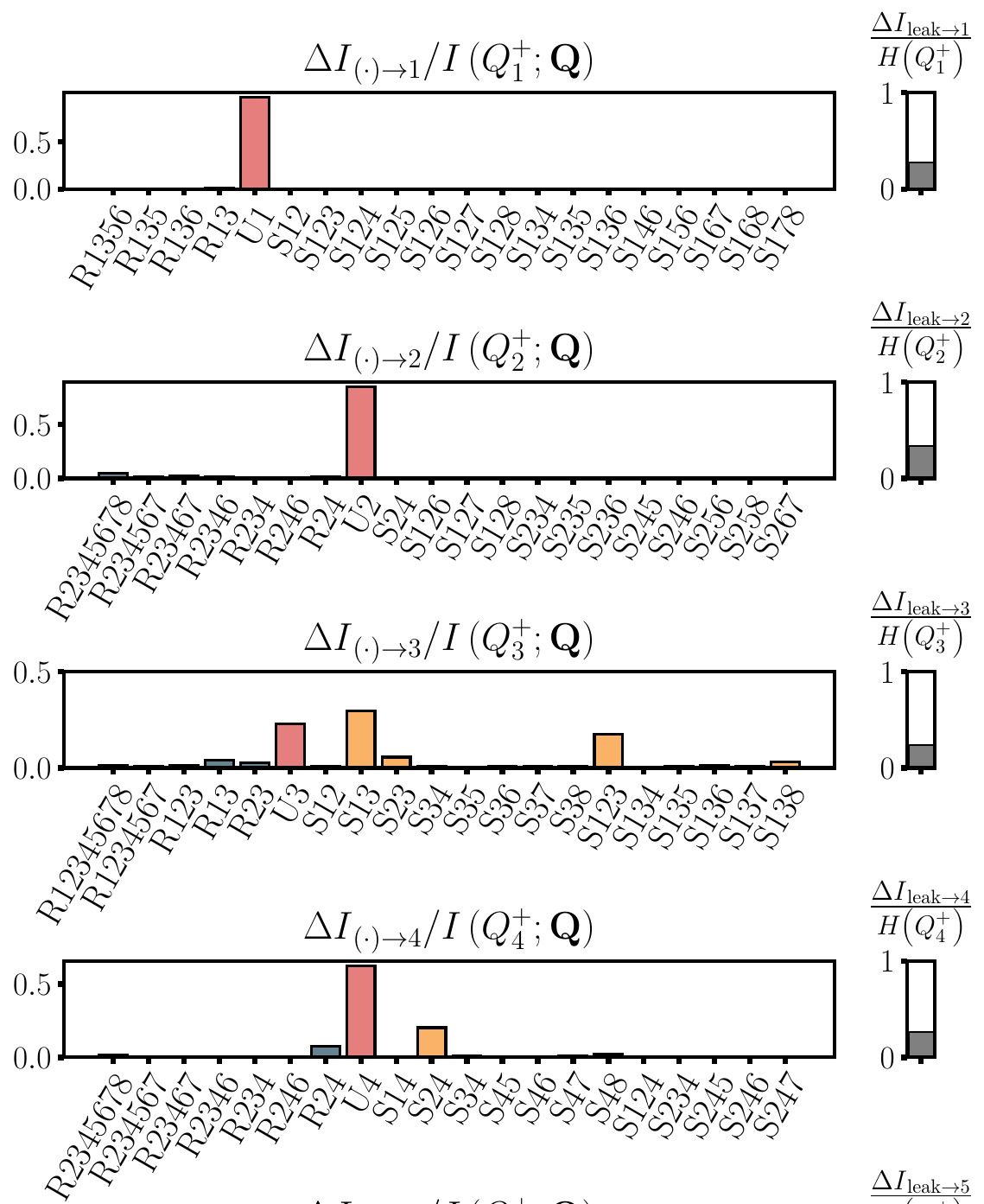}
        };
        \fill [white] (2,0) rectangle (9,0.5);
    \end{tikzpicture}
    \begin{tikzpicture}
        \node[anchor=south west,inner sep=0] (image) at (0,0) {
            \includegraphics[trim={0 0 0 23.cm},clip,width=0.485\textwidth]{supplementary/eight_species_noise_005_surd.pdf}
        };
        \fill [white] (0,11) rectangle (2,10);
    \end{tikzpicture}
    \caption{Eigth interacting species~\cite{pcm2020}. Redundant (R),
      unique (U) and synergistic (S) causalities for $Q_i$ with
      $i=1,\dots,8$. Only the top $N=20$ contributions, satisfying the
      condition ${\Delta I}_{(\cdot)\rightarrow j} / I(Q_j^+; \bQ)
      \geq 10^{-3}$, are represented for each variable. The gray bar
      is the causality leak.}
    \label{fig:eight}
\end{figure}

We test the effectiveness of our method in a network model
containing eight interacting species. This system was proposed by Leng
\textit{et al.}~\cite{pcm2020} to demonstrate the power of the partial
cross-mapping (PCM) method in accurately reconstructing the underlying
causal networks from multivariate time series of high
dimensionality. The equations of the system are given by:
\begin{subequations}
\begin{align}
Q_1(n+1) &= Q_1(n)\left[3.9 - 3.9Q_1(n)\right] + \eta_1(n),\\
Q_2(n+1) &= Q_2(n)\left[3.5 - 3.5Q_2(n)\right] + \eta_2(n),\\
Q_3(n+1) &= Q_3(n)\left[3.62 - 3.62Q_3(n) - 0.35Q_1(n)\right] + \eta_3(n),\\
Q_4(n+1) &= Q_4(n)\left[3.75 - 3.75Q_4(n) - 0.35Q_2(n)\right] + \eta_4(n),\\
Q_5(n+1) &= Q_5(n)\left[3.65 - 3.65Q_5(n) - 0.35Q_3(n)\right] + \eta_5(n),\\
Q_6(n+1) &= Q_6(n)\left[3.72 - 3.72Q_6(n) - 0.35Q_3(n)\right] + \eta_6(n),\\
Q_7(n+1) &= Q_7(n)\left[3.57 - 3.57Q_7(n) - 0.35Q_6(n)\right] + \eta_7(n),\\
Q_8(n+1) &= Q_8(n)\left[3.68 - 3.68Q_8(n) - 0.35Q_6(n)\right] + \eta_8(n),
\end{align}
\end{subequations}
where $\eta_i(n)$ terms for $i=1,\dots,8$, are white noise of zero
mean and standard deviation of 0.005. The results provided in
Figure \ref{fig:eight} show the robustness of SURD to reconstruct
all the causal links of the system according to their governing
equations. Additionally, SURD is able to identify those variables that
individually and synergistically cause the future of the target
variables through unique and synergistic causalities,
respectively. This validation case demonstrates that SURD is able to
identify causal relationships in a setting with a larger number of
variables.
%




\subsection{System with contemporaneous causal dependencies}
\label{sec:instantaneous}

We evaluate the performance of SURD in scenarios with contemporaneous causal links. In these cases, the vector of observed variables can also include variables at time $t+\Delta T$, excluding the target variable. For a system with $N$ variables and target $Q^+_i$, the vector of observables is
\begin{equation}
    \bQ = [Q_1, Q_2, \ldots, Q_N, Q_1^+, \ldots, Q_{i-1}^+, Q_{i+1}^+, \ldots, Q_N^+].
\end{equation}
Note that $Q_i^+$ cannot be included in the vector of observed variables, since doing so would already reveal all the information about the target variable. For example, to calculate the causalities to $Q_1^+$ in a system with three variables, the vector of observed variables is $\bQ = [Q_1, Q_2, Q_3, Q_2^+, Q_3^+]$. For simplicity, we use the notation $Q_4=Q_1^+$, $Q_5=Q_2^+$ and $Q_6=Q_3^+$.
\begin{figure}[t!]
    \centering
    \subfloat[]{
    \includegraphics[width=0.485\textwidth]{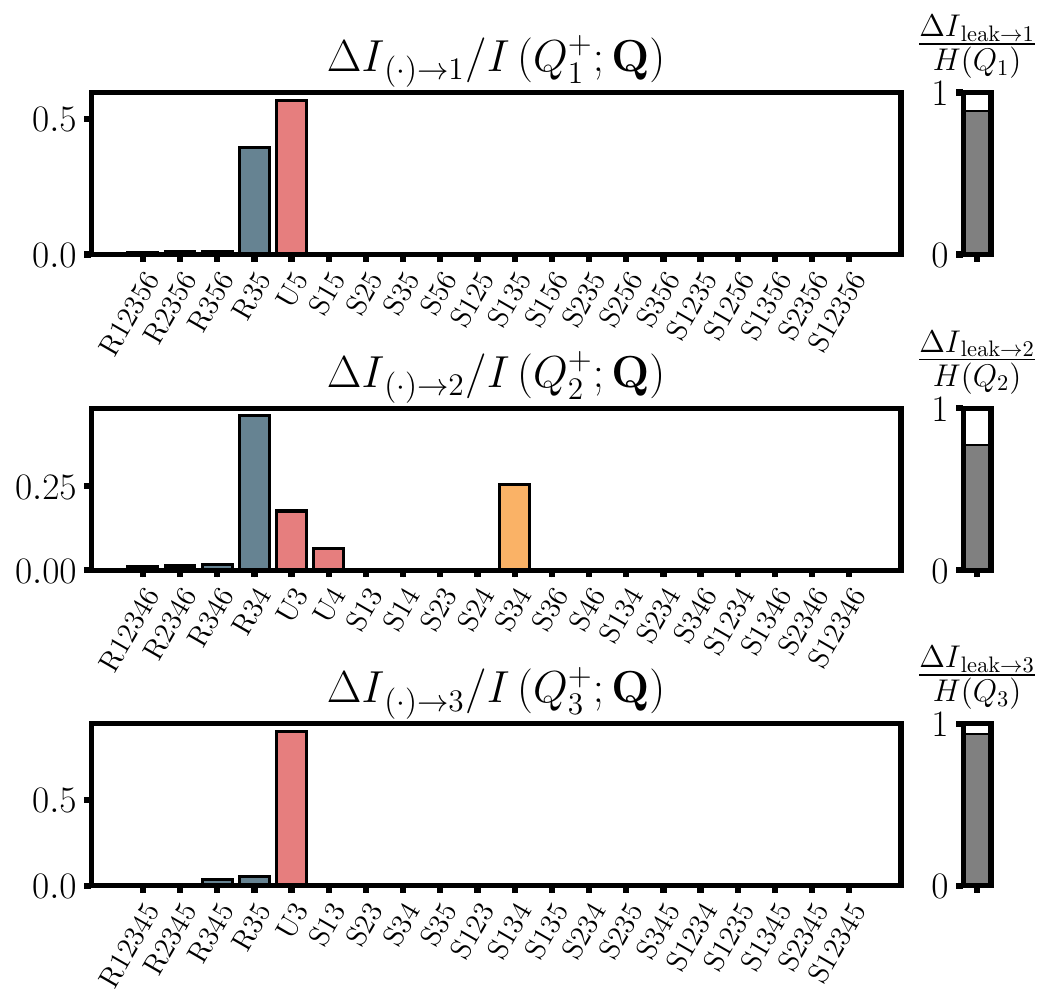}
    }
    \hfill
    \subfloat[]{
    \includegraphics[width=0.485\textwidth]{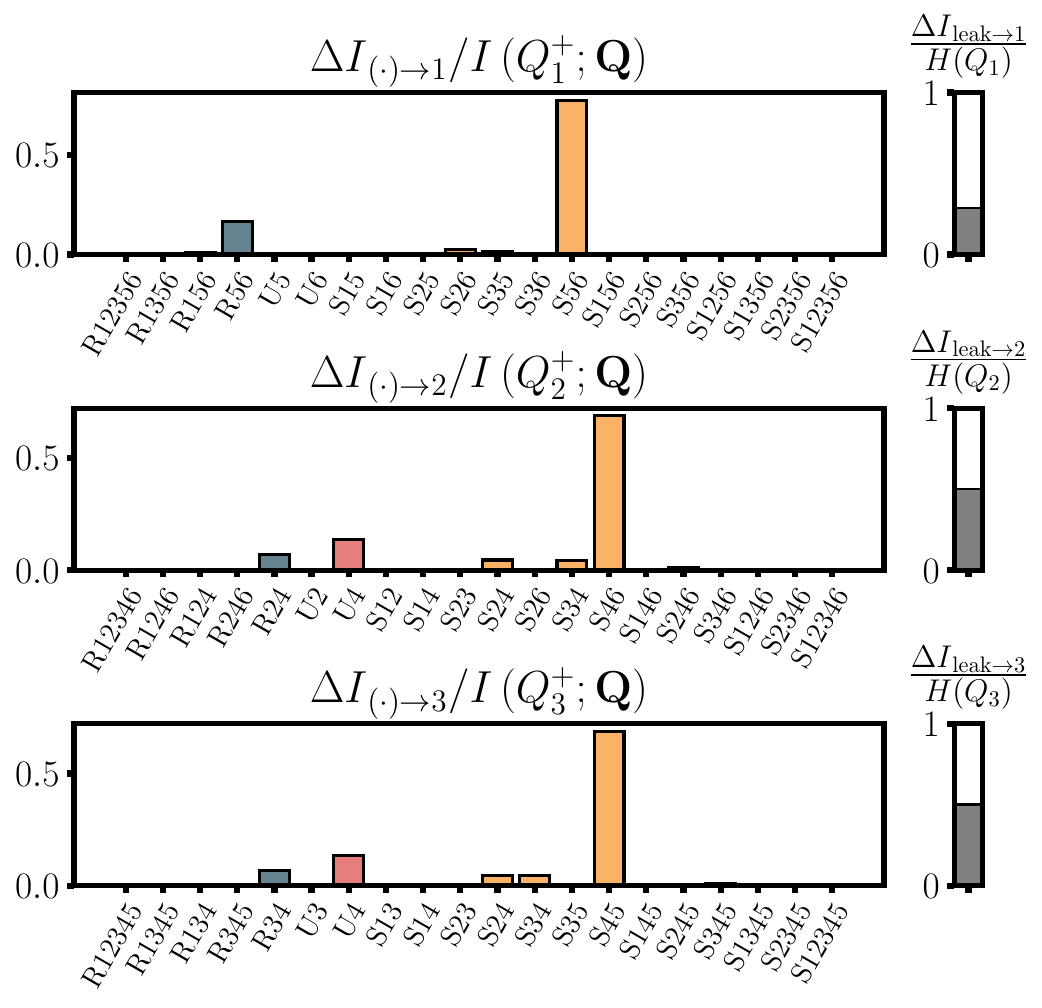}
    }
    \caption{System with instantaneous causal dependencies for (a) mediator and (b) synergistic collider variables. Results from SURD with redundant (R), unique (U)
      and synergistic (S) causalities in blue, red and yellow,
      respectively. The notation employed is such that the present variables {$Q_4=Q_1^+$, $Q_5=Q_2^+$ and $Q_6=Q_3^+$}. The gray bar is the
      causality leak.}
    \label{fig:mediator contemporaneous}
\end{figure}

We use as test beds the systems with mediator and synergistic variables depicted in Figures \ref{fig:mediator} and \ref{fig:synergistic}. For the system with mediator variables, we introduce an contemporaneous dependence of $Q_2$ on $Q_1$, i.e., $Q_2^+ \rightarrow Q_1^+$. The equations of the system with mediator variables can be defined as follows:
\begin{subequations}
\begin{align}
Q_1 (n+1) &= \sin \left[Q_2(n+1)\right] + 0.01W_1(n) \\
Q_2(n+1) &= \cos \left[Q_3(n)\right] + 0.01W_2(n) \\
Q_3(n+1) &= 0.5Q_3(n) + 0.1W_3(n).
\end{align}
\end{subequations}
Figure \ref{fig:mediator contemporaneous}(a) shows how SURD can identify the contemporaneous causal dependency of $Q_2^+$ on $Q_1^+$ through the unique causality $\Delta I ^ U _{5\rightarrow 1}$, {where the index 5 refers to variable $Q_2^+$}. For variables $Q_2^+$ and $Q_3^+$, the most relevant causalities are consistent with their dependencies on $Q_3$. However, a new synergistic causality, $\Delta I ^ S _ {34 \rightarrow 1}$, emerges in $Q_2^+$ due to its contemporaneous dependence on $Q_1^+$. 

We also test the system with synergistic collider modified to include contemporaneous links. In this case, the equations are given by:
\begin{subequations}
\begin{align}
Q_1 (n+1) &= \sin \left[Q_2(n+1)Q_3(n+1)\right] + 0.001W_1(n) \\
Q_2(n+1) &= 0.5 Q_2(n) + 0.1W_2(n) \\
Q_3(n+1) &= 0.5 Q_3(n) + 0.1W_3(n).
\end{align}
\end{subequations}
Figure \ref{fig:mediator contemporaneous}(a) shows that the most relevant causality for $Q_1^+$ is the component $\Delta I ^ S_{56 \rightarrow 1}$, {where indices $5$ and $6$ represent variables $Q_2^+$ and $Q_3^+$}, respectively. Furthermore, as the dependency between them is instantaneous and these variables are part of the observed vector, it is possible to write $Q_2^+ = f(Q_1^+, Q_3^+)$ and $Q_3^+ = f(Q_1^+, Q_2^+)$, indicating that the components $\Delta I ^ S_{45 \rightarrow 2}$ and $\Delta I ^ S_{45 \rightarrow 3}$ also play a significant role in determining the variables $Q_2^+$ and $Q_3^+$, respectively.
%

\subsection{\, Causality from combination of variables in mediator, confounder, and collider cases}
\label{sec:mutivariate}

\begin{figure}[t!]
    \centering
    \subfloat[$Q_3\rightarrow Q_2 \rightarrow Q_1$]{\includegraphics[width=0.23\textwidth]{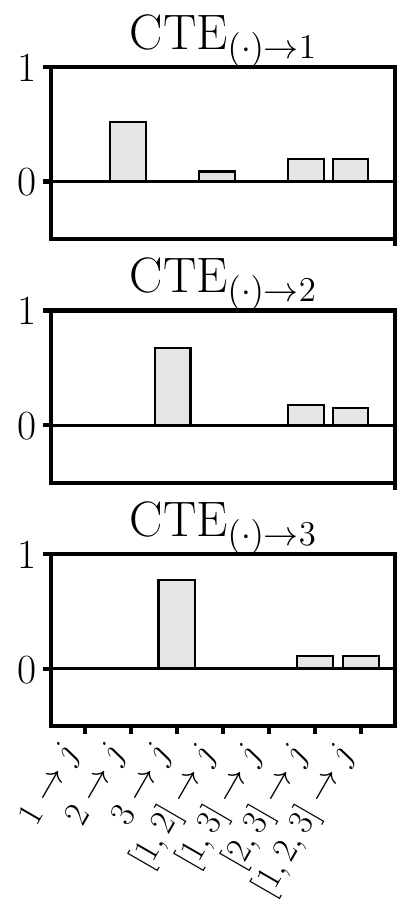}}
    \hspace{0.015\textwidth}
    \subfloat[$Q_1\leftarrow Q_3 \rightarrow Q_2$]{\includegraphics[width=0.23\textwidth]{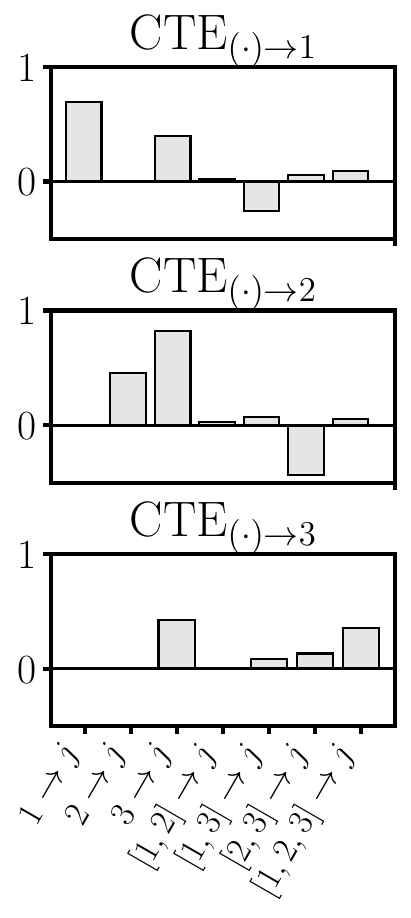}}
    \hspace{0.015\textwidth}
    \subfloat[$\left(Q_2,Q_3\right)\rightarrow Q_1$]{\includegraphics[width=0.23\textwidth]{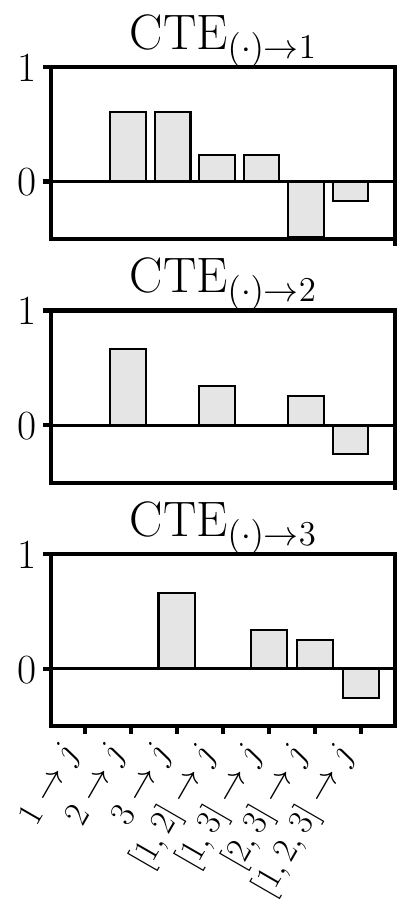}}
    \hspace{0.015\textwidth}
    \subfloat[$Q_2 \equiv Q_3 \rightarrow Q_1$]{\includegraphics[width=0.23\textwidth]{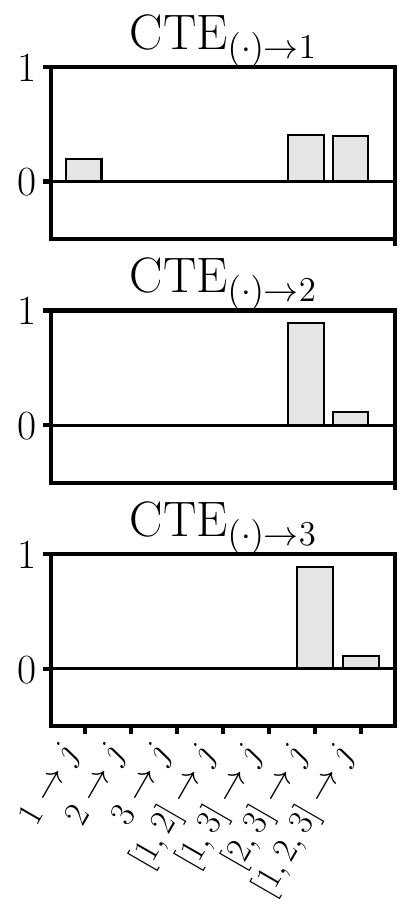}}
    \caption{Results from the multivariate version of the conditional
      transfer entropy (CTE)~\cite{Lozano2022} for the systems with
      (a) mediator, (b) confounder, (c) synergistic collider, and (d)
      redundant collider variables. Further details about these
      systems are provided in the main text.}
    \label{fig:info flux bblocks}
\end{figure}

\begin{figure}[t!]
    \centering
    \subfloat[$Q_3\rightarrow Q_2 \rightarrow Q_1$]{\includegraphics[width=0.23\textwidth]{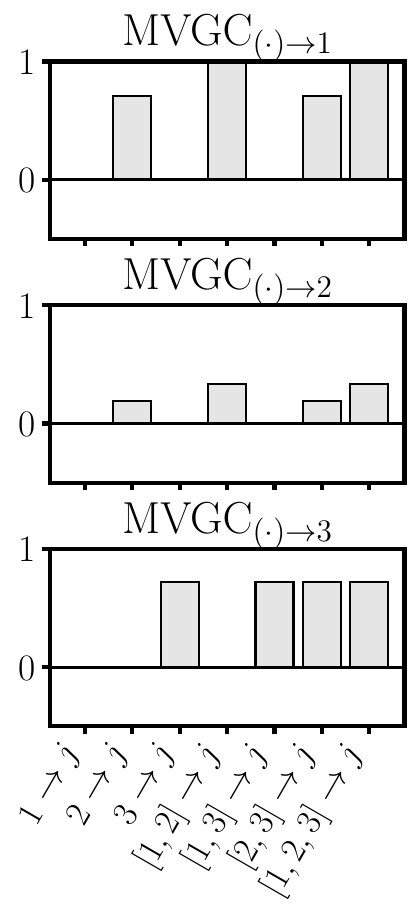}}
    \hspace{0.015\textwidth}
    \subfloat[$Q_1\leftarrow Q_3 \rightarrow Q_2$]{\includegraphics[width=0.23\textwidth]{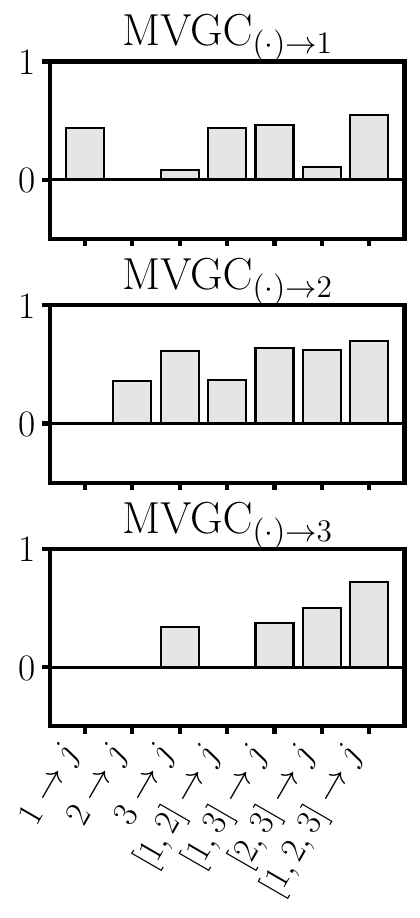}}
    \hspace{0.015\textwidth}
    \subfloat[$\left(Q_2,Q_3\right)\rightarrow Q_1$]{\includegraphics[width=0.23\textwidth]{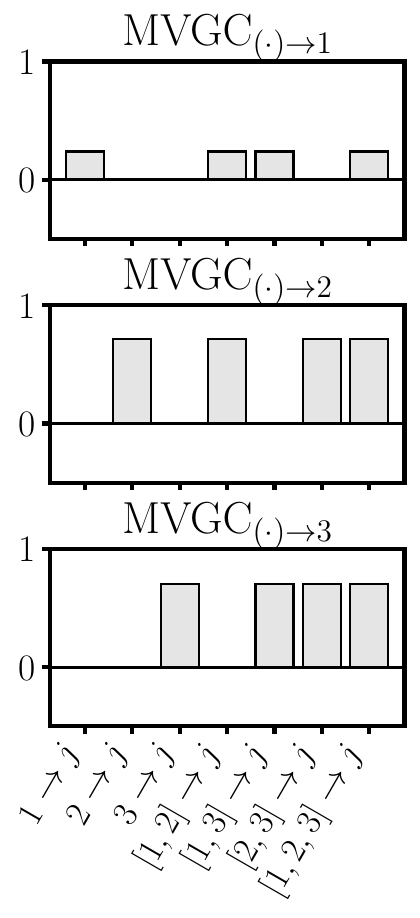}}
    \hspace{0.015\textwidth}
    \subfloat[$Q_2 \equiv Q_3 \rightarrow Q_1$]{\includegraphics[width=0.23\textwidth]{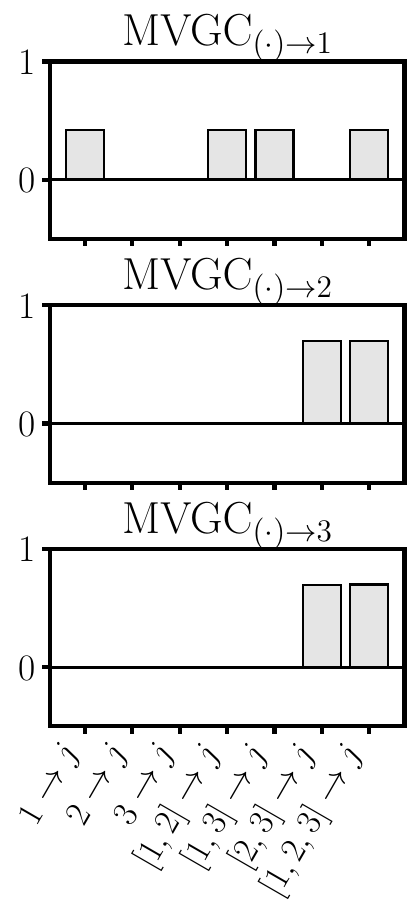}}
    \caption{Results from the multivariate version of the multivariate
      Granger causality (MVGC)~\cite{barnett2014} for the systems with (a)
      mediator, (b) confounder, (c) synergistic collider, and (d)
      redundant collider variables. Further details about these
      systems are provided in the main text.}
    \label{fig:mvgc bblocks}
\end{figure}

In this section, we compare the redundant and synergistic causalities
from SURD with the multivariate versions of CTE and CGC, referred to
as MCTE and MVCG, respectively. The results are calculated for systems
with mediator, confounder, synergistic collider, and redundant
collider variables. Although MCTE and MVCG can quantify causality from
combinations of variables (e.g., $[Q_1,Q_2,Q_3,\dots]\rightarrow
Q_j$), we show here that the results are not easily interpretable or
may be directly erroneous. Figure~\ref{fig:info flux bblocks} shows
how MCTE yields negative values of causality for cases in which
synergistic effects play a major role, such as systems with confounder
and synergistic collider variables. For example, in the synergistic
collider $(Q_2,Q_3)\rightarrow Q_1$, MCTE$_{23\rightarrow 1}$ is
strongly negative. The results from MVGC (Figure~\ref{fig:mvgc
  bblocks}) are constrained to be nonnegative; however, MVGC does not
provide information about the redundant or synergistic nature of the
interactions. This makes it very challenging to interpret whether the
identified links due to combined variables are consistent with the
relationships between variables. Returning to the case of the
synergistic collider $(Q_2,Q_3)\rightarrow Q_1$, SURD correctly
detects $\Delta I^S_{23\rightarrow 1}>0$ as the main causality to
$Q_1$, whereas MVCG identifies MVGC$_{123\rightarrow 1}$,
MVGC$_{12\rightarrow 1}$, and MVGC$_{13\rightarrow 1}$ as all equally
important, while MVGC$_{23\rightarrow 1}$ is zero.

\subsection{\, Causal graphs for PCMCI in multivariate systems}
\label{sec:graphs}

\begin{figure}[h!]
    \centering
    \subfloat[Mediator variable]{\includegraphics[trim={0 2.8cm 0 0},clip,width=0.245\textwidth]{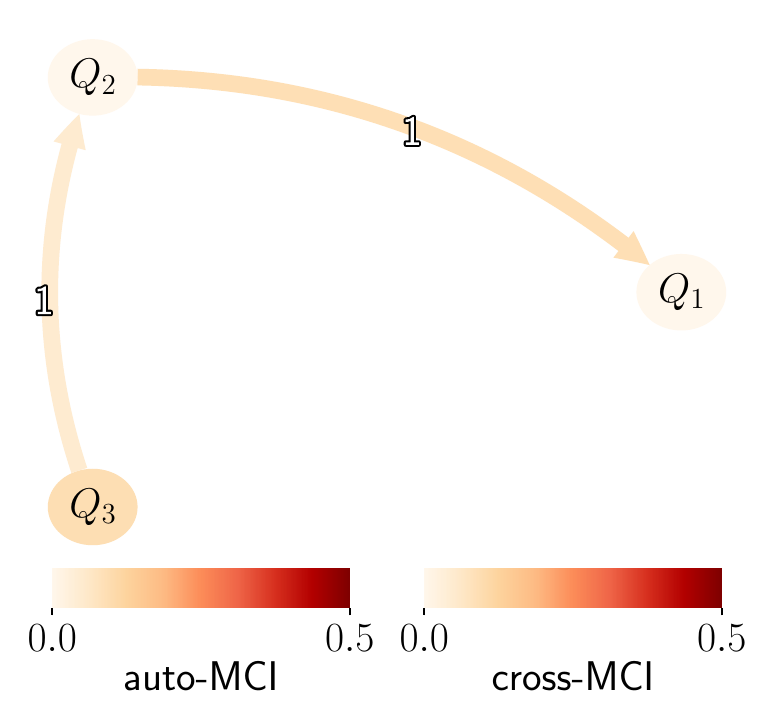}}\hfill
    \subfloat[Confounder variable]{\includegraphics[trim={0 2.8cm 0 0},clip,width=0.245\textwidth]{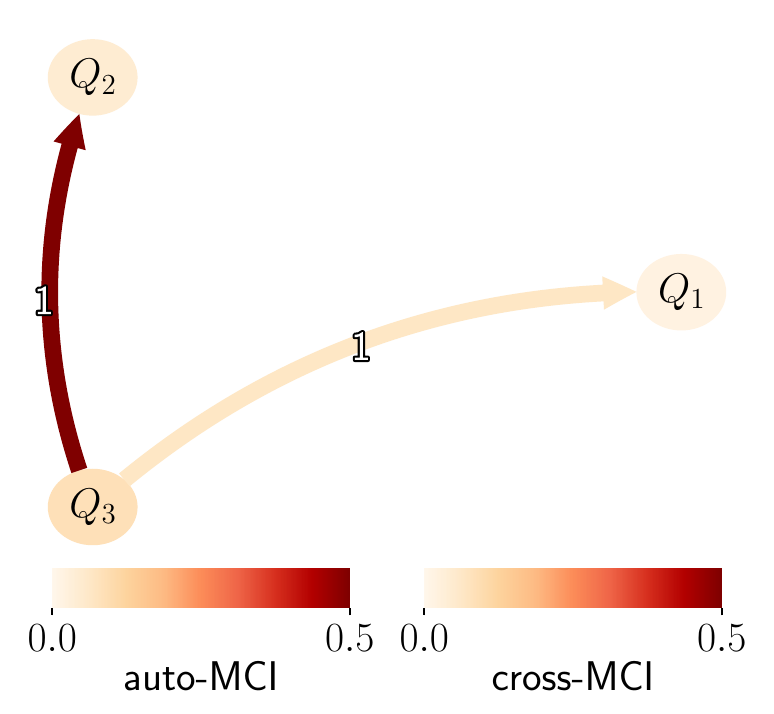}}\hfill
    \subfloat[Synergistic collider]{\includegraphics[trim={0 2.8cm 0 0},clip,width=0.245\textwidth]{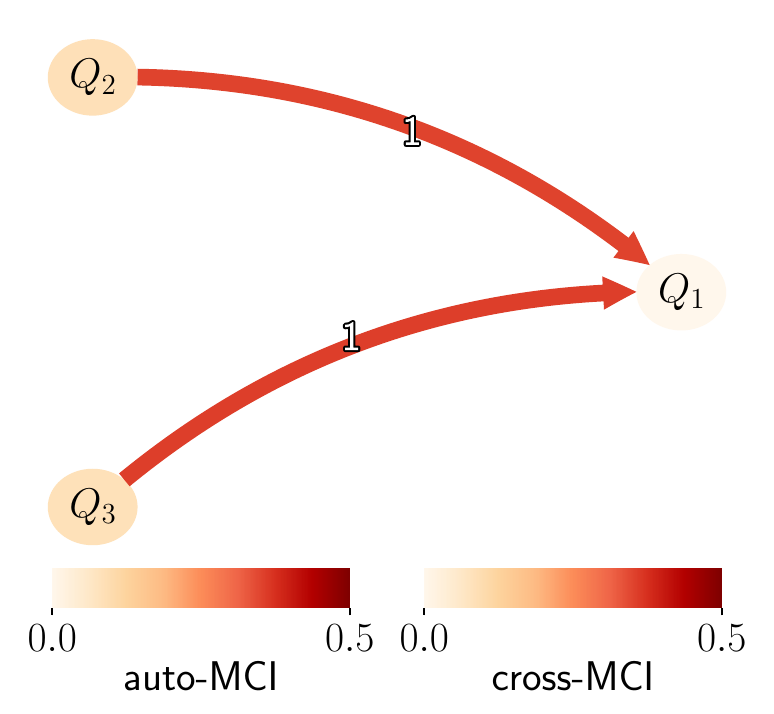}}\hfill
    \subfloat[Redundant collider]{\includegraphics[trim={0 2.8cm 0 0},clip,width=0.245\textwidth]{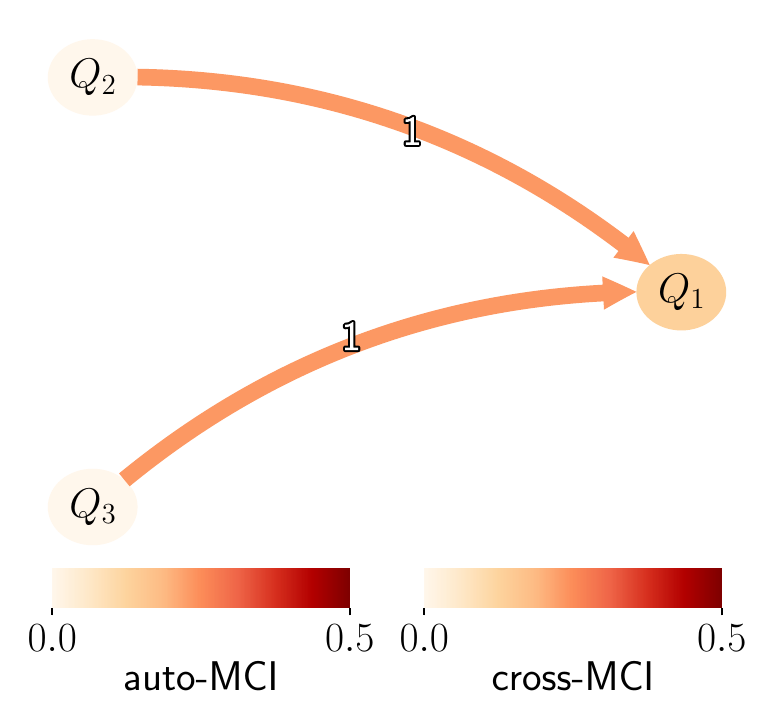}}
    
    \caption{Summary graph of the causality analysis performed using
      the PCMCI method for the systems with mediator, confounder, synergistic collider, and redundant collider variables. Cross-causality is quantified with the color intensity of the links,
      whereas self-causality is quantified with the color intensity of the variables nodes. Both quantities range from 0 (light orange) to 0.5 (dark orange). The indices in
      the arrows represent the time lag at which the causal links are
      identified.}
    \label{fig: pcmci results fundamental}
\end{figure}

The primary outcome of the PCMCI algorithm for the mediator,
confounder, and collider is illustrated through the causal graphs in
Figure~\ref{fig: pcmci results fundamental}. In this visual
representation, only the links that are statistically significant at a
specified threshold level are shown. Additionally, a measure of the
causal strength is included for both cross-causal and self-causal
connections. This visualization facilitates a clear evaluation of the
consistency of the causality analysis with respect to the functional
relationships of the systems. Among all cases, the redundant collider
stands out as the case where the summary graph lacks coherence. In
this particular scenario, the approach identifies a fully
interconnected graph, failing to correctly pinpoint the effects of the
duplicated variables $Q_2$ and $Q_3$ on $Q_1$. This issue arises from
incorporating duplicated variables into the conditional set of
variables. Additionally, in the case of the confounder variable,
although the causal connections across variables align with the
interdependencies within the system, the strength of the self-causal
connections for variables $Q_1$ and $Q_2$ is weak. 

\begin{figure}[h!]
    \centering
    \subfloat[Energy cascade]{\includegraphics[trim={0 2.5cm 0 0},clip,width=0.33\textwidth]{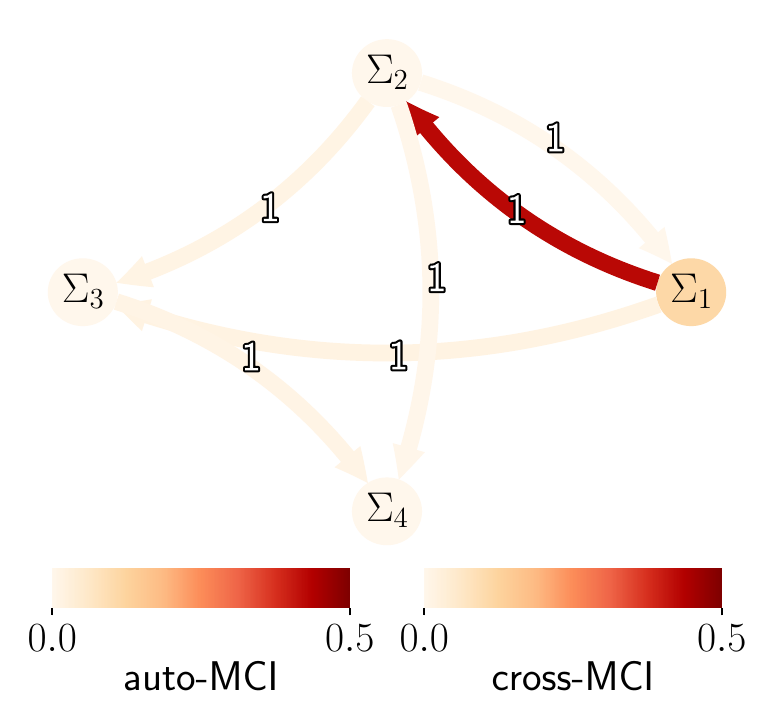}}\hfill
    \subfloat[Uncoupled R\"ossler-Lorenz system]{\includegraphics[trim={0 2.5cm 0 0},clip,width=0.33\textwidth]{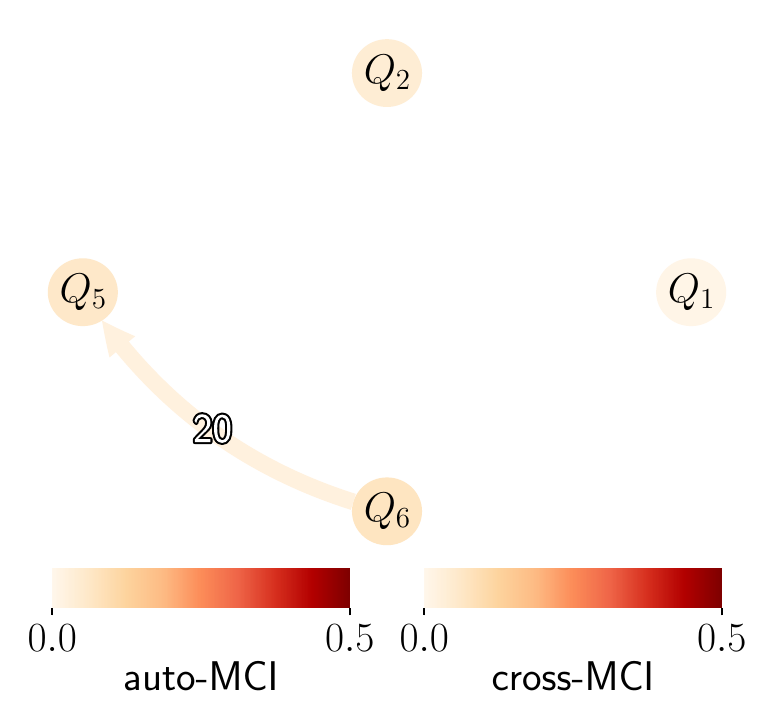}}\hfill
    \subfloat[Coupled R\"ossler-Lorenz system]{\includegraphics[trim={0 2.5cm 0 0},clip,width=0.33\textwidth]{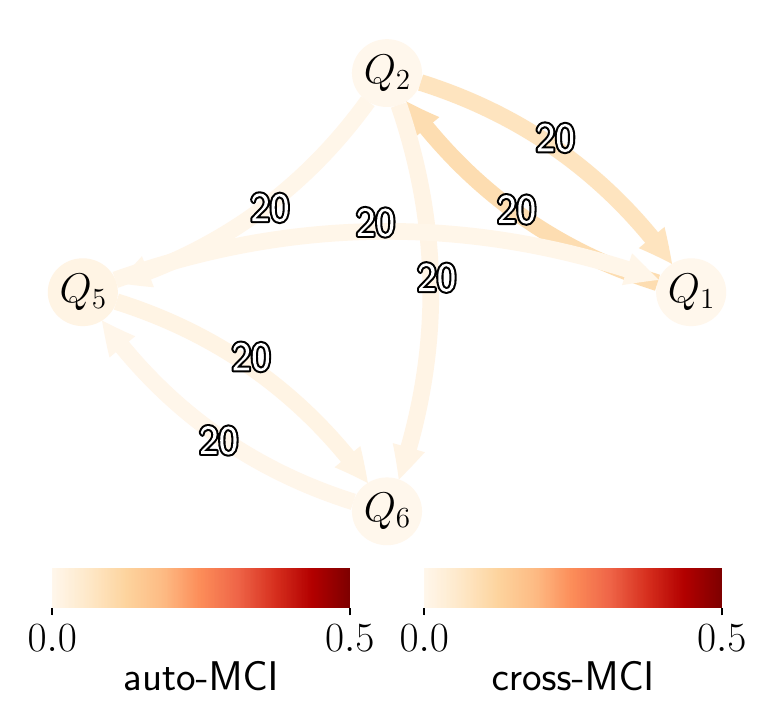}}
    
    \caption{Summary graph of the causality analysis performed using
      the PCMCI method for the turbulent energy cascade and the R\"ossler-Lorenz system. Cross-causality is quantified with the color intensity of the links,
      whereas self-causality is quantified with the color intensity of the variables nodes. Both quantities range from 0 (light orange) to 0.5 (dark orange). 
      The indices in the arrows represent the time lag at which the causal links are identified for panels (b) and (c). In panel (a), an index of one is shown for all links, although the time lag for identification of causality used in the analysis is different for each variable. These values are consistent with the ones provided in the main text.}
    \label{fig: pcmci results}
\end{figure}

Figure \ref{fig: pcmci results}(a) shows the summary graph of the causality analysis from PCMCI for the turbulent energy cascade (see Figure \ref{fig:cascade} in the main text). The most relevant causal links in the graph are $\langle \Sigma_1 \rangle \rightarrow \langle \Sigma_2 \rangle ^+$ and $\langle \Sigma_1 \rangle \rightarrow \langle \Sigma_1 \rangle ^+$. The other detected relationships, except for $\langle \Sigma_2 \rangle \rightarrow \langle \Sigma_1 \rangle ^+$, are consistent with the hypothesis of forward propagation of energy in turbulence, which have very weak causal strength. The weak causalities can be attributed to the high importance of redundant causalities for variables other than $\langle \Sigma_1\rangle$, as shown in the results from SURD in the main text. 

Additionally, the results for the R\"ossler-Lorenz system discussed in Figures \ref{fig:Rossler_c0} and \ref{fig:Rossler_c2} are provided in graph format in Figures \ref{fig: pcmci results}(b) and \ref{fig: pcmci results}(c). For both cases, PCMCI fails to identify causal links that are consistent with the equations of the systems. Specifically, in the uncoupled R\"ossler-Lorenz system (see Figure \ref{fig:Rossler_c0}), the method cannot identify a coupling between variables $Q_1$ and $Q_2$. In the coupled system (see Figure \ref{fig:Rossler_c2}), the links provided by PCMCI are more representative of the real connections between variables. However, the link $Q_5 \rightarrow Q_1^+$ is inconsistent with the coupling in the R\"ossler-Lorenz system, which occurs through $Q_2 \rightarrow Q_5^+$.

\subsection{\, PCMCI for different independence tests}

\begin{table}[t!]
\centering
\begin{tabular}{|l|c|c|c|c|c|c|c|}
\hline
\textbf{Case}  & \textbf{ParCorr} & \textbf{rParCorr} & \textbf{GPDC} & \textbf{CMI} \\
\hline
Mediator variable              & \xmark & \xmark & \cmark & \cmark \\
Confounder variable            & \cmark & \cmark & \cmark & \cmark \\ 
Synergistic collider variable  &  \xmark &  \xmark &  \cmark & \cmark \\
Redundant collider variable    &  \xmark &  \xmark &  \xmark & \xmark \\
Lotka-Volterra prey-predator model \cite{lotka1925,volterra1926} & \cmark & \cmark & \cmark & \xmark\\
Three-interacting species system \cite{pcm2020} & \xmark & \cmark & \xmark & \xmark\\
Moran effect model \cite{moran1953}                 & \cmark & \cmark & \xmark & \cmark\\
One-way coupling nonlinear logistic difference system \cite{sugihara2012} & \cmark & \xmark & \xmark & \xmark\\
Two-way coupling nonlinear logistic difference system \cite{sugihara2012} & \xmark & \xmark  & \cmark & \xmark\\
Stochastic system with linear dependencies \cite{ding2006} & \cmark & \cmark & \cmark  & \cmark\\
Stochastic system with non-linear dependencies \cite{bueso2020} & \cmark  & \xmark & \xmark & \cmark\\
Synchronization of two variables in logistic maps \cite{may1976}   & \cmark & \cmark & \xmark & \cmark\\
Synchronization of three variables in logistic maps \cite{may1976}  & \xmark & \xmark  & \xmark & \xmark\\
Uncoupled Rössler-Lorenz system \cite{lorenz1963, rossler1977} & \xmark &  \xmark & \xmark & \cmark\\
One-way coupled Rössler-Lorenz system \cite{lorenz1963, rossler1977} & \xmark & \xmark  & \xmark &  \cmark\\
\hline
\end{tabular}
\caption{{{Summary of the performance of different independence tests for
  PCMCI.}} The markers \cmark\ and \xmark\ denote
  consistent and inconsistent identification of causal links,
  respectively, according to the functional dependency of variables
  within the system.  The independence tests considered are partial correlation (ParCorr), robust partial correlation (rParCorr), Gaussian process regression and a distance correlation (GPDC) and conditional mutual information with a $k$-nearest neighbor estimator (CMI). }
\label{tab:summary pcmci}
\end{table}

We analyze the variability in the results of PCMCI for different independence tests. The optimal confidence interval $\alpha_{\rm{PC}}$ for PCMCI is selected during the initial condition selection phase (PC phase) based on the Akaike Information criterion~\cite{akaike2011} from a default list of values, i.e., $\alpha_{\rm{PC}} = [0.05, 0.1, 0.2, 0.3, 0.4, 0.5]$. Next,  the significance level for the  momentary conditional independence phase (MCI phase)
is set to $\alpha_{\rm{MCI}} = 0.01$ to obtain the causal graph and strengths. The independence tests analyzed are partial correlation (ParCorr), robust partial correlation (rParCorr), Gaussian process regression and a distance correlation (GPDC), and conditional mutual information with a $k$-nearest neighbor estimator (CMI). For a detailed discussion of these independence tests and their assumptions, the reader is referred to Ref.~\cite{runge2019pcmci}.

The summary of the results for the other tests is compiled in Table~\ref{tab:summary pcmci}. The results for the CMI test are the same as those reported in the main text. The table shows that there is strong variability in the conclusions depending on the independence test and the parameters used. The CMI test with the $k$-NN estimator might not be the best choice in systems with fully deterministic dynamics, such as the Lotka-Volterra prey-predator model and nonlinear logistic difference systems. In these cases, ParCorr and rParCorr offer a more consistent estimation of the causal graph. Moreover, the GPDC test is the only test that identifies the causal relationships in the system with confounder variables.

\section{Sensitivity of SURD}

\subsection{Sensitivity of SURD to sample size and partition refinement}
\label{sec:convergence}

\begin{figure}[t]
    \centering
    \subfloat[]{
    \begin{minipage}{0.45\textwidth}
        \includegraphics[trim={0 1.6cm 0 0},clip,width=\textwidth]{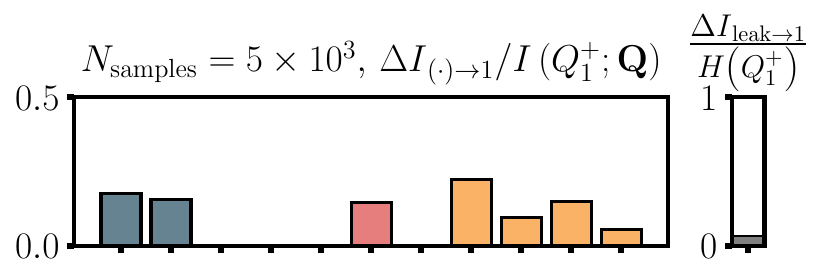}
        \includegraphics[trim={0 1.6cm 0 0},clip,width=\textwidth]{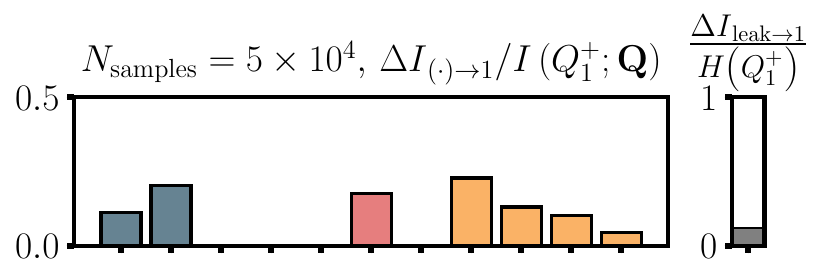}
        \includegraphics[trim={0 1.6cm 0 0},clip,width=\textwidth]{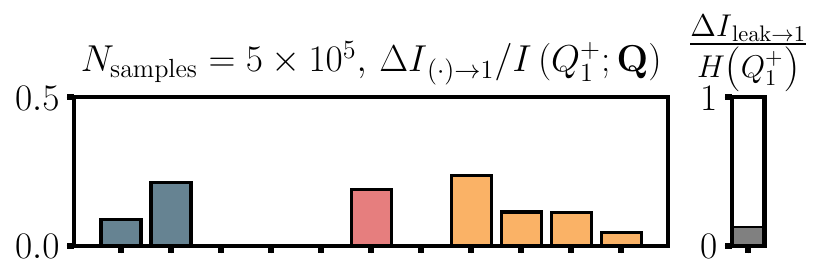}
        \includegraphics[trim={0 0cm 0 0},clip,width=\textwidth]{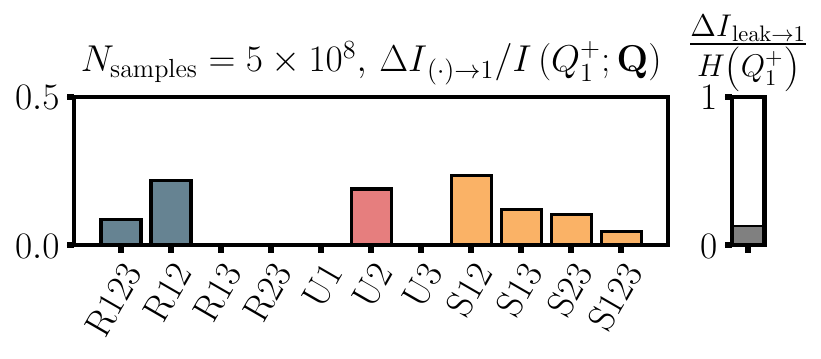}
    \end{minipage}
    }
    \hspace{0.03\textwidth}
    \subfloat[]{
    \begin{minipage}{0.45\textwidth}
        \includegraphics[trim={0 1.6cm 0 0},clip,width=\textwidth]{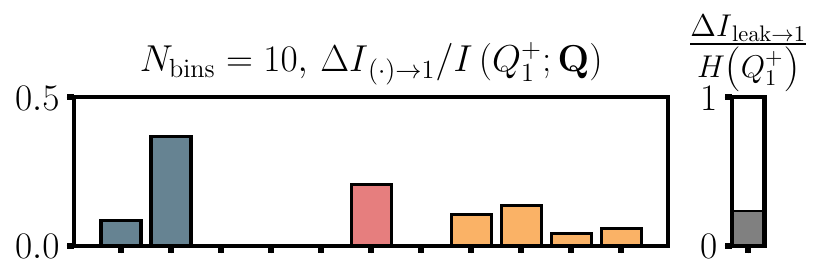}
        \includegraphics[trim={0 1.6cm 0 0},clip,width=\textwidth]{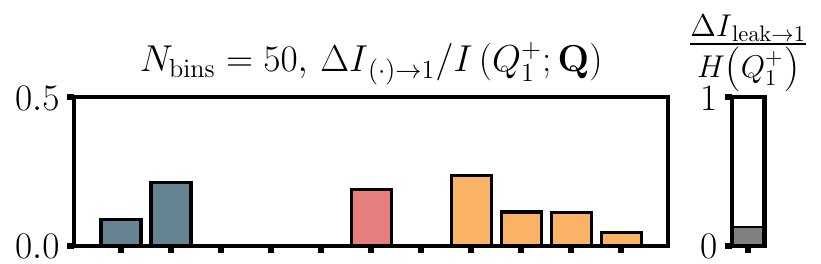}
        \includegraphics[trim={0 1.6cm 0 0},clip,width=\textwidth]{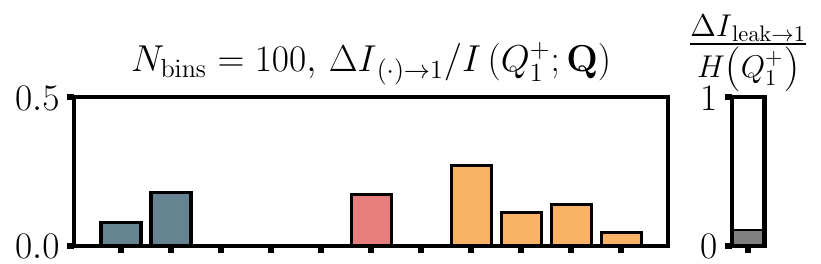}
        \includegraphics[trim={0 0cm 0 0},clip,width=\textwidth]{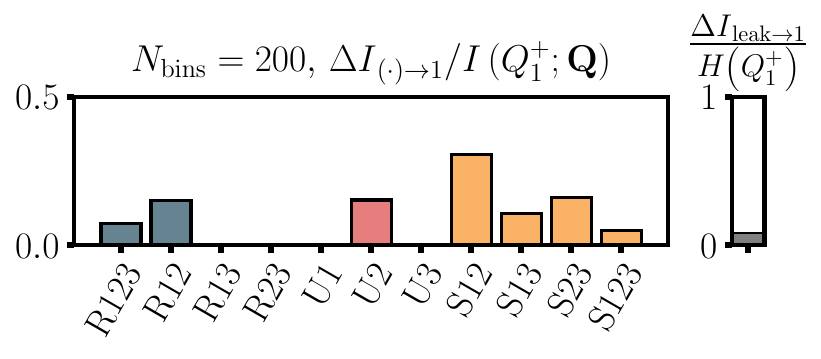}
    \end{minipage}
    }
    \caption{ Sensitivity of SURD in the Lorenz system for (a) number
      of samples $N_\text{samples}$ used to estimate the probability
      distributions using the binning method for $N_\text{bins}=50$
      held constant and (b) the number of bins $N_\text{bins}$ used to
      partition values of the variables for $N_\text{samples}=10^5$.}
    \label{fig:covergence}
\end{figure}

We investigate the sensitivity of SURD to the number of samples
($N_{\text{samples}}$) and number of bins ($N_{\text{bins}}$) used to
partition the variables in situations where the binning method is
employed to estimate the probability distributions. The Lorenz system
is used as a testbed:
\begin{subequations}
\begin{align}  
  \label{eq:L}
 \frac{\dd Q_1}{\dd t} &= 10[ Q_2 - Q_1 ], \\
 \frac{\dd Q_2}{\dd t} &= Q_1 [28 - Q_3 ] - Q_2, \\
 \frac{\dd Q_3}{\dd t} &= Q_1 Q_2 - \frac{8}{3}Q_3.
\end{align}
\end{subequations}
The system was integrated over time to collect
$N_{\text{samples}}=5\times 10^3, 5\times 10^4, 5\times 10^5,$ and
$5\times10^8$ events after transients. Probability distributions were
calculated using uniform bins with $N_{\text{bins}}=10, 50, 100,$ and
200 per variable. Our primary focus is on causalities to $Q_1$, but
the conclusions drawn also apply to $Q_2$ and $Q_3$.

The sensitivity to $N_{\text{samples}}$ is displayed in
Figure~\ref{fig:covergence}(a), where $N_{\text{samples}}$ varies
while maintaining $N_{\text{bins}}=50$ constant. For
$N_{\text{samples}}>5\times 10^3$, the changes in causality remain
within a few percentage points of difference. The sensitivity to the
size of the partition is assessed in Figure~\ref{fig:covergence}(b),
where $N_{\text{bins}}$ varies while $N_{\text{samples}}$ is held
constant at $N_{\text{samples}} = 5\times 10^5$. The causalities
exhibit quantitative resemblance for all partitions, with the
exception of $N_{\text{bins}}=10$, which may be too coarse to capture
the continuous dynamics of the variables.\\


\subsection{Sensitivity  of SURD to sample size and polynomia order}
\label{sec:convergence tmap}

\begin{figure}[t]
    \centering
    \subfloat[]{
    \begin{minipage}{0.45\textwidth}
        \includegraphics[trim={0 1.6cm 0 0},clip,width=\textwidth]{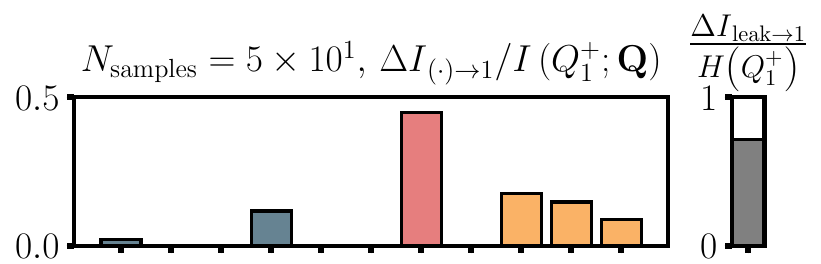}
        \includegraphics[trim={0 1.6cm 0 0},clip,width=\textwidth]{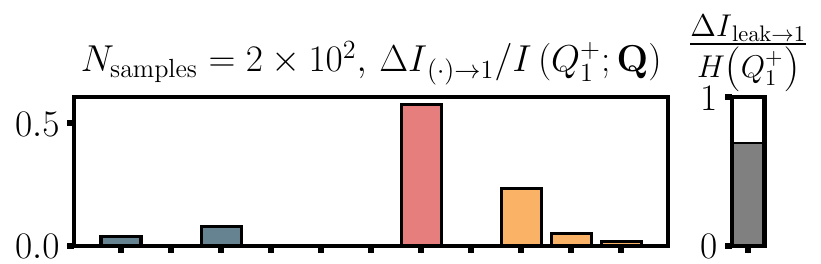}
        \includegraphics[trim={0 1.6cm 0 0},clip,width=\textwidth]{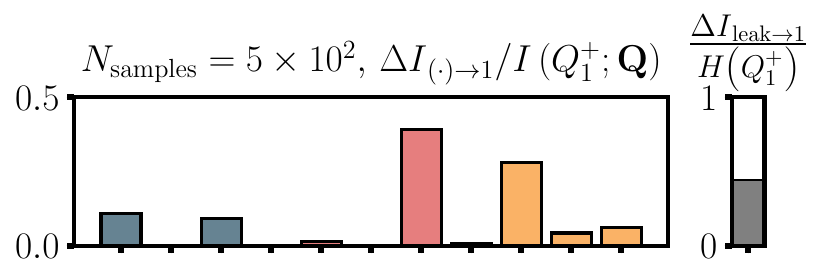}
        \includegraphics[trim={0 0cm 0 0},clip,width=\textwidth]{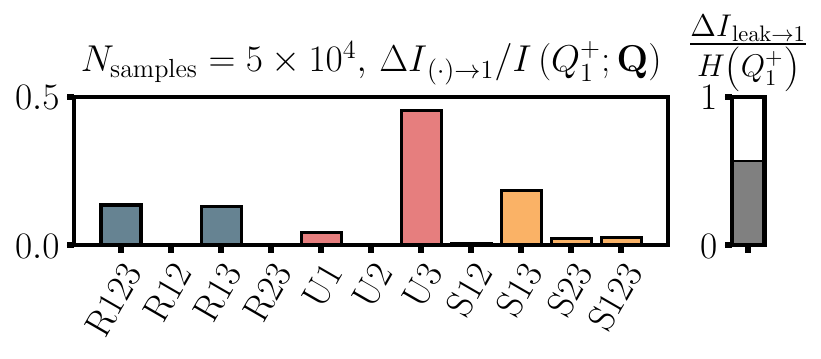}
    \end{minipage}
    }
    \hspace{0.03\textwidth}
    \subfloat[]{
    \begin{minipage}{0.45\textwidth}
        \includegraphics[trim={0 1.6cm 0 0},clip,width=\textwidth]{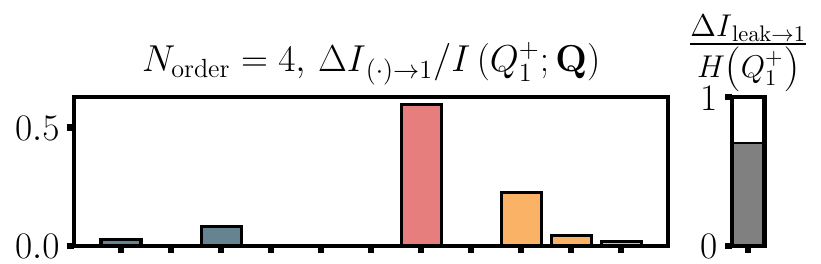}
        \includegraphics[trim={0 1.6cm 0 0},clip,width=\textwidth]{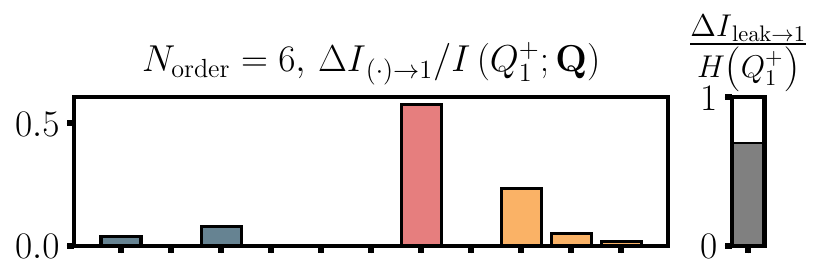}
        \includegraphics[trim={0 1.6cm 0 0},clip,width=\textwidth]{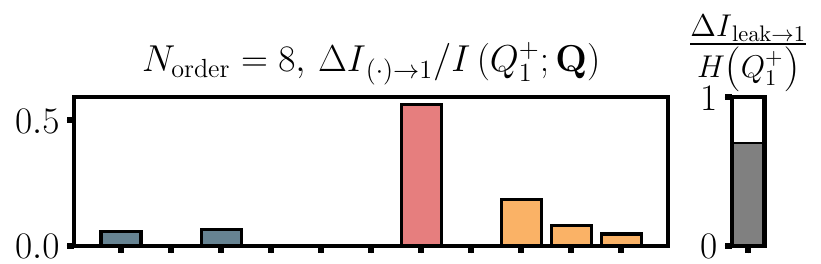}
        \includegraphics[trim={0 0cm 0 0},clip,width=\textwidth]{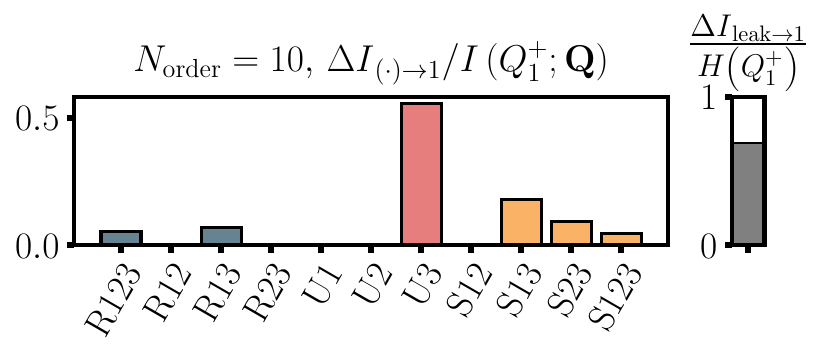}
    \end{minipage}
    }
    \caption{ Sensitivity of causality in the system with confounder
      variables and bimodal stochastic forcing for (a) number of
      samples $N_\text{samples}$ used to estimate the approximated
      probability distributions with the transport map method for
      $N_\text{order}=6$ held constant and (b) the order of
      polynomials $N_\text{order}$ used to partition values of the
      variables for $N_\text{samples}=200$. The causalities are
      ordered from left to right according to $N_{\bi \rightarrow
        1}^{\alpha}$.}
    \label{fig:covergence tmap}
\end{figure}

We investigate the sensitivity of SURD to the number of samples
($N_{\text{samples}}$) and the order of polynomials
($N_{\text{order}}$) used to estimate the probability distributions
using transport maps. The case considered for testing the results is
the system with a confounder variable, as defined in the main text,
but with bimodal stochastic forcing. The latter follows a bimodally
distributed random variable $W = \alpha X + (1-\alpha) Y$, where $X =
\mathcal{N}(-2,1)$ and $Y = \mathcal{N}(2,1)$ are unimodal random
variables following a Gaussian distribution and $\alpha = 0.3$
represents the mixture coefficient. The system was integrated over
time to collect $N_{\text{samples}}=5\times 10^1, 5\times 10^2,
5\times 10^3,$ and $5\times10^4$ events after transients. The
probability distributions were estimated using the transport map
method with $N_{\text{order}}=4, 6, 8,$ and 10. Our primary focus is
on causalities to $Q_1$, but the conclusions drawn also apply to $Q_2$
and $Q_3$.

The impact of the number of samples on the causalities estimated with
the transport map method is depicted in Figure~\ref{fig:covergence
  tmap}(a), where $N_{\text{samples}}$ is varied while keeping
$N_{\text{order}}=6$ constant. For $N_{\text{samples}}>5\times 10^2$,
the changes in causality only differ within a few percent and are
limited to synergistic contributions, which are linked to higher-order
probability distributions. Figure~\ref{fig:covergence tmap}(b)
assesses the sensitivity to polynomial order, with $N_{\text{order}}$
varying while $N_{\text{samples}}$ is fixed at $N_{\text{samples}} =
200$. The causalities show similar quantitative values for all
polynomial orders. Once again, the slight percentage differences
noticed when increasing the polynomial order are associated with
synergistic causalities. In conclusion, transport mapping theory
provides an accurate estimation of the probability distribution when
the number of samples is small or the number of variables is
high. However, these estimates are only approximations of the true
distribution, which might become more challenging in more complex
distributions.

\section{{Effect of non-separability of the variables}}

{One of the prevailing weaknesses in some of the
  previous methods for causal inference arises from the
  non-separability of the variables \cite{sugihara2012}. The issue is a
  consequence of Takens' embedding theorem, which states that under
  the right conditions, the dynamics of a system can be captured by
  embedding a sequence of past observations into a higher-dimensional
  space. In such cases, the future of a variable can be fully
  forecasted using only its own past, without the need for any other
  variables. Consequently, methods for causal inference based on
  predictability, such as Granger causality, might miss causal
  connections when including past observations into the model. For
  example, consider a system where $Q_1 \rightarrow Q_2$. By the
  Takens' embedding theorem, $Q_2$ could be forecasted using only its
  own past regardless of $Q_1$, leading to erroneous conclusions in
  Granger causality. }

{Takens' embedding theorem can also be interpreted
  within the framework of information theory~\cite{yuan2024}. If the
  variable $Q_1$ is causal to $Q_2$, then part of the past information
  from $Q_1$ is encoded into $Q_2$. Thus, from an
  information-theoretic viewpoint, non-separability arises from the
  flow of information among interacting variables. SURD is less
  susceptible to the issue of non-separability as it monitors all
  transfers of information among variables within the system, even if
  redundant.}

{Here, we employ the example introduced by
  Sugihara \textit{et al.} \cite{sugihara2012} to illustrate the robustness of SURD under the
  effect of multiple time lags with non-separable variables. The
  system is given by:
  \begin{subequations}
    \begin{align}
        \label{eq: sugihara1a}
      Q_1(n+1) &= Q_1(n) \left[r_1 - r_1 Q_1(n) - \beta_{2\rightarrow 1} Q_2(n) \right], \\
      \label{eq: sugihara1b}
      Q_2(n+1) &= Q_2(n) \left[r_2 - r_2 Q_2(n) - \beta_{1\rightarrow 2} Q_1(n)\right], 
    \end{align}
  \end{subequations}
  where the coupling from $Q_2$ to $Q_1$ is controlled through
  $\beta_{2\rightarrow 1}$ and the coupling from $Q_1$ to $Q_2$,
  through $\beta_{1\rightarrow 2}$. In this simple system, we can
  recover algebraically the influence of $Q_1$ on $Q_2$ using
  $Q_2(n+1)$ and $Q_2(n)$ (and vice versa):
  \begin{subequations}
    \begin{align}
        \label{eq: sugihara2a}
        \beta_{2\rightarrow 1} Q_2(n) &= 1 - Q_1(n)-\frac{Q_1(n+1)}{r_1 Q_1 (n)}, \\
      \label{eq: sugihara2b}
        \beta_{1\rightarrow 2} Q_1(n) &= 1 - Q_2(n)-\frac{Q_2(n+1)}{r_2 Q_2 (n)}.
    \end{align}
  \end{subequations}
We can substitute Equation (\ref{eq: sugihara2a}) into (\ref{eq:
  sugihara1b}) and obtain an expression for $Q_2(n)$ as a function of
$Q_1(n)$ and $Q_1(n-1)$:
  \begin{equation}
    \label{eq: sugihara3a}
      Q_2(n)= \frac{r_2}{\beta_{2\rightarrow 1}} \left[ \left(1-\beta_{1\rightarrow 2} Q_1(n-1)\right) \left( 1 - Q_1(n-1) - \frac{Q_1(n)}{r_1 Q_1(n-1)}\right) - \frac{1}{\beta_{2 \rightarrow 1}} \left(1 - Q_1(n-1) - \frac{Q_1(n)}{r_1 Q_1(n-1)} \right)^2 \right].
  \end{equation}
}
{Introducing Equation (\ref{eq: sugihara3a}) into
  (\ref{eq: sugihara1a}), we obtain an expression for $Q_1$ that is
  exclusively a function of its own past, i.e. $Q_1(n)$ and $Q_1(n-1)$:
  \begin{equation}
    \label{eq: sugihara4}
      Q_1(n+1)= f(Q_1(n),Q_1(n-1)).
  \end{equation}
  }
  
{Methods for causal inference based on the
  predictability of $Q_1$ might incorrectly conclude that $Q_2$ does
  not cause $Q_1$ if the values $Q_1(n)$ and $Q_1(n-1)$ are included
  in the predictive model. To address this, we assess the causal
  connections to $Q_1$ and $Q_2$ using SURD and a non-linear version
  of CGC. We use a non-linear implementation of CGC because its linear
  counterpart failed in all considered scenarios, which does not allow
  us to demonstrate the problem of non-separability. The non-linear
  CGC consists of an artificial neural network (ANN) trained to
  predict the target variables $Q_1(n+1)$ and $Q_2(n+1)$, given
  different sets of past instances of $Q_1$ and $Q_2$.  The model for
  $Q_1$ (similarly for $Q_2$) is
   \begin{subequations}
     \begin{align}
       Q_1(n+1) &= \text{ANN}_1(\bQ_1) + \hat{\varepsilon}(n+1), \\
       Q_1(n+1) &= \text{ANN}_{12}(\bQ_1, \bQ_2) + \varepsilon(n+1), 
     \end{align}
\end{subequations}  
   where vector of observables is defined as $\bQ = [\bQ_1, \bQ_2]$
   with $\boldsymbol{Q}_1 = \left[Q_1^n, Q_1^{n-1}, \cdots,
     Q_1^{n-\Delta n}\right]$ (similarly for $\boldsymbol{Q}_2$) and
   $\Delta n$ is the maximum lag considered. Note that, from the point
   of view of SURD, $\bQ$ only contains two variables (i.e., $\bQ_1$
   and $\bQ_2$), although these are vectors. This differs from the
   discussion in \S S1.3, where different time lags are considered as
   different variables. The network architecture includes three hidden
   layers with 1024, 512 and 256 neurons, respectively, and it is
   trained using an Adam optimizer with a maximum of 200 epochs and an
   initial learning rate of 0.01, which is reduced by a factor of 0.3
   with a period of 125 iterations.
  \begin{figure}[t!]
    \centering
    \subfloat[]{
    \includegraphics[height = 0.34\textwidth]{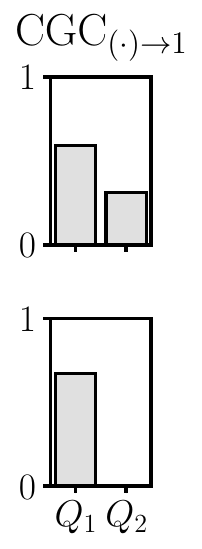}
    \includegraphics[height = 0.35\textwidth]{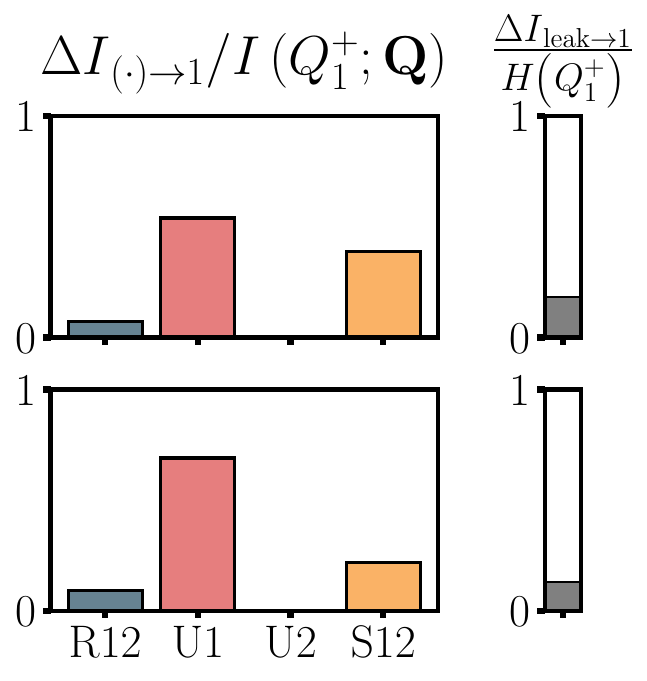}}
    \hspace{0.05\textwidth}
    \subfloat[]{
    \includegraphics[height = 0.34\textwidth]{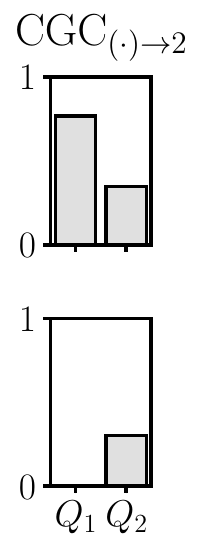}
    \includegraphics[height = 0.35\textwidth]{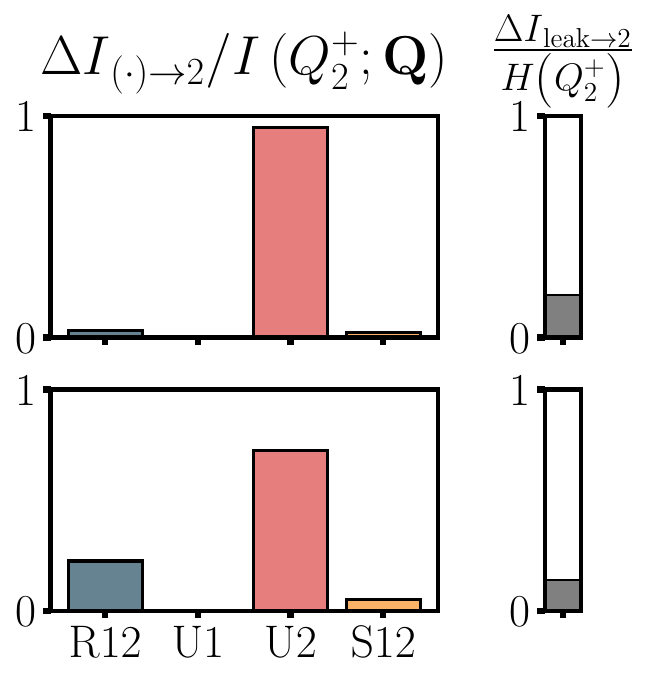}}
    \caption{Effect of non-separability. Performance of non-linear CGC
      and SURD for the target variables (a) $Q_1$ and (b) $Q_2$. The
      results in the top row are for $\Delta n = 0$ ($\mathbf{Q} =
      [Q_1(n), Q_2(n)]$) and in the bottom row for $\Delta n = 1$
      ($\mathbf{Q} = [\mathbf{Q}_1, \mathbf{Q}_2]$ where $\mathbf{Q}_1
      = [Q_1(n), Q_1(n-1)]$ and $\mathbf{Q}_2 = [Q_2(n),
        Q_2(n-1)]$). The system is simulated for the parameters
      $[r_1,r_2,\beta_{2\rightarrow 1}, \beta_{1\rightarrow 1}] =
      [3.8, 3.5, 0.2, 0.01]$.
    }
    \label{fig:non-separability}
\end{figure}
  }

{Figure \ref{fig:non-separability} displays the
  results from non-linear CGC and SURD using $\Delta n = 0$
  ($\mathbf{Q} = [Q_1(n), Q_2(n)]$) and $\Delta n = 1$ ($\mathbf{Q} =
  [\mathbf{Q}_1, \mathbf{Q}_2]$ where $\mathbf{Q}_1 = [Q_1(n),
    Q_1(n-1)]$ and $\mathbf{Q}_2 = [Q_2(n), Q_2(n-1)]$). For $\Delta n
  = 0$, both non-linear CGC and SURD identify the coupling between
  $Q_1$ and $Q_2$. However, with an additional time lag for both
  variables, non-linear CGC incorrectly determines that $Q_1$ does not
  influence $Q_2$ and vice versa, as these can be completely
  determined by their own past. In contrast, SURD continues to show
  the causal dependency between $Q_1$ and $Q_2$.  The improved
  robustness of SURD is attributable to the fact that, under a
  statistical steady state, the flow of information between variables
  remains unchanged. The main difference observed in SURD is an
  increase in redundant causality due to duplicated information from
  the inclusion of additional time lags.}

\section{{Application of SURD to predictive modeling}}

{SURD can also inform the development of predictive
  and/or reduced-order models of dynamical systems. By leveraging
  knowledge of the causal structure of the system, SURD enables the
  construction of minimal models by selecting the most effective input
  variables while disregarding those with irrelevant or duplicated
  information. This section illustrates an application of SURD to
  temporal forecasting of variables in the synergistic and redundant
  collider systems, as shown in Figures \ref{fig:synergistic} and
  \ref{fig:redundant}. The approach employs long-short-term memory
  (LSTM) artificial neural networks trained to predict $Q_1(n+1)$,
  using the exact values of $Q_1(n)$, $Q_2(n)$, and $Q_3(n)$. Several
  models are trained using different sets of input variables. The
  network architecture includes a sequence input layer with the
  corresponding number of input features, an LSTM layer with 200
  hidden units to capture temporal dependencies between the signals,
  and a fully connected layer to map the previous layer to the output
  variable. The network is trained using an Adam optimizer with a
  maximum of 200 epochs and an initial learning rate of 0.01, which is
  reduced by a factor of 0.3 with a period of 125 iterations.}
\begin{figure}[t!]
    \centering
    \subfloat[]{
    \includegraphics[height = 0.24\textwidth]{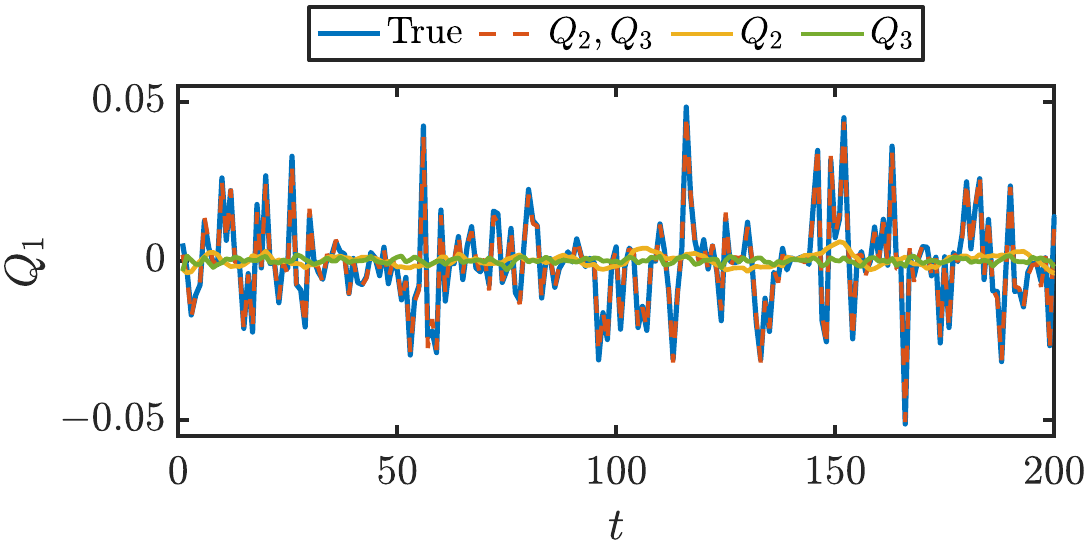}}
    \hfill
    \subfloat[]{
    \includegraphics[height = 0.24\textwidth]{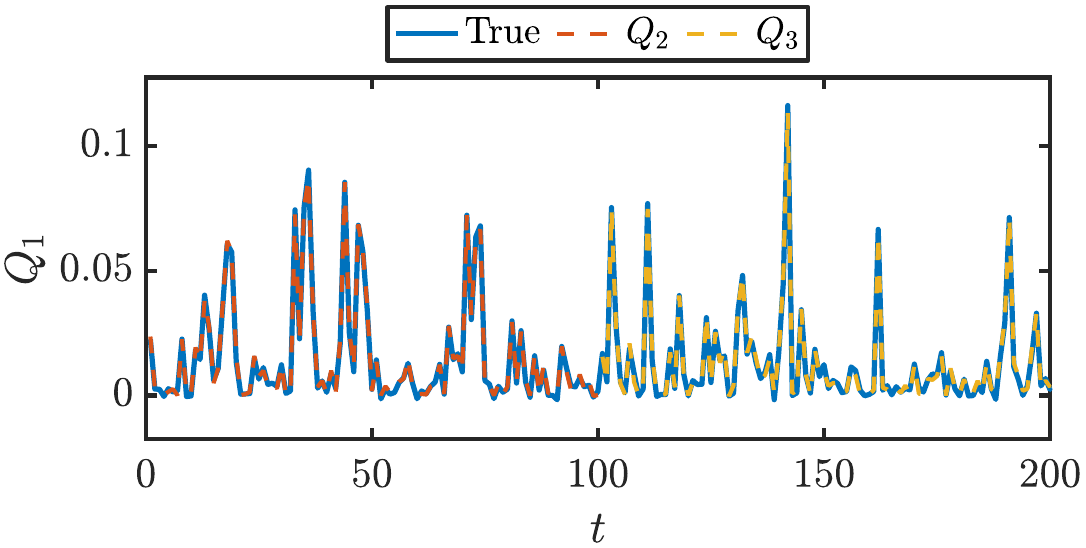}}
    \caption{Comparative performance of LSTM models for forecasting
      the future of $Q_1$ using different input variables for (a)
      system with a synergistic collider (where $Q_2$ and $Q_3$
      collectively influence the future of $Q_1$) and (b) system with
      redundant collider (where $Q_2$ and $Q_3$ contain the same
      information about the future of $Q_1$).  The legend indicates
      the variables used as input to the LSTM model.  In panel (b),
      the prediction is performed using $Q_2$ for the first half of
      the temporal sequence, while $Q_3$ is used for the second half.}
    \label{fig:forecasting}
\end{figure}

{In the first case (Figure \ref{fig:synergistic}),
  $Q_2$ and $Q_3$ synergistically influence $Q_1$, as previously
  indicated by $\Delta I^S_{23\rightarrow 1}$. Therefore, it is
  crucial for models to incorporate both variables as inputs to ensure
  accurate predictions. This is illustrated in Figure
  \ref{fig:forecasting}(a), where the forecasting performance of the
  models using $[Q_2, Q_3]$ significantly surpasses those that include
  either variable alone. This outcome is consistent with the
  synergistic causality detected by SURD, where $Q_2$ and $Q_3$
  collectively drive the future of $Q_1$. Generally, accurate
  forecasting of variables affected by synergistic causalities is
  achievable only when all synergistically interacting variables are
  incorporated into the model.}

{In the second case (Figure \ref{fig:redundant}),
  $Q_2$ and $Q_3$ exhibit redundant causality to $Q_1$, as revealed by
  $\Delta I^R_{23\rightarrow (\cdot)}$. Hence, predictive models can
  use either $Q_2$ or $Q_3$ without compromising their predictive
  accuracy as shown in Figure \ref{fig:forecasting}(b). In scenarios
  of high redundancy, minimal predictive models can be optimized by
  selecting the most convenient variable from the redundant set. This
  interchangeability provides a strategic advantage in model
  construction, allowing for the selection of variables based on
  practical considerations, such as measurement ease or data
  availability. For a more detailed discussion on
  information-theoretic causality for reduced-order modeling of
  chaotic dynamical systems, the reader is referred to
  Ref. \cite{Lozano2022}.}


\makeatletter\@input{xx.tex}\makeatother